\documentclass[twocolumn,superscriptaddress,twoside,rmp]{revtex4-1}

\usepackage[dvips]{graphicx}
\usepackage{amssymb,amsfonts,amsmath}
\usepackage{color}
\usepackage{ulem}
\usepackage[hidelinks]{hyperref}
\usepackage[dvipsnames]{xcolor}
\usepackage[paperwidth=210mm,paperheight=297mm,centering,hmargin=1.7cm,vmargin=2.89cm]{geometry}

\newcommand{\rv}{{\mathbf r}}
\newcommand{\rt}{({\mathbf r},t)}
\newcommand{\one}{(1)}
\newcommand{\rhoJ}{[\rho,{\mathbf J}]}
\newcommand{\rhoJJdot}{[\rho,{\mathbf J},\dot{\mathbf J}]}
\newcommand{\nJJdot}{[n,{\mathbf J},\dot{\mathbf J}]}
\newcommand{\ov}{{\boldsymbol{\omega}}}
\newcommand{\ev}{{\bf e}}

\newcommand{\Tr}{{\rm Tr}\,}

\newcommand{\e}{{\rm e}}
\newcommand{\J}{{\bf J}}
\newcommand{\Jv}{{\bf J}}
\newcommand{\Av}{{\bf A}}
\newcommand{\Bv}{{\bf B}}

\newcommand{\pv}{{\bf p}}

\newcommand{\Fv}{{\bf F}}
\newcommand{\fv}{{\bf f}}

\newcommand{\vel}{{\bf v}}

\newcommand{\xib}{{\boldsymbol \xi}}
\newcommand{\pib}{{\boldsymbol \pi}}
\newcommand{\sigmab}{{\boldsymbol \sigma}}

\newcommand{\qv}{{\bf q}}

\newcommand{\av}{{\bf a}}

\newcommand{\taub}{{\boldsymbol\tau}}

\newcommand{\Gv}{{\bf G}}
\newcommand{\etal}{{\it et al.}}

\newcommand{\bibitemORIGINAL}[2]{\bibitem[\color{black}#1]{#2}}

\newcommand{\bibitemESSENTIAL}[2]{\bibitem[\color{black}#1]{#2}}

\makeatletter
\def\l@subsubsection#1#2{}
\makeatother

\begin{document}

\title{Power functional theory for many-body dynamics}

\author{Matthias Schmidt}
\email{Matthias.Schmidt@uni-bayreuth.de}
\affiliation{Theoretische Physik II,
  Physikalisches Institut,
  Universit{\"a}t Bayreuth,
  D-95447 Bayreuth, Germany,
  \href{https://www.mschmidt.uni-bayreuth.de}{www.mschmidt.uni-bayreuth.de}}

\date{28 March 2022,
  \href{https://doi.org/10.1103/RevModPhys.94.015007}
       {Rev. Mod. Phys. {\bf 94}, 015007 (2022).}}

\begin{abstract}
The rich and diverse dynamics of particle-based systems ultimately
originates from the coupling of their degrees of freedom via internal
interactions. To arrive at a tractable approximation of such many-body
problems, coarse-graining is often an essential step. Power functional
theory provides a unique and microscopically sharp formulation of this
concept. The approach is based on an exact one-body variational
principle to describe the dynamics of both overdamped and inertial
classical and quantum many-body systems. In equilibrium, density
functional theory is recovered, and hence spatially inhomogeneous
systems are described correctly. The dynamical theory operates on the
level of time-dependent one-body correlation functions. Two- and
higher-body correlation functions are accessible via the dynamical
test-particle limit and the nonequilibrium Ornstein-Zernike route.  We
describe the structure of this functional approach to many-body
dynamics, including much background as well as applications to a broad
range of dynamical situations, such as the van Hove function in
liquids, flow in nonequilibrium steady states, motility-induced phase
separation of active Brownian particles, lane formation in binary
colloidal mixtures, and both steady and transient shear phenomena.
\end{abstract}

\maketitle

\tableofcontents

\section{Introduction}
\label{SECintroduction}

\subsection{Soft matter dynamics}
Soft matter science covers a broad range of diverse systems and their
phenomena.  \citet{nagel2017} and \citet{evans2019physicsToday}
described the gamut from colloids to polymers, from granulates to
active systems, from liquid crystals to biomolecular systems and
beyond those. Although the systems are typically out of true
equilibrium, in many instances the concepts of equilibrium statistical
physics can be fruitfully exploited in order to understand and predict
the behaviour observed in the lab. However, the genuine dynamical
behaviour of soft matter is varied and rich, and it often constitutes
the central focus of research.  \citet{balucani1994},
\citet{dhont1996}, \citet{zwanzig2001}, \citet{goetze2008}, and not
least \citet{hansen2013} provided accessible and thorough treatments
of soft matter and dynamical liquid state
theory. \citet{schilling2021} gave a recent comprehensive review of
dynamical coarse-graining strategies.

Specific recent studies were aimed at the dynamical structure of the
hard sphere liquid \cite{stopper2018dtpl} as well as of complex
ordered states \cite{bier2008prl}, microfluidics \cite{squires2005},
slow dynamics and the glass transition \cite{dyre2006}, gelation, and
the topical field of active systems, as reviewed by
e.g.\ \citet{liverpool2013} and by \citet{loewen2016}. Starting from a
microscopic point of view, one would expect phenomena such as these to
ultimately originate from the large number of degrees of freedom in
the system, which are coupled via the interparticle
interactions.\footnote{There are interesting counterexamples, where
  the dynamics of a single particle already are exceedingly rich, such
  as in magnetically driven topological transport
  \cite{loehr2016,loehr2018} and in active \cite{maes2020} and
  viscoeslastic \cite{berner2018} solvents.} On the many-body level of
description, computer simulation techniques offer in principle direct
access to the physical behaviour of a given system. In order to
rationalize the bare data that is output from simulations, however, a
theoretical framework is required, which (i) condenses the many-body
information into digestible and intelligible form, and (ii) formulates
interrelations between the simulated quantities. Furthermore, (iii)
both strong conceptual and practical reasons (such as computational
efficiency) speak for having a reliable and predictive stand-alone
theory.

For equilibrium properties classical density functional theory (DFT),
as established by \citet{evans1979}, satisfies the previously
described needs. Its basic (variational) variable is the density
profile, i.e., the position-resolved microscopic probability to find a
particle at the given spacepoint. The density profile determines all
physical properties of the system, at given thermodynamic statepoint,
and in the presence of a fixed external (one-body) potential. Hence,
the equation of state, the phase diagram, correlation functions,
solvation forces, interfacial tension etc.\ are all accessible. The
DFT framework was originally conceived for quantum systems at zero
temperature (i.e., for the ground state) by
\citet{HK1964}\footnote{Accessible and compact descriptions of
  electronic DFT were given by \citet{kohn1999nobel},
  \citet{jones1989}, and \citet{jones2015}.}. Only one year later,
\citet{mermin1965} accomplished the generalization to finite
temperatures.  \citet{evans1979} formulated the classical version of
DFT, and his approach has become textbook material
\cite{hansen2013}. The quantum and classical theories are very similar
in their formal structure, although the approximative functional forms
that are used in either field differ very substantially from each
other.\footnote{See the excellent reviews by \citet{roth2010review},
  \citet{lutsko2010review}, and by \citet{tarazona2008review} for
  descriptions of the state of the art of classical DFT; the foreword
  by \citet{evans2016specialIssue} of a special issue on classical DFT
  described recent progress.}

To give a sense of the breadth of classical DFT subject matters, we
enlist recent pivotal DFT studies. These have addressed
e.g.\ atomically resolved three-dimensional structures of electrolytes
near a solid surface
\cite{martinjimenez2017natCom,hernandez-munoz2019}, solvation
phenomena in water \cite{jeanmairet2013jpcl}, and water-graphene
capacitors \cite{jeanmairet2019capacitor}. Much work has been carried
out addressing hydrophobicity, where liquid water (or a more general
liquid) avoids contact with a substrate or solute. Here the density
fluctuations near the substrate were quantified \cite{evans2015prl}
and a unified description was obtained for hydrophilic and
superhydrophobic surfaces in terms of wetting and drying transitions
of liquids \cite{evans2019pnas,remsing2019pnas}. Furthermore critical
drying of liquids \cite{evans2016prl} and superhydrophobicity
\cite{giacomello2016,giacomello2019} were investigated. There is much
progress on the conceptual level, as exemplified by the recent
systematic incorporation of two-body correlation functions
\cite{tschopp2020,tschopp2021} and fluctuation profiles
\cite{eckert2020auxiliaryFields} into the one-body DFT framework.

A time-dependent version of classical DFT or ``dynamical DFT'' was
proposed by \citet{evans1979} and later much advocated for by
\citet{marconi1999ddft} and by \citet{archer2004ddft}.  Selected
examples of insightful dynamical DFT studies include the uptake
kinetics of molecular cargo into hollow hydrogels
\cite{monchojorda2019acsnano}, the particle-scale-resolved
non-equilibrium sedimentation of colloids
\cite{royall2007dynamicSedimentation}, the bulk dynamics of colloidal
Brownian hard disks \cite{stopper2018dtpl}, the pair dynamics in
inhomogeneous liquids and glasses
\cite{archer2007dtpl,hopkins2010dtpl}, and the growth of monolayers of
hard rods on planar substrates \cite{klopotek2017}.  The dynamical DFT
can be viewed as being based on the approximation that the
nonequilibrium dynamics are represented as a sequence of ``adiabatic
states'' that each are taken to be in equilibrium.  While the
adiabatic approximation for dynamical processes can be valid in
certain cases, important physical effects are absent
\cite{fortini2014prl}, such as drag forces \cite{krinninger2016prl},
viscosity \cite{delasheras2018velocityGradient}, and structural
nonequilibrium forces
\cite{stuhlmueller2018structural,delasheras2020fourForces,
  geigenfeind2019laning,treffenstaedt2021dtpl}.

Going beyond the somewhat {\it ad hoc} equation of motion of dynamical
DFT is facilitated by the formally exact power functional variational
framework of \citet{schmidt2013pft}. Power functional theory provides
a minimization principle for the description of the dynamics. The
internal force field, as arising from the interparticle interactions
of particles that undergo Brownian dynamics, consists of both
adiabatic and superadiabatic (above adiabatic) contributions. The
former are accounted for in dynamical DFT; the latter posses genuine
nonequilibrium character, as they are generated from a {\it kinematic}
functional of the density profile and of the microscopically resolved
flow.  The variational fields are dynamical one-body objects,
i.e.\ they depend on the position coordinate and the time. In
particular, the superadiabatic force field is a functional of these
kinematic fields, while the adiabatic force field is a functional of
the instantaneous density profile alone. The power functional theory
has been formulated for different types of underlying many-body
dynamics, such as overdamped Brownian motion \cite{schmidt2013pft},
including active systems
\cite{krinninger2016prl,krinninger2019jcp,hermann2019acif,hermann2019tension},
classical Hamiltonian dynamics as relevant for molecular dynamics
\cite{schmidt2018md}, and nonrelativistic quantum dynamics
\cite{schmidt2015qpft}. The last case promises to help to overcome
limitations of (adiabatic) time-dependent electronic DFT, and
furthermore to act as a conceptual bridge between the classical and
quantum worlds, due to the strong formal similarities between the
classical Hamiltonian and quantum versions of the power
functional. Developing time-dependent quantum DFT \cite{runge1984}
[see \citet{chan2005} for a classical analog] constitutes an active
field of research \cite{onida2002, nakatsukasa2016}. There is much
current interest in bringing the quantum and classical DFT communities
closer together \cite{CECAM2019}, and power functional theory provides
a concrete theoretical structure for making the corresponding
progress.

The review is organized as follows.
Section~\ref{SECforcesAsTheFundament} gives an overview of the central
power functional concept of working on the level of locally resolved
forces and exploiting functional dependencies.  Section
\ref{SECmanybodyDynamics} begins with a specification of the splitting
of the forces that act in typical many-body systems into internal and
external contributions (Sec.~\ref{SECinternalAndExternalForces}). The
splitting applies to a generic and broad class of systems and it forms
a primary motivation for the specific choice of one-body kinematic
fields as fundamental dynamical variables.  We then turn to the level
of one-body fields and derive corresponding equations of motion,
starting with underlying classical (Newtonian) dynamics of the
many-body system. This includes the equations of motion for the
one-body operators, i.e.\ for phase space functions that represent the
density and current. The form of these equations is that of a force
balance relationship, or Newton's second law, including transport
effects (Sec.~\ref{SEConebodyMD}). While transport effects are
familiar from a hydrodynamic description, the present treatment is
entirely microscopic and does not involve coarse graining in the sense
of averaging out microscopic length scales. The description is
microscopically ``sharp''. One-body distribution functions are
obtained by averaging over the many-body phase space probability
distribution function (of which the time evolution is governed by the
Liouville equation).

The one-body equations of motion for overdamped Brownian classical
motion (Sec.~\ref{SEConebodyBD}) and for nonrelativistic quantum
dynamics (Secs.~\ref{SEConebodyQM}) are similar to the previously
mentioned Newtonian case. Although the respective derivations are
elementary (to a certain degree) and the underlying dynamics are
apparently different from each other, it is surprising that the
one-body description possesses universal status. There are clear
differences though. The quantum one-body dynamics feature additional
genuine quantum contributions (dependent on $\hbar$), the appearance
of the quantum kinetic stress tensor distribution as well as of
different types of force densities. For overdamped classical dynamics,
the local force directly translates into an instantaneous particle
current, and not into its time derivative, as is the case for quantum
and Newtonian dynamics, which instead feature inertia. As we
demonstrate, in all considered cases of many-body time evolution, the
nontrivial coupling arises directly in the force density distribution
(along with the kinetic stress in the inertial cases). The details of
the definition of the internal force density differ amongst the three
types of dynamics, as do the different types of averages (phase space
average, quantum expectation value, and positional configuration space
integral). Nevertheless, one can view these differences as merely
technical and the internal force density as a universal and
fundamental physical object. At this stage, however, the internal
force density is only defined as a many-body average. Hence, the
one-body theory is not closed and does not form a stand-alone
framework. The one-body quantities merely constitute observables,
which, although being characteristic of the full dynamics, lack a
mechanism to restore the full information and evolve the system in
time.

In equilibrium, a closed theory on the one-body level is available
through the well-established framework of density functional theory,
which ascertains that the reduction of information that is inherent to
relying on one-body fields is perfectly compensated for, without any
principle loss, by the recognition and use of functional dependencies.
Section~\ref{SECdft} hence describes the adiabatic state and its
treatment via the classical version of density functional theory,
which is prominently used in the description of bulk and inhomogeneous
fluids, solids, liquid crystals and further self-organized states of
matter \cite{hansen2013,evans2016specialIssue}. The reasons for laying
out the framework are twofold. The first reason ist that DFT forms a
blueprint, or prototype, for the subsequent construction of the power
functional framework.  Both approaches share on an abstract, formal
level many similarities, such as a truly microscopic foundation, the
existence of a central functional object, which is minimized at the
physical solution, and the generation of meaningful averages
(correlators) via functional differentiation. Physically, however, the
frameworks are distinct as to whether equilibrium (DFT) or
nonequilibrium (power functional theory) situations are addressed. The
second, and possibly more important, reason for covering DFT is its
relevance in genuine nonequilibrium for the description of the
adiabatic state (Sec.~\ref{SECadiabaticState}). Briefly, any dynamical
theory (on the one-body level) that can account for spatial
inhomogeneity needs to reduce to DFT in the equilibrium limit. In
power functional theory, this reduction is generic. The adiabatic
contribution to the dynamics is unique and it forms the part that is
independent of the flow.

After an overview of the history of DFT (Sec.~\ref{SECoverviewDFT})
and its general structure (Sec.~\ref{SECsketchDFT}), we start from the
partition sum
(Sec.~\ref{SECstatisticalMechanicsAndFunctionalDerivatives}) and show
how its functional derivative(s) with respect to the external
potential are meaningful response functions; these are equivalent to
correlation functions in the classical case.  We cover several recent
developments that are crucial for the dynamical material to follow and
that are not covered in the above standard introductory DFT
literature. This includes the derivation of the Mermin-Evans
variational principle via the Levy constrained search method
(Sec.~\ref{SEClevysConstrainedSearch}). The intrinsic elegance and
prowess of this method are not only a boon for the equilibrium
framework, as a delicate {\it reductio ad absurdum} argument is
circumvented, but also vital for the construction of the power
functional.  We describe two-body correlation functions
(Sec.~\ref{SECstaticTwoBodyCorrelationFunctions}) and derive the
Ornstein-Zernike relation directly from the DFT minimization principle
(Sec.~\ref{SECornsteinZernikeRelation}). This derivation separates
cleanly the fundamental concept from the technicalities of defining
and manipulating the various response and correlation functions that
are involved. It is this type of derivation which is later generalized
to the dynamical functional calculus in order to obtain the
nonequilibrium version of the Ornstein-Zernike relation.  An overview
of approximate free energy functionals is presented
(Sec.~\ref{SECfexcSimple}). We conclude the section with an account of
dynamical density functional theory (Sec.~\ref{SECddft}).

Section~\ref{SECpft} describes power functional theory, starting with
an overview of the concept (Sec.~\ref{SECoverviewPFTboth}).  We cover
the formulations for classical inertial dynamics and for diffusive
overdamped Brownian dynamics, as well as for nonrelativistic many-body
quantum dynamics. In all these cases the power functional plays both
the role of a Gibbs-Appell-Gaussian that determines the dynamics via a
minimization principle but also constitutes a functional generator for
time correlation functions. The reduction to the one-body level is
performed using a dynamic generalization of the Levy search method,
where, in particular, the constraint of fixed one-body current creates
a one-body extremal principle with respect to the current (or the time
derivative of the current in the inertial cases). This concept allows
to formulate closed one-body equations of motion in all three cases of
microscopic dynamics considered (Sec.~\ref{SECpftMD}), i.e.\ for
molecular dynamics, overdamped Brownian motion and quantum mechanics.
It is shown that the natural splitting into intrinsic and external
(and additional transport effects, in the inertial cases) translates
into an analogous splitting of the functional generator. In all cases,
the internal force density plays a central role in coupling the
microscopic degrees of freedom.  The interrelated kinematic one-body
fields, i.e.\ the density profile, the local current or velocity, and
the local acceleration, play the role of order parameters.

The superadiabatic force contributions act on top of the adiabatic
force field. Although there is no exact solution for the
superadiabatic functional contribution available (similar to the
corresponding situation in equilibrium DFT, where the excess free
energy functional is unknown in general), the framework shows
existence and uniqueness. In contrast to the deterministic
Gibbs-Appell-Gaussian formulation in classical mechanics, the
constraint is a statistical one, as has been used in the Levy method
of classical DFT construction. The framework implies a fundamental
functional map from the density and current (and the current time
derivative, in the inertial cases) to the external force field that
generates these dynamics.  A discussion is given of simple approximate
forms of the superadiabatic free power functional
(Sec.~\ref{SECptexc}). We show how local and semi-local gradient
functionals describe important classes of physically distinct effects,
such as drag, viscous and structural nonequilibrium forces. Based on
the concept of functional differentiation of the Euler-Lagrange
equation, we describe the derivation of nonequilibrium
Ornstein-Zernike relations (Sec.~\ref{SECnoz}). This includes the
introduction of time direct correlation functions, which are
identified as functional derivatives of the superadiabatic free power
functional.

We then turn to several recent applications. The dynamical
test-particle limit (Sec.~\ref{SECdtpl}) constitutes an alternative,
formally exact route to the time-dependent two-body structure.  A
practical and simple explicit computational simulation scheme that
implements kinematic functional dependencies is provided by the custom
flow method, which we lay out for overdamped Brownian dynamics
(Sec.~\ref{SECcustomFlow}).  This method is vital in the study of
viscous and structural forces (Sec.~\ref{SECstructuralForces}), which
is based on splitting the Brownian dynamics into flow and structural
contributions.  Viscoelasticity, as originating from memory dependence
of the superadiabatic free power functional, is demonstrated to occur
for hard spheres under time-dependent step shear
(Sec.~\ref{SECviscoelasticity}).  Lane formation in counter driven
mixtures is shown to originate from a superadiabatic demixing force
contribution (Sec.~\ref{SECsuperdemixing}).  An overview of power
functional theory for active Brownian particles, including the
treatment of motility-induced phase separation, is described
(Sec.~\ref{SECactiveBrownianParticles}).  We draw conclusions and an
outlook on future work in Sec.~\ref{SECconclusions}.

The appendix contains an overview of Hamilton's action principle
(Appendix~\ref{SECHamiltonsActionPrinciple}), from which both the
Lagrangian and Hamiltonian formulations of classical mechanics are
derived. This familiar material serves to review the essentials of
functional calculus, which we spell out explicitly in spatio-temporal
and in time-slice forms in
Appendix~\ref{SECfunctionalDerivativeAppendix}.  As Hamilton's
principle only requires stationarity and not necessarily an extremum
of the functional, this case also constitutes a counterexample to
dynamical minimization, as performed in the Gibbs-Appell-Gaussian
formulation of classical mechanics (Appendix~\ref{SECgagOne}). Despite
the considerable fame of its originators, and its wide use both in the
nonequilibrium liquids computer simulation community
\cite{EvansMorriss}, as well as in mechanical applications of
classical dynamics with constraints, the method seems to be crucially
undervalued and very little known in a wider statistical physics
community. As the power functional performs a similar variation, we
lay out the (deterministic) Gibbs-Appell-Gaussian theory.

\subsection{Forces as the basis}
\label{SECforcesAsTheFundament}
The forces that govern the behaviour of typical many-body systems
naturally split into internal forces, which act between the
constituent particles, whether they be atoms, molecules or colloids,
and forces that are of external nature. Typically the external forces
depend on a single space coordinate only, i.e., the external force
that acts on a given particle $i$ depends only on its position
$\rv_i$, and possibly explicitly on time. If the particles possess
additional degrees of freedom, such as the orientations of anisotropic
particles, the external force field can also depend on these, as it
might on the type of particle in the case of multicomponent
systems. Thus, in general the external force will depend on the same
degrees of freedom that characterize a single particle (such as
position, orientation, and species). Hence, one refers to such forces
as {\it one-body} forces (of external nature in the present
case). Even in cases where no explicit external forces are present,
such as in a bulk fluid, one might regard one of the particles being
fixed, say, at the origin and consider the forces that this ``test''
particle exerts on the remaining system as external. This is Percus'
test-particle limit \cite{percus1962}, which relates inhomogeneous
one-body distribution functions to bulk two-body correlation
functions.

The external forces can be of various physical origins and hence model
a broad range of real-world experimental situations, such as gravity,
container walls, light, and electric and magnetic fields. The
mathematical description of forces via one-body fields allows for
systematic classification into conservative contributions, as derived
by the negative gradient of an external potential, and nonconservative
contributions, where such a potential does not exist. Both types of
forces might of course be simultaneously present, and they might, or
not, be time dependent. A mesoscopic example of the time-dependent
conservative case is the switching of a laser tweezer in strength
and/or position. Nonconservative forces can represent the influence of
e.g.\ shear flow in overdamped systems, such as sheared hard spheres
at a hard wall \cite{brader2011shear}.

Restricting ourselves to the simple case of a one-component system of
spheres, the external force field is
\begin{align}
  \fv_{\rm ext}(\rv,t) &=
  -\nabla V_{\rm ext}(\rv,t) +  \fv_{\rm nc}(\rv,t),
  \label{EQexternalForce}
\end{align}
where $V_{\rm ext}(\rv,t)$ is the external potential, and $\fv_{\rm
  nc}(\rv,t)$ is the nonconservative contribution to the force field;
here~$\rv$~indicates position, $t$~indicates time, and $\nabla$
denotes the derivative with respect to $\rv$.

In cases where the nonconservative forces vanish [$\fv_{\rm
    nc}(\rv,t)\equiv 0$] and the external potential is
time-independent [$V_{\rm ext}(\rv,t)\equiv V_{\rm ext}(\rv)$] a
well-defined equilibrium state exists (typically). Averaging over the
equilibrium time evolution of the system then provides a method to
calculate quantities of interest. On physical grounds one would be
interested in the response of the preferred particle positions to the
action of $V_{\rm ext}(\rv)$.  Valleys in the external potential
should be populated more likely by particles than peaks of the
external potential. As the external potential acts on single particles
individually, a meaningful corresponding observable is the one-body
density distribution (or ``density profile'')
\begin{align}
  \rho(\rv,t) &= \Big\langle \sum_i \delta(\rv-\rv_i) \Big\rangle,
  \label{EQdensityDistribution}
\end{align}
where the sum is over all particles $i=1,\ldots,N$, with~$N$ being the
total number of particles, $\delta(\cdot)$ is the three-dimensional
Dirac function, and the angles representing a statistical average (to
be specified in detail later) over microstates. For an equilibrium
system, the one-body density distribution will be time independent,
but in general be ``inhomogeneous'' in space, i.e.\ $\rho(\rv)\neq
{\rm const.}$ In practice, Eq.~\eqref{EQdensityDistribution} amounts
to ``counting'' the number of occurrences of any particle (hence the
sum) at a given position $\rv$ (see \citet{rotenberg2020} for an
excellent account of modern and more efficient force-sampling
simulation methods). Hence, Eq.~\eqref{EQdensityDistribution} can be
viewed as an idealized, infinitely sharply resolved histogram of
particle positions. Its normalization is $\int d\rv \rho(\rv,t)=N$,
due to the property of the Dirac distribution $\int d\rv
\delta(\rv)=1$ (for suitable integration domain).

Summarizing, in an equilibrium many-body system, it is natural to
consider the influence of a position-dependent external potential
$V_{\rm ext}(\rv)$ on the system. As a result it is plausible to
consider $\rho(\rv)$ as a meaningful response function to assess the
physical behaviour. One would view the relationship between the two
fields to be a causal one: i.e., $V_{\rm ext}(\rv)$ provides the
physical reason for the form of $\rho(\rv)$.  A primary example is the
barometric law of the isothermal atmosphere with an exponentially (in
height) decreasing density profile in response to gravity.  A
diffusive force field emerges in an inhomogeneous system,
$-k_BT\nabla\ln\rho(\rv)$, where $k_B$ is the Boltzmann constant, $T$
indicates temperature.  The diffusive force can counteract the
external force, e.g.\ the gravitational pull in the previous example.
This effect is already present in the ideal gas. In an interacting
system, however, the relationship between external potential and the
density profile is a much more subtle, and by far richer, one.

In equilibrium the system will on average not move. Hence, the
external forces need to be balanced by an (average) intrinsic force
field, which consists of the previously mentioned ideal diffusive
contribution and an interparticle interaction contribution, $\fv_{\rm
  int}(\rv)$.  Hence, in equilibrium the sum of all forces must
vanish,
\begin{align}
  - k_BT \nabla \ln \rho(\rv)
  + \fv_{\rm int}(\rv)
  + \fv_{\rm ext}(\rv)
  &= 0.
  \label{EQforceBalanceEquilibrium}
\end{align}
As a result of the force cancellation, no temporal changes occur in
the averaged quantities. Here the intrinsic force field $-k_BT \nabla
\ln\rho(\rv) + \fv_{\rm int}(\rv)$ consists of a sum of ideal and
excess (above ideal) contributions, and hence it contains all effects
that are not of external nature. The excess contribution $\fv_{\rm
  int}(\rv)$ arises from the internal interactions and is given by
\begin{align}
  \fv_{\rm int}(\rv)
  &= -\Big\langle
  \sum_i \delta(\rv-\rv_i)\nabla_i u(\rv^N)
  \Big\rangle\Big/\rho(\rv),
  \label{EQforceExcAsAverage}
\end{align}
where $u(\rv^N)$ is the interparticle interaction potential; the set
of all particle position coordinates is denoted by
$\rv^N\equiv\rv_1,\ldots,\rv_N$, and $\nabla_i$ is the derivative with
respect to $\rv_i$. Here $u(\rv^N)$ can be, but need not be, due to
only pairwise contributions.

The average in Eq.~\eqref{EQforceExcAsAverage} can again be viewed as
a histogram, but in contrast to the one-body density
\eqref{EQdensityDistribution} the entries are not just events that are
being counted but rather (vectorial) values ($-\nabla_iu$). Hence, the
``bin'' corresponding to $\rv$ can attain (say) large values due to
both a large number of events and due to large values of the local
force. The normalizing factor $1/\rho(\rv)$ scales out the first of
these effects (number of events).  While the force ``operator'' in
Eq.~\eqref{EQforceExcAsAverage} is entirely deterministic, the
statistical nature of the problem is prominently present in the
average over microstates.

It could be argued that the dependence of the positions on the forces
is a concept that dates back to Newton, with Gibbs's extension to a
statistical description. However, the precise nature of the
relationship between density and external potential is an equally
important and arguably more fundamental one, as established in the
1960s (and described in Sec.~\ref{SECdft}). In fact, for a given
system (as specified by its internal interactions), knowledge of the
one-body distribution function alone is sufficient to reconstruct the
corresponding external potential. This mathematical map is at the
heart of both quantum and classical DFT, see \citet{mermin1965} and
\citet{evans1979}, respectively.

Within classical DFT one expresses the equilibrium force field
\eqref{EQforceExcAsAverage} that arises due to the internal
interactions as the gradient of a functional derivative as follows:
\begin{align}
  \fv_{\rm int}(\rv) &= 
  -\nabla\frac{\delta F_{\rm exc}[\rho]}{\delta \rho(\rv)}.
  \label{EQforceExcDFT}
\end{align}
In Eq.~\eqref{EQforceExcDFT} $F_{\rm exc}[\rho]$ is a mathematical map
from the function $\rho(\rv)$ to the value of the excess (over ideal
gas) intrinsic Helmholtz free energy. As an intrinsic contribution,
this value is solely due to the internal interactions $u(\rv^N)$,
independent of the external potential.  Such a map constitutes a {\it
  functional}. The functional derivative $\delta/\delta\rho(\rv)$
creates the ``response'' of the value of $F_{\rm exc}[\rho]$ to
changes in density $\rho(\rv)$ at position $\rv$ (a pragmatic
introduction to functional calculus is given in
Appendixes~\ref{SECHamiltonsActionPrinciple} and
\ref{SECfunctionalDerivativeAppendix}).

The result of the functional derivative is hence position dependent,
and we have made this position-dependence explicit in the notation on
the left-hand side of Eq.~\eqref{EQforceExcDFT}. Recall that the
position-dependence in the ``probabilistic'' expression
\eqref{EQforceExcAsAverage} arises due to the presence of the
delta-function. The functional $F_{\rm exc}[\rho]$ depends also on $T$
(and on system volume $V$) and the functional is specific to the
choice of interparticle interaction potential $u(\rv^N)$. One highly
nontrivial feature is that $F_{\rm exc}[\rho]$ is independent of
$V_{\rm ext}(\rv)$. Recall that Eq.~\eqref{EQforceExcAsAverage} at
face value seems to depend on $V_{\rm ext}(\rv)$, as the external
potential enters the Boltzmann factor and hence determines the
statistical ensemble that defines the average.  However, the existence
of the unique relationship $\rho(\rv)\to V_{\rm ext}(\rv)$ frees
$F_{\rm exc}[\rho]$ of any dependence on $V_{\rm ext}(\rv)$, and hence
renders it an entirely {\it intrinsic} quantity.

It is instructive to use Eq.~\eqref{EQforceExcDFT} to re-write the
equilibrium force balance condition \eqref{EQforceBalanceEquilibrium}
as
\begin{align}
  -k_BT\nabla \ln \rho(\rv)
  -\nabla\frac{\delta F_{\rm exc}[\rho]}{\delta \rho(\rv)} 
  &=
  \nabla V_{\rm ext}(\rv),
  \label{EQequilibriumForceBalance}
\end{align}
and we recall that any nonconservative contribution to the external
force field~\eqref{EQexternalForce} needs to be absent and the
external potential must be independent of time in order for an
equilibrium state to exist.

Equation \eqref{EQequilibriumForceBalance} can be viewed as an overall
gradient of a scalar function, and upon spatial integration (and
multiplication by $-1$) one obtains
\begin{align}
  k_BT \ln[\rho(\rv)\Lambda^d]
  + \frac{\delta F_{\rm exc}[\rho]}{\delta\rho(\rv)}
  &= \mu-V_{\rm ext}(\rv),
  \label{EQequilibriumEulerLagrange}
\end{align}
where $\mu$ arises formally as an integration constant, which can be
identified with the chemical potential. Furthermore
$\Lambda=\sqrt{2\pi\beta\hbar^2/m}$ is the thermal de Broglie
wavelength \cite{hansen2013}, with particle mass $m$, inverse
temperature $\beta=1/(k_BT)$, and spatial dimensionality $d$ of the
system; note that $\nabla \ln[\rho(\rv)\Lambda^d]=\nabla\ln\rho(\rv)$
as $\Lambda$ is a constant. In practical applications of equilibrium
DFT, one typically solves Eq.~\eqref{EQequilibriumEulerLagrange}, or
its exponentiated version, numerically for $\rho(\rv)$ given $V_{\rm
  ext}(\rv)$. This is a nontrivial problem, as
Eq.~\eqref{EQequilibriumEulerLagrange} is an implict equation for
$\rho(\rv)$, due to the complex (in general) dependence of $F_{\rm
  exc}[\rho]$ on~$\rho(\rv)$.

Practical applications of DFT require making approximations for
$F_{\rm exc}[\rho]$.  A famous exception, where the exact solution has
been obtained by \citet{percus1976} [see \citet{robledo1981}], is the
one-dimensional system of hard rods. However, for certain realistic
systems, such as three-dimensional hard spheres, powerful
approximations are available
\cite{rosenfeld1989,tarazona2000,roth2002}; these can yield stunningly
accurate results relative even to large scale simulation results.

Even simple, mean-field-like approximations to $F_{\rm exc}[\rho]$
often yield physically correct qualitative and semi-quantitative
results. Here the accessibility of physical quantities goes far beyond
the one-body density profile, as thermodynamics, phase behaviour, two-
and higher-body correlation functions, etc.\ can be obtained. One of
the reasons for both the robust reliability of simple DFT
approximations and the width of the range of accessible quantities
lies in the fact that Eq.~\eqref{EQequilibriumEulerLagrange}
constitutes, within the calculus of variations, an Euler-Lagrange
equation corresponding to {\it minimization} of the grand potential
functional $\Omega[\rho]$. At the minimum of the functional
\begin{align}
  \frac{\delta \Omega[\rho]}{\delta \rho(\rv)} &= 0 \qquad {\rm (min)},
  \label{EQvariationalPrincipleDFT}
\end{align}
where $\Omega[\rho]$ consists of a sum of intrinsic and external
contributions,
\begin{align}
  \Omega[\rho] &= F_{\rm id}[\rho] + F_{\rm exc}[\rho]
  +\int d\rv \rho(\rv)(V_{\rm ext}(\rv)-\mu).
  \label{EQOmegaFunctionalOfDensity}
\end{align}
In Eq.~\eqref{EQOmegaFunctionalOfDensity} the intrinsic free energy
functional for the ideal gas is
\begin{align}
  F_{\rm id}[\rho] &= 
  k_BT\int d\rv \rho(\rv)(\ln(\rho(\rv)\Lambda^d)-1).
\end{align}
The functional derivative with respect to the density profile yields
$\delta F_{\rm id}[\rho]/\delta\rho(\rv)=k_BT\ln(\rho(\rv)\Lambda^d)$,
as appears in Eq.~\eqref{EQequilibriumEulerLagrange}. In carrying out
the derivative, as is typical in functional differentiation, the space
integral is cancelled by a Dirac delta function which arises from the
identity $\delta \rho(\rv)/\delta\rho(\rv')=\delta(\rv-\rv')$.

Given the many successes of equilibrium DFT, it is natural to use it
as a springboard for the formulation of dynamical theories. One way of
doing so is to start from a description of the forces that are present
in the system. Surely the challenge for such a formulation is to get
to grips with the internal force field \eqref{EQforceExcAsAverage},
where now the average is built over the nonequilibrium distribution of
microstates, at a given time $t$. Knowing the forces is crucial, as
this allows to progress in time and obtain the complete dynamics of
the system, as we demonstrate in Sec.~\ref{SECmanybodyDynamics}.

If the system is driven out of equilibrium, either because the
external potential changes in time, or by the addition of a
nonconservative contribution to the external force field, then a
non-vanishing average flow will result. The flow is quantified by the
average current distribution as follows:
\begin{align}
  \Jv(\rv,t) &= \Big\langle
  \sum_i \delta(\rv-\rv_i)\vel_i
  \Big\rangle,
  \label{EQcurrentDistribution}
\end{align}
where we recall that $\rv_i$ is the position of particle~$i$, its
velocity is $\vel_i$, and the average is performed at time $t$. As in
the case of the internal force field~\eqref{EQforceExcAsAverage}, the
average \eqref{EQcurrentDistribution} will acquire (say) large values
at position~$\rv$ due to frequent occurrences of particles, but also
due to large values of the many-body velocity $\vel_i$. Scaling out
the former effect leads to the definition of the local velocity field
\begin{align}
  \vel(\rv,t) = \Jv(\rv,t)/\rho(\rv,t),
  \label{EQvelocityField}
\end{align}
which is fully microscopically resolved (and hence different from a
hydrodynamic field as appearing in, say, the Navier-Stokes equation).
In a truly microscopic treatment, we need to specify the time
evolution of the positions $\rv^N$ on the many-body level. Several
choices exist; for simplicity, but also because of its practical
relevance in the description of colloidal systems, we focus first on
overdamped Brownian dynamics. Typical implementations in computer
simulations are based on the Euler algorithm\footnote{The benefits of
  using adaptive time-stepping in Brownian dynamics were described by
  \citet{sammueller2021}.} to perform the time evolution; here the
particle displacements are induced by (i) all deterministic forces
that act on particle $i$ at time $t$, and (ii) an additional random
(white noise) displacement which models diffusion, at constant
$T$. Hence, the time evolution is based on (stochastic) trajectories,
i.e., on the Langevin picture.  An equivalent, and for theoretical
purposes often more convenient and arguably more powerful, formulation
is based on the many-body probability distribution function
$\Psi(\rv^N,t)$ for finding microstate $\rv^N$ at time~$t$. Having
access to $\Psi(\rv^N,t)$ allows averages [such as the density profile
  \eqref{EQdensityDistribution}, the internal force field
  \eqref{EQforceExcAsAverage}, and the current distribution
  \eqref{EQcurrentDistribution}] to be explicitly specified, via
integration over all microstates, as follows:
$\langle\cdot\rangle=\int d\rv^N\cdot\Psi(\rv^N,t)$.  In the
overdamped limit considered here, there is no need to keep track of
the momentum part of classical phase space; only the position
(configuration) part is relevant.

The Smoluchowski equation \cite{dhont1996} is the dynamical equation
for $\Psi(\rv^N,t)$ for overdamped Brownian motion. This Fokker-Planck
equation can be viewed as the following many-body continuity equation
that expresses conservation of probability:
\begin{align}
  \frac{\partial \Psi(\rv^N,t)}{\partial t} &=
  -\sum_i \nabla_i\cdot\vel_i(\rv^N,t) \Psi(\rv^N,t),
  \label{EQsmoluchowski1}
\end{align}
where the expression on the right-hand side is the (negative)
divergence of the probability current $\vel_i(\rv^N,t) \Psi(\rv^N,t)$
in configuration space. Here the ``configurational'' many-body
velocity $\vel_i(\rv^N,t)$ of particle $i$ is given via
\begin{align}
  \gamma\vel_i(\rv^N,t) &= 
  -\nabla_i u(\rv^N) + \fv_{\rm ext}(\rv_i,t)\notag\\&\qquad
  -k_BT \nabla_i \ln \Psi(\rv^N,t),
  \label{EQsmoluchowski2}
\end{align}
where $\gamma$ is the friction constant against a static background.
The first and second terms on the right-hand side of
Eq.~\eqref{EQsmoluchowski2} are due to the internal and external
(deterministic) forces, respectively, and the third term represents
the thermal force that arises due to the diffusive Brownian motion.
For the present case of overdamped Brownian motion, it is the
configurational velocity $\vel_i$, as given by
Eq.~\eqref{EQsmoluchowski2}, that enters the averaged one-body current
distribution \eqref{EQcurrentDistribution}. We reiterate the
conceptual and practical difference of Eq.~\eqref{EQsmoluchowski2} to
the one-body velocity field $\vel(\rv,t)$, see
Eq.~\eqref{EQvelocityField}. The former is a configuration space
function, and hence constitutes an important formal object, whereas
the latter is the result of microscopically sharp
coarse-graining. Therefore, this is a more concrete and intuitively
accessible vector field in physical space.

Given an initial state of the system at time $t$, the time evolution
is fully determined by Eqs.~\eqref{EQsmoluchowski1} and
\eqref{EQsmoluchowski2}. This is a high-dimensional problem, and the
feasibility of direct solutions can be assessed along the reasoning by
\citet{kohn1999nobel} in equilibrium: If we were to attempt a
numerical solution in a one-dimensional problem of, say, 10 particles,
and restrict ourselves to a numerical grid containing 10 grid points,
with 10 bytes to represent the value at each grid point, we need 100GB
memory, in order to store a single instance of $\Psi$. (Optimists for
the development of computer resources are welcome to consider 20
particles.)  In the present (power functional) context, for very small
systems, both analytical \cite{hermann2018activeSedimentation} and
numerical \cite{stuhlmueller2018structural} solutions were obtained.
For conceptual purposes, it is very important to have specified a
concrete many-body dynamics. In practice trajectory-based Brownian
dynamics simulations (see e.g.~\citet{sammueller2021}) offer a
powerful alternative, based on importance sampling, that is
well-suited for tackling realistic, large systems.

Developing a stand-alone theoretical dynamical framework offers both
practical benefits of computational efficiency as well as providing a
conceptual framework for formulating fundamental physical questions,
analysing simulation data, and identifying physical mechanisms for
phenomena that are observed in simulation work and in experiment. As
the external forces \eqref{EQexternalForce} remain of one-body
character even if the system is no longer in equilibrium, we seek a
description on the basis of one-body correlation functions.

In a time-dependent situation, the sum of the external and internal
forces will not cancel in general, and will hence influence the
average motion. Hence, the sum of the terms on the left-hand side of
the force balance relation \eqref{EQforceBalanceEquilibrium} will no
longer vanish. Moreover having a nonconservative contribution to the
external force field is no longer forbidden, as was the case in
equilibrium. In the overdamped limit considered here, the resulting
driving force will be balanced by a friction force,
$-\gamma\vel(\rv,t)$. This plausibility argument leads to the correct
one-body equation of motion,
\begin{align}
  \gamma\vel(\rv,t) &= -k_BT\nabla\ln\rho(\rv,t)
  - \nabla\frac{\delta F_{\rm exc}[\rho]}{\delta \rho(\rv,t)} 
\notag\\&\quad  + \fv_{\rm sup}(\rv,t)
  + \fv_{\rm ext}(\rv,t),
  \label{EQofMotionNonequilibrium}
\end{align}
where the one-body density profile $\rho(\rv,t)$ and the velocity
field $\vel(\rv,t)$ are microscopically resolved in space and in time.
The forces on the right-hand side of
Eq.~\eqref{EQofMotionNonequilibrium} represent (i) ideal diffusion,
(ii) an internal ``adiabatic'' excess force that arises from the
excess free energy functional $F_{\rm exc}[\rho]$, (iii) an additional
internal superadiabatic force field $\fv_{\rm sup}(\rv,t)$ that is due
to the flow and occurs only in nonequilibrium, and iv) the external
(driving) force field $\fv_{\rm ext}(\rv,t)$. Here the superadiabatic
force field $\fv_{\rm sup}(\rv,t)$ accounts for all contributions, due
to internal interactions, that are of genuine nonequilibrium character
and hence are not contained in the adiabatic excess force field.

One might be surprised by the occurrence of a genuine equilibrium
object, the excess free energy functional $F_{\rm exc}[\rho]$, in an
out-of-equilibrium situation. As $F_{\rm exc}[\rho]$ requires an
underlying statistical ensemble and Boltzmann distributed microstates,
one might query its validity in
Eq.~\eqref{EQofMotionNonequilibrium}. However, this situation is
well-founded due to the {\it adiabatic construction}. Here one
considers a hypothetical ``adiabatic'' equilibrium system, which
possess the same interparticle interaction potential $u(\rv^N)$ as the
real system. Furthermore the adiabatic system possesses the same
one-body density distribution as the nonequilibrium system at a fixed
snapshot $t$ in time,
\begin{align}
  \rho_{{\rm ad},t}(\rv) &= \rho(\rv,t),
  \label{EQadiabaticDensityDefinition}
\end{align}
where $\rho_{{\rm ad},t}(\rv)$ is the density profile in the adiabatic
system. One can perform the adiabatic construction at each point in
time; hence, $\rho_{{\rm ad},t}(\rv)$ inherits an apparent time
dependence, although by construction the underlying many-body system
is in equilibrium with no explicit time dependence. As the adiabatic
system is in equilibrium, we may invoke the Mermin-Evans theorem of
DFT and conclude that there is a unique ``adiabatic'' external
potential $V_{{\rm ad},t}(\rv)$ which stabilizes the given density
$\rho_{{\rm ad},t}(\rv)$.

Hence, we can formulate the force balance
\eqref{EQforceBalanceEquilibrium} in the adiabatic system as
\begin{align}
  - k_BT \nabla\ln\rho_{{\rm ad},t}(\rv)
  + \fv_{\rm ad}(\rv,t) 
  &= \nabla V_{{\rm ad},t}(\rv),
  \label{EQforceBalanceAdiabatic}
\end{align}
where the adiabatic excess force field $\fv_{\rm ad}(\rv,t)$ is given
either (i) by the microscopic average \eqref{EQforceExcAsAverage} over
the equilibrium ensemble of the adiabatic system, or (ii) as the
density functional relationship \eqref{EQforceExcDFT}.  In the first
case, we may sample $\fv_{\rm ad}(\rv,t)$ directly with an equilibrium
method that offers access to the adiabatic system.  This task involves
finding $V_{{\rm ad},t}(\rv)$.  This constitutes an inverse problem
that requires computational effort. A brute force method consists of
guessing $V_{{\rm ad},t}(\rv)$ and sampling $\rho_{{\rm ad},t}(\rv)$
and then adjusting $V_{{\rm ad},t}(\rv)$ iteratively, such that the
external potential is, say, increased in regions with excessively high
density relative to the target density. Once a satisfactorily small
error in Eq.~\eqref{EQadiabaticDensityDefinition} is achieved, one can
directly solve Eq.~\eqref{EQforceBalanceAdiabatic} for $\fv_{\rm
  ad}(\rv,t)$. However, more direct methods based on the custom flow
method exist (as described in Sec.~\ref{SECcustomFlow}).

Within DFT the inverse problem has already been addressed implicitly
and, as a result, the adiabatic force field is directly available. We
can hence make the second term on the right-hand side of the equation
of motion \eqref{EQofMotionNonequilibrium} fully explicit as
\begin{align}
  \fv_{\rm ad}(\rv,t) &= -\nabla \left.
  \frac{\delta F_{\rm exc}[\rho]}
       {\delta \rho(\rv)}
       \right|_{\rho(\rv)=\rho(\rv,t)},
       \label{EQfadx}
\end{align}
which shows explicitly how the equilibrium free energy functional
enters the dynamical theory~\eqref{EQofMotionNonequilibrium} in a
well-defined and unambiguous way.

In the time evolution equation \eqref{EQofMotionNonequilibrium} the
genuine nonequilibrium contributions to the internal force field are
contained in $\fv_{\rm sup}(\rv,t)$. These forces do not occur in
equilibrium and cannot be obtained based on a free energy description.
Setting $\fv_{\rm sup}(\rv,t)=0$ can in specific cases be a reasonable
approximation, and the resulting dynamical theory is commonly referred
to as the dynamical density functional theory (DDFT).

\begin{figure*}
  \includegraphics[width=1.99\columnwidth,angle=0]{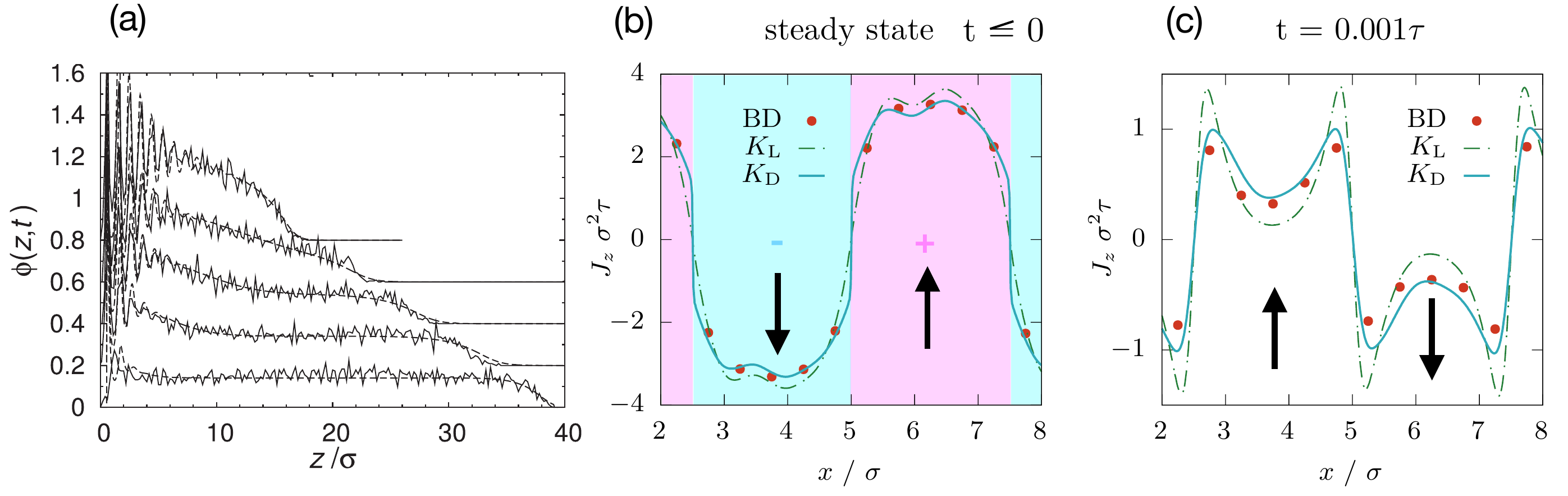}
  \caption{Dynamics of the density profile of colloidal hard spheres
    (a) in dynamical sedimentation, as primarily governed by adiabatic
    forces, and (b), (c) motion reversal under temporal switching of
    step shear, as a purely superadiabatic effect.  (a) Local packing
    fraction $\phi(z,t)=\rho(z,t)\pi\sigma^3/6$ as a function of the
    (scaled) height coordinate $z/\sigma$, where $\sigma$ is the hard
    sphere diameter. Results at increasing time are shifted upward by
    0.2 units. The system is initially almost homogeneous and over the
    course of time develops a strong density gradient, including
    layering at the bottom of the container.  Shown are results from
    confocal microscopy experiment (full lines) and from DDFT (dashed
    lines) using a density-dependent mobility.  (b),(c) Current
    profiles in the flow $\ev_z$-direction, $J_z \sigma^2\tau$, as a
    function of the position $x$ in the gradient $\ev_x$-direction of
    the (inhomogeneous) shear field.  The system is three-dimensional
    and it is homogeneous in the third direction $\ev_y$. Shear is
    induced by a square wave external force that acts alternating in
    the positive (light violet and $+$) and negative (light cyan and
    $-$) $\ev_z$-direction with strength $5k_BT/\sigma$ for times
    $t<0$. At time $t=0$ the external force is switched off. Due to
    the viscoelastic memory of the hard sphere fluid, the current
    immediately reverses its direction, as can be seen by comparing
    the down-up sequence of arrows in (b) with the up-down sequence in
    (c).  Results are obtained from event-driven BD simulations
    (symbols) and from power functional theory with spatially local
    ($K_L$) and non-local diffusing memory kernel ($K_D$).  Adapted
    from~\citet{royall2007dynamicSedimentation} (a) and from
    \citet{treffenstaedt2019shear} (b), (c).}
  \label{FIGsedimentationAndShear}
\end{figure*}

A simple counter-example, where the adiabatic approximation fails, is
steady shear of a homogeneous fluid, where $\rho(\rv,t)=\rm const$.
As the density is constant, no adiabatic effects occur on the one-body
level [the gradient in Eq.~\eqref{EQfadx} vanishes], although the
system can be driven arbitrarily far out-of-equilibrium by increasing
the shear rate. [This concept was carried much further by
  \citet{delasheras2020fourForces}, see their Supplemental Material
  for fully inhomogeneous flow patterns.]
Figure~\ref{FIGsedimentationAndShear} shows results from an adiabatic
treatment of sedimentation [Fig.~\ref{FIGsedimentationAndShear}(a)]
and superadiabatic effects in time-dependent shear
[Figs.~\ref{FIGsedimentationAndShear}(b) and
  \ref{FIGsedimentationAndShear}(c)] of Brownian hard spheres.

Hence, a complete dynamical theory needs to specify the superadiabatic
force field $\fv_{\rm sup}(\rv,t)$. This task is accomplished within
the power functional framework, where the superadiabatic force field
is expressed as a functional derivative \cite{schmidt2013pft},
\begin{align}
  \fv_{\rm sup}(\rv,t) &= -\frac{\delta P_t^{\rm exc}[\rho,\Jv]}
  {\delta \Jv(\rv,t)}.
  \label{EQfsupPFT}
\end{align}
In Eq.~\eqref{EQfsupPFT} the variation is performed at fixed density
distribution and at fixed time $t$, and the superadiabatic excess
power functional $P_t^{\rm exc}[\rho,\Jv]$ is a functional of both the
density and the current distribution. As $P_t^{\rm exc}[\rho,\Jv]$
originates from $u(\rv^N)$, it is in general both non-local in space
and non-local in time. The dependence is on the history of both
fields, i.e., on their values at times $<t$, where $t$ is the time at
which the variation \eqref{EQfsupPFT} is performed; the continuity
equation holds \cite{schmidt2013pft}.  The functional carries units of
energy per time, or power, $[P_t^{\rm exc}]={\rm J}/{\rm s}={\rm W}$.

Besides the occurrence of memory effects, the mathematical structure
is significantly richer than that of the DDFT, due to the fact that
the dependence on the current now occurs on both sides of
Eq.~\eqref{EQofMotionNonequilibrium}, on the left-hand side via
Eq.~\eqref{EQvelocityField} and on the right-hand side via
Eq.~\eqref{EQfsupPFT}. Hence, the current is defined by an implicit
relationship, which offers by far greater flexibility in describing
physical effects than an explicit theory, such as the DDFT. Recall
that the Euler-Lagrange equation \eqref{EQequilibriumEulerLagrange} of
equilibrium DFT is an implicit equation as well, albeit one for the
density profile. In equilibrium, it is precisely this structure, that
allows freezing, capillary behaviour, wetting,
etc.\ \cite{evans2016specialIssue} to be described.

For completeness, the temporal changes of the density profile
$\rho(\rv,t)$ are obtained from the current $\Jv(\rv,t)$ via the
continuity equation
\begin{align}
  \frac{\partial \rho(\rv,t)}{\partial t} = -\nabla\cdot\Jv(\rv,t).
\end{align}

The variational structure of power functional theory is analogous to
that of equilibrium DFT.  However, the similarity occurs on a deep,
structural level, as power functional theory is based on a variational
(extremal) principle, akin to the equilibrium minimization principle
with respect to the density distribution
\eqref{EQvariationalPrincipleDFT}. In the dynamical case, the
minimization is rather performed with respect to the current, at fixed
density distribution, and at fixed time,
\begin{align}
  \frac{\delta R_t[\rho,\Jv]}{\delta \Jv(\rv,t)}
  &= 0 \qquad \rm (min).
  \label{EQdynamicalVariationalPrinciple}
\end{align}
In Eq.~\eqref{EQdynamicalVariationalPrinciple} the total power
functional $R_t[\rho,\Jv]$ consists of a sum,
\begin{align}
  R_t[\rho,\Jv] = \dot F[\rho] + P_t[\rho,\Jv] - X_t[\rho,\Jv],
  \label{EQRt}
\end{align}
where $\dot F[\rho]$ is the time derivative of the total (ideal and
excess) intrinsic free energy functional $F[\rho]=F_{\rm
  id}[\rho]+F_{\rm exc}[\rho]$, the superadiabatic contribution
$P_t[\rho,\Jv]$ accounts for genuine nonequilibrium effects, and
$X_t[\rho,\Jv]$ is the external power. Both $\dot F[\rho]$ and
$P_t[\rho,\Jv]$ are of intrinsic nature, i.e., they depend on
$u(\rv^N)$, but not on the external force field. 

The genuine nonequilibrium power splits into ideal and excess
(superadiabatic) contributions ($P_t[\rho,\Jv]=P_t^{\rm
  id}[\rho,\Jv]+P_t^{\rm exc}[\rho,\Jv]$) where the exact ideal gas
dissipation contribution is local in space and time and given by
\begin{align}
  P_t^{\rm id}[\rho,\Jv] = \frac{\gamma}{2}
  \int d\rv \frac{\Jv^2(\rv,t)}{\rho(\rv,t)}.
\end{align}
The external power is the following sum of mechanical and motionless
contributions:
\begin{align}
  X_t[\rho,\Jv] = \int d\rv (\Jv(\rv,t)\cdot\fv_{\rm ext}(\rv,t)
  -\rho(\rv,t)\dot V_{\rm ext}(\rv,t)).
\end{align}
Inserting the decomposition \eqref{EQRt} into the dynamical extremal
principle \eqref{EQdynamicalVariationalPrinciple} and carrying out the
functional derivative yield an Euler-Lagrange equation that is
identical to the equation of motion \eqref{EQofMotionNonequilibrium}
with the superadiabatic excess force given by Eq.~\eqref{EQfsupPFT}.
The proof of this identity requires the derivative $\delta P_t^{\rm
  id}[\rho,\Jv]/\delta\Jv(\rv,t)=\gamma\Jv(\rv,t)/\rho(\rv,t)$.

Furthermore, using successively the functional chain rule, the
continuity equation, and spatial integration by parts, one finds the
total time derivative of the intrinsic free energy functional as
\begin{align}
  \dot F[\rho] = \frac{d}{dt} F[\rho] 
  &= \int d\rv \frac{\delta F[\rho]}{\delta\rho(\rv,t)}\dot\rho(\rv,t)
  \label{EQstarFdotZero}\\
  &=-\int d\rv \frac{\delta F[\rho]}{\delta \rho(\rv,t)}\nabla\cdot\Jv(\rv,t)\\
  &=\int d\rv \Jv(\rv,t)\cdot\nabla\frac{\delta F[\rho]}{\delta \rho(\rv,t)}.
  \label{EQstarFdot}
\end{align}
Due to the linear dependence on $\Jv(\rv,t)$, the form
\eqref{EQstarFdot} can be differentiated easily with respect to the
current with the density profile held fixed. The result is the total
(ideal and excess) adiabatic force field:
\begin{align}
  -\frac{\delta \dot F[\rho]}{\delta \Jv(\rv,t)} &= 
  -\nabla \frac{\delta F[\rho]}{\delta \rho(\rv,t)}
  \notag\\&
  = -k_BT\nabla \ln\rho(\rv,t)
  -\nabla \frac{\delta F_{\rm exc}[\rho]}{\delta\rho(\rv,t)},
\end{align}
where the last term is $\fv_{\rm ad}(\rv,t)$, see Eq.~\eqref{EQfadx}.
Lastly, the derivative of the external power is $\delta
X_t[\rho,\Jv]/\delta \Jv(\rv,t)=\fv_{\rm ext}(\rv,t)$. The equation of
motion \eqref{EQofMotionNonequilibrium} follows straightforwardly upon
the collection of all terms.

That the power functional \eqref{EQRt} exists is not an assumption.
Via a constructive proof it is derived from an underlying many-body
extremal principle. We do not reproduce the proof here (see
Sec.~\ref{SECpftBD}), but instead state only the starting point, which
is a many-body version of the one-body power functional \eqref{EQRt},
defined as
\begin{align}
  {\cal R}_t &= \int d\rv^N \Psi(\rv^N,t) \sum_i
  \left( \frac{\gamma \tilde\vel_i^2}{2} - \tilde\vel_i\cdot\fv_i^{\rm tot}
  +\dot V_{\rm ext}(\rv_i,t)
  \right).
  \label{EQRcalt}
\end{align}
In Eq.~\eqref{EQRcalt} $\tilde\vel_i(\rv^N,t)$ are configuration space
functions that represent trial velocities and $\fv_i^{\rm tot}$ is the
total force acting on particle $i$. The physical values of the
velocities are attained upon minimizing ${\cal R}_t$ with respect to
$\tilde\vel_i$ [which can easily be explicitly performed due to the
  simple quadratic structure of Eq.~\eqref{EQRcalt}]. Using a
dynamical version of Levy's constrained search method
\cite{levy1979,dwandaru2011levy} yields the one-body power functional
$R_t[\rho,\Jv]$ as given in Eq.~\eqref{EQRt}, with the one-body
minimization principle \eqref{EQdynamicalVariationalPrinciple}.  In
the following we flesh out this material and first turn to the
fundamentals.

\section{Many-body description}
\label{SECmanybodyDynamics}
\subsection{Internal and external forces}
\label{SECinternalAndExternalForces}

We consider particles (colloids, atoms, molecules, macromolecules, or
quantum particles) with position coordinates $\rv_i$, where the
particle index $i=1,\ldots,N$, and $N$ is the total number of
particles. As a short-hand notation $\rv^N\equiv
\rv_1,\ldots,\rv_N$. Position space is $d$-dimensional with $d=3$
being often the most relevant case, but important systems, such as
particles adsorbed at substrates or confined between plates have $d=2$
or even $d=1$ (confinement in channels). The case $d=1$ is also
important for conceptual purposes, as (some) exact results are
available.

The force on particle $i$ is also a $d$-dimensional vector, which
typically can be split into internal and external parts,
\begin{align}
  \fv_i(\rv^N,t) 
  &= \fv_{{\rm int},i}(\rv^N) + \fv_{\rm ext}(\rv_i,t).
  \label{EQftoti}
\end{align}
Here $\fv_{{\rm int},i}(\rv^N)$ is the internal force on particle $i$
which is exerted due to the cumulative effect of all other particles
in the system. There are typically no self-interactions and the
internal interactions do not depend on time.  The external force
field, however, is in general time dependent and characterized
(defined by) the property that it depends only on the position of
particle $i$, but not on the positions of all other particles $j\neq
i$. Hence, $\fv_{\rm ext}(\rv,t)$ can be viewed as a prescribed
external force field of a generic position coordinate $\rv$ and time
$t$. The external force field hence couples to the degrees of freedom
in the system, but there is no ``back action''; i.e., $\fv_{\rm
  ext}(\rv,t)$ is externally imposed, independent of the system
degrees of freedom.

We consider internal forces that are obtained from an interparticle
interaction potential $u(\rv^N)$ as the following negative gradient:
\begin{align}
  \fv_{{\rm int,}i}(\rv^N) = -\nabla_i u(\rv^N),
\end{align}
where, as before, $\nabla_i$ denotes the derivative with respect to
$\rv_i$. The total internal potential energy $u(\rv^N)$ can, but need
not, come from pairwise interparticle interactions.

\subsection{Hamiltonian dynamics}
\label{SEConebodyMD}

We consider classical particles first and start by deriving the
microscopic continuity equation by building the time derivative of the
density operator,
\begin{align}
  \frac{d}{dt}\hat\rho &=
  \frac{d}{dt}\sum_i\delta(\rv-\rv_i)\\
  &= \sum_i\Big(
  \frac{\partial}{\partial \rv_i}\delta(\rv-\rv_i)
  \Big)\cdot\dot\rv_i\\
  &= \sum_i\Big(
  \frac{\partial}{\partial(\rv_i-\rv)}\delta(\rv-\rv_i)
  \Big)\cdot\vel_i\\
  &= \sum_i \Big(
  -\frac{\partial}{\partial \rv}\delta(\rv-\rv_i)
  \Big)\cdot\vel_i
  \label{EQsignChangeContinuity}\\
  &= -\frac{\partial}{\partial \rv}\cdot
  \sum_i \delta(\rv-\rv_i)\vel_i
  \label{EQderivativeOutsideContinuity}\\
  &= -\nabla\cdot\hat\Jv,
  \label{EQcontinuityOperatorMD}
\end{align}
where the spatial derivative is $\nabla=\partial/\partial \rv$, the
microscopic density operator is defined as
$\hat\rho=\sum_i\delta(\rv-\rv_i)$, and the microscopic one-body
current operator is given as $\hat\Jv = \sum_i \delta(\rv-\rv_i)
\vel_i$.  Note the sign change in Eq.~\eqref{EQsignChangeContinuity}
from the change to the argument of the derivative. This substitution
enables one in Eq.~\eqref{EQderivativeOutsideContinuity} to move the
divergence operator outside of what becomes the current operator.

As the Newtonian dynamics are second order in time, we expect to
obtain a useful result when differentiating one more in time. Hence,
consider
\begin{align}
  \frac{d^2}{dt^2} \hat\rho&=
  \frac{d}{dt}(-\nabla\cdot\hat\Jv)
  =-\nabla\cdot \frac{d}{dt}\hat\Jv,
\end{align}
where we have used Eq.~\eqref{EQcontinuityOperatorMD} in the first
step. We can make progress with the following time derivative of the
current operator,
\begin{align}
  \frac{d\hat \Jv}{dt} &=
  \frac{d}{dt}\sum_i \delta(\rv-\rv_i)\vel_i\\
  &= \sum_i\Big(\frac{d}{dt}\delta(\rv-\rv_i)\Big)\vel_i
  +\sum_i \delta(\rv-\rv_i) \frac{d\vel_i}{dt}\\
  &= \sum_i \frac{\partial \delta(\rv-\rv_i)}{\partial \rv_i}
  \cdot\dot\rv_i\vel_i
  + \sum_i \delta(\rv-\rv_i)\frac{\fv_i}{m}\\
  &= -\nabla\cdot\sum_i \delta(\rv-\rv_i)\vel_i\vel_i
  +\frac{\hat\Fv}{m}\\
  &= \frac{\nabla\cdot \hat\taub}{m}   +\frac{\hat\Fv}{m},
\end{align}
where the kinetic stress operator is defined as $\hat\taub = -\sum_i
m\vel_i\vel_i\delta(\rv-\rv_i)$ and the force density operator is
$\hat\Fv = \sum_i \delta(\rv-\rv_i)\fv_i$. Hence, we have obtained the
operator equation of motion
\begin{align}
  m \frac{d}{dt}\hat\Jv &= \nabla\cdot\hat\taub + \hat\Fv,
  \label{EQforceDensityBalanceMD}
\end{align}
which expresses the total change in current (multiplied by mass) as
the sum of the divergence of the stress tensor (as a transport effect)
plus the force density. Equation~\eqref{EQforceDensityBalanceMD} can
be viewed as Newton's second law on the (classical) one-body operator
level.  Readers with a background in hydrodynamics will immediately be
familiar with the kinetic stress being a velocity-velocity dyadic
product, as this appears in standard derivations of the Navier-Stokes
equation, see \citet{hansen2013}. However, in contrast to a continuum
mechanical treatment, here $\hat\taub$ is resolved on a microscopic
scale (via the delta function in position). Note also that the trace,
$\Tr \hat\taub/2$, is the locally resolved kinetic energy density
operator.

\subsubsection*{Distribution functions as averages}
We obtain {\it one-body distribution functions} via averaging
according to
\begin{align}
  O(\rv,t) &= \langle \hat O(\rv,t;\rv^N,\pv^N) \rangle,
  \label{EQgenericOperatorAverage}
\end{align}
where $\hat O$ is a phase space function (``operator'') that
additionally depends on a generic position argument $\rv$ and
explicitly on time, in the most general case.  In
Eq.~\eqref{EQgenericOperatorAverage} the average is over the
probability distribution of microstates at time $t$. (This is now a
statistical description; an ensemble of systems is propagated forward
in time.)  Hence,
\begin{align}
  \left\langle\cdot\right\rangle &=
  \int d\rv^N d\pv^N \cdot \Psi(\rv^N,\pv^N,t),
  \label{EQphaseSpaceAverage}
\end{align}
where $\Psi$ is the many-body probability distribution function to
find microstate $\rv^N,\pv^N$ at time $t$. The differential phase
space volume element (which determines how to integrate over $\Psi$ in
order to obtain probabilities) is $d\rv^N d\pv^N$, and the
distribution function is normalized [$\int d\rv^N d\pv^N
  \Psi(\rv^N,\pv^N,t) = 1$] at all times~$t$.  For classical inertial
dynamics the time evolution of $\Psi$ is governed by the following
Liouville equation:
\begin{align}
  \frac{\partial \Psi}{\partial t} &=
  -\sum_i\Big(
  \frac{\pv_i}{m}\cdot\frac{\partial}{\partial \rv_i}
  +\fv_i\cdot\frac{\partial}{\partial \pv_i}
  \Big)\Psi.
\end{align}

\subsubsection*{One-body equation of motion}
The operator identities for the time derivative of density
\eqref{EQcontinuityOperatorMD} and current
\eqref{EQforceDensityBalanceMD} can be averaged over the phase space
distribution function according to
Eq.~\eqref{EQphaseSpaceAverage}. This yields reduced, yet
microscopically sharp one-body equations of motion:
\begin{align}
  \dot\rho\rt &= -\nabla\cdot\Jv\rt
  \label{EQcontinuityAverageMD},\\
  m\dot\Jv\rt &=  \nabla\cdot\taub\rt 
  + \Fv_{\rm int}\rt + \rho\rt\fv_{\rm ext}\rt.
  \label{EQmJdotFromManyBody}
\end{align}
In Eq.~\eqref{EQmJdotFromManyBody} the one-body distribution functions
are averaged according to Eq.~\eqref{EQphaseSpaceAverage}, i.e.,
$\rho(\rv,t)=\langle\hat\rho\rangle$,
$\Jv(\rv,t)=\langle\hat\Jv\rangle$,
$\taub(\rv,t)=\langle\hat\taub\rangle$,
$\Fv(\rv,t)=\langle\hat\Fv\rangle$, etc; the internal force density
operator is $\hat\Fv_{\rm int}=-\sum_i\delta(\rv-\rv_i)\nabla_i
u(\rv^N)$.  The kinematic fields are interrelated by time integration:
let the system be initially in equilibrium and $\rho(\rv,t\leq
0)=\rho(\rv,0)$, and $\Jv(\rv,t\leq 0)=0$.  At times $t>0$, then
\begin{align}
  \Jv(\rv,t) &= \int_0^t dt' \dot\Jv(\rv,t'),\\
  \rho(\rv,t) &= \rho(\rv,0) - \int_0^t dt' \nabla\cdot\Jv(\rv,t').
\end{align}
(Note that rotational contributions to the current leave the density
unchanged.) The one-body equations of motion are not closed as
$\rho\rt$, $\Jv\rt$, $\nabla\cdot\taub\rt$ and $\Fv_{\rm int}\rt$ are
unknown; only $\fv_{\rm ext}\rt$ is given, and we have only two
equations.

The nontrivial contribution due to the interparticle coupling is
\begin{align}
  & \nabla\cdot\taub\rt + \Fv_{\rm int}\rt \equiv
  \label{EQtotalInternalForceDensityMD}
  \\&\quad
   -\nabla\cdot
   \Big\langle
   \sum_i \delta(\rv-\rv_i)\frac{\pv_i\pv_i}{m}
   \Big\rangle
   -\Big\langle\sum_i \delta(\rv-\rv_i) \nabla_i u(\rv^N)
   \Big\rangle.\notag
\end{align}
We treat this force density field in Sec.~\ref{SECpftBD} using
dynamical functional methods.

\subsection{Brownian dynamics}
\label{SEConebodyBD}

We turn to the case of $N$ (classical) colloidal particles in
$d$-dimensional space, dispersed in a solvent at temperature $T$, and
undergoing overdamped Brownian motion with friction coefficient
$\gamma$, internal interaction potential $u(\rv^N)$ and under the
influence of an external force field $\fv_{\rm ext}(\rv,t)$. The
Langevin equations of motion are
\begin{align}
  \gamma \dot\rv_i &= \fv_i^{\rm det}(\rv^N,t) + \xib_i(t),
  \label{EQLangevinEquationOfMotion}
\end{align}
where the deterministic force acting on particle $i$ is a vector field
given by
\begin{align}
  \fv_i^{\rm det}(\rv^N,t) &= -\nabla_i u(\rv^N) 
  + \fv_{\rm ext}(\rv_i,t).
  \label{EQfdetDefinition}
\end{align}
The random contribution on the right-hand side of
Eq.~\eqref{EQLangevinEquationOfMotion} is a stochastic white noise
term with prescribed moments
\begin{align}
  \overline{\xib_i(t)} &= 0,
  \label{EQnoiseNoBias}\\
  \overline{\xib_i(t)\xib_j(t')} &= 
  2 k_BT\gamma \delta_{ij} {\bf 1} \delta(t-t'),
  \label{EQnoiseAutoCorrelator}
\end{align}
where the overline denotes an average over the noise realizations, the
left-hand side of Eq.~\eqref{EQnoiseAutoCorrelator} is a dyadic
product, $\delta_{ij}$ denotes the Kronecker symbol, and $\bf 1$
indicates the $d\times d$ unit matrix.

The Langevin scheme is well suited to carry out computer simulations,
via discretizing the equations of motion and using a simple Euler or
adaptive time-stepping algorithms \cite{sammueller2021} to integrate
the positions forward in time. The noise can be generated from pseudo
random number algorithms (e.g.\ the Box-Muller transform to generate
Gaussian distributed random numbers).  Building averages then requires
one, in principle, to average both over initial states (of which the
distribution function needs to be known and is in practice often
assumed to be equilibrated) and over different realizations of the
noise.  In the Langevin framework, the time-dependent probability
distribution function of microstates, $\Psi(\rv^N,t)$, does not appear
explicitly. This often makes calculations hard, as averages of
interest have to be reduced to the only known ones for the noise,
Eqs.~\eqref{EQnoiseNoBias} and \eqref{EQnoiseAutoCorrelator}.

Having $\Psi(\rv^N,t)$ is a powerful feature, with the microscopic
foundation of the concept of entropy resting upon it.  The explicit
introduction of $\Psi(\rv^N,t)$ into the framework is achieved by
complementing the Langevin picture by the corresponding Fokker-Planck
equation of motion for $\Psi(\rv^N,t)$. In the present case of
overdamped motion this is the Smoluchowski equation,
\begin{align}
  \frac{\partial \Psi}{\partial t} &=
  -\sum_i \nabla_i\cdot\vel_i \Psi.
  \label{EQSmoluchowski}
\end{align}
Here the many-body configurational velocity $\vel_i(\rv^N,t)$ of
particle~$i$ is a function (not a differential operator) defined via
\begin{align}
  \gamma\vel_i  &= \fv_i^{\rm det} - k_BT\nabla_i \ln \Psi \\
  &= -\nabla_i u(\rv^N) + \fv_{\rm ext}(\rv_i,t)
  -k_BT \nabla_i \ln \Psi,
\end{align}
where we have used Eq.~\eqref{EQfdetDefinition} to make $\fv_i^{\rm
  det}(\rv^N,t)$ explicit. The last term on the right-hand sides
corresponds to the noise contribution in the Langevin equation
\eqref{EQLangevinEquationOfMotion}; by differentiating the logarithm,
the term can be analogously rewritten as $-(k_BT/\Psi)\nabla_i
\Psi$. The position derivative in the Smoluchowski equation
\eqref{EQSmoluchowski} acts both on $\vel_i$ and on the distribution
function; hence, Eq.~\eqref{EQSmoluchowski} has the form of a
continuity equation for the local conservation of probability, as the
right-hand side expresses the negative divergence of a probability
current $\vel_i\Psi$.

It is instructive to rewrite the Smoluchowski equation in operator
form
\begin{align}
  \frac{\partial\Psi}{\partial t} &=
  -\sum_i \nabla_i \cdot \gamma^{-1}
  (\fv_i^{\rm det} - k_BT\nabla_i \ln\Psi)\Psi \\
  &=-\gamma^{-1}\sum_i
  \Big(
  (\nabla_i\cdot\fv_i^{\rm det}) 
  + \fv_i^{\rm det}\cdot\nabla_i - k_BT\nabla_i^2
  \Big) \Psi\\
  &\equiv \hat\Omega \Psi,
\end{align}
where the {\it Smoluchowski operator} is defined as
\begin{align}
\hat\Omega  &=-\gamma^{-1}\sum_i
  \Big(
  (\nabla_i\cdot\fv^{\rm det}_i) 
  + \fv_i^{\rm det}\cdot\nabla_i - k_BT\nabla_i^2
  \Big).
  \label{EQsmoluchowskiOperator}
\end{align}
Hence, the Smoluchowski equation is in compact notation simply
\begin{align}
  \frac{\partial \Psi}{\partial t} &= \hat\Omega \Psi,
\end{align}
which is a partial differential equation of first order in time and
second order in position. However, in contrast to the Schr\"odinger
equation, here $\Psi$ is real. Hence, there is no coupling of real and
imaginary parts, as occurs in quantum mechanics. The Smoluchowski
equation is instead a drift-diffusion equation for the many-body
distribution function. In particular, the diffusive effect is
generated by the Laplace operator~$\nabla_i^2$.

Again, one central purpose of $\Psi$ is to facilitate building
averages $O$ via
\begin{align}
 O &= \langle\hat O \rangle = \int d\rv^N \hat O \Psi(\rv^N,t),
\end{align}
where $\hat O$ is an operator that constitutes a physical
observable. Clearly, if $\hat O$ is a configuration space function,
$\hat O(\rv^N,t)$, then the order of terms in the integrand does not
matter, $\langle \hat O \rangle=\int d\rv^N \hat O \Psi=\int d\rv^N
\Psi \hat O$.

For the case of the density operator
$\hat\rho=\sum_i\delta(\rv-\rv_i)$ we obtain the one-body density
distribution
\begin{align}
  \rho(\rv,t) &= \langle \hat\rho \rangle
  = \int d\rv^N \sum_i\delta(\rv-\rv_i) \Psi(\rv^N,t).
\end{align}

We turn to the description of the one-body dynamics and are interested
in the time evolution of $\rho(\rv,t)$. Hence, we consider the time
derivative
\begin{align}
  \frac{\partial}{\partial t} \rho(\rv,t) &=
  \frac{\partial}{\partial t} 
  \int d\rv^N \sum_i \delta(\rv-\rv_i) \Psi\\
  & =\int d\rv^N \sum_i \delta(\rv-\rv_i) \frac{\partial\Psi}{\partial t}\\
  & =-\int d\rv^N \sum_i \delta(\rv-\rv_i) 
  \sum_j \nabla_j\cdot\vel_j \Psi
  \label{EQrhoTimeDerivativeBD1}\\
  & =\int d\rv^N \sum_i\sum_j \left(\nabla_j \delta(\rv-\rv_i) \right)
  \cdot\vel_j \Psi
  \label{EQrhoTimeDerivativeBD2}\\
  & =-\nabla\cdot\int d\rv^N \sum_i\delta(\rv-\rv_i)\vel_i \Psi
  \label{EQrhoTimeDerivativeBD3}\\
  & =-\nabla\cdot \Jv(\rv,t),
  \label{EQrhoTimeDerivativeBD4}
\end{align}
where we have used: the Smoluchowski equation \eqref{EQSmoluchowski}
for Eq.~\eqref{EQrhoTimeDerivativeBD1}, integration by parts for
Eq.~\eqref{EQrhoTimeDerivativeBD2}, and the identity
$\nabla_j\delta(\rv-\rv_i) = -\delta_{ij}\nabla\delta(\rv-\rv_i)$ for
Eq.~\eqref{EQrhoTimeDerivativeBD3}. In the last step
[Eq.~\eqref{EQrhoTimeDerivativeBD4}] we have defined the one-body
current distribution
\begin{align}
  \Jv(\rv,t) &= \langle\hat\Jv\rangle,\\
  \hat \Jv &= \sum_i\delta(\rv-\rv_i)\vel_i,
\end{align}
where $\hat \Jv$ is the current operator.  As an aside, the current
operator can alternatively be expressed, using the {\it velocity
  differential operator} $\hat\vel_i$, as
\begin{align}
  \hat\Jv &= \sum_i\delta(\rv-\rv_i)\hat\vel_i,\\
  \gamma \hat\vel_i &= \fv_i^{\rm det} -k_BT \nabla_i,
\end{align}
where $\fv_i^{\rm det}(\rv^N,t)$ is still given via
Eq.~\eqref{EQfdetDefinition}.  It is straightforward to show that
$\hat\vel_i\Psi=\vel_i\Psi$ and hence that both velocity
representations yield the same one-body current distribution. [The
  many-body velocity should not be confused with the average,
  microscopically resolved {\it velocity field}
  $\vel(\rv,t)=\Jv(\rv,t)/\rho(\rv,t)$.]

It remains to express the current distribution via the forces that act
in the system. As the dynamics are overdamped, no further time
derivative is required. We rather rewrite as follows,
\begin{align}
  \gamma \Jv\rt &= \gamma \int d\rv^N \sum_i\delta(\rv-\rv_i)\vel_i \Psi\\
   &= \int d\rv^N
  \sum_i\delta(\rv-\rv_i)(\fv_i^{\rm det}-k_BT\nabla_i\ln\Psi) \Psi\\
   &= \int d\rv^N
    \sum_i\delta(\rv-\rv_i)
   \notag\\&\qquad\qquad
   \times(-(\nabla_i u)+\fv_{\rm ext}(\rv_i,t)-k_BT\nabla_i) \Psi\\
   &= -\int d\rv^N
    \sum_i\delta(\rv-\rv_i)(\nabla_i u)\Psi\notag\\
    & \qquad +\int d\rv^N\sum_i\delta(\rv-\rv_i)
    \fv_{\rm ext}(\rv_i,t) \Psi\label{EQcurrentBDintermediate}\\
    &\qquad -\int d\rv^N\sum_i\delta(\rv-\rv_i)
    k_BT\nabla_i \Psi
    \notag\\
    &\equiv \Fv_{\rm int}\rt + \rho\rt \fv_{\rm ext}\rt - k_BT\nabla\rho\rt.
    \label{EQcurrentBDfinal}
\end{align}
In Eq.~\eqref{EQcurrentBDfinal} we have defined the first integral in
Eq.~\eqref{EQcurrentBDintermediate} as the internal force density
distribution
\begin{align}
  \Fv_{\rm int}(\rv,t) &= -\int d\rv^N
  \sum_i\delta(\rv-\rv_i)(\nabla_i u)\Psi.
  \label{EQinternalForceDensityDefinition}
\end{align}
In the second integral in Eq.~\eqref{EQcurrentBDintermediate} we have
replaced $\fv_{\rm ext}(\rv_i,t)$ by $\fv_{\rm ext}(\rv,t)$ due to the
presence of the delta function. In the third integral in
Eq.~\eqref{EQcurrentBDintermediate} we have integrated by parts, and
have once more used
$\nabla_i\delta(\rv-\rv_i)=-\nabla\delta(\rv-\rv_i)$.

The equations of motion of motion follow as
\begin{align}
  \gamma \Jv\rt  &= \Fv_{\rm int}\rt 
  + \rho\rt \fv_{\rm ext}\rt - k_BT\nabla\rho\rt,
  \label{EQforceDensityBalanceBD}\\
  \frac{\partial\rho\rt}{\partial t} &= -\nabla\cdot\Jv\rt.
\end{align}
The current can be eliminated to obtain a single equation for the time
evolution of the one-body density
\begin{align}
  \frac{\partial\rho\rt}{\partial t} &= 
  -\gamma^{-1}\nabla\cdot\Fv_{\rm int}\rt
  -\gamma^{-1}\nabla\cdot\rho\rt\fv_{\rm ext}\rt
  \notag\\&\quad
  + D \nabla^2 \rho\rt,
\end{align}
where $D=k_BT/\gamma$ is the diffusion constant according to
Einstein's relation.

It is instructive to scale Eq.~\eqref{EQforceDensityBalanceBD} by the
density profile. We first define the internal microscopic one-body
force field by normalizing as follows:
\begin{align}
  \fv_{\rm int}(\rv,t) &= \Fv_{\rm int}(\rv,t)/\rho(\rv,t).
  \label{EQfintAsRatio}
\end{align}
The microscopic velocity field is obtained as before as the ratio
\begin{align}
  \vel(\rv,t) &= \Jv(\rv,t)/\rho(\rv,t).
\end{align}
We can now rewrite the force density balance
\eqref{EQforceDensityBalanceBD} by dividing with the density profile,
which yields a {\it force balance} relationship:
\begin{align}
  \gamma \vel\rt &= 
  \fv_{\rm int}\rt + \fv_{\rm ext}\rt - k_BT\nabla\ln\rho\rt.
  \label{EQforceFieldBalanceBD}
\end{align}
Note that there are no transport contributions (kinetic stress is
absent), as the motion is overdamped. However, diffusive effects do
occur.  The equations of motion are not closed on the one-body level,
as the internal force density distribution $\Fv_{\rm int}(\rv,t)$ is
unknown at this stage, and only defined via the many-body average
\eqref{EQinternalForceDensityDefinition}.

There are three possible ways out. 

\begin{itemize}

\item[(i)] Solve the many-body dynamics numerically, either using
  trajectory-based BD or, for small number of degrees of freedom, the
  Smoluchowski equation.

\item[(ii)] Relate $\Fv_{\rm int}(\rv,t)$ to higher-body (two-body,
  three-body, etc.) correlation functions and formulate closure
  relations. This is both technically and conceptually difficult.

\item[(iii)] Express $\Fv_{\rm int}(\rv,t)$ in a variational way via a
  generator (generating functional). This is also technically and
  conceptually difficult, but it is complementary to (i) and
  (ii). (Hybrid forms of (i) and (iii) could be imagined.)  We
  describe the power functional for overdamped BD below in
  Sec.~\ref{SECpftBD}.

\end{itemize}

\subsection{Quantum dynamics}%
\label{SEConebodyQM}
  
Besides its relevance for a broad variety of systems, the importance
of quantum dynamics in the present context lies not least in its
formal similarities with the classical Hamiltonian dynamics
(Sec.~\ref{SEConebodyMD}).  As we demonstrate in the following, a case
can be made for the universality of the dynamical one-body point of
view.\footnote{See e.g.\ \citet{tarantino2021} and
  \citet{tchenkoue2019} for recent work addressing the force balance
  in the context of time-dependent density functional theory.}
Readers who are primarily interested in classical systems may directly
proceed to Sec.~\ref{SECdft}, where we cover classical density
functional theory. Besides the significant importance in its own
right, this approach also acts both as a blueprint and an integral
component for the dynamical theory. The connection is via the
adiabatic construction, as described in Sec.~\ref{SECadiabaticState}.

We consider $N$ spinless nonrelativistic quantum particles that are
coupled by an internal interaction potential $u(\rv^N)$. The particles
have an electrical charge $q$ and mass $m$. The particles are exposed
to a magnetic vector potential $\Av(\rv,t)$ and an external potential
energy $V_{\rm ext}(\rv,t)$. The form of $u(\rv^N)$ is general,
possibly including a Coulombic contribution.  We use the position
representation of the Schr\"odinger equation
\begin{align}
  i\hbar\frac{\partial}{\partial t} \Psi(\rv^N,t) &= 
  \hat H \Psi(\rv^N,t),
  \label{EQSchroedingerEquation}
\end{align}
with $\Psi(\rv^N,t)$ being the quantum mechanical wave function and
$\hat H$ the Hamiltonian. The wave function is normalized at all times
as $\int d\rv^N \Psi^*\Psi = \int d\rv^N|\Psi|^2 = 1$, $\forall t$;
the asterisk denotes the complex conjugate.  The Hamiltonian has the
following form of kinetic energy plus potential energy:
\begin{align}
  \hat H &= \sum_i \frac{\hat\pv_i^2}{2m}
  +u(\rv^N) + \sum_i V_{\rm ext}(\rv_i,t),
\end{align}
where the (kinematic) momentum operator is
\begin{align}
  \hat\pv_i = -i\hbar\nabla_i - q \Av(\rv_i,t),
  \label{EQkinematicMomemtumOperator}
\end{align}
with the first term acting via differentiation and the second term
acting via multiplication (on the wave function in position
representation).

Our goal is to obtain the reduced one-body dynamics. Consider the
general Heisenberg equation of motion for an operator $\hat O$,
\begin{align}
  \frac{d\hat O}{dt} &= 
  \frac{i}{\hbar}[\hat H,\hat O] + \frac{\partial \hat O}{\partial t},
  \label{EQHeisenbergOfMotion}
\end{align}
where the brackets denote the commutator of two operators.  Quantum
mechanical averages $O$ are built by the bra-ket sandwich,
\begin{align}
  O = \langle\hat O\rangle = 
  \langle\Psi|\hat O|\Psi\rangle
  =\int d\rv^N \Psi^*\hat O\Psi,
\end{align}
where, depending on the form of the operator $\hat O$, its expectation
value can have both explicit and implicit (due to the dynamics) time
dependence.

Applying Eq.~\eqref{EQHeisenbergOfMotion} to the position operator
yields the result
\begin{align}
  \frac{d\rv_i}{dt} &= \frac{\hat\pv_i}{m},
\end{align}
which shows that calling Eq.~\eqref{EQkinematicMomemtumOperator} the
kinematic momentum is justified.

We next differentiate $\hat\pv_i$ in time. The calculation (which is
omitted) is lengthier, but straightforward.  We define the force
operator for particle $i$ as
\begin{align}
  \hat\fv_i &= -(\nabla_i u(\rv^N)) - (\nabla_i V_{\rm ext}(\rv_i,t))
  -q\dot\Av(\rv_i,t)\notag\\
  &\quad + \frac{q}{2m}(\hat\pv_i\times {\bf B}(\rv_i,t)
  -{\bf B}(\rv_i,t)\times\hat\pv_i).
\end{align}
where the magnetic field is $\Bv\rt = \nabla\times \Av(\rv,t)$.

Then the equation of motion \eqref{EQHeisenbergOfMotion} for
$\hat\pv_i$ attains the compact form
\begin{align}
  \frac{d\hat\pv_i}{dt} &= \hat\fv_i,
\end{align}
which is Newton's second law on the operator level. 

We next summarize the relevant one-body operators, for density $\hat
n$, current $\hat \Jv$, kinetic stress $\hat\taub$ and internal force
density $\hat\Fv_{\rm int}$. These are defined, respectively, by
\begin{align}
  \hat n &= \sum_i  \delta(\rv-\rv_i),\\
  \hat\Jv &= \frac{1}{2m}
  \sum_i\big(\hat\pv_i
  \delta(\rv-\rv_i) + \delta(\rv-\rv_i)
  \hat\pv_i\big),
  \label{EQcurrentOperatorQM}\\
  \hat\taub &= -\frac{1}{2m}
  \sum_i\big( \hat\pv_i\delta_i\hat\pv_i + \hat\pv_i\delta_i\hat\pv_i^{\sf T}
  \big),
  \label{EQkineticStressOperatorQM}\\
  \hat\Fv_{\rm int} &= -\sum_i(\nabla_i u(\rv^N))\delta(\rv-\rv_i),
  \label{EQinternalForceDensityOperatorQM}
\end{align}
where we have used the short-hand notation
$\delta_i=\delta(\rv-\rv_i)$ in Eq.~\eqref{EQkineticStressOperatorQM},
and the superscript $\sf T$ indicates the transpose of a $d\times d$
matrix (here a dyadic product). Both the density and internal force
density are multiplication operators. All occurring kinematic momentum
operators act on all arguments to their right.

We now derive the corresponding equations of motion, beginning with
the density operator. (The change of notation from $\hat\rho$ to $\hat
n$ is cosmetic, done in order to conform to quantum convention.)  We
consider the density operator for particle $i$, and apply the
Heisenberg equation of motion \eqref{EQHeisenbergOfMotion}, which
yields
\begin{align}
  \frac{d}{dt} \delta(\rv-\rv_i) &=
  \frac{i}{\hbar}[\hat H, \delta(\rv-\rv_i)] 
  +\frac{\partial}{\partial t}\delta(\rv-\rv_i)
  \label{EQdndtStart}\\
  &= \frac{i}{2m\hbar}
  \sum_j[\hat\pv_j^2, \delta(\rv-\rv_i)],
  \label{EQdndtIntermediate}
\end{align}
where the commutator of the potential energy contributions and the
density operator (delta function) vanishes, as does the partial time
derivative of the delta function.

To address Eq.~\eqref{EQdndtIntermediate}, we consider the general
form of the commutator of $\hat\pv_i^2\equiv p^2$ with a function
$g(\rv^N)$,
\begin{align}
  [p^2,g(\rv^N)] &= p^2g-pgp+pgp-gp^2\\
  &= p(pg-gp)+(pg-gp)p\\
  &= p[p,g]+[p,g]p\\
  &= p(-i\hbar\nabla_i g)-i\hbar(\nabla_i g)p,
\end{align}
where we have used $[p,g]=-i\hbar (\nabla_i g)$. Returning to the full
notation, hence, we have
\begin{align}
  [\hat\pv_i^2,g(\rv^N)]=
  -i\hbar\big(
  \hat\pv_i\cdot(\nabla_i g) + (\nabla_i g)\cdot\hat\pv_i
  \big).
\end{align}
The application to Eq.~\eqref{EQdndtIntermediate} yields zero for the
case $i\neq j$. For $i=j$ we obtain
\begin{align}
&  \frac{i}{2m\hbar}[\hat\pv_i^2,\delta(\rv-\rv_i)]
    \\&=
    \frac{-i^2\hbar}{2m\hbar}
    \Big[\hat\pv_i\cdot\big(\nabla_i \delta(\rv-\rv_i)\big)
      +\big(\nabla_i\delta(\rv-\rv_i)\big)\cdot\hat\pv_i
      \Big]\\
    &=-\nabla\cdot \frac{1}{2m}
    \big(\hat\pv_i\delta(\rv-\rv_i)+\delta(\rv-\rv_i)\hat\pv_i
    \big),
\end{align}
where we have used
$\nabla_i\delta(\rv-\rv_i)=-\nabla\delta(\rv-\rv_i)$. Recalling that
the left-hand side of Eq.~\eqref{EQdndtStart} is the time derivative
of $\delta(\rv-\rv_i)$ and summing over all particles yields
\begin{align}
  \frac{d}{dt} \hat n &= -\nabla\cdot\hat\Jv,
  \label{EQcontinuityOperatorQM}
\end{align}
which can rightfully be called the operator continuity equation. Here,
the anticipated form \eqref{EQcurrentOperatorQM} of the current
operator $\hat\Jv$ applies.  Building quantum mechanical expectation
values yields the one-body density distribution and the one-body
current distribution, which are defined, respectively, by
\begin{align}
  n(\rv,t) &= \langle\hat n\rangle = \int d\rv^N\Psi^*\hat n \Psi,\\
  \Jv(\rv,t) &= \langle\hat\Jv\rangle = \int d\rv^N\Psi^*\hat \Jv \Psi.
\end{align}
Building the quantum average over the operator continuity equation
\eqref{EQcontinuityOperatorQM} yields the continuity equation
\begin{align}
  \frac{\partial}{\partial t} n(\rv,t) &= -\nabla\cdot \Jv(\rv,t),
  \label{EQcontinuityAverageQMAppendix}
\end{align}
where we have changed the notation from total to partial time
derivative. This is purely cosmetic; the character of the time
derivative has not changed. In both cases the time derivative is with
respect to the real dynamics, and at fixed position $\rv$.  Clearly
the form \eqref{EQcontinuityAverageQMAppendix} is identical to the
classical result \eqref{EQcontinuityAverageMD}.

\subsubsection*{Current operator dynamics} 

We turn to the time evolution of the current operator
\eqref{EQcurrentOperatorQM}. Our hope, if not expectation, is to be
able to identify a relationship to the transport contribution
represented by the kinetic stress tensor
\eqref{EQkineticStressOperatorQM} and to the internal force density
operator \eqref{EQinternalForceDensityOperatorQM}. Hence, we are
seeking an analogue of the classical MD force density relationship
[Eq.~\eqref{EQforceDensityBalanceMD}]. This can indeed be established,
albeit not without a certain level of engagement in the quantum
formalism; however, all manipulations are straightforward in
principle. We start by considering the time evolution of the current
operator of particle $i$, defined as $\Jv_i =
(\delta_i\hat\pv_i+\hat\pv_i\delta_i)/(2m)$, where
$\delta_i=\delta(\rv-\rv_i)$, such that the (total) current operator
is $\hat \Jv=\sum_i \hat \Jv_i$. Hence,
\begin{align}
  \frac{d}{dt}\hat\Jv_i &=
  \frac{i}{\hbar}[\hat H,\hat\Jv_i] + \frac{\partial}{\partial t} \hat\Jv_i.
  \label{EQdJdtQM}
\end{align}
The last term can be simplified as
\begin{align}
  \frac{\partial}{\partial t}\hat\Jv_i &=
  \frac{1}{2m}\frac{\partial}{\partial t}
  \big(\delta_i\hat\pv_i+\hat\pv_i\delta_i \big)\\
  &=\frac{1}{2m}
  \big(\delta_i(-q\dot\Av(\rv_i,t))+(-q\dot\Av(\rv_i,t))\delta_i \big)\\
  &=-\frac{q}{m}\delta_i\dot\Av(\rv_i,t),
  \label{EQdJdtQMpartialTimeDerivative}
\end{align}
which is a multiplication operator, and an expected part of the force
density balance. We hence still need to consider the first
(commutator) term in Eq.~\eqref{EQdJdtQM}. We first address the
kinetic energy contribution to the Hamiltonian,
\begin{align}
  & \frac{i}{\hbar}
  \big[\sum_j \frac{\hat\pv_j^2}{2m}, 
    \frac{\hat\pv_i\delta_i+\delta_i\hat\pv_i}{2}
    \big]
  =\frac{i[\hat\pv_i^2,\hat\pv_i\delta_i+\delta_i\hat\pv_i]}{4m\hbar}
  \label{EQkineticEnergyContributionToHamiltonianCommutator}
  \\
  &=\frac{i}{4m\hbar}\Big(
  \hat p_i^\alpha[\hat p_i^\alpha,\hat\pv_i\delta_i+\delta_i\hat\pv_i]
  +[\hat p_i^\alpha,\hat\pv_i\delta_i+\delta_i\hat\pv_i]\hat p_i^\alpha
  \Big),\notag
\end{align}
where $\hat p_i^\alpha$ is the $\alpha$-th Cartesian component of
$\hat\pv_i$ and the Einstein summation convention is implied.
Contributions with $i\neq j$ vanish as there is no coupling between
$\nabla_j$ and~$\rv_i$. Hence, we need the following commutator
identity (which can explicitly be proven):
\begin{align}
  &\frac{i}{4m\hbar}\sum_i\big[\hat\pv_i^2,
    \hat\pv_i\delta_i+\delta_i\hat\pv_i)\big]
  \notag\\\quad  &=
  \nabla\cdot\hat\taub
  +\frac{q}{2m}\sum_i[\delta_i
    (\hat\pv_i\times{\bf B}_i-{\bf B}_i\times\hat\pv_i)\notag\\&
    \qquad\qquad\qquad
  +(\hat\pv_i\times{\bf B}_i-{\bf B}_i\times\hat\pv_i)\delta_i]
  +\frac{\hbar^2}{4m}\nabla\nabla^2\hat n,
  \label{EQcanExplicitlyBeProven}
\end{align}
where the kinetic one-body stress operator is a second rank tensor
given by
\begin{align}
  \hat\taub &= -\frac{1}{2m}\sum_i(\hat\pv_i\delta_i\hat\pv_i
  +\hat\pv_i\delta_i\hat\pv_i^{\sf T}).
\end{align}

It remains to consider the potential energy contribution to the
commutator in Eq.~\eqref{EQdJdtQM}. Defining the total potential
energy as $V(\rv^N)=u(\rv^N) + \sum_i V_{\rm ext}(\rv_i,t)$, we have
\begin{align}
  \frac{i}{\hbar}
  \big[ V,
    \frac{\hat\pv_i\delta_i + \delta_i\hat\pv_i}{2m}
    \big]
  &=-\frac{i}{2m\hbar}[\hat\pv_i\delta_i+\delta_i\hat\pv_i,V]\\
  &=-\frac{i}{2m\hbar}(-i\hbar)[\nabla_i\delta_i+\delta_i\nabla_i,V],
  \label{EQpotentialEnergyCommutator}
\end{align}
as clearly the magnetic contribution, $-2q\Av_i\delta_i$, commutes
with the potential energy. Hence, we can rewrite
Eq.~\eqref{EQpotentialEnergyCommutator} as
\begin{align}
&-\frac{1}{2m}\big([\nabla_i\delta_i, V] + [\delta_i\nabla_i,V]
  \big)\\
  &=-\frac{1}{2m}\big(
  \nabla_i\delta_i V- V\nabla_i\delta_i + \delta_i\nabla_i V - V \delta_i \nabla_i
  \big)\\
  &=-\frac{1}{2m}\big(
  (\nabla_i\delta_i)V + \delta_i(\nabla_i V) + \delta_i V \nabla_i
 -V(\nabla_i\delta_i)
\notag\\&
\qquad \quad - V \delta_i\nabla_i
+\delta_i(\nabla_i V) + \delta_i V \nabla_i
-V \delta_i\nabla_i
  \big)\\
  &= \frac{1}{m}\delta_i(-\nabla_i V)\\
  &= \frac{1}{m}\delta(\rv-\rv_i)(-\nabla_i V)\\
  &= \frac{1}{m}\big[
    \delta_i (-\nabla_i u(\rv^N)) + \delta_i (-\nabla_i V_{\rm ext}(\rv_i,t))
    \big],
  \label{EQdJdtQpotentialEnergyContribution}
\end{align}
where in the last step we have split the total potential energy into
internal and external contributions. Equation
\eqref{EQdJdtQpotentialEnergyContribution} is the (maybe expected)
contribution to the force density due to potential forces.

Collecting all terms, i.e.,
Eqs.~\eqref{EQdJdtQMpartialTimeDerivative},
\eqref{EQcanExplicitlyBeProven} and
\eqref{EQdJdtQpotentialEnergyContribution}, yields
\begin{align}
  m\frac{d}{dt} \hat\Jv &=
  \nabla\cdot\hat\taub 
  +\frac{\hbar^2}{4m} \nabla\nabla^2\hat n
  + \hat\Fv,
  \label{EQoperatorNewtonSecondLawQM}
\end{align}
where the total force density operator is defined as
\begin{align}
  \hat \Fv &= \frac{1}{2}\sum_i(\hat\fv_i\delta_i+\delta_i\hat\fv_i),
\end{align}
with the force operator of particle $i$ given by
\begin{align}
  \hat\fv_i &= -(\nabla_i u(\rv^N)) - (\nabla_i V_{\rm ext}(\rv_i,t))
  -q\dot\Av(\rv_i,t)\notag\\
  &\quad+\frac{q}{2m} (\hat\pv_i\times{\bf B}_i-{\bf B}_i\times\hat\pv_i).
\end{align}

Building the quantum average $\langle\cdot\rangle=\int
d\rv^N\Psi^*\cdot\Psi$ of Eq.~\eqref{EQoperatorNewtonSecondLawQM}
yields the force density balance in the form of Newton's second law
for one-body current and force density distributions,
\begin{align}
  m \frac{d}{dt} \Jv\rt &= 
  \frac{\hbar^2}{4m}\nabla\nabla^2 n\rt
  +\nabla\cdot\taub\rt
  +\Fv\rt.
  \label{EQNewtonsSecondLawQMAverageAppendix}
\end{align}
Here the total one-body force density is given by
\begin{align}
  \Fv\rt &= \Fv_{\rm int}\rt 
  - n\rt(\nabla V_{\rm ext}\rt + q\dot\Av\rt)
  \notag\\&\quad
  +q\Jv\rt\times{\bf B}\rt,
  \label{EQtotalInternalForceDensityQMAppendix}
\end{align}
and the internal force density given by the quantum average $\Fv_{\rm
  int}(\rv,t)=-\langle\sum_i (\nabla_i
u(\rv^N))\delta(\rv-\rv_i)\rangle$.

As a corollary, for a single particle: $N=1$, $\rv^N\equiv\rv_1$,
$\Fv_{\rm int}\rt=0$, and the resulting equation of motion takes on
the following form (leaving away arguments $\rv,t$):
\begin{align}
  m\frac{d\Jv}{dt} &= (-q\dot\Av-\nabla V_{\rm ext})n
  +q \Jv\times{\bf B}\notag\\&\quad
  +\nabla\cdot\taub_{\rm id} 
  + \frac{\hbar^2}{4m} \nabla\nabla^2 n,\\
  \taub_{\rm id} &= - m\frac{\Jv\Jv}{n} 
  - \frac{\hbar^2}{4m}\frac{(\nabla n)(\nabla n)}{n},
\end{align}
which is exact, i.e., equivalent to the Schr\"odinger equation for a
single quantum particle. For $N\geq 2$ internal interactions will be
relevant and the one-body description is no longer closed as both
$\Fv_{\rm int}\rt$ and $\taub\rt$ are unknown.  In order to address
this issue, we return to the quantum dynamical case in
Sec.~\ref{SECquantumPFT}, where we introduce functional generators for
these fields, which then allow to construct a formally closed one-body
theory.

The classical and quantum force balance relationships,
Eqs.~\eqref{EQmJdotFromManyBody} and
\eqref{EQNewtonsSecondLawQMAverageAppendix}, bear striking
similarities to each other; note that the external force field is
$\fv_{\rm ext}\rt =-q\dot\Av\rt + q\vel\rt\times\Bv\rt-\nabla V_{\rm
  ext}\rt$. Sometimes the first term on the right-hand side of
Eq.~\eqref{EQNewtonsSecondLawQMAverageAppendix} is subsumed into a
modified kinetic stress tensor, $\taub_{\rm QM}\rt=\taub\rt +
\hbar^2\nabla\nabla n\rt/(4m)$, which then renders
Eqs.~\eqref{EQmJdotFromManyBody} and
\eqref{EQNewtonsSecondLawQMAverageAppendix} formally identical.

\section{The adiabatic state}
\label{SECdft}

\subsection{The adiabatic construction}
\label{SECadiabaticState}
\begin{figure}
  \includegraphics[width=0.9\columnwidth,angle=0]
                  {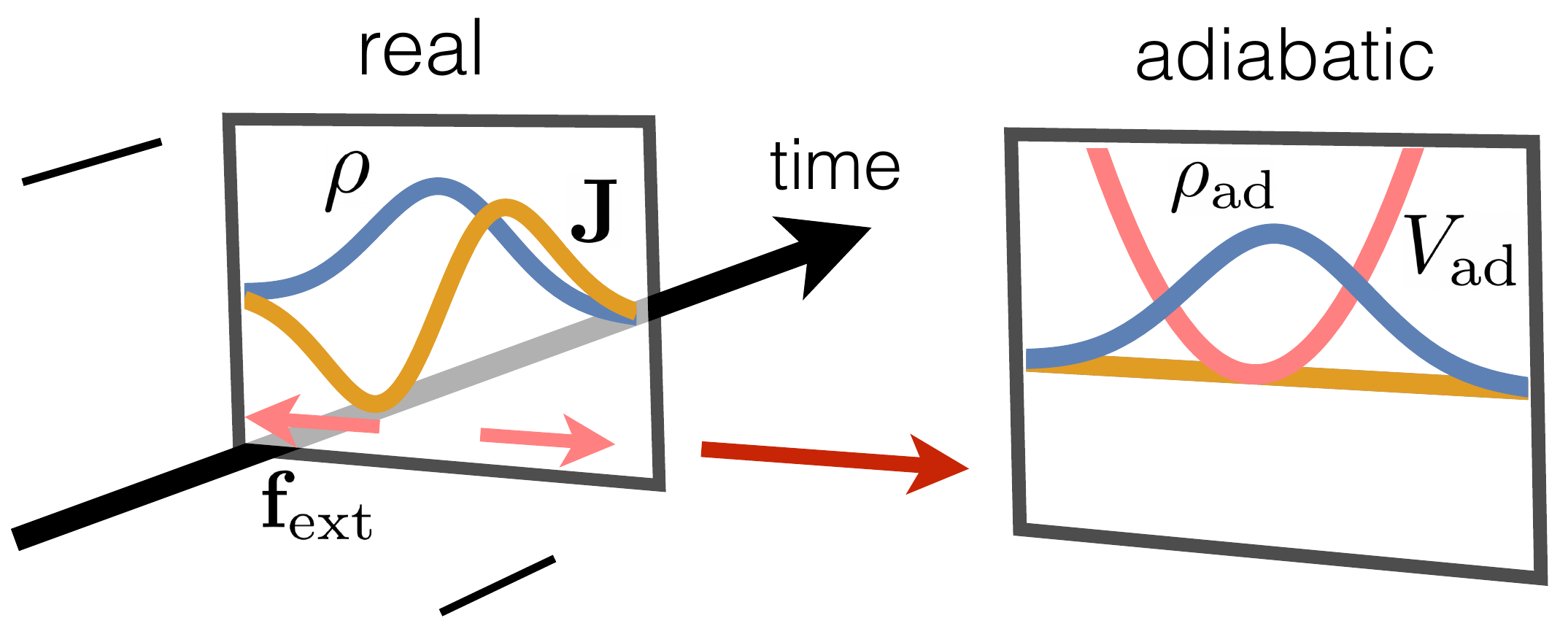}
  \caption{Illustration of the adiabatic construction. Left panel: The
    real system evolves in time according to BD. It is characterized
    by an in general inhomogeneous density distribution $\rho(\rv,t)$
    and an inhomogeneous one-body current $\Jv(\rv,t)\neq 0$. The
    adiabatic state is constructed at (each) fixed time, such that
    $\rho(\rv,t)$ is identical to the density profile $\rho_{{\rm
        ad},t}(\rv)$ in the adiabatic system at the time
    considered. Right panel: The adiabatic system is in equilibrium,
    and hence there is no average current. The inhomogeneous adiabatic
    density profile is stabilized by the action of an external
    potential~$V_{{\rm ad},t}(\rv)$, which acts solely in the
    adiabatic system, but not in the real system. In the real system
    it is the external force field $\fv_{\rm ext}(\rv,t)$ that drives
    the dynamics.}
  \label{FIGadiabaticConstruction}
\end{figure}

In the following we describe the concept of splitting the internal
force field into an adiabatic and an additional superadiabatic
contribution. We restrict ourselves to the case of (classical)
overdamped Brownian many-body dynamics.  The adiabatic construction,
illustrated in Figure~\ref{FIGadiabaticConstruction}, was explicitly
demonstrated on the basis of computer simulation results by
\citet{fortini2014prl} using a one-dimensional hard core system. A
range of subsequent studies were aimed at the Gaussian core model
\cite{bernreuther2016gcm,stuhlmueller2018structural}, the
Lennard-Jones liquid \cite{schindler2016dynamicPairCorrelations},
Weeks-Chandler-Andersen repulsive particles
\cite{delasheras2020fourForces}, and hard disks
\cite{delasheras2018velocityGradient,jahreis2019shear}. An elegant and
computationally straightforward implementation in simulation work is
via the custom flow method of \citet{delasheras2019customFlow}, as
described in Sec.~\ref{SECcustomFlow}.

We start by recalling the BD one-body force field balance
\eqref{EQforceFieldBalanceBD} where the time-dependent internal force
field $\fv_{\rm int}(\rv,t)$ is defined via the correlator
\eqref{EQinternalForceDensityDefinition} and the ratio
\eqref{EQfintAsRatio}, i.e.,
\begin{align}
  \fv_{\rm int}(\rv,t)\rho(\rv,t) &= -\Big\langle
  \sum_i (\nabla_i u(\rv^N))\delta(\rv-\rv_i)
  \Big\rangle.
  \label{EQadiabaticStateFint}
\end{align}
The time-dependent density profile is
$\rho(\rv,t)=\langle\hat\rho\rangle$, where the average is carried out
over the nonequilibrium many-body probability distribution
$\Psi(\rv^N,t)$.  We compare Eq.~\eqref{EQforceFieldBalanceBD} at
time~$t$ with the force balance relationship in a second ``adiabatic''
system, which is at rest (on average) and in equilibrium. Hence, in
the adiabatic system
\begin{align}
  0 &= -k_BT\nabla\ln\rho_{{\rm ad},t}(\rv) + \fv_{{\rm ad},t}(\rv) 
  - \nabla V_{{\rm ad},t}(\rv),
  \label{EQadsupForceBalanceEquilibrium}
\end{align}
where $\rho_{{\rm ad},t}(\rv)$ is the density profile and $V_{{\rm
    ad},t}(\rv)$ is the external potential in the adiabatic system;
$-\nabla V_{{\rm ad},t}(\rv)$ is the external force field, which is
necessarily of gradient nature as the adiabatic system is in
equilibrium. The internal force field $\fv_{{\rm ad},t}(\rv)$ in the
adiabatic system, expressed as an average, is given via
\begin{align}
  \fv_{{\rm ad},t}(\rv) \rho_{{\rm ad},t}(\rv) &= -\Big\langle
  \sum_i (\nabla_i u(\rv^N))\delta(\rv-\rv_i)
  \Big\rangle_{\rm eq},
  \label{EQadsupInternalForceAsCorrelator}
\end{align}
which is similar in form to the nonequilibrium internal force field
[Eq.~\eqref{EQadiabaticStateFint}], with the sole distinction (and an
important one) that an equilibrium average is carried out (at fixed
$N,V,T$, i.e., canonically, indicated by the subscript eq). The
density profile in the adiabatic system is the equilibrium average
$\rho_{{\rm ad},t}(\rv)=\langle\hat\rho\rangle_{\rm eq}$.

Assume the most general case of a spatially inhomogeneous system that
evolves in time, i.e., consider $\rho(\rv,t)$ as having nontrivial
dependence on both of its arguments. At (each) time~$t$ we choose the
adiabatic system in such a way that its density profile coincides with
that in the nonequilibrium system. This amounts to the density
matching condition
\begin{align}
  \rho_{{\rm ad},t}(\rv) &= \rho(\rv,t),
  \label{EQadsupDensityMatch}
\end{align}
where the adiabatic density profile has acquired a parametric
dependence on time but is itself stationary as the adiabatic system is
in equilibrium at the same temperature $T$ of the nonequilibrium
system. (More precisely, were one to evolve the adiabatic system
according to its own time evolution, i.e., along a new ``adiabatic'
time axis $t_{\rm ad}$, then with respect to $t_{\rm ad}$ no changes
in the adiabatic density profile occur.)

The many-body distributions in the real system and in the adiabatic
system will in general differ from each other
[$\Psi(\rv^N,t)\neq\Psi_{{\rm ad},t}(\rv^N)$].  However,
\begin{align}
  \int d\rv^N \hat\rho \Psi_{{\rm ad},t}(\rv^N) &= 
  \int d\rv^N \hat\rho \Psi(\rv^N,t),
\end{align}
which is the density matching condition \eqref{EQadsupDensityMatch}
written in explicit average form.

Per construction $\Psi_{{\rm ad},t}(\rv^N)$ needs necessarily to be of
normalized Boltzmann form, as is appropriate for the canonical
ensemble at fixed $N,V$, and $T$. One might wonder whether such a
distribution is guaranteed to exist. Note that we are dealing with a
potentially complex situation, as $\rho_{{\rm ad},t}(\rv)$ can have a
virtually arbitrary shape (as long as it is one that occurs in a real
time evolution of the system). The answer to the question is
affirmative, based on a Hamiltonian with an unchanged interparticle
interaction potential $u(\rv^N)$, i.e., the adiabatic system is being
composed of, say, Lennard-Jones particles, when the real system under
investigation is a time-dependent process in the Lennard-Jones
system. The freedom that we need to introduce in the adiabatic system
is the presence of an ``adiabatic'' external potential, which is, from
the standpoint of the real system, of entirely virtual nature, and in
particular different from the real force field $\fv_{\rm ext}(\rv,t)$
that drives the time evolution. Mathematically, for a given
$u(\rv^N)$, there is a unique map in equilibrium, from the density
distribution to the external potential, $\rho_{\rm ad}\to V_{\rm
  ad}$. This is indeed ensured by the theorem due to
\citet{mermin1965} and \citet{evans1979}.

It is clearly of interest to study the internal force field in the
adiabatic reference system. In order to gain access to $\fv_{{\rm
    ad},t}(\rv)$, there are two obvious routes.
\begin{itemize}
\item[(i)] We can use the correlator expression
  \eqref{EQadsupInternalForceAsCorrelator} and carry out the
  average. The Mermin-Evans theorem ensures that $V_{{\rm ad},t}(\rv)$
  is unique. Hence, the Hamiltonian is fully and uniquely specified as
  is the canonical equilibrium probability distribution, which is
  required to carry out the average.
\item[(ii)] The second route takes a shortcut, based directly on the
  external potential $V_{{\rm ad},t}(\rv)$ (which again is determined
  in principle from the Mermin-Evans theorem) and a trivial
  rearranging of the equilibrium force balance relationship
  \eqref{EQadsupForceBalanceEquilibrium} into the form
\begin{align}
  \fv_{{\rm ad},t}(\rv) &= 
  k_BT\nabla\ln\rho_{{\rm ad},t}(\rv) + \nabla V_{{\rm ad},t}(\rv).
  \label{EQadsupInternalAdiabaticForceField}
\end{align}
\end{itemize}
Both routes are directly accessibly in many-body simulation work and
they can be equally useful. The density profile in the adiabatic
system is known [recall the density matching condition
  \eqref{EQadsupDensityMatch}]. Hence, using either method in practice
requires one to have an explicit representation of the Mermin-Evans
map $\rho_{\rm ad}\to V_{\rm ad}$. For computer simulation work,
custom flow \cite{delasheras2019customFlow} delivers this task.

An important point concerns higher-order correlation functions, i.e.,
those beyond the one-body density profile. While the density profile
is per construction guaranteed to be the same in the dynamical and the
adiabatic system, higher-body correlation functions in general will
differ. This is a straightforward consequence of the differences in
underlying many-body distributions; recall the Boltzmann form in the
adiabatic system versus the result of the Smoluchowski dynamics in the
real system. In practice, the respective two-body density correlation
functions are accessible in simulation work, see
\citet{fortini2014prl}. Moreover, recent conceptual advances in DFT
have demonstrated the relevance of two-body correlations, e.g., in the
quest for systematically incorporating interparticle attraction; see
the pioneering work by \citet{tschopp2020} and \citet{tschopp2021}.

\begin{figure}
  \includegraphics[width=0.9\columnwidth,angle=0]
                  {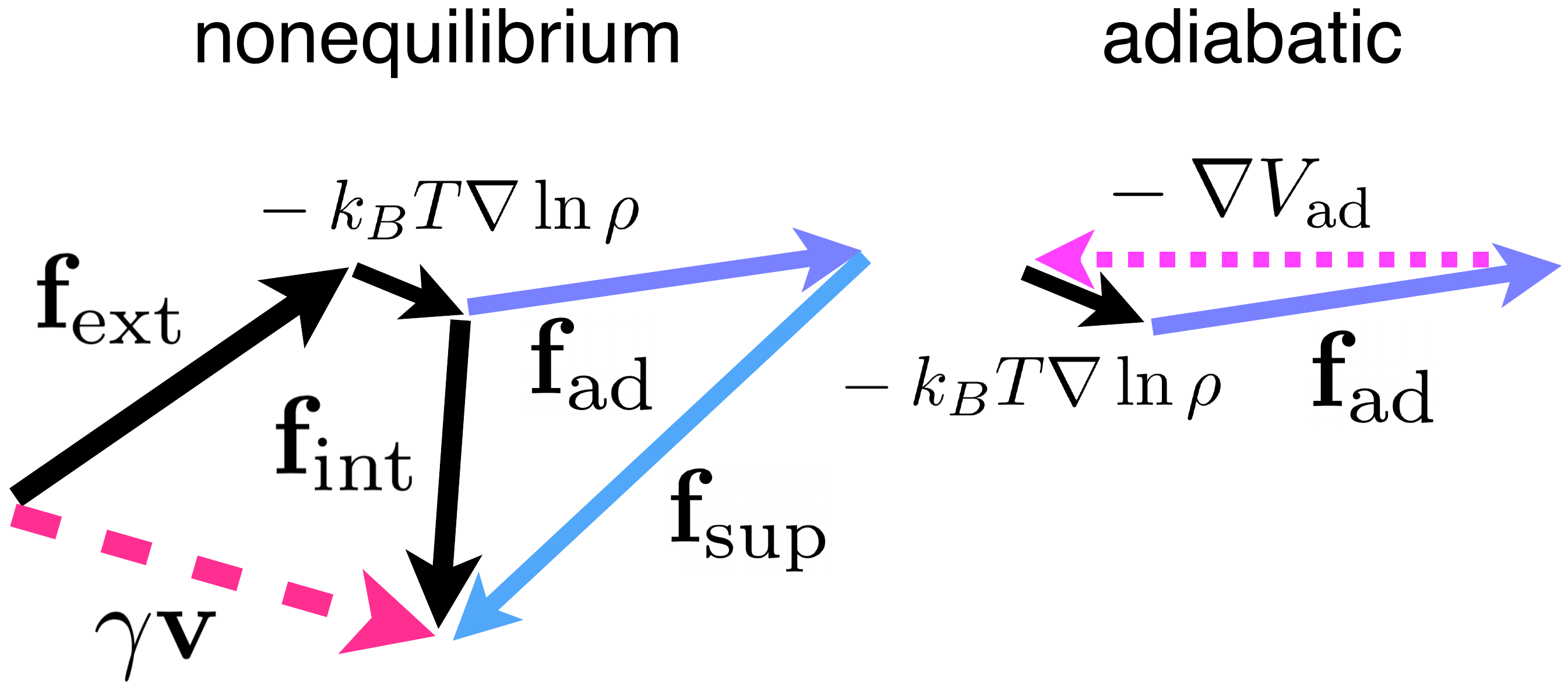}
  \caption{Force balance in nonequilibrium (left panel) and in the
    adiabatic system (right panel).  In nonequilibrium the sum of the
    external force field $\fv_{\rm ext}\rt$, the diffusive force field
    $-k_BT\nabla\ln\rho\rt$, and the internal force field $\fv_{\rm
      int}\rt$ add up and generate the (scaled) flow $\gamma\vel\rt$.
    The internal force field $\fv_{\rm int}\rt$ consists of a sum of
    adiabatic and superadiabatic contributions [$\fv_{\rm
        ad}\rt+\fv_{\rm sup}\rt$]. In the adiabatic system the sum of
    the diffusive force field $-k_BT\nabla\ln\rho\rt$, the internal
    adiabatic force field $\fv_{\rm ad}\rt$, and the adiabatic
    external force field $-\nabla V_{{\rm ad},t}(\rv)$ vanishes, as
    there is no flow (average one-body motion) in equilibrium.  The
    superadiabatic force field $\fv_{\rm sup}\rt$ constitutes the
    genuine nonequilibrium contribution to the real dynamics.}
  \label{FIGforceBalance}
\end{figure}

As is the case for the higher-correlation functions, there is hence no
reason to expect that the internal force density in the adiabatic
system will be identical to the counterpart in the real system, and
therefore $\fv_{\rm int}(\rv,t)\neq \fv_{{\rm ad},t}(\rv)$ in general.
Note that for the case of pair forces the pair distribution function
determines the local force density.  Nevertheless, as the
interparticle interaction potential is the same in the real and the
adiabatic system, and the one-body density distribution has the same
form, we might want $\fv_{{\rm ad},t}(\rv)$ to capture some of the
properties of $\fv_{\rm int}(\rv,t)$. (This is made rigorous in
Sec.~\ref{SECpft}.) As a consequence the difference of the real force
density and that in the adiabatic system might be a simpler object
than the bare $\fv_{\rm int}(\rv,t)$ itself. Hence, we define the
superadiabatic force field $\fv_{\rm sup}(\rv,t)$ as the difference
\begin{align}
  \fv_{\rm sup}(\rv,t) &= \fv_{\rm int}(\rv,t) 
  - \fv_{\rm ad}(\rv,t),
  \label{EQadsupInternalForceSplitting}
\end{align}
where we have changed the notation $\fv_{{\rm ad},t}(\rv)$ to
$\fv_{\rm ad}(\rv,t)$. Here the term superadiabatic refers to the
contribution above adiabatic, or more accurately, the contribution
that acts in addition to the adiabatic force field. [This implies no
  simple relationship of the relative sign, the direction, or the
  magnitude of $\fv_{\rm sup}(\rv,t)$, as compared to either $\fv_{\rm
    int}(\rv,t)$ or $\fv_{\rm ad}(\rv,t)$.] We cover the behaviour of
these fields in model setups with several simplifying geometries when
discussing power functional applications in Sec.~\ref{SECpft}.  Figure
\ref{FIGforceBalance} shows an illustration.

We insert the adiabatic-superadiabatic internal force splitting
\eqref{EQadsupInternalForceSplitting} into the nonequilibrium force
balance relationship \eqref{EQforceFieldBalanceBD} and (trivially)
obtain the equation of motion in the form
\begin{align}
  \gamma\vel(\rv,t) &=
  -k_BT\nabla\ln\rho(\rv,t)\notag \\&\quad+ \fv_{\rm ad}(\rv,t)
  + \fv_{\rm sup}(\rv,t) + \fv_{\rm ext}(\rv,t),
  \label{EQadsupEquationOfMotion}
\end{align}
with the adiabatic construction \cite{fortini2014prl} implied (i.e.,
the density matching condition \eqref{EQadsupDensityMatch} that
uniquely specifies the adiabatic system).  The functional dependencies
of $\fv_{\rm ad}(\rv,t)$ and $\fv_{\rm sup}(\rv,t)$ are fundamentally
very different from each other. In the adiabatic system, owing to the
Mermin-Evans map $\rho_{\rm ad}\to V_{\rm ad}$ and
Eq.~\eqref{EQadsupInternalAdiabaticForceField} we have a
density-functional dependence
\begin{align}
  \fv_{\rm ad}(\rv,t) &=
  \fv_{\rm ad}(\rv,t,[\rho]) = \fv_{{\rm ad},t}(\rv,[\rho_{\rm ad}]),
\end{align}
where the adiabatic and dynamic density profiles are identical [see
  Eq.~\eqref{EQadsupDensityMatch}] by construction.  Hence, $\fv_{\rm
  ad}(\rv,t)$ is an instantaneous (Markov-type) density functional
with neither memory nor dependence on other kinematic variables. Its
complexity lies entirely in the spatially nonlocal dependence on the
density distribution. We see in Sec.~\ref{SECpft} that the
superadiabatic force field depends functionally on density and flow as
follows:
\begin{align}
  \fv_{\rm sup}(\rv,t) &= \fv_{\rm sup}(\rv,t,[\rho,\Jv]) =
  \fv_{\rm sup}(\rv,t,[\rho,\vel]),
\end{align}
i.e., with an additional dependence on the current distribution or,
equivalently, on the microscopic velocity field. In general the
functional dependence will again be non-local in space, but also
non-local in time (in the form of history-dependence, i.e., dependence
on $\rho(\rv,t')$ and $\vel(\rv,t')$ at times $t'$ that lie not in the
future, i.e., $t'\leq t$).

In general $\fv_{\rm sup}(\rv,t)\neq 0$ only if the system is in
motion, i.e., $\vel(\rv,t')\neq 0$. The superadiabatic force field
vanishes [$\fv_{\rm sup}(\rv,t)=0$] if the system is at rest at all
prior times [$\vel(\rv,t')=0$]. Hence, in a system with no flow the
equation of motion \eqref{EQadsupEquationOfMotion} reduces to
\begin{align}
  0 &= -k_BT\nabla\ln\rho(\rv,t) + \fv_{\rm ad}(\rv,t) 
  + \fv_{\rm ext}(\rv,t),
\end{align}
where the density profile and hence the adiabatic forces field are
both invariant in time. Necessarily the external force field is also
invariant in time and of the form $-\nabla V_{\rm ext}(\rv)$. We hence
revover the exact static equilibrium limit from the time-dependent
theory. (This still is a highly nontrivial many-body problem,
encompassing a broad range of relevant physical phenomena, from phase
behaviour in bulk and at interfaces, structural correlations, etc.)
The adiabatic construction hence allows to systematically split the
problem of determining $\fv_{\rm int}(\rv,t)$ into the problem of
separately determining (and hence modeling and rationalizing) both the
adiabatic and superadiabatic contributions.

As it turns out, the adiabatic-superadiabatic splitting is not merely
a formal one. Important and prominent physical effects, such as drag
against a dense surrounding, both bulk and shear viscosity, and
nonequilibrium structural forces that are of genuine nonequilibrium
character are solely accounted for by the superadiabatic effects,
i.e., they can be understood only if $\fv_{\rm sup}(\rv,t)$ is
correctly accounted for (Sec.~\ref{SECpft} presents the corresponding
theoretical development as well as concrete applications.) In contrast
$\fv_{\rm ad}(\rv,t)$ is free of any of the previously mentioned
effects.

Nevertheless, $\fv_{\rm ad}(\rv,t)$ is in general neither a negligible
nor a small contribution (although there are special cases where
$\fv_{\rm ad}(\rv,t)=0$ or a small number, such as in strong shear
flow of a nearly homogeneous system \cite{jahreis2019shear} or the
specifically tailored systems of \citet{delasheras2020fourForces}). We
later demonstrate the functional structure, which involves a
superadiabatic current-density functional $P_t^{\rm exc}[\rho,\Jv]$,
which generates the superadiabatic force field via functional
differentiation as follows:
\begin{align}
  \fv_{\rm sup}(\rv,t,[\rho,\Jv]) &= 
  -\frac{\delta P_t^{\rm exc}[\rho,\Jv]}{\delta \Jv(\rv,t)}.
\end{align}
The magnitude and direction of $\fv_{\rm ad}(\rv,t)$ and those of
$\fv_{\rm sup}(\rv,t)$ are in general decoupled from each other. As a
rule of thumb, $\fv_{\rm ad}(\rv,t)$ is more prominent the more the
density profile deviates from a homogeneous profile, and $\fv_{\rm
  sup}(\rv,t)$ grows large with increased driving.

In the following, we first address the adiabatic force profile, then
describe the theoretical (equilibrium density-functional) structure
that one can associate with it. The Mermin-Evans theorem has the
important feature that the adiabatic force field is obtained from a
generating (intrinsic excess Helmholtz) free energy functional $F_{\rm
  exc}[\rho]$ via
\begin{align}
  \fv_{\rm ad}(\rv,t) 
  &= -\nabla\frac{\delta F_{\rm exc}[\rho]}{\delta \rho(\rv)}
  \Big|_{\rho(\rv)=\rho(\rv,t)}.
\end{align}
$F_{\rm exc}[\rho]$ is an intrisic object in the sense that it is
independent of the external potential, and characteristic for (and
dependent on) the internal interaction potential $u(\rv^N)$. This
might come as a surprise given the coupled nature of the many-body
problem behind the equilibrium force balance relationship
\eqref{EQadsupForceBalanceEquilibrium}, but this property can be made
entirely rigorous. Moreover, a functional minimization principle lies
behind this beautiful mathematical structure, and powerful physical
theory, which is the density functional framework, to which we turn in
the following and which we lay out in some detail.

As a final remark, it is worthwhile to point out that the concept of
integrating out degrees of freedom, or partial noise averaging in the
nonequilibrium system in order to arrive at effective internal
interactions [which differ in general from the bare $u(\rv^N)$] is
entirely different in character to the adiabatic construction. See
e.g.\ \citet{farage2015} for an insightful study of how
self-propulsion of active Brownian particles generates an effective
attractive tail of the pair potential, which originally was purely
repulsive. \citet{turci2021} recently addressed many-body
contributions to the effective attraction.

\subsection{Timeline of density functional theory}
\label{SECoverviewDFT}

The free gas-liquid interface, as treated by \citet{vanderWaals1893}
via a square-gradient approximation, can be viewed as the historically
first DFT. He concluded correctly that the interface between the
coexisting bulk fluid phases has finite width and is hence not a sharp
two-dimensional mathematical object. The theory extends the work
performed in his Ph.D thesis of 1873 [see the reissue
  \cite{vanderWaals1873}], which itself was dedicated to gas-liquid
bulk phase coexistence in bulk.\footnote{\citet{maxwell1874} reviewed
  that work a year after its publication in the journal {\it Nature}.}

The theory by \citet{onsager1949} of the isotropic-nematic phase
transition of long and thin hard rods is based on the virial (i.e.,
low density) expansion together with a geometrical scaling argument
that involves the particle aspect ratio;
\citet{vanroij2005pedagogical} gave a clear account of this. The phase
transition is of first order and the treatment is exact in the scaling
limit.  While neither Onsager nor van der Waals knew of free energy
density functionals, each of them was able to deduce a
self-consistency equation that with hindsight can be viewed as the
Euler-Lagrange equation of an underlying density functional.

The following decades saw much progress in the description of the
liquid state. Particular highlights include the formulation of the
integral equation closure by \citet{percus1958} and of scaled-particle
theory for hard spheres by \citet{reiss1959}. \citet{percus1976}
presented the exact solution for one-dimensional hard rods
\cite{tonks1936} that are exposed to an arbitrary external
potential. His solution has the form of an exact and closed
self-consistency equation for the density profile, with no
higher-order correlators being involved. These by then classic
approaches (see \citet{hansen2013}) formed the grassroots upon which
\citet{rosenfeld1989} later built his formidable fundamental-measure
density functional theory for hard sphere systems.

The birth of modern DFT is the treatment given by \citet{HK1964} of
the ground state properties of the electron gas. Their work
established that the ground state energy of a quantum system is a
functional of the one-body density distribution $n(\rv)$, and a unique
map exists $n(\rv)\to V_{\rm ext}(\rv)$. Only one year later,
\citet{mermin1965} generalized the theory to finite temperatures. At
$T>0$ entropy becomes relevant, and the framework that he developed
applies to quantum statistical physics. In the same year, Kohn and
Sham reintroduced orbitals into DFT, see e.g.\ \citet{kohn1999nobel}
for a well-accessible and compact description of the essentials of
electronic DFT.

In far-reaching work, \citet{evans1979} laid out the structure and the
foundation of the present-day use of classical DFT.
[\citet{evans2016specialIssue} discussed prior work.]  The paper also
contains the first formulation of the DDFT equation of motion.  The
DDFT approach lay virtually dormant for twenty years, until
\citet{marconi1999ddft} put it at the center of a new research
activity; see also \citet{archer2004ddft}.

Three important innovations were put forward in the same year:
\citet{levy1979} formulated the constrained search proof of the
Hohenberg-Kohn theorem.  \citet{rosenfeld1979mhnc} formulated the
modified hypernetted chain theory, including the hypothesis of
universality of the short-range structure in liquids.
\citet{ramakrishnan1979} developed their first-principle
order-parameter theory of freezing, which was based on a functional
Taylor expansion of the excess free energy functional around the
homogeneous bulk liquid.

\citet{tarazona1984} used weight functions in classical DFT to smooth
the density profile via spatial convolution. They take the weight
function to be proportional to the Mayer function,
$\exp(-\beta\phi(r))-1$, where $\phi(r)$ is the pair potential. The
method allowed them to incorporate nonlocal interparticle correlations
into DFT. \citet{rosenfeld1988} formulated his scaled-field particle
approach, which unifies the hitherto distinct scaled-particle and
Percus-Yevick theories.  \citet{vanderlick1989} formulated the exact
solution for mixtures of one-dimensional (polydisperse) hard rods in
an arbitrary external potential.

\citet{rosenfeld1989} constructed fundamental measure theory (FMT) for
hard sphere mixtures. His density-functional approach is geometric in
nature and was at that time (and still is) different in theoretical
structure than all other existing DFTs approximations.
\citet{kierlik1990} give an alternative and elegant formulation of
FMT, based on scalar weight functions. In a noteworthy extension of
his hard sphere functional, \citet{rosenfeld1994nonspherical} obtained
an initial generalization of FMT to nonspherical hard bodies; he
identified a relationship to the Gauss-Bonnet theorem of integral
geometry and related the Mayer bond to topological properties of the
system. This approach was carried further for specific systems such as
hard needle-sphere \cite{schmidt2001rsf} and hard plate-sphere
mixtures \cite{esztermann2006}.  More generally shaped bodies were
addressed by \citet{hansengoos2009prl} and by \citet{wittmann2015}.

The hard sphere FMT functional recevied a boost in popularity through
the version by \citet{rosenfeld1997RSLT}, which is based on respecting
the properties of the free energy functional upon dimensional
reduction. This version of FMT cured the initial defect of FMT, which
yielded the fluid unstable with respect to the crystal.
\citet{tarazona2000} introduced a new tensorial weight function into
FMT from considering cavitylike, one-dimensional density
distributions. His functional predicts freezing from first principles
in excellent quantitative agreement with simulation
benchmarks. Remaining inaccuracies are due to the description of the
fluid rather than the solid. To go beyond the Percus-Yevick
(compressibility) equation of state, seemingly inherent to FMT,
\citet{roth2002} formulated the White Bear version of FMT, which they
based on the Carnahan-Starling equation of state;
\citet{hansengoos2006} generalized this approach to multicomponent
mixtures of hard spheres.  \citet{davidchack2016} compared the
performance of different versions of FMT against benchmark simulation
data.  Minimization in three dimensions was performed by
\citet{levesque2012jcp}. Recently \citet{lutsko2020stable}
reconsidered the original deficiency of the \citet{rosenfeld1989}
functional, and obtained a class of what he referred to as explicitly
stable functionals.

Progress was made at overcoming the hard sphere paradigm and hence to
arrive at a first-principles version of FMT for a wider range of
microscopic models. This includes the penetrable step function pair
potential \cite{schmidt1999ps}, as an example of a non-hard core
model, used to test the universality of the bridge functional
\cite{rosenfeld2000ps}. The FMT for the Asakura-Oosawa model
colloid-polymer mixture \cite{schmidt2000ao} of hard sphere colloids
and ideal effective polymer spheres proved to be a valuable tool for
the study of adsorption and confinement phenomena in such systems. See
\citet{schmidt2001wr,schmidt2004nahs,schmidt2011tnas} for work on free
energy functionals for more general nonadditive hard sphere mixtures.

On a more conceptual level, classical DFT was generalized to
quenched-annealed mixtures \cite{schmidt2002qa}, where the quenched
component forms a random matrix and the annealed component represents
an equilibrated fluid that is adsorbed in the resulting pore
structure. The theory predicts quenched-annealed fluid structure with
accuracy comparable to liquid integral equation theory
\cite{schmidt2002versus}. \citet{delasheras2014fullCanonical}
demonstrated how to practically obtain canonical information from
grand canonical DFT results. This proved to be a crucial step
clarifying the role of ensembles in DDFT
\cite{delasheras2016particleConservation}.  Recently \citet{lin2019ml}
and \citet{lin2020ml} constructed a DFT using nonlocal functional
ideas combined with machine learning; see also the approach of
\citet{cats2021}.

Today there is a broad range of applications of DFT, from fundamental
(toy) situations to very applied, relevant problems, such as the
calculation of solvation free energies of complex molecules, where DFT
performs orders of magnitude faster than simulations as shown by
\citet{jeanmairet2013jcp,jeanmairet2013jpcl,sergiievskyi2014} on the
basis of their classical molecular density functional model for
water. The behaviour of patchy colloids has been addressed using DFT
\cite{delasheras2011patchy}, as well as complex capillary phase
behaviour in model liquid crystals \cite{delasheras2005capillary}.
DFT is applied to complex problems such as nucleation of crystals and
polymorphic behaviour \cite{lutsko2018} and the hard sphere
crystal-fluid interface \cite{haertel2012}.  FMT was formulated for
lattice models by \citet{lafuente2004}; see \citet{oettel2016} for an
insightful application.

\subsection{Sketch of classical DFT}
\label{SECsketchDFT}

For an introduction to classical DFT see \citet{evans1979},
\citet{hansen2013}, and the reviews by \citet{tarazona2008review},
\citet{roth2010review}, and \citet{lutsko2010review}.  We consider
systems of particles in $d$ space dimensions, where all forces are
time-independent gradient fields. The total force that acts on
particle $i$ is
\begin{align}
  \fv_i(\rv^N) &= -\nabla_i u(\rv^N) - \nabla_i V_{\rm ext}(\rv_i).
\end{align}
The total potential energy is $u(\rv^N)+\sum_i V_{\rm ext}(\rv_i)$ and
the Hamiltonian has no explicit time dependence. We are interested in
equilibrium states and (typically) work in the grand ensemble at
chemical potential $\mu$, absolute temperature $T$, and system volume
$V$. The grand potential is expressed as
\begin{align}
  \Omega([\rho], \mu,V,T) &=
  F([\rho],V,T) + \int d\rv\rho(\rv)(V_{\rm ext}(\rv)-\mu),
  \label{EQdftSplittingIntrinsicAndExternal}
\end{align}
where the square brackets indicate a functional dependence, $F[\rho]$
is the intrinsic Helmholtz free energy density functional, which
crucially is independent of $V_{\rm ext}(\rv)$. The space integral in
the external contribution runs over~$V$. The dependence on $V$ is
often disregarded, as it can be subsumed into an appropriate form of
$V_{\rm ext}(\rv)$ that models system walls.

The minimization principle states that $\Omega$ is minimized at fixed
values of $\mu,V,T$ by the true equilibrium density distribution
$\rho_0(\rv)$. Hence,
\begin{align}
  \frac{\delta \Omega[\rho]}{\delta\rho(\rv)}\Big|_{\rho(\rv')=\rho_0(\rv')}
  &= 0 \qquad \rm (min).
  \label{EQdftMinimizationOfOmega}
\end{align}
The value of the functional at the minimum is the (equilibrium) value
of the grand potential $\Omega_0$ itself,
\begin{align}
  \Omega_0(\mu,V,T) &= \Omega([\rho_0],\mu,V,T).
\end{align}
The intrinsic free energy functional $F[\rho]$ can be split into ideal
and excess (over ideal gas) contributions, according to
\begin{align}
  F[\rho] &= F_{\rm id}[\rho] + F_{\rm exc}[\rho],
  \label{EQdftSplittingIdealAndExcess}
\end{align}
where the dependence on the thermodynamic parameters~$T,V$ has been
suppressed in the notation; no dependene on $\mu$ occurs, as this is
accounted for soley by the second term in
Eq.~\eqref{EQdftSplittingIntrinsicAndExternal}.  The ideal gas free
energy functional is given by
\begin{align}
  F_{\rm id}[\rho] &= k_BT \int d\rv 
  \rho(\rv)(\ln(\rho(\rv)\Lambda^d))-1);
\end{align}
recall that $\Lambda=\sqrt{2\pi\beta\hbar^2/m}$ is the thermal de
Broglie wavelength, with $\beta=1/(k_BT)$. Changing the value of
$\Lambda$ only adds a constant to $F_{\rm id}[\rho]$, which has no
effect on the minimization \eqref{EQdftMinimizationOfOmega}.  The
excess free energy functional~$F_{\rm exc}[\rho]$ is due to the
internal interaction potential $u(\rv^N)$, and typically
approximations are required to proceed towards application to actual
physical problems (freezing, adsorption, etc.).

Inserting the intrinsic-external splitting
\eqref{EQdftSplittingIntrinsicAndExternal} and the ideal-excess
decomposition \eqref{EQdftSplittingIdealAndExcess} of the intrinsic
free energy into the minimization condition
\eqref{EQdftMinimizationOfOmega} yields
\begin{align}
  0&=\frac{\delta\Omega[\rho]}{\delta\rho(\rv)}\label{EQEulerLagrangeZero}\\
  &=\frac{\delta}{\delta\rho(\rv)}(F_{\rm id}[\rho]+F_{\rm exc}[\rho])
  \notag\\&\quad
  +\frac{\delta}{\delta\rho(\rv)}\int d\rv'\rho(\rv')[V_{\rm ext}(\rv')-\mu]\\
  &= k_BT\ln(\rho(\rv)\Lambda^d)
  +\frac{\delta F_{\rm exc}[\rho]}{\delta\rho(\rv)}\notag\\
  &\quad +\int d\rv'\frac{\delta\rho(\rv')}{\delta\rho(\rv)}[V_{\rm ext}(\rv)-\mu]
  \\
  &= k_BT\ln[\rho(\rv)\Lambda^d]
  +\frac{\delta F_{\rm exc}[\rho]}{\delta\rho(\rv)}+V_{\rm ext}(\rv)-\mu,
  \label{EQEulerLagrangeFirstForm}
\end{align}
where we have used in the last step that
$\delta\rho(\rv')/\delta\rho(\rv)=\delta(\rv-\rv')$. Solving for the
first term on the right-hand side of
Eq.~\eqref{EQEulerLagrangeFirstForm} and exponentiating gives
\begin{align}
  \rho(\rv) &= \Lambda^{-d}\exp\Big(
  -\frac{\delta\beta F_{\rm exc}[\rho]}{\delta\rho(\rv)}
  -\beta V_{\rm ext}(\rv)
  +\beta\mu
  \Big),
  \label{EQEulerLagrange}
\end{align}
which forms a self-consistency equation for the determination of the
equilibrium density profile $\rho_0(\rv)$. (Recall that the
minimization equation \eqref{EQdftMinimizationOfOmega} holds at
$\rho(\rv)=\rho_0(\rv)$). In the case of the ideal gas $F_{\rm
  exc}[\rho]=0$ and the Euler-Lagrange equation
\eqref{EQEulerLagrange} reduces to the generalized barometric law
$\rho(\rv)=\Lambda^{-d}\exp(-\beta V_{\rm ext}(\rv)+\beta\mu)$.

We can alternatively rearrange the Euler-Lagrange equation
\eqref{EQEulerLagrangeFirstForm} in the following form:
\begin{align}
  V_{\rm ext}(\rv) &= \mu-k_BT\ln[\rho(\rv)\Lambda^d]
  -\frac{\delta F_{\rm exc}[\rho]}{\delta\rho(\rv)},
  \label{EQdftVextFromDensityProfile}
\end{align}
which makes the functional map $\rho(\rv)\to V_{\rm ext}(\rv)$
explicit: the right-hand side of
Eq.~\eqref{EQdftVextFromDensityProfile} is independent of $V_{\rm
  ext}(\rv)$, as it depends solely on $\rho(\rv)$ (and of course on
the form of the functional $F_{\rm exc}[\rho])$. Hence, knowing the
density profile is enough, in principle, to evaluate the right-hand
side of Eq.~\eqref{EQdftVextFromDensityProfile} and obtain the
corresponding external potential. One hence obtains, formally, a
corresponding pair of functions $V_{\rm ext}(\rv)$ and $\rho(\rv)$
that minimize the grand potential functional, i.e., satisfy
Eq.~\eqref{EQdftMinimizationOfOmega}. Physically, it is this hence
identified external potential, which then leads in equilibrium to the
prescribed target density profile.

The DFT framework is well-suited for addressing phase behaviour, where
multiple macrostates can coexist. This applies to general situations
with nonvanishing $V_{\rm ext}(\rv)$, such as in capillaries. At
coexistence, we have multiple stable phases, labeled by an index
$\alpha$, with corresponding density profiles $\rho_\alpha(\rv)$. Then
the map
\begin{align}
  \rho_\alpha(\rv) \to V_{\rm ext}(\rv)
\end{align}
is unique, as it should be. The external potential is the same in the
coexisting phases, as is external force field $-\nabla V_{\rm
  ext}(\rv)$. On the other hand, $V_{\rm ext}(\rv)\to\rho(\rv)$ is not
unique, due to the multiplicity of the density profile(s). This is of
course a real effect e.g.\ at phase coexistence. Typically, for
discontinuous (first order) phase transitions, the location of the
interface between the two coexisting phases constitutes further
freedom in the construction of a valid density profile.

\subsection{Statistical mechanics and variations}
\label{SECstatisticalMechanicsAndFunctionalDerivatives}

We work in the grand ensemble (or ``grand canonical'' ensemble), where
the particle number $N$ fluctuates and its mean is controlled by the
chemical potential $\mu$, which renders $\mu,V,T$ the
macrovariables. The grand partition sum is defined as
\begin{align}
  \Xi(\mu,V,T) &=   \Tr
       {\rm e}^{-\beta (H-\mu N)},
  \label{EQpartitionSumGrandEnsembleLive}
\end{align}
where the Hamiltonian is for $N=\rm const$ particles, and the
classical ``trace'' is defined as the sum over all particles and
integral over each, $N$-specific, phase space:
\begin{align}
  {\rm Tr} &= \sum_{N=0}^\infty 
  \frac{1}{h^{3N}N!}
  \int d\rv^N d\pv^N.
  \label{EQtraceGrandEnsembleBasicLive}
\end{align}

The corresponding thermodynamic potential is the grand potential,
which is given by
\begin{align}
  \Omega(\mu,V,T) = -k_B T \ln \Xi(\mu,V,T) 
  \label{EQgrandPotentialViaPartitionSumLive}.
\end{align}

The microstates now encompass all $\rv^N,\pv^N$ with $N=0,1,2,\ldots$,
distributed according to
\begin{align}
  \Psi_{\mu VT}(\rv^N,\pv^N) &= \frac{{\rm e}^{-\beta(H-\mu N)}}{\Xi},
  \label{EQpsiMuVT}
\end{align}
where $\mu,V,T$ are control parameters, see their occurrence on the
right-hand side of Eq.~\eqref{EQpsiMuVT}. Averages are built according
to
\begin{align}
  \langle \cdot \rangle_{\mu VT} 
  &= {\rm Tr} \cdot \Psi_{\mu VT}.
  \label{EQgrandAverageDefinitionLive}
\end{align}
It is then elementary to see that thermodynamic identities are
generated as parametric derivatives, such as
  \begin{align}
  -\frac{\partial \Omega}{\partial \mu}\Big|_{VT}
  &= \langle N \rangle_{\mu VT}.
  \label{EQmeanNumberOfParticleFromDerivativeLive}
\end{align}

Thus far everything has been general and applicaple to arbitrary forms
of $N$-body Hamiltonians. Consider now the specific form
\begin{align}
  H = \sum_i \frac{\pv_i^2}{2m} + u(\rv^N) 
  + \sum_i V_{\rm ext}(\rv_i),
  \label{EQHamiltonianManyBody}
\end{align}
which has no explicit time dependence, splits into internal and
external one-body contributions, and generates potential forces
only. Clearly, and trivially, $H$ depends on the function $V_{\rm
  ext}(\rv)$ as a (time-independent) one-body field. When $H$ is input
into Eq.~\eqref{EQgrandPotentialViaPartitionSumLive} the dependence on
$V_{\rm ext}(\cdot)$ persists and renders $\Omega$ a {\it functional}
of $V_{\rm ext}(\cdot)$. We spell out the dependence explicitly:
\begin{align}
  &  \Omega(\mu,V,T,[V_{\rm ext}]) =\notag\\&
  -k_BT \ln {\rm Tr}
  \exp\Big[-\beta\Big(
  \sum_i \frac{\pv_i^2}{2m} + u(\rv^N) \notag\\
  &\qquad\qquad\qquad\qquad\qquad
  + \sum_i V_{\rm ext}(\rv_i) -\mu N
  \Big)\Big],
  \label{EQOmegaFunctionalOfVext}
\end{align}
where the (only) dependence on the external potential is made explicit
in the notation.  Hence, any input field $V_{\rm ext}(\cdot)$ is
converted to a number (the value of $\Omega$ with units of energy) by
in principle carrying out the high-dimensional integrals that
constitute the classical trace. In particular the space integrals are
coupled via $u(\rv^N)$ and there is no hope in general of finding an
exact result. Nevertheless Eq.~\eqref{EQOmegaFunctionalOfVext} is
important as a meaningful starting point for an exact microscopic
formal description as the basis of Statistical Mechanics in
equilibrium. Recognizing the apparently trivial functional dependence
on $V_{\rm ext}(\rv)$, and the consquences that this has, is an
important modern achievement, see the wealth of research carried out
on the basis of \citet{evans1979}.

To study and understand the functional relationship better, it is
useful to consider functional derivatives of the grand potential with
respect to $V_{\rm ext}(\rv)$. All the usual reasons for studying
derivatives, as a means to study an object itself, apply here.  The
parameters $\mu,V,T$ are kept constant upon building the functional
derivative and hence
 \begin{align}
  \frac{\delta \Omega}{\delta V_{\rm ext}(\rv)}
  \Big|_{\mu VT} &=
  -\frac{k_BT}{\Xi} \Tr
  \frac{\delta}{\delta V_{\rm ext}(\rv)}
  {\rm e}^{-\beta(H-\mu N)}
  \label{EQdeltaOmega1}\\
  &= \Xi^{-1} \Tr
  {\rm e}^{-\beta(H-\mu N)}
  \frac{\delta H}{\delta V_{\rm ext}(\rv)}
  \label{EQdeltaOmega2}\\
  &= \Big\langle
  \frac{\delta H}{\delta V_{\rm ext}(\rv)} 
  \Big\rangle_{\mu VT}.
  \label{EQdeltaOmega3}
\end{align}
In Eqs.~\eqref{EQdeltaOmega1}-\eqref{EQdeltaOmega3} $\rv$ is a generic
position coordinate (which is in general different from $\rv_i$). The
functional derivative commutes with the classical trace operation
\eqref{EQtraceGrandEnsembleBasicLive}, and hence operates only on the
Boltzmann factor in Eq.~\eqref{EQdeltaOmega1}. The functional chain
rule then reproduces the exponential and generates the derivative of
the Hamiltonian, which is
\begin{align}
  \frac{\delta H}{\delta V_{\rm ext}(\rv)} &=
  \frac{\delta}{\delta V_{\rm ext}(\rv)} \sum_i V_{\rm ext}(\rv_i)
  \\ &
  = \sum_i \frac{\delta V_{\rm ext}(\rv_i)}{\delta V_{\rm ext}(\rv)}
  \\ &
  = \sum_i \delta(\rv-\rv_i) \equiv \hat\rho(\rv).
\end{align}
Insertion into Eq.~\eqref{EQdeltaOmega3} yields
\begin{align}
  \frac{\delta \Omega}{\delta V_{\rm ext}(\rv)}
  \Big|_{\mu VT}
  = \rho(\rv),
  \label{EQrhoFromOmegaDerivative}
\end{align}
where $\rho(\rv) = \langle \hat\rho(\rv) \rangle_{\mu VT}$ is the
(equilibrium) one-body density distribution in the grand
ensemble. Equation \eqref{EQrhoFromOmegaDerivative} is a powerful
generalization of the much more elementary
Eq.~\eqref{EQmeanNumberOfParticleFromDerivativeLive}, which relates
only the mean particle number to the negative partial derivative of
the grand potential with respect to the chemical potential.

As an aside, sometimes one chooses to define a ``local'' chemical
potential via
\begin{align}
  \mu_{\rm loc}(\rv) &= \mu - V_{\rm ext}(\rv),
\end{align}
where $\mu=\rm const$ is the ``true'' chemical potential.\footnote{The
  concept of a species-dependent local chemical potential has been
  exploited in work on sedimentation in binary colloidal mixtures
  \cite{delasheras2012sciRep,delasheras2013phaseStacking,
    delasheras2014sedimentationLiquid2014}.} This allows to rewrite
Eq.~\eqref{EQrhoFromOmegaDerivative} in a form that is even more
inline with Eq.~\eqref{EQmeanNumberOfParticleFromDerivativeLive},
namely
\begin{align}
  -\frac{\delta \Omega}{\delta \mu_{\rm loc}(\rv)}
  \Big|_{VT}
  = \rho(\rv).
\end{align}
Several remarks are in order.
\begin{itemize}
\item[(i)] The relationship \eqref{EQrhoFromOmegaDerivative} applies
  to the elementary concept of the (grand ensemble) grand
  potential. No further functional needs to be established, apart from
  recognizing that the (standard) partition sum is already
  functionally dependent on $V_{\rm ext}(\rv)$.
\item[(ii)] The mean density $\rho(\rv)$ is microscopically sharp,
  i.e., it can resolve inhomogeneities on (small) length scales, as
  determined by the interaction forces. In practice, this involves
  e.g.\ packing effects on the lengthscale of the particle size.  
\item[(iii)] Although functional calculus has certainly proved to be
  powerful, at this stage it is not obvious at all how deep the result
  \eqref{EQrhoFromOmegaDerivative} is, and whether it holds by mere
  accident.
\end{itemize}

Having had a useful outcome of applying functional calculus to the
partition sum, it is natural to consider its second functional
derivative. Using Eq.~\eqref{EQrhoFromOmegaDerivative} this can
immediately be rewritten as
\begin{align}
  \frac{\delta^2 \Omega}
       {\delta V_{\rm ext}(\rv') \delta V_{\rm ext}(\rv)}
       \Big|_{\mu VT}
       &= \frac{\delta \rho(\rv)}{\delta V_{\rm ext}(\rv')}
       \Big|_{\mu VT},
       \label{EQfunctionalDerivativesDensityResponse}
\end{align}
which is a surprising result, as it relates a very abstract object
(left-hand side) to the physical response of the density distribution
at space point $\rv$ upon changing the external potential at point
$\rv'$. Hence, the right-hand side of
Eq.~\eqref{EQfunctionalDerivativesDensityResponse} constitutes a {\rm
  density-response} function. We return to this point for an in-depth
study in Sec.~\ref{SECornsteinZernikeRelation} of the Ornstein-Zernike
relation.

Before doing so we further investigate
Eq.~\eqref{EQfunctionalDerivativesDensityResponse}. At fixed
thermodynamic parameters, consider
\begin{align}
  \frac{\delta\rho(\rv)}{\delta V_{\rm ext}(\rv')} &=
  \frac{\delta}{\delta V_{\rm ext}(\rv')}
  \frac{1}{\Xi} \Tr \e^{-\beta(H-\mu N)} \hat\rho(\rv)\\
  &= \Big(
  -\frac{1}{\Xi^2}
  \frac{\delta}{\delta V_{\rm ext}(\rv')}\Xi
  \Big)\Tr \e^{-\beta(H-\mu N)} \hat\rho(\rv)\notag\\
  & \quad + \frac{1}{\Xi} \Tr
  \frac{\delta}{\delta V_{\rm ext}(\rv')}
  \e^{-\beta(H-\mu N)}\hat\rho(\rv)
    \label{EQdensityResponseIntermediate}\\
  &\equiv \text{(i) + (ii)}.
\end{align}
We consider the two contributions separately.
\begin{itemize}
\item[(i)] We rearrange the first term in
  Eq.~\eqref{EQdensityResponseIntermediate} as
\begin{align}
  &  \Big(
  -\frac{1}{\Xi} \frac{\delta}{\delta V_{\rm ext}(\rv')} \Xi
  \Big)
  \Tr \frac{1}{\Xi} \e^{-\beta(H-\mu N)}\hat\rho(\rv)\notag\\
  &=\Big(
  -\frac{1}{\Xi}
  \Tr \e^{-\beta(H-\mu N)}   (-\beta)
  \frac{\delta H}{\delta V_{\rm ext}(\rv')}
  \Big)\rho(\rv)\\
  &= \beta\Big(
  \Tr \frac{1}{\Xi} \e^{-\beta(H-\mu N)} \hat\rho(\rv')
  \Big)\rho(\rv)\\
  &= \beta\rho(\rv')\rho(\rv),
  \label{EQdensityResponseIntermediate1}
\end{align}
where we have used the functional derivative of the Hamiltonian
$\delta H/\delta V_{\rm ext}(\rv')=\sum_i\delta(\rv'-\rv_i)\equiv
\hat\rho(\rv')$.

\item[(ii)] The second term in
  Eq.~\eqref{EQdensityResponseIntermediate} can be transformed as
\begin{align}
  & \frac{1}{\Xi} \Tr \e^{-\beta(H-\mu N)}
  (-\beta) \frac{\delta H}{\delta V_{\rm ext}(\rv')} \hat\rho(\rv)\notag\\
  &= \frac{-\beta}{\Xi} \Tr \e^{-\beta(H-\mu N)}
  \hat\rho(\rv')\hat\rho(\rv)\\
  &= -\beta\langle \hat\rho(\rv) \hat\rho(\rv') \rangle.
  \label{EQdensityResponseIntermediate2}
\end{align}
\end{itemize}

Adding Eqs.~\eqref{EQdensityResponseIntermediate1} and
\eqref{EQdensityResponseIntermediate2} together yields
\begin{align}
  -\frac{\delta \rho(\rv)}{\delta \beta V_{\rm ext}(\rv')}
  &= \langle \hat\rho(\rv)\hat\rho(\rv') \rangle
  -\rho(\rv)\rho(\rv'),
  \label{EQdensityVariance}
\end{align}
which relates a density response function (left-hand side) with a
density-density correlation function, i.e., the covariance of
$\hat\rho(\rv)$ and $\hat\rho(\rv')$ (right-hand side). We return to
static two-body correlation functions later, when we summarize their
standard definition (Sec~\ref{SECstaticTwoBodyCorrelationFunctions})
and derive the static Ornstein-Zernike relation
(Sec.~\ref{SECornsteinZernikeRelation}). Dynamic correlations
functions are described in Sec.~\ref{SECpft}. We next turn to proving
the existence of the free energy density functional.

\subsection{Levy's constrained search}
\label{SEClevysConstrainedSearch}

The standard Mermin-Evans derivation of classical DFT \cite{evans1979}
was described by \citet{hansen2013}, see {\it Two Theorems in Density
  Functional Theory} in Appendix B of that work.  The first step
consists of constructing a {\it many-body} variational theory on the
level of many-body distribution functions, using the (Mermin) grand
potential functional
\begin{align}
  \Omega_M
        [\Psi]
 &= {\rm Tr}\, \Psi (H-\mu N + k_BT\ln\Psi)
  \label{EQOmegaMdefinition}
\end{align}
and then, via reductio ad absurdum, obtaining the functional
dependence on the density profile.

For the quantum case, \citet{levy1979} developed and used his method
as an alternative, and arguably more explicit, derivation of the
Hohenberg-Kohn theorem.  \citet{dwandaru2011levy} applied the method
to classical DFT and they argue that it has similar advantages over
the conventional Mermin-Evans proof.

The starting point of Levy's search is to consider a function space
$\{\Psi(\rv^N,\pv^N)\}$ of normalized many-body distribution functions
[$\Tr \Psi(\rv^N,\pv^N)=1$] where the grand canonical trace
\eqref{EQtraceGrandEnsembleBasicLive} is defined as above. Each
many-body distribution implies a corresponding density profile.
Hence, we have a functional map
\begin{align}
  \Psi \to \rho(\rv)=\Tr \Psi\hat\rho,
  \label{EQpsiToRho}
\end{align}
which allows to build subspaces of distribution functions $\Psi$ that
generate the same (given) $\rho(\rv)$. Hence, within one subspace
$\{\Psi_1,\Psi_2,\ldots\}$, all
$\Psi_1\to\rho(\rv),\Psi_2\to\rho(\rv)$, etc.  In a different (primed)
subspace $\{\Psi_1',\Psi_2',\ldots\}$ all
$\Psi_1'\to\rho'(\rv),\Psi_2'\to\rho'(\rv)$, etc., with a unique
$\rho'(\rv)$, but $\rho'(\rv)\neq\rho(\rv)$.

The Mermin-Evans form of the intrinsic many-body Helmholtz free energy
functional is given by
\begin{align}
  F[\Psi_0]= \Tr \Psi_0\Big(
  \sum_i\frac{\pv_i^2}{2m} + u(\rv^N) + k_BT\ln\Psi_0
  \Big),
  \label{EQintrinsicFMerminEvansManyBodyAgain}
\end{align}
which resembles the Mermin functional \eqref{EQOmegaMdefinition},
where the external energy and chemical potential contributions have
been split off. The first two terms in the integrand of
Eq.~\eqref{EQintrinsicFMerminEvansManyBodyAgain} represent the
internal (kinetic and potential) energy $U_{\rm int}$, and the third
term involving the logarithm is entropy $S$ multiplied by negative
temperature. Hence, overall the structure is indeed that of an
(intrinsic) free energy, $U_{\rm int}-TS$. In
Eq.~\eqref{EQintrinsicFMerminEvansManyBodyAgain}, $\Psi_0$ is an
equilibrium many-body probability distribution function, associated
with a (Mermin) external potential $V_M(\rv)$.  In the proof by
contradiction one shows that any density distribution $\rho_0(\rv)$
determines uniquely a corresponding Mermin potential $V_M(\rv)$, which
renders $\Psi_0$ known. This implies functional dependence
$\Psi_0[\rho]$, which leads to the free energy
\eqref{EQintrinsicFMerminEvansManyBodyAgain} also being functionally
dependent on $\rho(\rv)$.

Here we argue differently, and rather operate on the function space of
general many-body phase space distribution functions $\Psi$.  The Levy
definition of the intrinsic Helmholtz free energy functional
\cite{dwandaru2011levy} is
\begin{align}
  F_L[\rho] &= \min_{\Psi\to\rho}
  \Tr \Psi\Big(
  \sum_i \frac{\pv_i^2}{2m} + u(\rv^N)
  +k_BT\ln\Psi
  \Big),
  \label{EQFLevy}
\end{align}
where the minimization is performed in the subspace
$\{\Psi|\Psi\to\rho(\rv)\}$, i.e., it is a search for the minimum
under the constraint of a given one-body density, as expressed in
Eq.~\eqref{EQpsiToRho}.  It is this constraint that makes the value of
the integral \eqref{EQFLevy} functionally dependent on
$\rho(\rv)$. This works for any normalized, non-negative trial form of
$\Psi$. [The integral in Eq.~\eqref{EQpsiToRho} can be carried out
  regardless of whether $\Psi$ is a valid equilibrium distribution.]

The Levy version of the {\it grand potential functional} is defined as
\begin{align}
  \Omega_L[\rho] &= F_L[\rho] + \int d\rv\rho(\rv)(V_{\rm ext}(\rv)-\mu),
  \label{EQomegaL}
\end{align}
where $F_L[\rho]$ is given via Eq.~\eqref{EQFLevy}. The functional
$\Omega_L[\rho]$ forms the basis of DFT as follows.

{\it Theorem.}---The Levy form \eqref{EQomegaL} of the grand potential
functional has the properties
\begin{align}
  \Omega_L[\rho_0] = \Omega_0,\\
  \Omega_L[\rho] \geq \Omega_0,
\end{align}
with the equilibrium density profile $\rho_0(\rv)$ and any trial
density profile $\rho(\rv)$.

{\it Proof.}---The idea for the proof is based on Levy's argument of a
double minimization. The first step consists of the constrained
(search) minimization. Then the constraint is relaxed and the overall
minimum is identified. We show this explicitly in the following.

For completeness we spell out $\rho_0(\rv)=\Tr \Psi_0\hat\rho$, where
$\Psi_0=\Xi^{-1}\exp( -\beta(\sum_i\pv_i^2/(2m) + u(\rv^N) + \sum_i
V_{\rm ext}(\rv_i)-\mu N ))$, the grand potential is
$\Omega_0=-k_BT\ln\Xi$, and the grand partition sum is
$\Xi=\Tr\exp(-\beta(H-\mu N))$.  It is a standard exercise
\cite{hansen2013} to show via the Gibbs-Bogoliubov inequality that the
Mermin functional \eqref{EQOmegaMdefinition} satisfies
$\Omega_M[\Psi_0]=\Omega_0\leq \Omega_M[\Psi]$. Rephrasing this, we
can obtain from $\Omega_M$, as defined in
Eq.~\eqref{EQOmegaMdefinition}, the value of the grand potential via
minimization in the space of many-body distribution functions,
\begin{align}
  \Omega_0 &= \min_\Psi
  \Tr \Psi \Big(
  \sum_i\frac{\pv_i^2}{2m} + u(\rv^N)
  +k_BT\ln\Psi\notag\\&\qquad\qquad\qquad
  +\sum_i V_{\rm ext}(\rv_i)
  -\mu N
  \Big).
  \label{EQOmega0asMinimum}
\end{align}
We next decompose the overall minimization into two steps,
\begin{align}
  \min_\Psi (\cdot) \, &= \min_\rho \min_{\Psi\to\rho} (\cdot),
\end{align}
where the inner (right) minimization on the right-hand side is a
search under the constraint of prescribed $\rho(\rv)$ and the outer
(left) minimization then finds the minimum upon varying~$\rho(\rv)$.

Applying this general concept to Eq.~\eqref{EQOmega0asMinimum} yields
\begin{align}
  \Omega_0 &= \min_\rho \min_{\Psi\to\rho}
  \Tr \Psi \Big(
  \sum_i\frac{\pv_i^2}{2m} + u(\rv^N)
  +k_BT\ln\Psi
  \notag\\&\qquad\qquad\qquad\qquad
  +\sum_i V_{\rm ext}(\rv_i)
  -\mu N
  \Big).
\end{align}
Clearly, inside of the inner minimization $\rho(\rv)$ is fixed and hence
\begin{align}
  \Omega_0 &= \min_\rho \min_{\Psi\to\rho}
  \Tr \Psi \Big(
  \sum_i\frac{\pv_i^2}{2m} + u(\rv^N)
  +k_BT\ln\Psi\notag
  \\&\qquad\qquad\qquad\quad
  +\int d\rv V_{\rm ext}(\rv)\sum_i \delta(\rv-\rv_i)
  -\mu N
  \Big)\\
  &=\min_\rho\Big[
    \min_{\Psi\to\rho} \Tr\Psi
    \Big(
    \sum_i\frac{\pv_i^2}{2m} + u(\rv^N)
    +k_BT\ln\Psi\Big)\notag
    \\&\qquad\qquad
    +\int d\rv(V_{\rm ext}(\rv)-\mu)
    \min_{\Psi\to\rho}\Tr\Psi\sum_i\delta(\rv-\rv_i)
    \Big]\\
  &=\min_\rho\Big[
    F_L[\rho]+\int d\rv(V_{\rm ext}(\rv)-\mu)\rho(\rv)
    \Big],
\end{align}
where the final form is written using Eq.~\eqref{EQFLevy} for
$F_L[\rho]$.  This proves the Theorem and identifies the Levy form
\eqref{EQFLevy} with the aforementioned intrinsic free energy
functional $F[\rho]$ [Eq.~\eqref{EQdftSplittingIdealAndExcess}].

Levy's constrained search method is flexible. \citet{dwandaru2011levy}
used it to formulate classical DFT in the canonical ensemble. Here the
constraint of fixed particle number is implemented straightforwardly
by setting up the previously mentioned reasoning in the canonical
ensemble. The canonical intrinsic free energy functional formally
resembles the grand ensemble form \eqref{EQFLevy}, but with the trace
and many-body probability distribution function expressed canonically,
see \citet{dwandaru2011levy} for details.
\citet{schmidt2011internalEnergyFunctional} has used Levy's method for
the construction of an internal-energy functional, which depends both
on the density profile and on a microscopically resolved entropy
density (both act as constraints). An extended set of closely related
fluctuation profiles in inhomogeneous fluids have been systematically
studied by \citet{eckert2020auxiliaryFields}.  The fluctuation
profiles include the local compressibility.  Based on early work
\cite{stewart2012pre,stewart2014jcp}, this one-body function was shown
to be a highly useful indicator for a range of important phenomena
from solvent-mediated interactions \cite{chacko2017}, solvophobicity
and hydrophobicity \cite{evans2015jpcm}, drying and wetting
\cite{evans2016prl,evans2017} to the physical mechanism behind
hydrophobicity \cite{coe2022}.

In practice, approximations for canonical functionals are
scarce. There are alternative methods to obtain canonical information,
see the method by \citet{gonzalez1997} and the framework of
\citet{white2000prl} and \citet{white2002jpcm}. Starting from grand
ensemble data requires one, in principle, to carry out an inverse
Laplace transform, which is a numerically difficult task.  The direct
decomposition method by \citet{delasheras2014fullCanonical}
circumvents this problem by rather solving a linear set of equations
that is numerically tractable. Obtaining canonical information can be
crucial, in particular for small systems, and when comparing to
results from experiment or simulation. For ensemble differences to
vanish, typically the thermodynamic limit is required. The
decomposition method allows to obtain canonical information from grand
canonical results, as are characteristic for numerical DFT
applications.  \citet{delasheras2016particleConservation} used this
approach to formulate particle-conserving adiabatic dynamics in order
to avoid erroneous particle number fluctuations of dynamical
density-functional theory; see also \citet{schindler2019} and
\citet{wittmann2021}.

\subsection{Static two-body correlation functions}
\label{SECstaticTwoBodyCorrelationFunctions}

We recall the fundamental property of the grand potential $\Omega$, as
expressed via the grand partition sum and viewed as a functional of
the external potential, to generate correlation functions from
functional derivatives [Eqs.~\eqref{EQrhoFromOmegaDerivative},
  \eqref{EQfunctionalDerivativesDensityResponse} and
  \eqref{EQdensityVariance} in
  Sec.~\ref{SECstatisticalMechanicsAndFunctionalDerivatives}]. Explicitly,
\begin{align}
  \frac{\delta\Omega[V_{\rm ext}]}{\delta V_{\rm ext}(\rv)} &=
  \rho(\rv),\\
  \frac{\delta^2\Omega[V_{\rm ext}]}
       {\delta V_{\rm ext}(\rv')\delta V_{\rm ext}(\rv)}
       &= \frac{\delta\rho(\rv)}{\delta V_{\rm ext}(\rv')}
       \label{EQsecondDerivativeOmega}
       \\
       &=-\beta\big(
       \langle\hat\rho(\rv)\hat\rho(\rv')\rangle
       -\rho(\rv)\rho(\rv')
       \big).
       \label{EQHcorrelatorDefinition}
\end{align}
From the interchangeability of the order of the second derivatives in
Eq.~\eqref{EQsecondDerivativeOmega}, the symmetry
$\delta\rho(\rv)/\delta V_{\rm ext}(\rv')= \delta\rho(\rv')/\delta
V_{\rm ext}(\rv)$ follows.

We summarize the definitions of several closely related two-body
functions \cite{hansen2013}. The {\it density-density correlation
  function} $H_2(\rv,\rv')$ is defined as
\begin{align}
  H_2(\rv,\rv')
  &= \langle\hat\rho(\rv)\hat\rho(\rv')\rangle
    -\rho(\rv)\rho(\rv'),
    \label{EQdensityAutoCorrelator}
\end{align}
where the symmetry $H_2(\rv,\rv')=H_2(\rv',\rv)$ holds. One can
alternatively express Eq.~\eqref{EQdensityAutoCorrelator} as the
auto-correlator of density fluctuations around the mean density
profile, 
\begin{align}
  H_2(\rv,\rv') &=
  \big\langle
  (\hat\rho(\rv)-\rho(\rv))(\hat\rho(\rv')-\rho(\rv'))
  \big\rangle
  \label{EQHAsDensityDensityCorrelator}
\end{align}
To show the equivalence of Eqs.~\eqref{EQdensityAutoCorrelator} and
\eqref{EQHAsDensityDensityCorrelator}, we omit the position arguments
and indicate dependence on $\rv'$ by a prime.  Mulitplying out
Eq.~\eqref{EQHAsDensityDensityCorrelator} we obtain
$\langle\hat\rho\hat\rho'\rangle - \langle\hat\rho\rho'\rangle -
\langle\rho\hat\rho'\rangle + \langle\rho\rho'\rangle =
\langle\hat\rho\hat\rho'\rangle - \langle\hat\rho\rangle\rho' -
\rho\langle\hat\rho'\rangle + \rho\rho' =
\langle\hat\rho\hat\rho'\rangle - \rho\rho' - \rho\rho' + \rho\rho' =
\langle\hat\rho\hat\rho'\rangle - \rho\rho'$, which is
Eq.~\eqref{EQdensityAutoCorrelator}.

The {\it total correlation function} $h(\rv,\rv')$ is defined via
\begin{align}
  H_2(\rv,\rv') &= \rho(\rv)\rho(\rv')h(\rv,\rv')
  +\rho(\rv)\delta(\rv-\rv'),
\end{align}
where rearranging gives
\begin{align}
  h(\rv,\rv') &= \frac{H_2(\rv,\rv')}{\rho(\rv)\rho(\rv')}
  -\frac{\delta(\rv-\rv')}{\rho(\rv)}.
  \label{EQtotalCorrelationFunctionDefinition}
\end{align}
The correlation function $h(\rv,\rv')$ carries no units (the the delta
function carries units of inverse volume).

The {\it pair correlation function} $g(\rv,\rv')$ is defined via
\begin{align}
  g(\rv,\rv') &= 1 + h(\rv,\rv') \\
  &= \frac{H_2(\rv,\rv')+\rho(\rv)\rho(\rv')}{\rho(\rv)\rho(\rv')}
  -\frac{\delta(\rv-\rv')}{\rho(\rv)}\\
  &= \frac{\langle\hat\rho(\rv)\hat\rho(\rv')\rangle}
  {\rho(\rv)\rho(\rv')}
  -\frac{\delta(\rv-\rv')}{\rho(\rv)}.
  \label{EQpairCorrelationFunctionFromDensityOperators}
\end{align}

Alternatively and equivalently, one can express the pair correlation
function as
\begin{align}
  g(\rv,\rv') &= \frac{1}{\rho(\rv)\rho(\rv')}
  \Big\langle
  \sum_i{\sum_j}' \delta(\rv-\rv_i)\delta(\rv'-\rv_j)
  \Big\rangle,
  \label{EQpairCorrelationFunctionFromDistinctPart}
\end{align}
where the primed sum indicates that the terms with $i=j$ have been
omitted. The equivalence with the prior definition
\eqref{EQpairCorrelationFunctionFromDensityOperators} can be seen by
considering the product of the two density operators,
\begin{align}
  \hat\rho(\rv)\hat\rho(\rv') &=
  \sum_i \delta(\rv-\rv_i) \sum_j \delta(\rv'-\rv_j)\\
  &= \sum_i{\sum_j}'\delta(\rv-\rv_i)\delta(\rv'-\rv_j)\notag\\
  &\quad  +\delta(\rv-\rv')\sum_i\delta(\rv-\rv_i),
\end{align}
where the last term is $\delta(\rv-\rv')\hat\rho(\rv)$. Averaging
yields the sum of a distinct part (from different particles) and a
self part (considering the same particle twice), according to
\begin{align}
  \langle\hat\rho(\rv)\hat\rho(\rv')\rangle &=
 \Big\langle \sum_i{\sum_j}'\delta(\rv-\rv_i)\delta(\rv'-\rv_j)\Big\rangle
 \notag\\
  &\qquad +\delta(\rv-\rv')\langle\hat\rho(\rv)\rangle,
 \label{EQaveragedRhoRhoOperators}
\end{align}
where of course $\rho(\rv)=\langle\hat\rho(\rv)\rangle$. Input of the
result \eqref{EQaveragedRhoRhoOperators} into
Eq.~\eqref{EQpairCorrelationFunctionFromDensityOperators} yields
Eq.~\eqref{EQpairCorrelationFunctionFromDistinctPart}.

Finally, the {\it density-response} function is defined as
\begin{align}
  \chi(\rv,\rv') &= 
  \frac{\delta\rho(\rv)}{\delta V_{\rm ext}(\rv')},
\end{align}
which is often formulated in the form of a response relationship. Here
the density change $\delta \rho(\rv)$ at position $\rv$ in response to
a change in external potential $\delta V_{\rm ext}(\rv')$ at position
$\rv'$ is expressed as
\begin{align}
  \delta\rho(\rv) &= 
  \int d\rv'\chi(\rv,\rv')\delta V_{\rm ext}(\rv').
\end{align}
A priori the density-response function is very different in character
from the density-density correlation functions above. However, in the
present classical context, we can identify these conceptually very
different objects:
\begin{align}
  \chi(\rv,\rv') &= -\beta H_2(\rv,\rv').
\end{align}
We have so far used explicit correlator expressions and functional
derivatives of the partition sum (in the form of the grand potential)
with respect to the external potential.  We next turn to the density
functional structure.

One defines the {\it one-body direct correlation function} as
\begin{align}
  c_1([\rho],\rv;T,V) &= -\beta
  \frac{\delta F_{\rm exc}[\rho]}{\delta\rho(\rv)},
  \label{EQc1functional}
\end{align}
where $F_{\rm exc}[\rho]$ is the excess (over ideal gas) contribution
of the Helmholtz excess free energy functional; recall the
ideal-excess splitting \eqref{EQdftSplittingIdealAndExcess}.  Strictly
speaking, Eq.~\eqref{EQc1functional} defines a functional, and the
direct correlation function $c_1(\rv)$ is obtained by evaluating this
functional, for given thermodynamic parameters, at the physical
equilibrium density profile $\rho_0(\rv)$, i.e.,
\begin{align}
  c_1(\rv) &= c_1([\rho_0], \rv; T,V).
\end{align}

We can of course build higher than first derivatives of the excess
free energy density functional. Going to the second derivative, i.e.,
one order higher than in Eq.~\eqref{EQc1functional}, gives the
two-body direct correlation ``function'' $c_2(\rv,\rv')$, defined as
\begin{align}
  c_2([\rho],\rv,\rv';T,V) &= 
  -\beta 
  \frac{\delta^2 F_{\rm exc}[\rho]}{\delta\rho(\rv)\delta\rho(\rv')},
  \label{EQc2Definition}
\end{align}
where we have again made the functional dependence on $\rho(\rv)$
explicit in the notation. Evaluating at the equilibrium density
profile, $\rho(\rv)=\rho_0(\rv)$, gives the two-body direct
correlation function, central to liquid integral equation theory
\cite{hansen2013}. Of course, in a bulk fluid
$c_2(\rv,\rv')=c_2(|\rv-\rv'|)$ due to global translational and
rotational symmetry; see
\citet{hermann2021noether,hermann2021noetherPopular} for the
consequences that arise from symmetries according to Nother's theorem.

\subsection{Static Ornstein-Zernike relation}
\label{SECornsteinZernikeRelation}

The Ornstein-Zernike relation connects the two-body direct correlation
function with the density-response function (which can be expressed
and rewritten in various forms, as described above in
Sec.~\ref{SECstaticTwoBodyCorrelationFunctions}).  Here we give a
derivation that identifies the underlying physical concept and
separates this from the more technical points that arise from the use
of the different form of correlators and response functions $H_2,h,g$,
and $\chi$.  This type of derivation was used recently in equilibrium
by \citet{eckert2020auxiliaryFields} in their derivation of
Ornstein-Zernike relations for fluctuation profiles and by
\citet{tschopp2021} in their fundamental measure theory for
inhomogeneous two-body correlation functions. The dynamical
generalization formed the basis of the nonequilibrium Ornstein-Zernike
relation \cite{brader2013nozOne,brader2014nozTwo}.

We address the general, inhomogeneous case in the following. Consider
the Euler-Lagrange equation \eqref{EQdftVextFromDensityProfile} of
DFT, which we make fully explicit as
\begin{align}
 k_BT\ln(\rho_0(\rv)\Lambda^d) &= 
 -\frac{\delta F_{\rm exc}[\rho]}{\delta\rho(\rv)} \Big|_{\rho(\rv)=\rho_0(\rv)}
 -V_{\rm ext}(\rv) +\mu,
 \label{EQEulerLagrangeDFTexplicit}
\end{align}
where $\rho_0(\rv)$ is the equilibrium density profile in the presence
of the external potential $V_{\rm ext}(\rv)$. Hence,
Eq.~\eqref{EQEulerLagrangeDFTexplicit} is valid for any corresponding
pair of fields $\rho_0(\rv)$ and $V_{\rm ext}(\rv)$.  As
Eq.~\eqref{EQEulerLagrangeDFTexplicit} stays true upon changing
$V_{\rm ext}(\rv)$ and correspondingly changing $\rho_0(\rv)$, its
derivative with respect to the external potential is also true.
Introducing a new primed spatial position variable, from
Eq.~\eqref{EQEulerLagrangeDFTexplicit} we hence obtain
\begin{align}
&  \frac{\delta k_BT\ln(\rho_0(\rv)\Lambda^d)}{\delta V_{\rm ext}(\rv')}=
  \notag\\& \quad  
  -\frac{\delta}{\delta V_{\rm ext}(\rv')}
  \frac{\delta F_{\rm exc}}{\delta\rho(\rv)}\Big|_{\rho(\rv)=\rho_0(\rv)}
  -\frac{\delta (V_{\rm ext}(\rv)-\mu)}{\delta V_{\rm ext}(\rv')}.
  \label{EQprotoOZ}
\end{align}
As the variation is performed at a constant thermodynamic statepoint,
the second term on the right-hand side is simply
\begin{align}
  -\frac{\delta(V_{\rm ext}(\rv)-\mu)}{\delta V_{\rm ext}(\rv')}
  & =-\frac{\delta V_{\rm ext}(\rv)}{\delta V_{\rm ext}(\rv')}
  =-\delta(\rv-\rv'),
  \label{EQprotoOZdeltaFunction}
\end{align}
as no dependence on the density profile is involved. In order to
calculate the remaining terms, we need (i) the standard rules of
functional calculus and (ii) to recognize the relationship with the
previously mentioned correlators. The left-hand side of
Eq.~\eqref{EQprotoOZ} can be related to
Eq.~\eqref{EQHcorrelatorDefinition}, i.e.,
$H_2(\rv,\rv')=-k_BT\delta\rho(\rv)/\delta V_{\rm ext}(\rv')$, by
rewriting it as
\begin{align}
  k_BT\frac{\delta\ln(\rho(\rv)\Lambda^d)}{\delta V_{\rm ext}(\rv')} &=
 \frac{k_BT}{\rho(\rv)}
 \frac{\delta\rho(\rv)}{\delta V_{\rm ext}(\rv')}
 = -\frac{H_2(\rv,\rv')}{\rho(\rv)},
 \label{EQprotoOZleftHandSide}
\end{align}
where we have dropped the subscript 0. To obtain the remaining (first)
term on the right-hand side of Eq.~\eqref{EQprotoOZ}, we carry out the
functional derivative, by using the functional chain rule and once
more the definition \eqref{EQHcorrelatorDefinition},
\begin{align}
  -&\frac{\delta}{\delta V_{\rm ext}(\rv')}
  \frac{\delta F_{\rm ext}[\rho]}{\delta \rho(\rv)}\Big|_{\rho=\rho_0}
  \notag\\
  &= -\int d\rv'' \frac{\delta^2 F_{\rm exc}[\rho]}
  {\delta\rho(\rv'')\delta\rho(\rv)}
  \frac{\delta\rho(\rv'')}{\delta V_{\rm ext}(\rv')}\\
  &= -\int d\rv'' \frac{\delta^2 F_{\rm exc}[\rho]}
  {\delta\rho(\rv'')\delta\rho(\rv)}
  \Big(-\frac{1}{k_BT} H_2(\rv'',\rv')\Big)\\
  &= \int d\rv'' \frac{\delta^2 \beta F_{\rm exc}[\rho]}
      {\delta\rho(\rv'')\delta\rho(\rv)}
      H_2(\rv'',\rv')\\
  &= -\int d\rv'' c_2(\rv'',\rv)
      H_2(\rv'',\rv')\\
  &= -\int d\rv'' c_2(\rv,\rv'')
      H_2(\rv'',\rv').
      \label{EQprotoOZpartialRightHandSide}
\end{align}
We can now restore the starting equality \eqref{EQprotoOZ} by equating
Eq.~\eqref{EQprotoOZleftHandSide} with the sum of
Eq.~\eqref{EQprotoOZpartialRightHandSide} and the delta function
\eqref{EQprotoOZdeltaFunction}.  The result is
\begin{align}
  H_2(\rv,\rv') &= \rho(\rv)\int d\rv'' c_2(\rv,\rv'') H_2(\rv'',\rv')
  +\rho(\rv)\delta(\rv-\rv'),
  \label{EQOZwithH}
\end{align}
where we have multiplied by $-\rho(\rv)$. Equation \eqref{EQOZwithH}
can already be viewed as the static Ornstein-Zernike relation. Its
standard form is expressed in terms of the total correlation function
$h(\rv,\rv')$ where according to
Eq.~\eqref{EQtotalCorrelationFunctionDefinition} we have
$H_2(\rv,\rv')=\rho(\rv)\rho(\rv')h(\rv,\rv')+\rho(\rv)\delta(\rv-\rv')$.
We insert this identity into the integral in Eq.~\eqref{EQOZwithH},
\begin{align}
  &\int d\rv'' c_2(\rv,\rv'') H_2(\rv'',\rv')\\
  &=
  \int d\rv'' c_2(\rv,\rv'') \big( 
  \rho(\rv'')\rho(\rv')h(\rv'',\rv')+\rho(\rv'')\delta(\rv''-\rv')\big)\notag\\
  &=\int d\rv'' c_2(\rv,\rv'') 
  \rho(\rv'')\rho(\rv')h(\rv'',\rv')\notag\\&\quad +
  \int d\rv'' c_2(\rv,\rv'')\rho(\rv'')\delta(\rv''-\rv')\\
  &=\rho(\rv') \Big(\int d\rv'' c_2(\rv,\rv'') 
  \rho(\rv'')h(\rv'',\rv')+ c_2(\rv,\rv')\Big).
\end{align}
We use this result in Eq.~\eqref{EQOZwithH}, divide by
$\rho(\rv')\rho(\rv)$, and obtain the standard form of the {\it
  inhomogeneous Ornstein-Zernike relation} as
\begin{align}
  h(\rv,\rv') &= c_2(\rv,\rv') +\int d\rv''
  c_2(\rv,\rv'')\rho(\rv'')h(\rv'',\rv'),
  \label{EQinhomogeneousOZ}
\end{align}
where the total correlation function $h(\rv,\rv')$ is a probabilistic
object, defined via Eqs.~\eqref{EQHAsDensityDensityCorrelator} and
\eqref{EQtotalCorrelationFunctionDefinition}, and $c_2(\rv,\rv')$ is
the total correlation function, defined as the second functional
density derivative \eqref{EQc2Definition} of the excess free energy
functional.

The Ornstein-Zernike relation is a fundamental sum rule, different in
character from hierarchies that relate two-body to three-body (and/or
higher) correlation functions. The Ornstein-Zernike relation is closed
on the two-body level. (It also involves the one-body density
profile.)  Three- and higher-body versions exist and can be
systematically derived.  Alternatively, without the density functional
context, the Ornstein-Zernike relation can be viewed as the definition
of the direct correlation function $c_2(\rv,\rv')$. (This is the
original concept by Ornstein and Zernike.) The combination with a
``closure'' relation, i.e., an approximate additional relation between
$h$ and $c_2$, forms the basis of liquid state integral equation
theory.  In a bulk fluid $\rho(\rv)=\rho_b=\rm const$ and the spatial
dependence is only on $r_{\alpha\beta}=|\rv_\alpha-\rv_\beta|$, where
$\alpha,\beta=1,2,3$ labels space points. Then
\begin{align}
  h(r_{13}) = c_2(r_{13}) + \rho_b \int d\rv_2 c_2(r_{12})h(r_{23}).
\end{align}
One can visualize the integrals via diagrammatic notation, which gives
deep insights into the mathematical structure and forms a useful
calculation device.  There are generalizations to mixtures and to
anisotropic interparticle interactions. While this is conceptually
(rather) straightforward, in actual applications the use can be highly
challenging.  We demonstrate in Sec.~\ref{SECnoz} how the power
functional permits to generalize to time-dependent correlation
functions.

\subsection{Approximate free energy functionals}
\label{SECfexcSimple}

In practical applications of DFT an approximation for the nontrivial
excess part $F_{\rm exc}[\rho]$ of the density functional is required.
Carrying out such work, an example being to investigate the behaviour
of a given fluid in the presence of an external potential, requires
solving the Euler-Lagrange equation \eqref{EQEulerLagrange}, which is
typically performed numerically.  Owing to decades of fundamental
research efforts, a wide range of useful concrete prescriptions for
$F_{\rm exc}[\rho]$ is available. The different functionals vary both
in the underlying concepts to perform the approximation and in the
resulting mathematical complexity. We refer the reader to the
pertinent literature\footnote{Good starting points are
  \citet{hansen2013, evans1992, tarazona2008review, roth2010review,
    lutsko2010review, evans2016specialIssue}.} and here only describe
briefy several basic concepts that are relevant for the construction
of power functional approximations.

The arguably simplest model form for $F_{\rm exc}[\rho]$ is the
local-density approximation (LDA). This scheme requires as input the
bulk fluid equation of state, which determines the excess free energy
density per unit volume, $\psi_{\rm exc}(\rho_b)$ as a function of the
bulk density $\rho_b$. The LDA density functional then sums up local
contributions from all space points via $F_{\rm exc}[\rho]=\int d\rv
\psi_{\rm exc}(\rho(\rv))$, thereby ignoring all spatial correlations
that the internal interactions generate. Nevertheless, for slowly
varying spatial inhomogeneities (as measured on the lengthscale of
interparticle correlations in the system) the LDA can be a very good
approximation. (See the studies of colloidal sedimentation by
\citet{delasheras2012sciRep,geigenfeind2017sampleHeight,eckert2021}).

Taking simple account of some correlation effects is possible via the
following square gradient approximation:
\begin{align}
  F_{\rm exc}[\rho] &= \int d\rv \Big[
  \psi_{\rm exc}(\rho(\rv)) + \frac{m}{2}(\nabla\rho(\rv))^2
  \Big].
  \label{EQFexcSquareGradient}
\end{align}
where $m$ determines the strength of square gradient
contribution. Microscopically, $m$ is related to the second moment of
the bulk direct correlation function, $m = k_BT \int d\rv r^2
c_2(r)/6$ (for $d=3$), which allows to make a connection with the
underlying model fluid.

Building the functional derivative of Eq.~\eqref{EQFexcSquareGradient}
with respect to the density profile gives
\begin{align}
  \frac{\delta F_{\rm exc}[\rho]}{\delta \rho(\rv)} &=
  \frac{\partial \psi_{\rm exc}(\rho)}{\partial \rho(\rv)}
  -m\nabla^2\rho(\rv),
\end{align} 
which can directly input into the Euler-Lagrange equation
\eqref{EQEulerLagrange}.  The resulting theory is along the lines of
van der Waals' historical treatment of the free gas-liquid interface
and it is akin to a Landau theory, when identifying $\rho(\rv)$ as the
local order parameter.  Approximations of the form
\eqref{EQFexcSquareGradient} are sometimes referred to as semilocal.

A better accounting of the microscopic correlations that arise from
interparticle forces and, in particular, from short-ranged repulsion,
requires genuine nonlocal approximations. Most nonlocal functionals
rely on introducing one or several weighted densities that are
obtained by convolution of the density profile with suitable weight
function(s). Arguably the most successful scheme of this form is
Rosenfeld's fundamental-measure theory for hard sphere mixtures (as
well as for certain further model fluids). In short, the weighted
densities are built according to
\begin{align}
  n_\alpha(\rv) &= \int d\rv' \rho(\rv') w_\alpha(\rv-\rv'),
\end{align}
where $\alpha$ is an index that labels the different weight functions.
The \citet{rosenfeld1989} functional has the form
\begin{align}
  F_{\rm exc}[\rho] &=
  k_BT \int d\rv \Phi(\{n_\alpha(\rv)\}),
\end{align}
where the (scaled) excesse free energy density per volume,
$\Phi(\{n_\alpha(\rv)\})$, depends on all weighted densities. It is
given by simple rational expression
\begin{align}
  \Phi(\{n_\alpha(\rv)\}) = -n_0\ln(1-n_3) + \frac{n_1n_2}{1-n_3}
  +\frac{n_2^3}{24\pi(1-n_3)^2},
\end{align}
in the form by \citet{kierlik1990}; we have omitted the position
argument of $n_\alpha(\rv)$ for clarity.  For the one-dimensional hard
core system (``hard rods'') $\Phi_{\rm
  1d}(\{n_\alpha\})=-n_0\ln(1-n_1)$ and the fundamental-measure
functional is identical to Percus' exact functional. In three
dimensions a range of improved FMTs exist: these successfully describe
freezing and crossover to reduced dimensionality; the White-Bear
version incorporates the quasi-exact Carnahan-Starling equation of
state, see \citet{roth2010review}.  We recall
Sec.~\ref{SECoverviewDFT}, which gives an overview of further
developments in constructing free energy density functionals.

\subsection{Dynamical density functional theory}
\label{SECddft}

The original proposal of \citet{evans1979} for the dynamical DFT was
subsequently reconsidered by \citet{marconi1999ddft},
\citet{archer2004ddft}, and \citet{espanol2009}. In the following
their dynamical theory is described on the basis of the adiabatic
construction, which was laid out in Sec.~\ref{SECadiabaticState}.

We recall the Euler-Lagrange equation \eqref{EQEulerLagrangeFirstForm}
of equilibrium DFT:
\begin{align}
  \frac{\delta F[\rho]}{\delta\rho(\rv)} 
  + V_{\rm ext}(\rv) - \mu &= 0,
  \label{EQddftEulerLagrange}
\end{align}
where for compactness of notation, we have left away the fact that
equality holds for $\rho(\rv)=\rho_0(\rv)$, where $\rho_0(\rv)$
indicates the equilibrium density profile.  Building the negative
gradient of Eq.~\eqref{EQddftEulerLagrange} yields
\begin{align}
  -\nabla \frac{\delta F[\rho]}{\delta\rho(\rv)}
  -\nabla V_{\rm ext}(\rv) &= 0,
  \label{EQddftForceBalance}
\end{align}
which has the clear physical interpretation of a force balance
relationship of vanishing sum of intrinsic and external forces.

The aim is to formulate a dynamical one-body theory that drives the
time evolution in nonequilibrium, based on the equilibrium intrinsic
force term in Eq.~\eqref{EQddftForceBalance}. We recall from
Sec.~\ref{SECmanybodyDynamics} the one-body continuity equation
\begin{align}
  \dot\rho(\rv,t) &= -\nabla\cdot\Jv(\rv,t),
  \label{EQddftContinuityEquation}
\end{align}
which links changes in the time-dependent density profile to the
divergence of the microscopic current distribution $\Jv(\rv,t)$. We
have seen that the exact equation of motion
\eqref{EQforceDensityBalanceBD} for the case of overdamped Brownian
motion represents the current as being instantaneously generated by
the sum of all force densities that act in the system,
\begin{align}
  \gamma\Jv(\rv,t) &= 
  -k_BT\nabla\rho(\rv,t) + \Fv_{\rm int}(\rv,t) + \rho \fv_{\rm ext}(\rv,t).
\end{align}
As before $\gamma$ is the friction constant against the static
background solvent.  In equilibrium, we know that the internal force
density and the external force field satisfy, respectively,
\begin{align}
  \Fv_{\rm int}(\rv) &=
 -\Big\langle\sum_i\delta(\rv-\rv_i)\nabla_i u
  \Big\rangle_{\rm eq}
  \label{EQddftFintEquilibrium}
  = -\rho\nabla\frac{\delta F_{\rm exc}[\rho]}{\delta\rho(\rv)},\\
  \fv_{\rm ext}(\rv) &= -\nabla V_{\rm ext}(\rv).
\end{align}
To use Eq.~\eqref{EQddftFintEquilibrium} in a dynamical context we use
the concept of the {\it adiabatic state}, which as described in
Sec.~\ref{SECadiabaticState} consists of considering, at each time
$t$, a true equilibrium (``adiabatic'') system, with its genuine
one-body density profile $\rho_{{\rm ad},t}(\rv)$. As the adiabatic
system is in equilibrium, its density distribution is independent of
time. However, per construction, the nonequilibrium system has at each
time $t$ a corresponding adiabatic state. Both, the nonequilibrium
system and the adiabatic system share the same internal interaction
potential $u(\rv^N)$ and they are related by the identification
\begin{align}
  \rho(\rv,t) &= \rho_{{\rm ad},t}(\rv),
\end{align}
where the dependence on time is real in the nonequilibrium system and
parametric only in the adiabatic system (where $t$ instead ``selects''
the fitting adiabatic state in a sequence of equilibrium systems
indexed by~$t$). In the adiabatic system, via the
Hohenberg-Kohn-Mermin-Evans map, we can identify a unique external
one-body potential $V_{{\rm ad},t}(\rv)$, which stabilizes the given
$\rho_{{\rm ad},t}(\rv)$. Again the dependence on time of the
adiabatic external potential is merely parametric. The adiabatic
system is in equilibrium and hence its external potential $V_{{\rm
    ad},t}(\rv)$ is static. (This point is of mere conceptual
inportance; in practice one treats the adiabatic system using a
corresponding equilibrium ensemble that relieves one of a secondary
time evolution in the adiabatic system.)

In the adiabatic system the external force field needs to balance the
intrinsic force field: $-k_BT\nabla\ln\rho_{{\rm ad},t}(\rv)+\fv_{{\rm
    ad},t}(\rv)-\nabla V_{{\rm ad},t}(\rv)=0$, where $\fv_{{\rm
    ad},t}(\rv)$ is the one-body force field in the adiabatic system
that arises due to internal interactions. Hence, we have a chain of
functional relationships
\begin{align}
  \rho(\rv,t) \to \rho_{{\rm ad},t}(\rv)
  \to V_{{\rm ad},t}(\rv)
  \to \fv_{{\rm ad},t}(\rv).
  \label{EQddftAdiabaticMap}
\end{align}
Dynamical DFT amounts to approximating the real internal one-body
force field by that in the adiabatic system,
\begin{align}
  \fv_{\rm int}(\rv,t) &\approx \fv_{{\rm ad},t}(\rv).
  \label{EQddftApproximationInternalForceField}
\end{align}
As a result of the approximation all forces are known in the
nonequilibrium system [as $\rho(\rv,t)$ is known at time $t$ and
  $\fv_{\rm ad, t}(\rv)$ is a density functional]. As the force
balance is known via the approximation
\eqref{EQddftApproximationInternalForceField}, the dynamical theory is
closed. [The continuity equation \eqref{EQddftContinuityEquation}
  forms the supplemental, secondary relation]. Hence, we have the
instantaneous relationship
\begin{align}
  \gamma\Jv(\rv,t) &= -k_BT \nabla\rho(\rv,t)
  +\rho \fv_{\rm ad}(\rv,t)
  +\rho \fv_{\rm ext}(\rv,t),
\end{align}
where we have dropped the subscript $t$ and $\fv_{\rm ext}(\rv,t)$
needs no longer be restricted to gradient form. Note that in cases
where it is restricted the real external potential, which generates
the instantaneous external force field via $\fv_{\rm
  ext}(\rv,t)=-\nabla V_{\rm ext}(\rv,t)$, will in general be
significantly different than the external potential that acts in the
adiabatic system [$V_{{\rm ad},t}(\rv)$]. Hence, $V_{\rm
  ext}(\rv,t)\neq V_{{\rm ad},t}(\rv)$, possibly strikingly so.  To
see this, first consider first e.g.\ a switching process that changes
$V_{\rm ext}(\rv,t)$ abruptly, but that had not allowed enough time to
pass to generate a noticable effect on $\rho(\rv,t)$, and hence
leaving $V_{{\rm ad},t}(\rv)$ virtually intact. It is important to
appreciate the difference between both external potentials, $V_{\rm
  ext}(\rv,t)$ drives the time evolution in the real system, $V_{{\rm
    ad},t}(\rv)$ rather stops the time evolution in the adiabatic
system.  When the real system is in equilibrium then both potentials
are identical. As a second illustrative example, consider free
expansion of an initially confined density distribution, where at any
time the adiabatic potential needs to stabilize the broadening density
profile, although after the initial time $V_{\rm ext}(\rv,t)\equiv 0$,
see e.g.\ \cite{schmidt2013pft}.

Using the approximation \eqref{EQddftApproximationInternalForceField}
in a practical application requires one to have access to the
adiabatic map \eqref{EQddftAdiabaticMap}, of which the nontrivial part
is the map from the density profile to the external potential in the
adiabatic system,
\begin{align}
  \rho_{{\rm ad},t}(\rv) \to V_{{\rm ad},t}(\rv).
\end{align}
Of course, DFT as an approximative computational scheme is perfectly
suited to perform this task. Consider the following Euler-Lagrange
equation in the adiabatic system:
\begin{align}
  k_BT \ln\rho_{{\rm ad},t}(\rv) +
  \frac{\delta F_{\rm exc}[\rho_{{\rm ad},t}]}
       {\delta\rho_{{\rm ad},t}(\rv)}
       =\mu_{\rm ad} - V_{{\rm ad},t}(\rv),
\end{align}
where $\mu_{\rm ad}$ is the chemical potential that controls the
density in the adiabatic system, and $F_{\rm exc}[\rho]$ is the
intrinsic excess free energy functional, which arises from
$u(\rv^N)\neq 0$. Hence, the internal force field in the adiabatic
system is available as a density functional as
\begin{align}
  \fv_{{\rm ad},t}(\rv) &= -\nabla
  \frac{\delta F_{\rm exc}[\rho]}{\delta \rho(\rv)}
  \Big|_{\rho(\rv)=\rho_{{\rm ad},t}(\rv)},
\end{align}
which is a directly accessible quantity (recall
$\rho(\rv,t)=\rho_{{\rm ad},t}(\rv)$) once the excess free energy
functional is known (as an approximation, as is typical for
equilibrium DFT applications).

In summary, the DDFT equations of motion are
\begin{align}
  \dot\rho(\rv,t) &= -\nabla\cdot\Jv(\rv,t),\\
  \gamma\Jv(\rv,t) &= 
  -k_BT\nabla \rho(\rv,t)
  -\rho(\rv,t)\nabla\frac{\delta F_{\rm exc}[\rho]}{\delta \rho(\rv,t)}
  \notag\\&\quad
  +\rho(\rv,t) \fv_{\rm ext}(\rv,t),
  \label{EQcurrentDDFT}
\end{align}
Eliminating the current (which is a useful object in its own right, as
we demonstrate in Sec.~\ref{SECpft}) yields a standard form of DDFT:
\begin{align}
  \dot \rho(\rv,t)
  &= D\nabla^2 \rho(\rv,t)
  + \gamma^{-1}\nabla\cdot\rho(\rv,t)
  \nabla\frac{\delta F_{\rm exc}[\rho]}{\delta\rho(\rv,t)}
  \notag\\&\quad
  - \gamma^{-1}\nabla\cdot\rho(\rv,t)\fv_{\rm ext}(\rv,t),
  \label{EQddftOfMotionDensityOnly}
\end{align}
where $D=k_BT/\gamma$ is the (single-particle) diffusion constant, and
$\nabla^2=\nabla\cdot\nabla$ is the Laplace operator.  For the ideal
gas $F_{\rm exc}[\rho]=0$, as there are no interparticle interactions
[$u(\rv^N)\equiv 0$]. Hence, the second term on the right-hand side of
Eq.~\eqref{EQddftOfMotionDensityOnly} vanishes. The leaves over the
sum of the first term (diffusion) and the third term (drift), which
then constitutes the correct drift-diffusion equation for the ideal
gas.

The contribution due to the internal interactions in the system [the
  second term on the right-hand side of
  Eq.~\eqref{EQddftOfMotionDensityOnly}] will in general have
spatially nonlocal dependence on the density distribution. Recall that
$F_{\rm exc}[\rho]$ describes in equilibrium all correlation effects,
from the particle scale to macroscopic scales (say near a gas-liquid
critical point or in complete wetting situations).  The temporal
dependence of the equation of motion \eqref{EQddftOfMotionDensityOnly}
remains simplistic though, and it is virtually unchanged over the
ideal drift-diffusion equation: The time dependence is local (i.e.,
Markovian) and hence memory effects are absent.  A further, and
related problem is the value of $\gamma$ (and hence of the diffusion
constant $D=k_BT/\gamma$). If $\gamma$ has the value of the free
single-particle motion, how can slowing down, as is typical at high
densities, occur? --The theory seems to lack a corresponding
mechanism.  Several ways out to remedy this seeming absence of
essential physics have been proposed (some are described later) and
are primarily based on an empirical footing. As a recent
representative investigation of the differences of intrinsic
time-scales obtained from dynamical DFT as compared to BD simulation
work, we mention the studies of the van Hove pair correlation function
in liquids by \citet{treffenstaedt2021dtpl} and
\citet{treffenstaedt2021asymptotic}. We give an overview of the
dynamical test-particle limit, which underlies their treatment, below
in Sec.~\ref{SECdtpl}.

Nevertheless, per construction the equilibrium limit of the
interacting many-body system with arbitrary spatial inhomogeneity is
incorporated in an, in principle, exact fashion. This is typically not
the case in approaches that are developed genuinely in nonequilibrium
(where the assumption of a homogeneous bulk fluid is made). It is
difficult to conceive how an entirely different dynamical approach
would be able to reduce naturally to DFT when applied to time
evolution in equilibrium.

We show below that power functional theory delivers this feat, and
that the internal force density field exactly splits into $\fv_{\rm
  int}(\rv,t)=\fv_{\rm ad}(\rv,t) + \fv_{\rm sup}(\rv,t)$, where the
adiabatic force field $\fv_{\rm ad}(\rv,t)$ is identical to that in
DDFT, and the superadiabatic force field is generated from a kinematic
excess free power functional $P_t^{\rm exc}[\rho,\Jv]$ via functional
differentiation, $\fv_{\rm sup}(\rv,t)=-\delta P_t^{\rm exc}/\delta
\Jv(\rv,t)$. Power functional theory elevates the microscopic current
distribution $\Jv(\rv,t)$ from the status of a mere book-keeping
device to that of a genuine degree of freedom (an order parameter) of
the physical system.

Our presentation of the DDFT, based on
Eq.~\eqref{EQddftApproximationInternalForceField} is similarly {\rm ad
  hoc} as the original proposal of the theory \cite{evans1979}. There
has been much refined reasoning, based on the Langevin picture and
Dean's equation \cite{marconi1999ddft}, on the Smoluchowski equation
\cite{archer2004ddft}, and on the projection operator formalism
\cite{espanol2009}. While these studies shed some light on deep
connections with the many-body dynamics, and each one of the
derivations has also gained wide-spread recognition, they have thus
far not provided a systematic basis of assessing the fundamental
approximation that is involved. This step remained {\rm ad hoc}, in
the sense that no systematic way for improvement is implied. In
contrast, we later see that the adiabatic state arises naturally in
the power functional framework, as a formal exact one-body treatment
of the dynamics, which allows one to formalize and build concrete
approximations for the superadiabatic force contributions.

Applications of the DDFT framework are numerous; an exhaustive list
was given by \citet{tevrugt2020}. Here we mention selected insightful
DDFT studies. \citet{royall2007dynamicSedimentation} presented an
investigation of sedimentation of model hard-sphere-like colloidal
dispersions confined in horizontal capillaries, based on a use of
laser scanning confocal microscopy, Brownian dynamics computer
simulations, and DDFT (additional details were given by
\citet{schmidt2008dynamicSedimentation}). The researchers could obtain
quantitative agreement of the results from the respective approaches
for the time evolution of the one-body density distribution and the
osmotic pressure on the walls. In order to match the theoretical
results to the experimental data, a density-dependent mobility
$\gamma^{-1}$ was empirically introduced.

\citet{dzubiella2003mfddft} formulated the DDFT concept based on the
mean-field (quadratic in density) free energy functional.  DDFT has
been used to describe protein adsorption on polymer-coated
nanoparticles \cite{angioletti2014,angioletti2018} and the uptake
kinetics of molecular cargo into hollow hydrogels
\cite{monchojorda2019acsnano}.  DDFT has been applied to lane
formation in oppositely driven binary mixtures
\cite{chakrabarti2003epl,chakrabarti2004pre}.  DDFT has been used for
lattice models for problems such as growth of hard-rod monolayers via
deposition \cite{klopotek2017}.  \citet{bleibel2016} derived a DDFT
including two-body hydrodynamic interactions.  \citet{menzel2016}
established a DDFT for active microswimmer suspensions.  A DDFT for
translational Brownian dynamics which includes hydrodynamic
interactions was described by \citet{rex2009epje}.
\citet{scacchi2018} investigated the formation of a cavitation bubble
as a local phase transition.

\citet{goddard2012prl} derived a DFT for colloidal fluids including
inertia and hydrodynamic interactions.  \citet{wittkowski2012}
formulated an extended DDFT for colloidal mixtures with temperature
gradients.  Using DDFT \citet{scacchi2017pre} investigated the laning
instability of a sheared colloidal suspension.  \citet{waechtler2016}
performed a stability analysis based on DDFT in order to investigate
nonequilibrium lane formation in a two-dimensional Lennard-Jones fluid
composed of two particle species driven in opposite directions.
\citet{anero2013} constructud an approach that they call functional
thermo-dynamics, which represents a generalization of dynamic density
functional theory to non-isothermal situations.  DDFT has also been
used to describe polymeric systems \cite{qi2017}.
\citet{archer2004rauscher} aimed to clarify confusions in the
literature as to whether or not dynamical density functional theories
for the one-body density of a classical Brownian fluid should contain
a stochastic noise term.

\section{Power functional theory}
\label{SECpft}
\subsection{Dynamic minimization principle}
\label{SECoverviewPFTboth}

Power functional theory is based on a formally exact minimization
principle on the one-body level of dynamic correlation functions. The
theory was formulated originally for Brownian dynamics by
\citet{schmidt2013pft}, and subsequently generalized to
nonrelativistic quantum dynamics \cite{schmidt2015qpft}, and classical
Hamiltonian dynamics \cite{schmidt2018md}. Here we provide an overview
of the central concepts. Key ideas of the microscopic foundation are
described in Sec.~\ref{SECmicroscopicFoundation}. For the full
treament, see the original papers.

The kinematic fields, i.e., the density $\rho(\rv,t)$ [referred to as
  $n(\rv,t)$ in the quantum case], the current $\Jv(\rv,t)$, and in
case of inertial dynamics also the time derivative of the current,
$\dot\Jv(\rv,t)$, are the relevant functional variables.  Two
continuity equations interrelate these fields. The variational
principle is instantaneous in time, involving minimization with
respect to the highest relevant time derivative, i.e., with respect to
$\Jv(\rv,t)$ in the overdamped Brownian case, and with respect to
$\dot\Jv(\rv,t)$ in both the classical and the quantum inertial
cases. Integration in time then determines the current (in the
inertial cases), as well as the density according to
\begin{align}
  \Jv(\rv,t) &= \Jv(\rv,0) + \int_0^t dt' \dot\Jv(\rv,t'),
  \label{EQcontinuityEquationCurrentMD}\\
  \rho\rt &= \rho(\rv,0)-\int_0^t dt'\nabla\cdot \Jv(\rv,t').
  \label{EQcontinuityEquationBD}
\end{align}
In practice, one proceeds in discrete time steps such that
minimization at time $t$ allows to proceed in time by one step, then
update according to Eqs.~\eqref{EQcontinuityEquationCurrentMD} and
\eqref{EQcontinuityEquationBD}, and proceed to the next time step; see
\citet{treffenstaedt2019shear}.

We first collect from Sec.~\ref{SECmanybodyDynamics} the one-body
equations of motion for the three different types of dynamics. For
overdamped Brownian dynamics the relationship of the current and the
force densities \eqref{EQforceDensityBalanceBD} is
\begin{align}
  \gamma \Jv\rt &= -k_BT \nabla\rho\rt + \Fv_{\rm int}\rt 
  + \rho\rt \fv_{\rm ext}\rt,
  \label{EQofMotionBDdensityForm}
\end{align}
where the internal force density distribution is $\Fv_{\rm
  int}(\rv,t)=-\langle\sum_i\delta(\rv-\rv_i)\nabla_i
u(\rv^N)\rangle$, with the average being taken over the instantaneous
configuration space probability distribution.

In molecular dynamics the equations of motion
\eqref{EQmJdotFromManyBody} and \eqref{EQtotalInternalForceDensityMD}
are
\begin{align}
  m\dot \Jv\rt &= \nabla\cdot\taub\rt + \Fv_{\rm int}\rt 
  + \rho\rt \fv_{\rm ext}\rt,
  \label{EQmdEquationOfMotion}
  \\ 
  \taub(\rv,t) &= -\frac{1}{m}\Big\langle
  \sum_i\delta(\rv-\rv_i)\pv_i\pv_i \Big\rangle,
\end{align}
where the kinetic stress distribution $\taub(\rv,t)$ captures
transport effects. The averages in $\Fv_{\rm int}(\rv,t)$
$\taub(\rv,t)$ are here over the many-body phase space distribution
function.

The quantum dynamics in
Eqs.~\eqref{EQNewtonsSecondLawQMAverageAppendix} and
\eqref{EQtotalInternalForceDensityQMAppendix} are similar to the
classical inertial case, but incorporate additional wave-like, genuine
quantum effects as follows:
\begin{align}
  m\dot\Jv\rt &= \nabla\cdot\taub\rt + \Fv_{\rm int}\rt
  +\frac{\hbar^2}{4m}\nabla\nabla^2n\rt\notag\\
  &\quad + n\rt\fv_{\rm ext}\rt, 
  \\ 
  \taub(\rv,t) &= -\frac{1}{2m}\Big\langle
  \sum_i\Big[
    \hat\pv_i\delta_i\hat\pv_i
    +\hat\pv_i\delta_i\hat\pv_i^{\sf T}\Big]
  \Big\rangle,
\end{align}
where $\delta_i=\delta(\rv-\rv_i)$, and all averages are bra-kets
using the instantaneous wave function. The continuity equations
\eqref{EQcontinuityEquationCurrentMD} and
\eqref{EQcontinuityEquationBD} apply (with the symbol $\rho$ replaced
by $n$ in the notation).

The central object of power functional theory is the free power
functional $R_t[\rho,\Jv]$ (BD) or free power {\it rate} functional,
$G_t[\rho,\Jv,\dot\Jv]$ (MD) and $G_t[n,\Jv,\dot\Jv]$ (QM).  Staying
with BD, the exact minimization principle states that at (each) time
$t$:
\begin{align}
  \frac{\delta R_t[\rho,\Jv]}
       {\delta\Jv(\rv,t)}\Big|_{\rho,\Jv=\Jv_0} &= 0
  \qquad \rm (min),
  \label{EQminimizationPrincipleBDOverview}
\end{align}
where $\Jv_0(\rv,t)$ is the real, physically realized current
distribution of the physical dynamics. Hence, $\Jv_0(\rv,t)=\langle
\sum_i \vel_i \delta_i \rangle$, averaged over the actual many-body
phase space distribution function at time~$t$.

\begin{figure}
  \includegraphics[width=0.9\columnwidth,angle=0]
                  {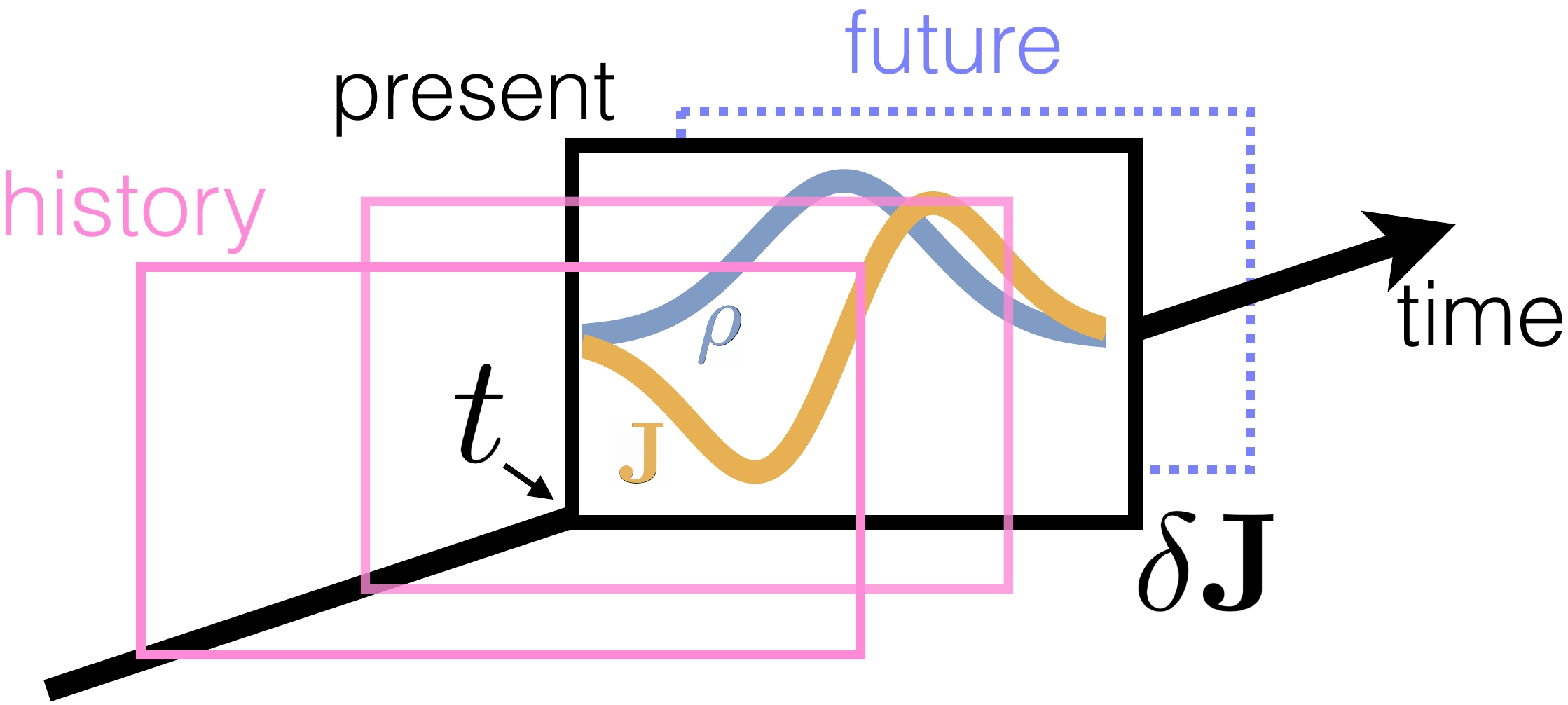}
  \caption{Time slice variational principle for overdamped BD. The
    position-resolved current varation $\delta\Jv(\rv,t)$ is performed
    at fixed time $t$. The values at times $<t$ (history) and at $>t$
    (future) are unaffected by the purely spatial variation in a time
    slice at the present time~$t$.}
  \label{FIGtimeSlice}
\end{figure}

The derivative in Eq.~\eqref{EQminimizationPrincipleBDOverview} is
performed as a spatial variation, instantaneously at fixed time $t$;
the same time argument occurs in both $R_t[\rho,\Jv]$ and in
$\Jv\rt$. This constitutes a ``time-slice'' variation where the
spatial argument can be chosen freely, but time is prescribed; see
Fig.~\ref{FIGtimeSlice} for a graphical illustration of the concept
and Appendix~\ref{SECfunctionalDerivativeAppendix} for background.
The dependence of $R_t[\rho,\Jv]$ on its functional arguments is in
general nonlocal in space, but it is causal in time, such that the
value of the density and the current contribute only at times $\leq t$
(i.e., there is no unphysical dependence on future times $>t$).  The
physical units of $R_t[\rho,\Jv]$ are those of energy per time, i.e.,
power, $[R_t]={\rm J}/{\rm s}={\rm W}$.  The density distribution
$\rho(\rv,t)$ is kept fixed under the variation
\eqref{EQminimizationPrincipleBDOverview}. Hence, the variation can be
viewed as a partial functional derivative with respect to the current,
with the density distribution being kept constant. This is not an
uncommon situation in functional calculus, see Hamilton's principle
(Appendix~\ref{SECHamiltonsActionPrinciple}).

For MD and QM the power functional minimization principle is
\begin{align}
  \frac{\delta G_t[\rho,\Jv,\dot\Jv]}
       {\delta \dot\Jv(\rv,t)}\Big|_{\rho,\Jv,\dot\Jv=\dot\Jv_0} &= 0
  \qquad \rm (min),
  \label{EQminimizationPrincipleInertial}
\end{align}
where the subscript 0 indicates again the physically realized
dynamics, $\dot\Jv_0(\rv,t)=\langle d\hat\J/dt\rangle$, with the
average taken over the state of the system at time $t$. The derivative
\eqref{EQminimizationPrincipleInertial} is taken in a time slice, and
the density and the current distributions are held constant under the
variation. The functional dependence on time is again causal, i.e., on
the value of the argument fields at times~$\leq t$. Together with the
continuity equations \eqref{EQcontinuityEquationCurrentMD} and
\eqref{EQcontinuityEquationBD}, one has formally exact equations of
motion that are closed on the one-body level. The many-body problem is
entirely encapsulated in the functional dependence of $G_t\rhoJJdot$
on its arguments.

As with the splitting of the grand potential density functional in
equilibrium, in the dynamical case a splitting of the functional into
intrinsic and external contributions holds. We return to BD, where the
total power functional splits according to
\begin{align}
  R_t\rhoJ &= P_t\rhoJ + \dot F[\rho] - X_t\rhoJ.
  \label{EQoverviewRTSplitting}
\end{align}
In Eq.~\eqref{EQoverviewRTSplitting} the superadiabatic free power
$P_t\rhoJ$ consists of ideal and excess contributions, $P_t\rhoJ =
P_t^{\rm id}\rhoJ + P_t^{\rm exc}\rhoJ$ with $P_t^{\rm id}\rhoJ$ the
exact dissipation functional of the ideal gas and $P_t^{\rm exc}\rhoJ$
accounting for excess superadiabatic effects (above excess free energy
changes). Free energy changes emerge via the time derivative $\dot
F[\rho] \equiv d F[\rho]/dt$ of the Helmholtz excess free energy
functional $F[\rho]$ of an equilibrium system with unchanged
interparticle interaction potential $u(\rv^N)$. The remaining term
$-X_t\rhoJ$ in Eq.~\eqref{EQoverviewRTSplitting} is the negative
external power. The total power functional $R_t\rhoJ$ constitutes free
power in the sense that not only energetic, but also entropic effects
are accounted for, analogously to the free energy in equilibrium.

The intrinsic contribution to the total free power is the sum
$W_t\rhoJ\equiv P_t\rhoJ+\dot F[\rho]$, which depends on the internal
interactions, but is independent of the external forces (which in
general are of time-dependent one-body form).  A splitting into ideal
and excess (above ideal) parts holds according to
\begin{align}
  W_t\rhoJ &= W_t^{\rm id}\rhoJ + W_t^{\rm exc}\rhoJ,
\end{align}
where $W_t^{\rm id}[\rho,\Jv]$ is due to the diffusive motion, and
$W_t^{\rm exc}[\rho,\Jv]$ arises from the interparticle interactions.
Both terms consist of a sum of adiabatic and superadiabatic
contributions
\begin{align}
  W_t^{\rm id}\rhoJ &= \dot F_{\rm id}[\rho] + P_t^{\rm id}\rhoJ ,
  \label{EQWtidSplitting}\\
  W_t^{\rm exc}\rhoJ &= \dot F_{\rm exc}[\rho] + P_t^{\rm exc}\rhoJ.
  \label{EQWtexcSplitting}
\end{align}
As one might expect for the non-interacting system, the two ideal
terms \eqref{EQWtidSplitting} are temporally local (as later given
explicitly). The excess part \eqref{EQWtexcSplitting} consists of an
instantaneous contribution, which one can identify with the time
derivative $\dot F_{\rm exc}[\rho]$ (also given explicitly later), and
a temporally nonlocal, i.e., memory-dependent superadiabatic term
$P_t^{\rm exc}[\rho,\Jv]$. The latter in general is also spatially
nonlocal, as is $\dot F_{\rm exc}[\rho]$, due to the coupling via the
interparticle interaction potential. We recall the total time
derivative of the (equilibrium) free energy functional as
\begin{align}
  \dot F[\rho] = \frac{d}{dt} F[\rho] 
  &= \int d\rv 
  \Big(\nabla\frac{\delta F[\rho]}{\delta\rho(\rv,t)}\Big)\cdot\Jv\rt,
  \label{EQFtimeDerivative}
\end{align}
which follows from the (functional) chain rule of differentiation, the
continuity equation, and integration by parts, see
Eqs.~\eqref{EQstarFdotZero}--\eqref{EQstarFdot}. Here the functional
derivative of $F[\rho]$ with respect to the time-dependent density is
defined via the adiabatic construction (Sec.~\ref{SECadiabaticState})
as
\begin{align}
  \frac{\delta F[\rho]}{\delta\rho(\rv,t)} &=
  \frac{\delta F[\rho_{{\rm ad},t}]}{\delta \rho_{{\rm ad},t}(\rv)}
  \Big|_{\rho_{{\rm ad},t}(\rv)=\rho(\rv,t)},
\end{align}
where $\rho_{{\rm ad},t}(\rv)$ is a trial density distribution in the
adiabatic system.

The ideal contribution to the time derivative
\eqref{EQFtimeDerivative} can be made more explicit as
\begin{align}
  \dot F_{\rm id}[\rho] &=
  \int d\rv \Big(\nabla\frac{\delta F_{\rm id}[\rho]}{\delta \rho\rt}\Big)
  \cdot \Jv\rt\\
  &= k_BT\int d\rv [ \nabla\ln\rho\rt]\cdot \Jv\rt,
  \label{EQoverviewFiddot}
\end{align}
where we recall the ideal gas free energy functional as $F_{\rm
  id}[\rho]=k_BT\int d\rv\rho(\rv)\{\ln[\rho(\rv)\Lambda^d]-1\}$.

The external power is the instantaneous expression
\begin{align}
  X_t\rhoJ &= \int d\rv
  \Big(
    \Jv(\rv,t)\cdot\fv_{\rm ext}(\rv,t) 
    - \rho(\rv,t) \dot V_{\rm ext}(\rv,t)
    \Big),
    \label{EQpftOverviewXt}
\end{align}
where the first contribution in the integral is the mechanical power
due to motion along the external force field. The second contribution
is static with respect to the particle coordinates and describes a
``charging'' effect due to temporal changes in the external potential
landscape at fixed particle positions.

The ideal dissipation functional is 
\begin{align}
  P_t^{\rm id}\rhoJ &= \frac{\gamma}{2}\int d\rv
  \frac{[\Jv(\rv,t)]^2}{\rho(\rv,t)},
  \label{EQpftOverviewPtid}
\end{align}
where, as previously, $\gamma$ is the friction constant and the
expression is spatially local and temporally Markovian, as one might
expect for ideal diffusive motion.

Inserting the splitting \eqref{EQoverviewRTSplitting} into the
minimization principle \eqref{EQminimizationPrincipleBDOverview}
yields
\begin{align}
  0 &= 
  \frac{\delta P_t^{\rm id}\rhoJ}{\delta \Jv\rt}
  +\frac{\delta P_t^{\rm exc}\rhoJ}{\delta \Jv\rt}\notag\\&\quad
  +\frac{\delta \dot F_{\rm id}[\rho]}{\delta \Jv\rt}
  +\frac{\delta \dot F_{\rm exc}[\rho]}{\delta \Jv\rt}
  -\frac{\delta X_t\rhoJ}{\delta \Jv\rt}.
  \label{EQminimizationPrincipleBDsplitting}
\end{align}
Three of the individual contributions can be obtained explicitly,
using Eqs.~\eqref{EQoverviewFiddot}, \eqref{EQpftOverviewXt}, and
\eqref{EQpftOverviewPtid}, which gives
\begin{align}
  \frac{\delta P_t^{\rm id}}{\delta \Jv} = \frac{\gamma\Jv}{\rho},
   \quad
   \frac{\delta \dot F_{\rm id}}{\delta \Jv} = k_BT \nabla
   \ln\rho,  \quad
   \frac{\delta X_t}{\delta \Jv} = \fv_{\rm ext}.
\end{align}
where arguments have been omitted for brevity. Use of the microscopic
velocity field $\vel\rt=\Jv\rt/\rho\rt$ then gives upon rearrangement
of Eq.~\eqref{EQminimizationPrincipleBDsplitting} a force balance
relationship
\begin{align}
  \gamma \vel\rt &=  -k_BT \nabla\ln\rho\rt
  -\nabla \frac{\delta F_{\rm exc}[\rho]}{\delta\rho\rt}
  \notag\\&\quad
  -\frac{\delta P_t^{\rm exc}[\rho,\Jv]}{\delta\Jv\rt}
  +\fv_{\rm ext}\rt,
 \label{EQpftOverviewForceBalanceVariational}
\end{align}
where the (negative) friction force (left-hand side) equals the sum of
all driving forces (right-hand side).  From averaging over the
microscopic dynamics (Sec.~\ref{SEConebodyBD}) we also know the
equation of motion \eqref{EQofMotionBDdensityForm},
\begin{align}
 \gamma \vel\rt &=
 -k_BT\nabla\ln\rho\rt + \fv_{\rm int}\rt + \fv_{\rm ext}\rt,
 \label{EQpftOverviewForceBalanceManyBody}
\end{align}
which is here divided by $\rho\rt$; recall that the internal force
field and the internal force density are related via $\fv_{\rm int}\rt
= \Fv_{\rm int}\rt/\rho\rt$.  From comparing
Eqs.~\eqref{EQpftOverviewForceBalanceVariational} and
\eqref{EQpftOverviewForceBalanceManyBody} we can hence identify the
internal one-body force field as
\begin{align}
  \fv_{\rm int}\rt = 
  -\nabla \frac{\delta F_{\rm exc}[\rho]}{\delta \rho\rt}
  -\frac{\delta P_t^{\rm exc}\rhoJ}{\delta \Jv\rt}.
  \label{EQinternalForceIdentification}
\end{align}
The right-hand side of Eq.~\eqref{EQinternalForceIdentification}
depends functionally on $\rho\rt$ and on $\Jv\rt$, as both $F_{\rm
  exc}[\rho]$ and $P_t^{\rm exc}[\rho,\Jv]$ inherit their functional
dependence from $R_t[\rho,\Jv]$. Hence, the left-hand side of
Eq.~\eqref{EQinternalForceIdentification} is also a functional of
these fields, i.e.,
\begin{align}
  \fv_{\rm int}(\rv,t) &= \fv_{\rm int}(\rv,t,[\rho,\vel]),
\end{align}
where the pairs $\rho\rt, \vel\rt$ and $\rho\rt, \Jv\rt$ are two
alternative sets of functional arguments
\cite{delasheras2018velocityGradient}.  We split into adiabatic and
superadiabatic force fields $\fv_{\rm int}\rt = \fv_{\rm ad}\rt +
\fv_{\rm sup}\rt$, where the two contributions are
\begin{align}
  \fv_{\rm ad}\rt &= 
  -\nabla \frac{\delta F_{\rm exc}[\rho]}{\delta \rho\rt},\quad
  \fv_{\rm sup}\rt = - \frac{\delta P_t^{\rm exc}[\rho,\Jv]}{\delta\Jv\rt}.
\end{align}
The adiabatic force field $\fv_{\rm ad}(\rv,t,[\rho])$ is an
instantaneous density functional. The superadiabatic force field
$\fv_{\rm sup}(\rv,t,[\rho,\vel])$ depends both on density and the
velocity field (or, equivalently, on the current distribution); we
refer to such an object as a {\it kinematic functional}.

Hence, the total internal force field is
\begin{align}
  \fv_{\rm int}(\rv,t,[\rho,\vel]) &= 
  \fv_{\rm ad}(\rv,t,[\rho])
  +\fv_{\rm sup}(\rv,t,[\rho,\vel]),
\end{align}
where the adiabatic force field depends instantaneously, at time $t$,
on the density distribution. The superadiabatic force field depends
also on the microscopic velocity field, and it does so via causal
dependence on time, i.e., the values of density and velocity at all
times $\leq t$ contribute, and they determine the internal force field
at time $t$. The internal one-body force density field $\fv_{\rm
  int}\rt$ plays a crucial role in the power functional formulation of
the dynamics, as it contains explicitly the interparticle coupling
that generates the many-body effects.

We can express the adiabatic force field in correlator form
\eqref{EQadsupInternalForceAsCorrelator} as
\begin{align}
  \fv_{\rm ad}(\rv,t,[\rho]) &=
  \frac{-1}{\rho\rt}\Big\langle
  \sum_i \delta(\rv-\rv_i)\nabla_i u(\rv^N)
  \Big\rangle_{\rm eq},
\end{align}
where the average is performed in an equilibrium system with density
profile $\rho_{{\rm ad},t}(\rv)=\rho(\rv,t)$. This density is
generated via an appropriate external (Mermin) potential $V_{{\rm
    ad},t}(\rv)$, which acts only in the adiabatic system, not in the
real dynamical system; see Sec.~\ref{SECadiabaticState}. If a real
external potential $V_{\rm ext}\rt$ is present, then in general
$V_{\rm ext}(\rv,t)\neq V_{{\rm ad},t}(\rv,t)$.

Recall that within classical DFT we have
\begin{align}
  \fv_{\rm ad}(\rv,t,[\rho]) &=
  -\nabla 
  \frac{\delta F_{\rm exc}[\rho_{\rm ad}]}{\delta \rho_{{\rm ad},t}(\rv)}
  \Big|_{{\rho_{\rm ad},t}(\rv)=\rho(\rv,t)}.
\end{align}

Considering the force balance
\eqref{EQpftOverviewForceBalanceManyBody}, we can conclude that the
external force field only appears explicitly; the internal force field
is independent thereof via the kinematic functional dependence.  It is
hence instructive to rearrange
Eq.~\eqref{EQpftOverviewForceBalanceManyBody} as
\begin{align}
  \fv_{\rm ext}(\rv,t) &= \gamma \vel(\rv,t)
  +k_BT\nabla\ln\rho(\rv,t)
  \notag\\&\quad
  -\fv_{\rm int}(\rv,t,[\rho,\vel]),
  \label{EQpftOverviewForceBalanceExternal}
\end{align}
which in this form constitutes a balance relationship of external
forces (left-hand side) with friction due to the flow, ideal diffusive
forces, and internal forces (three contributions on the right-hand
side).  Notably, the right-hand side of
Eq.~\eqref{EQpftOverviewForceBalanceExternal} is independent of
$\fv_{\rm ext}(\rv,t)$. Hence, if the kinematics, i.e., the history of
$\rho\rt$ and $\vel\rt$, are known, then one can determine the
external force field $\fv_{\rm ext}(\rv,t)$ that generates the
dynamics.  This implies the following functional map:
\begin{align}
  {\rm kinematics} & \quad \rightarrow \quad {\rm external\; force\; field}
 \\  
 \{\rho\rt,\vel\rt\} & \quad \rightarrow \quad \fv_{\rm ext}\rt.
\end{align}
This nonequilibrium map can be viewed as a generalization of the
following equilibrium Hohenberg-Kohn-Mermin-Evans map:
\begin{align}
  \rho(\rv) & \quad \rightarrow \quad V_{\rm ext}(\rv).
\end{align}
Given an initial equilibrium state at $t=0$, one can prescribe target
kinematic fields $\rho(\rv,t>0)$, $\vel(\rv,t>0)$ (which satisfy
physical constraints such as the continuity equation) and determine
the external force field $\fv_{\rm ext}\rt$ that generates the
prescribed dynamics. This requires access to the kinematic functional
dependence of the internal force field, as realized in the custom flow
method by \citet{delasheras2019customFlow}, which is described in
Sec.~\ref{SECcustomFlow}.

Two simple special cases are worth spelling out.
\begin{itemize}
\item[(i)] For ideal motion $P_t^{\rm exc}\rhoJ=\dot F_{\rm
  exc}[\rho]=0$, and hence
\begin{align}
\;\;\qquad  \dot\rho\rt &= \nabla\cdot \gamma^{-1}\rho\rt
  [k_BT \nabla\ln\rho\rt - \fv_{\rm ext}\rt]\\
  &= D \nabla^2 \rho\rt
  - \gamma^{-1} \nabla\cdot\rho\rt\fv_{\rm ext}\rt,
\end{align}
which is the free drift-diffusion equation with diffusion constant
$D=k_BT/\gamma$.

\item[(ii)] Neglecting only the superadiabatic excess contribution, $P_t^{\rm
  exc}\rhoJ=0$, leads to
\begin{align}
  \qquad\gamma \vel\rt &=
  -k_BT\nabla\ln\rho\rt
  -\nabla\frac{\delta F_{\rm exc}[\rho]}{\delta\rho(\rv,t)}
  + \fv_{\rm ext}\rt,
\end{align}
which is the equation of motion \eqref{EQcurrentDDFT} according to
DDFT by \citet{evans1979} and \citet{marconi1999ddft}.
\end{itemize}

We return to the description of the general framework and address
Molecular Dynamics next. The kinematic fields are
$\rho\rt,\Jv\rt,\dot\Jv\rt$.\footnote{An alternative and equivalent
  set of kinematic fields is $\rho\rt,\vel\rt,\av\rt$, where $\av\rt$
  is the local acceleration field. Rather than the bare microscopic
  acceleration $\av\rt=\dot\Jv\rt/\rho\rt$
  \citet{renner2022acceleration} recently argued that for constructing
  approximations it is advantageous to remove transport effects and
  instead use $\av(\rv,t)=\partial \vel(\rv,t)/\partial t
  =\dot\Jv(\rv,t)/\rho(\rv,t)+\vel(\rv,t)[\nabla\cdot\Jv(\rv,t)]/\rho(\rv,t)$.
} Consider Hamiltonians that contain an internal interaction potential
$u(\rv^N)$ and an external potential $V_{\rm ext}(\rv,t)$ and the
contributions from a magnetic field ${\bf B}\rt=\nabla\times{\bf
  A}\rt$. The power rate functional satisfies the minimization
principle
\begin{align}
  \frac{\delta G_t[\rho,\Jv,\dot \Jv]}{\delta \dot\Jv(\rv,t)}
  \Big|_{\rho,\Jv,\dot\Jv=\dot\Jv_0} &= 0 \quad \rm (min).
  \label{EQpftOverviewGtMinimal}
\end{align}
The minimum is attained at the physically realized form of the
acceleration density, $\dot\Jv_0\rt$.

The total power rate (with units ${\rm J}/{\rm s^2}={\rm W}/{\rm s}$)
splits into the following intrinsic and external contributions:
\begin{align}
  G_t[\rho,\Jv,\dot\Jv] &= 
  G_t^{\rm int}[\rho,\Jv,\dot\Jv] 
  - \int d\rv \dot\Jv\rt\\&
\cdot
  (q\vel\rt\times{\bf B}\rt-q\dot\Av\rt-\nabla V_{\rm ext}\rt),\notag
\end{align}
where the intrinsic contribution $G_t^{\rm int}\rhoJJdot$ solely
depends on the interparticle interaction potential. An insertion into
the minimization principle \eqref{EQpftOverviewGtMinimal} gives an
Euler-Lagrange equation of the form
\begin{align}
  \frac{\delta G_t^{\rm int}\rhoJJdot}{\delta \dot \Jv\rt}
  &=q\vel\rt\times\Bv\rt\notag\\&\quad
  -q\dot\Av\rt
  -\nabla V_{\rm ext}\rt.
  \label{EQpftOverviewMDEulerLagrange}
\end{align}
We split the intrinsic contribution into an approximate ideal and an
excess contribution
\begin{align}
  G_t^{\rm int}\rhoJJdot &= G_t^{\rm id}\rhoJJdot + G_t^{\rm exc}\rhoJJdot,
  \label{EQpftOverviewGtsplitting}
\end{align}
where the approximate ideal contribution is given by
\begin{align}
  G_t^{\rm id}[\rho,\Jv,\dot\Jv] &= \int d\rv
  \frac{\dot\Jv\rt}{\rho\rt}\Big(
  \frac{m\dot\Jv\rt}{2} - \nabla\cdot\taub^{\rm id}\rt
  \Big),
  \\
  \taub^{\rm id}\rt &= - m \frac{\Jv\rt\Jv\rt}{\rho\rt},
\end{align}
with $\taub^{\rm id}\rt$ being a factorized form of the kinetic stress
tensor. Calculating the derivative,
\begin{align}
  \frac{\delta G_t^{\rm id}\rhoJJdot}{\delta \dot\Jv\rt} &=
  m\frac{\dot\Jv\rt}{\rho\rt} 
  - \frac{\nabla\cdot\taub^{\rm id}\rt}{\rho\rt},
\end{align}
and inserting Eq.~\eqref{EQpftOverviewGtsplitting} into the
Euler-Lagrange equation \eqref{EQpftOverviewMDEulerLagrange} yields
upon rearranging the force balance relationship
\begin{align}
  \frac{m\dot\Jv\rt}{\rho\rt} &=
  \frac{\nabla\cdot\taub^{\rm id}\rt}{\rho\rt}
  -\frac{\delta G_t^{\rm exc}\rhoJJdot}{\delta\dot\Jv\rt}
  -\nabla V_{\rm ext}\rt.
  \label{EQpftOverviewMDforceBalanceVariational}
\end{align}

We also know the exact force balance directly from the many-body
dynamics \eqref{EQmdEquationOfMotion} as
\begin{align}
  \frac{m\dot\Jv\rt}{\rho\rt} &=
  \frac{\nabla\cdot\taub\rt}{\rho\rt} 
  +\fv_{\rm int}\rt
  -\nabla V_{\rm ext}\rt,
  \label{EQpftOverviewMDforceBalanceManyBody}
\end{align}
where the internal force field is the phase space average
\begin{align}
   \fv_{\rm int}\rt &=
  -\Big\langle
  \sum_i\delta(\rv-\rv_i) \nabla_i u(\rv^N)\Big\rangle\Big/ \rho\rt.
\end{align}
The kinetic stress is given by
\begin{align}
  \taub\rt &= -\frac{1}{m}\Big\langle
  \sum_i\delta(\rv-\rv_i)\pv_i\pv_i\Big\rangle.
\end{align}

By comparing Eq.~\eqref{EQpftOverviewMDforceBalanceVariational} with
Eq.~\eqref{EQpftOverviewMDforceBalanceManyBody} we can identify
\begin{align}
  -\frac{\delta G_t^{\rm exc}\rhoJJdot}{\delta\dot\Jv\rt} &=
  \fv_{\rm int}\rt
  + \frac{\nabla\cdot(\taub\rt-\taub^{\rm id}\rt)}{\rho\rt}.
\end{align}
Hence, $G_t^{\rm exc}\rhoJJdot$ is a functional generator of the
nontrivial part of the transport plus the internal force field. [Also
  here the functional derivative is taken while $\rho\rt$ and $\Jv\rt$
  are kept fixed, and the physical field values need to be inserted
  after the derivative has been taken.]

Together with the continuity equations, the Euler-Lagrange equation
forms a closed set of equations on the one-body level. The quantum
version is very similar in structure, but contains important
additional wave contributions, as described at the end of
Sec.~\ref{SECquantumPFT}.

\subsection{Microscopic foundation}
\label{SECmicroscopicFoundation}

\subsubsection*{Power functional for molecular dynamics}
\label{SECpftMD}
We return to (classical) Hamiltonian dynamics, and briefly describe
the key concepts of the many-body functional description that
underpins the power functional.  For the full presentation we refer
the reader to \citet{schmidt2018md}.

The microscopic many-body power rate functional is defined as
\begin{align}
  {\cal G}_t &= \int d\rv^N d\pib^N
  \sum_i \frac{(\fv_i-m\av_i)^2}{2m}\Psi(\rv^N,\pib^N,t)
  \notag\\&\quad
  -\int d\rv\frac{m}{2\langle\hat\rho\rangle}
  \Big\langle\frac{d\hat\Jv}{dt}\Big\rangle^2,
  \label{EQcurlyGt}
\end{align}
In Eq.~\eqref{EQcurlyGt} the particle-labelled acceleration fields
$\av_i(\rv^N,\pib^N,t)$ are (trial) variational fields on phase space;
we use the notation $\av^N=\av_1,\ldots,\av_N$ in the following.  At
the physical dynamics $m\av_i=\fv_i$, where $\fv_i$ is the force on
particle $i$.  Furthermore ${\cal G}_t$ is an instantaneous functional
at time $t$ (time slice derivatives need to be taken).  The units of
${\cal G}_t$ are power per time, i.e., ``power rate,'' which can be
seen by observing that $[\int d\rv^N d\pib^N\Psi]=1$, and
$[\fv_i^2/(2m)]=[\fv\cdot\av_i]={\rm Nm}/{\rm s}^2={\rm J}/{\rm
  s^2}={\rm W}/{\rm s}$.  The second term in Eq.~\eqref{EQcurlyGt} is
independent of the $\av_i$.

Minimizing ${\cal G}_t$ at a fixed time with respect to the $\av^N$
implies that
\begin{align}
  \frac{\delta {\cal G}_t}{\delta\av_i(\rv^N,\pib^N,t)} &= 0
  \quad \rm (min).
\end{align}

Explicitly carrying out the derivative yields
\begin{align}
  &  \frac{\delta {\cal G}_t}{\delta \av_i(\rv^N,\pib^N,t)}
  = [-\fv_i(\rv^N,\pib^N,t) + m\av_i(\rv^N,\pib^N,t)]\Psi.
\end{align}
Hence, ${\cal G}_t$ acts like a Gibbs-Appell-Gauss function (see
Appendix \ref{SECgagOne}) in that it uniquely determines the physical
dynamics by minimization. However, beyond this role, it is also a
generator of dynamical correlators, via
\begin{align}
  \frac{\delta{\cal G}_t}{\delta q\dot\Av(\rv',t)}
  &= \dot\Jv(\rv',t),
\end{align}
where the derivative acts both on the explicit appearance of $\fv_i$
and the ``hidden'' appearance in $d\hat\Jv/dt$ in
Eq.~\eqref{EQcurlyGt}.

In order to connect the many-body variational principle with the
one-body level, a constrained search is performed:
\begin{align}
  G_t[\rho,\Jv,\dot\Jv] &=
  \min_{\av^N\to\rho,\Jv,\dot\Jv} {\cal G}_t.
\end{align}
Hence, $G_t\rhoJJdot$ is a one-body functional, which is minimized by
$\dot\Jv\rt$ at the physical dynamics, see
Eq.~\eqref{EQpftOverviewGtMinimal}.  We recall the Levy method's use
in classical equilibrium density functional
(Sec.~\ref{SEClevysConstrainedSearch}), and refer the reader to
\citet{schmidt2018md} for the details of the present dynamical
treatment.

\subsubsection*{Power functional for Brownian dynamics}
\label{SECpftBD}
We return to overdamped Brownian many-body dynamics, as described in
Sec.~\ref{SEConebodyBD}, and present the key ideas of
\citet{schmidt2013pft}. They introduce trial velocity fields
$\tilde\vel_i(\rv^N,t)$, $i=1,\ldots, N$ on configuration space and
define the free power as an operator (phase space function) as
\begin{align}
  \hat{\cal R}_t
  &= 
  \sum_i\Big(
  \frac{\gamma}{2}\tilde\vel_i(\rv^N,t)  -\fv_i^{\rm tot}(\rv^N,t)
  \Big)\cdot\tilde\vel_i(\rv^N,t)
  \notag\\ &\quad
  +\sum_i\dot V_{\rm ext}(\rv_i,t).
\end{align}
For Brownian motion the total force on particle $i$ consists of
deterministic and diffusive contributions and is given by
\begin{align}
  \fv_i^{\rm tot} &= -\nabla_i u(\rv^N)
  -\nabla_i V_{\rm ext}(\rv_i,t) \notag\\
  &\quad + \fv_{\rm nc}(\rv_i,t)
  -k_BT\nabla_i\ln\Psi,
\end{align}
where $\fv_{\rm nc}(\rv,t)$ is a nonconservative external force field.
Averaging over configuration space creates the following functional
dependence on the trial velocities:
\begin{align}
  {\cal R}_t  &=
  \int d\rv^N \Psi(\rv^N,t)\hat{\cal R}(\rv^N,\tilde\vel^N,t).
  \label{EQRtManyBodyBD}
\end{align}
Due to its quadratic structure, ${\cal R}_t$ is minimized by the 
true velocity
\begin{align}
  \frac{\delta{\cal R}_t}{\delta\tilde\vel_i(\rv^N,t)}
    &= 0 \quad \rm (min),
\end{align}
and hence at the minimum $\tilde\vel_i = \vel_i \equiv
\gamma^{-1}\fv_i^{\rm tot}$, i.e., the true dynamics is recovered.
This can be seen by calculating the functional (time-slice) derivative
as follows:
\begin{align}
  \frac{\delta{\cal R}_t}{\delta\tilde\vel_i} &=
  (\gamma\tilde\vel_i-\fv_i^{\rm tot})\Psi,
\end{align}
where arguments $\rv^N,t$ have been left away for clarity As
$\Psi(\rv^N,t)\neq 0$ in general, the proposition follows.

A constrained search for the minimum yields the one-body power
functional
\begin{align}
  R_t [\rho,\Jv]
  &= \min_{\tilde\vel^N\rightarrow\rho,\Jv}{\cal R}_t,
  \label{EQRtViaConstrainedSearch}
\end{align}
see \citet{schmidt2013pft} for the full treatment, as well as for the
relationship to the time derivative of the many-body Mermin
functional; see also \citet{chan2005} and \citet{lutsko2021}. Below we
describe in Sec.~\ref{SECptexc} progress in formulating concrete
approximations for the power functional.

\subsubsection*{Quantum power functional theory}
\label{SECquantumPFT}
The quantum case is somewhat similar in mathematical structure to the
previously described classical Hamiltonian power functional
treatment. The additional quantum effects both arise in explicit,
$\hbar$-dependent terms and affect the structure of the nontrivial
parts of the functional generator. We follow \citet{schmidt2015qpft}.

We introduce complex-valued trial acceleration fields $\av^N\equiv
\av_1(\rv^N,t),\ldots,\av_N(\rv^N,t)$ and define a many-body power
rate functional
\begin{align}
  {\cal G}_t &= \int d\rv^N
  \sum_i \frac{|(\hat\fv_i-m\av_i)\Psi|^2}{2m}
  -\int d\rv \frac{m}{2\langle\hat n\rangle}
  \Big\langle
  \frac{d\hat\Jv}{dt}
  \Big\rangle^2,
\end{align}
where in the second term the trial fields $\av^N$ do not enter. Due to
the quadratic structure of ${\cal G}_t$, at the minimum the modulus
squared expression vanishes, and hence
\begin{align}
  m\av_i(\rv^N,t)\Psi(\rv^N,t) &= \hat\fv_i(\rv^N,t)\Psi(\rv^N,t)
  \label{EQquantumAtMinimum}
\end{align}
for the specific set $\av^N$ at the minimum. (This fixes the dynamics,
if Eq.~\eqref{EQquantumAtMinimum} is known at all times.) Hence, the
time-slice derivative satisfies
\begin{align}
  \frac{\delta {\cal G}_t}{\delta \av_i(\rv^N,t)} &= 0
  \quad\rm (min).
\end{align}
Furthermore ${\cal G}_t$ is a one-body generator via
\begin{align}
  \frac{\delta {\cal G}_t}{\delta q\dot\Av(\rv,t)} &=
    \dot\Jv(\rv,t).
\end{align}
We introduce a one-body constraint
\begin{align}
  \dot\Jv(\rv,t) &=
  \Big\langle
  \sum_i\Big(
  \frac{\av_i+\av_i^*}{2}\delta_i
  +\frac{\nabla\cdot\hat\taub_i}{m}
  +\frac{\hbar^2}{4m^2}\nabla\nabla^2 \hat n_i
  \Big)
  \Big\rangle,
\end{align}
where the left-hand side constitutes a prescribed target and the
$\av^N$ are trial fields in position representation. The constrained
search is
\begin{align}
  G_t[n,\Jv,\dot\Jv] &=
  \min_{\av^N\to n,\Jv,\dot\Jv} {\cal G}_t.
\end{align}
The true time evolution is still at the global minimum, and hence
\begin{align}
  \frac{\delta G_t[n,\Jv,\dot\Jv]}{\delta\dot\Jv(\rv,t)}
  \Big|_{n,\Jv} &= 0
  \quad \rm (min),
  \label{EQminimizationQM}
\end{align}
where the derivative is functional in position and at fixed time (time
slice). We split the total power rate functional into intrinsic and
external contributions according to
\begin{align}
  G_t\nJJdot &= G_t^{\rm int}\nJJdot
  \\&\quad - \int d\rv \dot\Jv\cdot
  \Big(
  \frac{q\Jv\times{\bf B}}{n}
  -q\dot\Av
  -\nabla V_{\rm ext}
  \Big),\notag
\end{align}
where the intrinsic contribution $G_t^{\rm int}\nJJdot$ is independent
of the external forces; we have omitted the arguments $\rv,t$ for
compactness of notation.  From the minimization condition
\eqref{EQminimizationQM} one obtains
\begin{align}
  \frac{\delta G_t^{\rm int}\nJJdot}{\delta \dot\Jv} &=
  \frac{q\Jv\times{\bf B}}{n} 
  - q\dot\Av
  -\nabla V_{\rm ext},
  \label{EQforceBalanceQM}
\end{align}
where the left-hand side is intrinsic and the right-hand side
constitutes the external force field. We split further into ideal and
excess contributions according to
\begin{align}
  G_t^{\rm int}\nJJdot &= G_t^{\rm id}\nJJdot + G_t^{\rm exc}\nJJdot,
  \label{EQGtSplittingIdealAndExcess}
\end{align}
where the ideal contribution \cite{bruetting2019viscosity} is
\begin{align}
  G_t^{\rm id}\nJJdot &=
  \int d\rv \frac{\dot\Jv}{n}\cdot
  \Big(
  \frac{m\dot\Jv}{2}
  -\nabla\cdot\taub_{\rm id} 
  - \frac{\hbar^2}{4m}\nabla\nabla^2 n
  \Big),
\end{align}
with the factorized dyadic form
\begin{align}
  \taub_{\rm id} &= -m\frac{\Jv\Jv}{n}
  -\frac{\hbar^2}{4m} \frac{(\nabla n)(\nabla n)}{n}.
\end{align}
The term $G_t^{\rm exc}[n,\Jv,\dot\Jv]$ in
Eq.~\eqref{EQGtSplittingIdealAndExcess} contains effects due to
internal interactions and possibly further transport terms. The
derivative of the ideal term is
\begin{align}
  \frac{\delta G_t^{\rm id}\nJJdot}{\delta \dot\Jv}\Big|_{n,\Jv} &=
  m\frac{\dot\Jv}{n}
  -\frac{1}{n}\nabla\cdot\taub_{\rm id}
  -\frac{1}{n}\frac{\hbar^2}{4m}\nabla\nabla^2 n,
\end{align}
which we insert into the force balance equation
\eqref{EQforceBalanceQM}. This gives the final equation of motion
\begin{align}
  m \dot\Jv &= -n \frac{\delta G_t^{\rm exc}\nJJdot}{\delta \dot\Jv}
  +\nabla\cdot\taub_{\rm id}
  +\frac{\hbar^2}{4m}\nabla\nabla^2 n\notag\\
  &\quad +q\Jv\times{\bf B} - n(q\dot\Av + \nabla V_{\rm ext}).
  \label{EQofMotionQuantum}
\end{align}
Together with the continuity equation, Eq.~\eqref{EQofMotionQuantum}
forms a closed dynamical theory on the one-body level. This is a
formal (yet important) result, as $G_t^{\rm exc}$ is unknown in
practice, as this would require solution of the coupled many-body
dynamics under the action of arbitrary external fields $V_{\rm ext}$
and $\Av(\rv,t)$. However, (i) approximations can be found (searched
for) and (ii) the functional relationship $-(\delta G_t^{\rm
  exc}/\delta\dot\Jv)[n,\Jv,\dot\Jv]$ is established.

The connection to time-dependent DFT is via the ground state energy
functional
\begin{align}
  E[n] &= \min_{\Psi\to n}
  \langle\Psi|\hat H|\Psi\rangle
  -\int d\rv n(\rv,t) V_{\rm ext}(\rv,t).
  \label{EQgroundStateFunctional}
\end{align}
In Eq.~\eqref{EQgroundStateFunctional} $E[n]$ is the intrinsic
(kinetic and internal interaction) contribution; often the interaction
part is split further into Hartree, exchange and correlation
terms. The first and second time derivatives are
\begin{align}
  \frac{d}{dt}E[n] &= \int d\rv \Jv\cdot\nabla\frac{\delta E[n]}{\delta n},\\
  \frac{d^2}{dt^2}E[n] &=
  \int d\rv\dot\Jv\cdot\nabla\frac{\delta E[n]}{\delta n}
  +\int d\rv d\rv' \Jv'\Jv:\nabla\nabla'
  \frac{\delta^2 E[n]}{\delta n\delta n'},
\end{align}
where $n'=n(\rv',t)$ and $\Jv'=\Jv(\rv',t)$.  The corresponding force
field is alternatively obtained via
\begin{align}
  -\frac{\delta \ddot E[n]}{\delta \dot\Jv\rt}\Big|_{n,\Jv} &=
  -\frac{\delta \dot E[n]}{\delta \Jv\rt}\Big|_{n}
  = -\nabla\frac{\delta E[n]}{\delta n\rt}.
\end{align}
Splitting $G_t^{\rm exc}[n,\Jv,\dot\Jv]=\ddot E[n]+G_t^{\rm
  sup}[n,\Jv,\dot\Jv]$ and insertion into the equation of motion
yields
\begin{align}
  m\dot\Jv &=
  -n\frac{\delta G_t^{\rm sup}\nJJdot}{\delta\dot\Jv}
  -n\nabla\frac{\delta E[n]}{\delta n}
  +\nabla\cdot\taub_{\rm id}
  \notag\\ &\quad
  +\frac{\hbar^2}{4m}\nabla\nabla^2 n
  +q\Jv\times{\bf B}
  -n(q\dot\Av + \nabla V_{\rm ext}),
\end{align}
where $G_t^{\rm sup}[n,\Jv,\dot\Jv]$ describes nonequilibrium effects,
beyond the adiabatic ground state, see \citet{bruetting2019viscosity}
for an explicit model calculation.

\subsection{Superadiabatic free power approximations}
\label{SECptexc}

Recall that the one-body equation of motion
\eqref{EQpftOverviewForceBalanceVariational} for overdamped BD, as
formulated by \citet{schmidt2013pft}, taken together with the
continuity equation, provides a formally exact description of the
dynamics, provided that the internal interaction contributions $F_{\rm
  exc}[\rho]$ and $P_t^{\rm exc}[\rho,\Jv]$ are known. The equation of
motion is closed; i.e., no further higher order correlators are
required. The adiabatic force field stems from the equilibrium excess
free energy density functional, $F_{\rm exc}[\rho]$; the
superadiabatic force field is generated from the superadiabatic free
power functional $P_t^{\rm exc}[\rho,\Jv]$. Both functionals depend
(only) on the internal interaction potential $u(\rv^N)$, and they are
unknown in practice. $F_{\rm exc}[\rho]$ is, however, a well-studied
object (although many mysteries remain).

What can we say about $P_t^{\rm exc}\rhoJ$? It certainly needs to
provide mechanisms to slow down the dynamics in typical situations, as
DDFT (where $P_t^{\rm exc}\rhoJ\equiv 0$) is often too fast. The
superadiabatic free power functional hence should describe both
dissipative, irreversible effects, but also provide genuine
structure-forming mechanism that occur in nonequilibrium.
--Reversible effects are already accounted for by the adiabatic
contribution (via the total time derivative $\dot F[\rho]$).

A series of studies have demonstrated that the superadiabatic free
power functional $P_t^{\rm exc}[\rho,\Jv]$ is indeed amenable to
analytical approximations
\cite{delasheras2018velocityGradient,delasheras2020fourForces,
  stuhlmueller2018structural,treffenstaedt2019shear,treffenstaedt2021dtpl}.
\citet{delasheras2018velocityGradient} have shown that it is possible
to use the local velocity gradient instead of the current distribution
as the relevant kinematic variable; their central ideas are presented
later. By considering higher than quadratic contributions to the power
functional, the velocity gradient concept was shown by
\citet{stuhlmueller2018structural} to also describe structural
nonequilibrium forces, i.e., nonequilibrium force contribution that
sustain density gradients. This approach was fully developed by
\citet{delasheras2020fourForces} in their splitting of the force
balance into flow and structural components.
\citet{treffenstaedt2019shear} demonstrated how to describe the
spatially and temporally nonlocal nature of viscous
forces. \citet{treffenstaedt2021dtpl} applied this approach to the
dynamical two-body structure of the bulk hard sphere fluid, i.e., its
van Hove function.

{\it Microscopic stress tensor.}---Let $\boldsymbol\sigma\rt$ be the
total interaction stress (we do not need to consider the kinetic
stress in overdamped BD). Then
\begin{align}
  \gamma\Jv\rt &= \nabla\cdot\boldsymbol\sigma\rt.
  \label{EQstressTensorGeneral}
\end{align}
In order to be fully explicit, the right-hand side of
Eq.~\eqref{EQstressTensorGeneral} expresses the divergence of a tensor
field with components $(\nabla\cdot\boldsymbol\sigma)_\beta =
\sum_{\alpha=1}^d\partial \sigma_{\alpha\beta}/\partial r_\alpha$,
where $\alpha,\beta=1,\ldots, d$ labels the Cartesian components, and
$r_\alpha$ is the $\alpha$-th component of $\rv$.  Note that the
interaction stress is very different from the kinetic stress
$\taub(\rv,t)$ described in Sec.~\ref{SEConebodyMD} in the context of
Molecular Dynamics. Rather than the transport mechanism that
$\taub(\rv,t)$ provides, the interaction stress $\boldsymbol\sigma\rt$
arises from the forces that act in the the system; for much background
see e.g.\ the accounts by \citet{balucani1994} and (in a polymer
context) by \citet{bird1987}.

Very much inspired by the formulation of mode-coupling theory on the
level of the stress tensor and the strain rate tensor [see the review
  by \citet{brader2010}], \citet{delasheras2018velocityGradient} have
shown that
\begin{align}
  \frac{\delta {\cal R}_t}{\delta\nabla\vel_{\rm sol}(\rv,t)}
  &= \sigmab(\rv,t),
\end{align}
where $\gamma {\vel}_{\rm sol}(r,t)$ is the external force field that
is induced by solvent flow.  Hence, building the divergence yields
\begin{align}
  \nabla\cdot\frac{\delta{\cal R}_t}{\delta \nabla\vel_{\rm sol}(\rv,t)}
  &= \nabla\cdot\sigmab\rt = \gamma \Jv\rt,
\end{align}
where we have used Eq.~\eqref{EQstressTensorGeneral}.  We choose an
inverse operator to $\nabla$ of ``electrostatic form'', defined as
operating on some test function $f(\rv)$ via the convolution
\begin{align}
  \nabla^{-1} f(\rv) &=
  \int d\rv' \frac{\rv-\rv'}{4\pi|\rv-\rv'|^3}f(\rv').
  \label{EQinverseNabla}
\end{align}
The convolution kernel is a radial, inverse square distance vector
field (equivalent to the electric field of a point charge). The
application of $\nabla^{-1}$ creates a vectorial dependence, via the
distance vector $\rv-\rv'$ on the right-hand side of
Eq.~\eqref{EQinverseNabla}. Note that
$\nabla\cdot\nabla^{-1}f(\rv)=f(\rv)$, which can readily be seen by
observing that $\delta(\rv)=\nabla\cdot\rv/(4\pi|\rv|^3)$.

A very well-known alternative to Eq.~\eqref{EQinverseNabla} is the
Irving-Kirkwood form of the stress tensor, as applicable for pairwise
forces. It is important to realize that the stress tensor is a
nonunique quantity \cite{SchofieldHenderson}, as only the
corresponding force density, i.e., the divergence of the stress tensor
is an observable quantity, see Eq.~\eqref{EQstressTensorGeneral}.  The
presence of the derivative allows significant freedom in the
particular choice of definition of the stress tensor.

Using the electrostatic form, we obtain the specific expression
\begin{align}
  \sigmab(\rv,t) &= \int d\rv^N
  \Psi(\rv^N,t)\sum_i
  \frac{(\rv-\rv_i)\fv_i^{\rm tot}(\rv^N,t)}{4\pi|\rv-\rv_i|^3},
  \label{EQstressElectrostaticForm}
\end{align}
where the numerator is a dyadic product of relative distance and the
force on particle $i$. Further significance for the form
\eqref{EQstressElectrostaticForm} comes from considering the
integrated stress:
\begin{align}
  \boldsymbol\Sigma(t) &= \int d\rv\sigmab(\rv,t)\\
  &= -\frac{1}{3}\int d\rv^N\Psi(\rv^N,t)
  \sum_i\rv_i \fv_i^{\rm tot}(\rv^N,t).
\end{align}
Hence, the (averaged) Clausis virial is then simply $-{\rm
  tr}\,\boldsymbol\Sigma(t)$.

{\it One-body level.}---We start using the splitting of the power
functional into ideal dissipative, superadiabatic, reversible and
external contributions, $R_t = P_t^{\rm id} + P_t^{\rm exc} + \dot F -
X_t$.  Here the external power is
\begin{align}
  X_t &= 
  \int d\rv(\Jv\rt\cdot\fv_{\rm ext}\rt 
  - \rho\rt \dot V_{\rm ext}\rt)\\
  &= -\int d\rv
  (\sigmab_{\rm ext}(\rv,t):\nabla\vel(\rv,t) + \rho\rt\dot V_{\rm ext}\rt),
  \label{EQXtViaStressTensor}
\end{align}
where we have introduced the external stress tensor field
\begin{align}
  \sigmab_{\rm ext}(\rv,t) &= \nabla^{-1}\rho(\rv,t)\fv_{\rm ext}(\rv,t).
\end{align}
The colon in Eq.~\eqref{EQXtViaStressTensor} indicates a double tensor
contraction; for two matrices $\sf A$ and $\sf B$ this is defined as
${\sf A}:{\sf B}=\sum_{ij}A_{ij}B_{ji}={\rm tr}\,{\sf A}\cdot{\sf B}$.

Hence, we can generate the velocity gradient tensor field via
\begin{align}
  \frac{\delta R_t}{\delta\sigmab_{\rm ext}(\rv,t)} &= \nabla\vel(\rv,t).
\end{align}
We can express the ideal dissipation and the adiabatic power
contributions via
\begin{align}
  P_t^{\rm id}[\rho,\vel] &= -\frac{1}{2}\int d\rv \sigmab\rt:\nabla\vel\rt,\\
  \dot F[\rho] &= \int d\rv \sigmab_{\rm ad}\rt:\nabla\vel\rt,
\end{align}
where $\sigmab\rt$, as before, is the total stress distribution, see
Eq.~\eqref{EQstressTensorGeneral}. The adiabatic stress distribution
is
\begin{align}
  \sigmab_{\rm ad}\rt &= 
  -\nabla^{-1}\rho\rt\nabla\frac{\delta F[\rho]}{\delta\rho\rt}.
\end{align}
We can now reformulate the variational principle $\delta R_t/\delta
\Jv=0$ in tensor form:
\begin{align}
  \nabla\cdot\frac{\delta R_t}{\delta\nabla\vel\rt}\Big|_\rho &= 0
  \quad \rm (min),
\end{align}
at the physical dynamics. An equivalent form is obtained by
integration,
\begin{align}
  \frac{\delta R_t}{\delta\nabla\vel\rt}\Big|_\rho
  &= \sigmab_{\rm stat}\rt,
\end{align}
where $\sigmab_{\rm stat}(\rv,t)$ is a static (artificial) stress with
vanishing divergence, $\nabla\cdot\sigmab_{\rm stat}=0$.

In order to make this framework more explicit, consider first
\begin{align}
  \frac{\delta P_t^{\rm id}}{\delta\nabla\vel\rt} &=
  -\sigmab\rt.
\end{align}
We can now collect all stress tensor contributions, $\sigmab =
\sigmab_{\rm ad} + \sigmab_{\rm sup} +\sigmab_{\rm ext}+\sigmab_{\rm
  stat}$, where the superadiabatic stress tensor distribution is
\begin{align}
  \sigmab_{\rm sup}\rt &= \nabla^{-1} \Fv_{\rm sup}\rt\\
  &= -\nabla^{-1}\rho\rt\frac{\delta P_t^{\rm exc}}{\delta \Jv\rt}
  \Big|_\rho \\
  &\equiv \frac{\delta P_t^{\rm exc}}{\delta\nabla\vel\rt}\Big|_{\rho}.
\end{align}
Hence, we have alternative forms of dependence on $\rho\rt$ and on
either $\Jv\rt$, $\vel\rt$ or $\nabla\vel\rt$, and hence $P_t^{\rm
  exc}[\rho,\Jv] \equiv P_t^{\rm exc}[\rho,\vel] \equiv P_t^{\rm
  exc}[\rho,\nabla\vel]$. In particular the velocity gradient form is
useful as a starting point for introducing approximations, as this
ensures consistency with spatial translational invariance according to
Noether's theorem; we refer the reader to \citet{hermann2021noether}.

The most general bilinear form (assuming the existence of a power
series) is
\begin{align}
  P_t^{\rm exc}[\rho,\vel] &=
  k_BT \int d\rv\int d\rv'\int_0^t dt'
  \rho(\rv,t)\nabla\vel(\rv,t):\notag\\&\quad\qquad
  {\sf M}(\rv-\rv',t-t'):
  \nabla\vel(\rv',t')\rho(\rv',t').
\end{align}
Note that terms linear in $\nabla\vel\rt$ are already accounted for in
the adiabatic term, and local contributions are contained in the ideal
dissipation functional. The kernel ${\sf M}(\rv-\rv', t-t')$ is a
dimensionless fourth-rank tensor that depends on the internal
interaction potential $u(\rv^N)$. We can approximate further by a
spatially local and Markovian form. Owing to rotational symmetry, this
is
\begin{align}
  P_t^{\rm exc}[\rho,\vel] &=
  \frac{1}{2}\int d\rv\rho 
  \Big[
    n_{\rm rot}(\nabla\times\vel)^2 + n_{\rm div}(\nabla\cdot\vel)^2
    \Big],
  \label{EQPtexcViscous}
\end{align}
where the constants $n_{\rm rot}$ and $n_{\rm div}$ possess units of
energy $\times$ time. The dynamic shear viscosity is $\eta=\rho n_{\rm
  rot}$ and the bulk (or volume) viscosity is $\zeta=\rho n_{\rm
  div}$.  For cases where $\rho=\rm const$ the resulting
superadiabatic force density is
\begin{align}
  \Fv_{\rm sup}\rt &= -\frac{\delta P_t^{\rm exc}[\rho,\vel]}
     {\delta \vel\rt}\\
  &=\eta (\nabla^2\vel\rt
  - \nabla\nabla\cdot\vel\rt)+\zeta \nabla\nabla\cdot\vel\rt,
\end{align}
which is identical to the Stokes form of hydrodynamic friction.

{\it Higher-order terms.}---Consider only rotational (shear)
components, and a spatially local form
\cite{stuhlmueller2018structural}
\begin{align}
  &P_t^{\rm exc}[\rho,\vel] = \int d\rv\Big[
    \int_0^t dt' n_{tt'}(\nabla\times\vel)\cdot(\nabla\times\vel')
    \label{EQPtexcStructural}
    \\ &\quad\;\;
    -\int_0^t dt' \int_0^t dt'' m_{tt't''}
    (\nabla\cdot\vel)(\nabla\times\vel')\cdot(\nabla\times\vel'')
    \Big],\notag 
  \label{EQPtexcHigherOrder}
\end{align}
where further terms involving $\nabla\cdot\vel$ have been omitted. The
temporal convolution kernels $n_{tt'}$ and $m_{tt't''}$ only depend on
the time differences $t-t'$ and $t-t''$ (and hence $t'-t''$).  The
resulting superadiabatic force density is
\begin{align}
  \Fv_{\rm sup}\rt &= \int_0^t dt'\nabla\cdot n_{tt'} \nabla\vel'
  \\
  &\quad-\int_0^t dt' \int_0^t dt'' \nabla m_{tt't''}
  (\nabla\times\vel')\cdot(\nabla\times\vel'')\notag
  \\
  &= \eta\nabla^2\vel
  - \chi \nabla(\nabla\times\vel)^2,
  \label{EQFsupShearSteadyState}
\end{align}
where the form \eqref{EQFsupShearSteadyState} holds in steady state,
with coefficients given by
\begin{align}
  \eta &= \lim_{t\to\infty} \int_0^t dt' n_{tt'},\\
  \chi &= \lim_{t\to\infty} \int_0^t dt' \int_0^t dt'' m_{tt't''},
\end{align}
Here $\eta$ is the coefficient of shear viscosity, and $\chi$ is the
coefficient of the migration force, which is a structural
(non-dissipative) force field, that can sustain and generate density
gradients in nonequilibrium, both in steady state and in
time-dependent situations. See \citet{stuhlmueller2018structural} for
explicit numerical results for a fluid under inhomogeneous shear
flow. That novel type of transport coefficients, such as the migration
coefficient $\chi$ arise is quite natural and it follows naturally
from the kinematic point of view. Obtaining a quantitative and
systematic understanding of how $\chi$ depends on density, temperature
etc.\ is an interesting topic for future work. We return to the
physics under shear flow below in Sec.~\ref{SECstructuralForces}.

\subsection{Nonequilibrium Ornstein-Zernike relation}
\label{SECnoz}
We give an abrideged version of the dynamical Ornstein-Zernike theory
by \citet{brader2013nozOne}.  The derivation of the full (tensorial)
version of the nonequilibrium Ornstein-Zernike equation can be found
in \citet{brader2014nozTwo}.  We recall the static Ornstein-Zernike
theory (Sec.~\ref{SECornsteinZernikeRelation}) as a template for
relating probabilistic and direct correlation function hierarchies to
each other. The power functional concept provides a time-dependent
analog.

In nonequilibrium it is natural to go from the pair correlation
function $g(\rv,\rv')$ to the van Hove function $G_{\rm
  vH}(\rv_1,t_1,\rv_2,t_2)\equiv G_{\rm vH}(1,2)$, where we have used
compact notation for spacetime points $1\equiv\rv_1,t_1$ and $2\equiv
\rv_2,t_2$. The van Hove function measures the probability of finding
a particle at point 2, given that a particle is at point 1. Even in a
bulk fluid at equilibrium the van Hove function is nontrivial, due to
the time lag between the two events, see
e.g.\ \citet{treffenstaedt2021dtpl} for recent work. The requirements
for a nonequilibrium Ornstein-Zernike relation are as follows.
\begin{itemize}
\item[(i)] It should determine $G_{\rm vH}(1,2)$.
\item[(ii)] It is not a hierarchy involving
higher (three-body, etc.) correlators.
\item[(iii)] An analogue of the direct correlation function
  $c_2(\rv,\rv')$ should occur.
\end{itemize}

We resort to the microscopic dynamics as specified via the
Smoluchowski equation. Averages are built according to $O(t)=\langle
\hat O(\rv^N,t)\rangle=\int d\rv^N \hat
O(\rv^N,t)\Psi(\rv^N,t)$. Examples include
$\rho\rt=\langle\hat\rho\rangle$ and $\Jv\rt=\langle\hat\Jv\rangle$
with $\dot\rho\rt=-\nabla\cdot\Jv\rt$.  The van Hove function is
defined as
\begin{align}
  G_{\rm vH}(1,2) &= \rho(1)^{-1} \langle \hat\rho(1)\hat\rho(2) \rangle,
  \label{EQvanHoveDefinition}
\end{align}
where we take the two times to be ordered ($t_1\geq t_2$). The
two-time average is taken over the distribution at the earlier time
$t_2$ with the conditional probability of finding the state at the
later time $t_1$.  For a quiescent bulk fluid,
Eq.~\eqref{EQvanHoveDefinition} reduces to the standard form [see
  \citet{hansen2013}] such that the dependence is only on the moduli
$|t_1-t_2|$ and $|\rv_1-\rv_2|$. Besides introducing the inhomogeneous
general form \eqref{EQvanHoveDefinition}, \citet{brader2013nozOne}
also considered the {\it front van Hove current}, defined as
\begin{align}
  \Jv_{\rm vH}^{\rm f}(1,2) &=
  \langle \hat\Jv(1)\hat\rho(2)\rangle,
\end{align}
where the first, ``front'', factor in the correlator is the current
operator.  \citet{brader2014nozTwo} also consider a corresponding van
Hove current-current correlator $\langle\hat\Jv(1)\hat\Jv(2)\rangle$.

The two-body continuity equation relates the two-body correlators
according to
\begin{align}
  \frac{\partial}{\partial t_1}
  \rho(1)G_{\rm vH}(1,2) = -\nabla_1\cdot \Jv_{\rm vH}^{\rm f}(1,2),
\end{align}
where $\nabla_1$ indicates the derivative with respect to $\rv_1$.  In
formal analogy to the static case, here we consider dynamical
functional derivatives.  We rewrite the Smoluchowski equation as
\begin{align}
  \frac{\partial}{\partial t} \Psi &= \hat\Omega(t) \Psi,
  \label{EQsmoluchwskiEquationWithOperator}
\end{align}
with the Smoluchowski time evolution operator
[Eq.~\eqref{EQsmoluchowskiOperator}]
\begin{align}
  \hat\Omega(t) &=
  -\sum_i \nabla_i \cdot \hat\vel_i,
\end{align}
where both $\nabla_i$ and $\hat\vel_i$ act via differentiation, and
the velocity operator is $\gamma\hat\vel_i = -(\nabla_i u) - k_BT
\nabla_i -(\nabla_i V_{{\rm ext},i}) + \fv_{{\rm nc},i}$. The formal
solution of Eq.~\eqref{EQsmoluchwskiEquationWithOperator} is
\begin{align}
  \Psi(\rv^N,t) &=
  {\rm e_+}^{\int_{t_0}^t ds\hat\Omega(s)} \Psi(\rv^N,t_0),
\end{align}
where $t_0$ is an initial time and the time-ordered exponential is
defined via its power series
\begin{align}
  {\rm e}_+^{\int_{t_0}^t ds \hat\Omega(s)} &=
  1+\int_{t_0}^t ds \hat\Omega(s)
  +\int_{t_0}^t ds_1\int_{t_0}^{s_1} ds_2 \hat\Omega(s_1)\hat\Omega(s_2)\notag\\
  +&\int_{t_0}^t ds_1
  \int_{t_0}^{s_1} ds_2
  \int_{t_0}^{s_2} ds_3 \hat\Omega(s_1) \hat\Omega(s_2) \hat\Omega(s_3)+\ldots.
\end{align}
Hence, the time arguments build a succession $t_0,s_3,s_2,s_1,t$ along
increasing time.  This order of labels allows to write the nested time
integrals in a natural way; note that times with increasing subscripts
$s_1,s_2,\ldots$ have the natural interpretation to be ordered
according to increasing temporal distance into the past, viewed from
the time $t$ at present. An excellent, accessible account of the
calculus of time-ordered exponentials was given by
\citet{brader2012pre}.

We use the following three ingredients.
\begin{itemize}
\item[(i)] Time-dependent functional derivatives satisfy $\delta
  \tilde u(\rv,t)/\delta \tilde
  u(\rv',t')=\delta(\rv-\rv')\delta(t-t')$, where $\tilde u\rt$ is
  some test function, {see also
    Appendix~\ref{SECfunctionalDerivativeAppendix}}.

\item[(ii)] The chain rule is
\begin{align}
  &  \frac{\delta}{\delta \tilde u(\rv,t)}
  {\rm e}_+^{\int_{t_1}^{t_2} ds \hat\Omega(s)}\\
&\quad =  \int_{t_1}^{t_2} ds
  {\rm e}_+^{\int_{s}^{t_2}ds'\hat\Omega(s')}
  \frac{\delta\hat\Omega(s)}{\delta \tilde u(\rv,t)}
  {\rm e}_+^{\int_{t_1}^s ds'\hat\Omega(s')}.\notag
\end{align}

\item[(iii)] The general definition of the two-time correlator between
operators $\hat a(1)$ and $\hat b(2)$ is
\begin{align}
  \quad\quad & \langle \hat a(1) \hat b(2) \rangle 
  \\ &\quad =
  \int d\rv^N \hat a(1)
  {\rm e}_+^{\int_{t_2}^{t_1}ds \hat\Omega(s)}
  \hat b(2)
  {\rm e}_+^{\int_{t_0}^{t_2}ds \hat\Omega(s)}
  \Psi(\rv^N,t_0).\notag
\end{align}
\end{itemize}

Using (i)--(iii) one can show the following relations, for which we
define the field
\begin{align}
  {\cal V}(2) &= \int_{t_0}^{t_2} dt_2' D \nabla_2^2 V_{\rm ext}(2'),
\end{align}
where $2'\equiv \rv_2,t_2'$, the free diffusion constant is
$D=k_BT/\gamma$, and ${\cal V}(2)$ has units of energy. The
relationships are
\begin{align}
  \frac{\delta \Jv(1)}{\delta \beta{\cal V}(2)} &=
  I(1,2) + \frac{\partial}{\partial t_2} \Jv_{\rm vH}^{\rm f}(1,2),
  \label{EQnozOne}\\
  \frac{\delta \rho(1)}{\delta \beta{\cal V}(2)}
  &= \rho(1)\frac{\partial}{\partial t_2} G_{\rm vH}(1,2).
  \label{EQnozTwo}
\end{align}
The instantaneous contribution is
\begin{align}
  I(1,2) &= -\gamma^{-1}\rho(1) 
  \frac{\delta \nabla V_{\rm ext}(1)}{\delta \beta{\cal V}(2)}.
\end{align}
The consistency with the equilibrium result from DFT can be seen by
integrating Eq.~\eqref{EQnozTwo} in time and assuming that
decorrelation happens at long times,
\begin{align}
  \int_{-\infty}^{t_2} dt_2'
  \frac{\delta \rho(1)}{\delta \beta{\cal V}(2')}
  &= \int_{-\infty}^{t_2} dt_2' \rho(1)
  \frac{\partial}{\partial t_2'} G_{\rm vH}(1,2')\\
  &= \langle \hat\rho(\rv_1)\hat\rho(\rv_2)\rangle
  -\rho(\rv_1)\rho(\rv_2)\\
  &= \frac{\delta\rho(\rv_1)}{\delta\beta V_{\rm ext}(\rv_2)}\Big|_{\rm eq}.
\end{align}
Note that here the equilibrium result is obtained via a dynamical
mechanism, which is very different from the standard static route, see
Sec.~\ref{SECornsteinZernikeRelation}.

To proceed, we first consider the adiabatic contribution to the
current (DDFT approximation), which is
\begin{align}
  \Jv_{\rm DDFT}(1) &=
  \frac{\rho(1)}{\gamma}
  \Big(
  -\nabla\frac{\delta F[\rho]}{\delta \rho(1)}
  -\nabla V_{\rm ext}(1) + \fv_{\rm nc}(1)
  \Big).
\end{align}
We calculate the derivative $\delta\Jv(1)/\delta\beta{\cal V}(3)$ and
use Eqs.~\eqref{EQnozOne} and \eqref{EQnozTwo} to obtain
\begin{align}
&  \Jv_{\rm vH}^{\rm f,DDFT}(1,3) =
  \Jv(1) G_{\rm vH}(1,3)
  -D\rho(1)\nabla_1\Big[
    G_{\rm vH}(1,3)\notag\\&\quad
    -\int d\rv_2 c_2(1,2_1) \rho(2_1)
    \left(G_{\rm vH}(2_1,3)-\rho(3_{-\infty})\right)
    \Big],
\end{align}
where the notation is $\rho(3_{-\infty})=\rho(\rv_3,-\infty)$, and
$2_1\equiv\rv_2,t_1$. Hence, $c_2(1,2_1)=c_2(\rv_1,\rv_2,t_1)$ is an
equal-time object (as is appropriate for adiabatic correlations). We
then have
\begin{align}
  c_2(\rv_1,\rv_2,t_1) &=
  -\beta \frac{\delta^2 F_{\rm exc}[\rho]}
  {\delta\rho(\rv_1)\delta\rho(\rv_2)}
  \Big|_{\rho=\rho(\rv,t_1)}.
\end{align}
Equilibrium is obtained as a special limit $\Jv(1)=0$, $\forall t$,
and the equal time limit $t_1=t_3$. Use that at equal times $G_{\rm
  vH}(1,3_1)=\rho(3_1)(h(1,3_1)+1)+\rho(1)\delta(\rv_1-\rv_3)$. Then
\begin{align}
  \Jv_{\rm vH}^{\rm f, DDFT}(1,3) &= -D\rho(1)\nabla_1
  \Big[ \delta(\rv_1-\rv_3)\\
    &\quad
    +\rho(3_1)
    \Big(
    h(1,3_1) - c_2(1,3_1)\notag\\&\quad
    -\int d\rv_2 c_2(1,2_1)\rho(2_1)h(2_1,3_1)
    \Big)\Big].\notag
\end{align}
For short times, the first term in the square brackets alone already
gives the exact result for the decay.  Hence, the term in parenthese
needs to vanish, which proves the equilibrium OZ relation
\eqref{EQinhomogeneousOZ}. This dynamical method hence provides an
alternative way to derive the static identity.

Considering next the superadiabatic contribution, we split the total
front van Hove current according to
\begin{align}
  \Jv_{\rm vH}^{\rm f}(1,3) &= 
  \Jv_{\rm vH}^{\rm f, DDFT}(1,3)
  +\Jv_{\rm vH}^{\rm f, sup}(1,3).
\end{align}
The superadiabatic contribution satisfies
\begin{align}
  &\Jv_{\rm vH}^{\rm f,sup}(1,3) = 
  \notag\\&\quad
  \Jv_{\rm vH}^{\rm f,sup}(1,3_{-\infty})
  -\rho(1)\int_{-\infty}^{t_3} dt_3'\nabla_3\cdot{\sf M}(1,3')\rho(3')\notag\\
  &\quad
  +\rho(1)\int d2\Big[ {\sf M}(1,2)\cdot
    (\Jv_{\rm vH}^{\rm f}(2,3)-\Jv(2)\rho(3_{-\infty}))\notag\\
    &\qquad\qquad
    +{\bf m}(1,2)\rho(2)(G_{\rm vH}(2,3)-\rho(3_{-\infty}))
    \Big],
\end{align}
with ${\bf m}(1,2)$ being the vectorial and ${\sf M}(1,3)$ the
tensorial time direct correlation functions.

From the general equation of motion, 
\begin{align}
  \Jv(1) &= \Jv_{\rm DDFT}(1) - \frac{\rho(1)}{\gamma}
  \frac{\delta P_{t_1}^{\rm exc}[\rho,\Jv]}{\delta \Jv(1)},
\end{align}
one can identify
\begin{align}
  {\bf m}(1,2) &= -\gamma^{-1}
  \frac{\delta}{\delta\rho(2)}
  \frac{\delta P_{t_1}^{\rm exc}[\rho,\Jv]}{\delta \Jv(1)},\\
  {\sf M}(1,2)^{\sf T} &= -\gamma^{-1}
  \frac{\delta}{\delta \Jv(2)}
  \frac{\delta P_{t_1}^{\rm exc}[\rho,\Jv]}{\delta \Jv(1)}.
\end{align}
This forms a connection from the dynamic pair structure to the
superadiabatic power functional. A more general relationship can be
obtained by considering sourced dynamics, as introduced by
\citet{brader2014nozTwo}.  Furthermore, functional line integration
\cite{brader2015functionalLineIntegration} provides a systematic means
to obtain nonequilibrium identities and, in particular, formulate
dynamical versions of common liquid state perturbation techniques.
\citet{hermann2021noether,hermann2021noetherPopular} formulated exact
sum rules for forces and correlations on the basis of Noether's
theorem, which allows one to exploit symmetries in variational
calculus. These Noether sum rules involve time direct correlation
functions, and they express their interdependence with and
relationship to averaged dynamical correlators.

\subsection{Dynamical test-particle limit and mixtures}
\label{SECdtpl}

The dynamical test-particle limit provides a formally exact route to
the time-dependent pair structure. It hence constitutes an alternative
to the nonequilibrium Ornstein-Zernike route, see
Sec.~\ref{SECnoz}. The concept relies on identifying the van Hove
function with a correspondingly constructed one-body density profile
and suitable initial conditions. The van Hove current is related to a
nonequilibrium one-body current. The dynamical test-particle limit
generalizes the static test particle of \citet{percus1962} to both
equilibrium pair dynamics (say, in a homogeneous fluid), as well as to
the time-dependent nonequilibrium pair structure.

The dynamical test-particle concept was first introduced by
\citet{archer2007dtpl} on the basis of dynamical density functional
theory and exemplified in a system of Gaussian core particles; this
model has become central to the study of interpenetrable soft matter
\cite{archer2001,archer2002}. \citet{hopkins2010dtpl} carried out a
thorough test-particle study for the hard sphere
fluid. \citet{brader2015dtpl} have overcome the the DDFT limitations
by providing the formally exact closed equations of motion for the van
Hove function, based on power functional theory
\cite{schmidt2013pft}. \citet{schindler2016dynamicPairCorrelations}
have shown, by analyzing BD computer simulation results, that the
superadiabatic contributions that determine the dynamics of the van
Hove function are comparable in magnitude to the adiabatic
contributions (i.e., those that are in principle accounted for in
dynamical DFT).

For the hard sphere fluid, \citet{treffenstaedt2021dtpl} recently
specified the superadiabatic forces that drive the equilibrium van
Hove function as consisting of drag, viscous, and structural
contributions. These force types are relevant in active Brownian
particles, in liquids under shear and in lane forming mixtures,
respectively. The explicit power functional approximation reproduces
these universal force fields in quantitative agreement with Brownian
dynamics simulation results. \citet{treffenstaedt2021dtpl} argue that
these findings demonstrate the existence of close interrelationships
between equilibrium and nonequilibrium hard sphere properties, as
expected from the general power functional point of view. We give an
outline of the dynamical test-particle theory in the following.

We use the following splitting into so-called self and distinct parts:
\begin{align}
  G_{\rm vH}(1,2) &= G_{\rm vH}^{\rm s}(1,2) + G_{\rm vH}^{\rm d}(1,2),\\
  \Jv_{\rm vH}(1,2) &= \Jv_{\rm vH}^{\rm s}(1,2) + \Jv_{\rm vH}^{\rm d}(1,2),
\end{align}
where the self part (superscript s) refers to the auto-correlation of
particle $i=j$, and the distinct part (suberscript d) refers to pairs
of different particles, $i\neq j$. Hence,
\begin{align}
  G_{\rm vH}^{\rm s}(1,2) &= \rho(1)^{\rm -1}
  \sum_i\langle\hat\rho_i(1)\hat\rho_i(2)
  \rangle,\\
  G_{\rm vH}^{\rm d}(1,2) &= \rho(1)^{\rm -1}
  {\sum_{ij}}'\langle\hat\rho_i(1)\hat\rho_j(2)\rangle,
\end{align}
where space-time points are indicated as $1\equiv \rv,t$ and
$2\equiv\rv',t'$, the particle-labelled density operator is
$\hat\rho_i(1)=\delta(\rv-\rv_i)$, and the primed sum indicates that
the case $i=j$ has been omitted.

The dynamical test-particle method applies to general nonequilibrium;
here we limit ourselves to a description of the equilibrium dynamics
of a bulk fluid of density $\rho_{\rm b}$. We introduce two
time-dependent {\it one-body} density distributions, $\rho_{\rm
  s}(\rv,t)$ and $\rho_{\rm d}(\rv,t)$ that respectively represent the
self and the distinct part of the van Hove function. Here we reuse the
symbols $\rv,t$ to indicate difference of the bare variables,
$\rv-\rv'\to\rv$, $t-t'\to t$, as is appropriate for a bulk fluid.
Consider the initial state ($t=0$) to be such that
\begin{align}
  \rho_{\rm s}(\rv,0) &= \delta(\rv),
  \label{EQtplInitialConditionSelf}\\
  \rho_{\rm d}(\rv,0) &= \rho_{\rm b}g(r),
  \label{EQtplInitialConditionDistinct}
\end{align}
where $g(r)$ is the static pair correlation function (see
Sec.~\ref{SECornsteinZernikeRelation}). The self initial condition
\eqref{EQtplInitialConditionSelf} describes the ``tagged'' particle as
being located at the origin at the initial time (or equivalently, the
origin of the coordinate system as being moved to the position of the
tagged particle at time $t=0$). The distinct initial condition
\eqref{EQtplInitialConditionDistinct} is the density profile of all
other particles in the fluid, according to the static test-particle
limit \cite{percus1962}; see \citet{rosenfeld1993testparticle} for
enforcing self-consistency with the Ornstein-Zernike route and
\citet{thorneywork2014} for a comparison to experimental results.

In the dynamical test-particle limit, we identify the time evolution
of the self and distinct one-body fields with that of the self and
distinct part of the van Hove function:
\begin{align}
  G_{\rm vH}^{\rm s}(\rv,t) &= \rho_{\rm s}(\rv,t),\\
  G_{\rm vH}^{\rm d}(\rv,t) &= \rho_{\rm d}(\rv,t),
\end{align}
where calculating the right-hand sides constitutes a dynamical
one-body problem.

To perform this task, we need to generalize power functional theory to
mixtures.  We sketch in the following the presentation by
\citet{brader2015dtpl}.  The generalization to orientational degrees
of freedom by \citet{krinninger2016prl} is closely related;
\citet{krinninger2019jcp} gave a comprehensive account. Applications
of the rotational version were presented by
\citet{hermann2019acif,hermann2019tension,landgraf2019torques}.

The many-body Smoluchowski dynamics for general mixtures is obtained
by keeping the continuity equation \eqref{EQsmoluchowski1} $\dot\Psi =
-\sum_i\nabla_i\cdot\vel_i\Psi$, but introducing $\gamma_i$ as the
friction constant of particle $i$, and relating the configurational
velocity of particle $i$ to the forces via
\begin{align}
  \gamma_i\vel_i &= -k_BT\nabla_i\ln\Psi
  -\nabla_i u(\rv^N) + \fv_{{\rm ext},i}(\rv_i,t),
  \label{EQdtplSmoluchowski2}
\end{align}
where $d$ indicates the spatial dimensionality, and $\fv_{{\rm
    ext},i}(\rv,t)$ is the external field that acts individually on
particle $i$. We recover the one-component version,
Eq.~\eqref{EQsmoluchowski2}, when setting $\gamma_i=\gamma$ and
$\fv_{{\rm ext},i}\rt=\fv_{\rm ext}\rt$, $\forall i$. In the
one-component system, we have the further requirement that $u(\rv^N)$
is invariant under permutations of the particle indices, which is not
necessarily implied in Eq.~\eqref{EQdtplSmoluchowski2}.

From this most general dynamics of having $N$ particles with distinct
properties, we specialize to mixtures via introducing sets ${\cal
  N}_\alpha$ of particle labels $i$ that contain identical particles
of the same species $\alpha$. We can then obtain the species-resolved
density and current operators, respectively, by restricting the
particle summation to those particles of the same species,
\begin{align}
  \hat\rho_\alpha &= \sum_{i\in{\cal N}_\alpha}\delta(\rv-\rv_i),
  \qquad
  \hat\Jv_ \alpha =
  \sum_{i\in{\cal N}_\alpha} \delta(\rv-\rv_i) \vel_i.
\end{align}
The species-resolved density and the current profile are obtained by
standard averages, $\rho_\alpha\rt=\langle\hat\rho_\alpha\rangle$ and
$\Jv_\alpha\rt=\langle\hat\Jv_\alpha\rangle$. And as the particle
identities are fixed in the course of time, the continuity equation is
$\dot\rho_\alpha\rt=-\nabla\cdot\Jv_\alpha\rt$. We also bin the
friction constants, such that there is a unique friction constant
$\gamma_\alpha$ for each species; formally one can express this as
$\gamma_i=\gamma_\alpha$, $\forall i\in{\cal N}_\alpha$. Similarly,
for the external potential $\fv_{{\rm ext},i}\rt=\fv_{\rm
  ext}^\alpha\rt, \forall i\in{\cal N}_\alpha$, where $\fv_{\rm
  ext}^\alpha\rt$ is the external force field acting on species
$\alpha$, with conservative contribution $-\nabla V_{\rm
  ext}^\alpha\rt$.

The power functional framework generalizes the generating functional
$R_t$ to remain a single object
\cite{brader2015dtpl,krinninger2019jcp}, but one that depends on the
set of all species-resolved profiles
$\{\rho_{\alpha'}\rt,\Jv_{\alpha'}\rt\}$, where $\alpha'$ labels the
species of the set. The extremal principle is
\begin{align}
  \frac{\delta R_t[\{\rho_{\alpha'},\Jv_{\alpha'}\}]}
  {\delta \Jv_\alpha(\rv,t)} &= 0
  \qquad {\rm (min)}, \quad \forall \alpha.
  \label{EQRtminimalMixtures}
\end{align}
The power functional for a mixture splits into intrinsic and external
contributions according to
\begin{align}
  R_t[\{\rho_{\alpha'},\Jv_{\alpha'}\}] &=
  \sum_\alpha(\dot F_{\rm id}[\rho_\alpha]
  +P_t^{\rm id}[\rho_\alpha,\Jv_\alpha] - X_t[\rho_\alpha,\Jv_\alpha])\notag\\
  &\quad + \dot F_{\rm exc}[\{\rho_{\alpha'}\}]
  + P_t^{\rm exc}[\{\rho_{\alpha'},\Jv_{\alpha'}\}],
  \label{EQRtdecompositionMixture}
\end{align}
where the ideal adiabatic, ideal dissipative, and external
contributions are given, respectively, by
\begin{align}
  \dot F_{\rm id}[\rho_\alpha] &= k_BT \int \!\! d\rv 
  \Jv_\alpha\rt\cdot\nabla\ln\rho_\alpha\rt,\\
  P_t^{\rm id}[\rho_\alpha,\Jv_\alpha] &= \frac{\gamma_\alpha}{2}
  \int d\rv \frac{\Jv_\alpha^2\rt}{\rho_\alpha\rt},\\
  X_t[\rho_\alpha,\Jv_\alpha] &= \!\!\int\!\! d\rv(
  \Jv_{\alpha}\rt\cdot\fv_{\rm ext}^\alpha\rt
  -\dot V_{\rm ext}^\alpha\rt \rho_\alpha\rt).
\end{align}
Inserting these forms into the free power decomposition
\eqref{EQRtdecompositionMixture} and using the minimization principle
\eqref{EQRtminimalMixtures} yields the equations of motion
\begin{align}
  \gamma_\alpha\vel_\alpha\rt 
  &= -k_BT\nabla\ln\rho_\alpha\rt
  + \fv_{\rm ad}^\alpha\rt
  \notag\\&\quad
  +\fv_{\rm sup}^\alpha\rt
  + \fv_{\rm ext}^\alpha\rt,
\end{align}
where the microscopic velocity profile of species $\alpha$ is
$\vel_\alpha\rt=\Jv_\alpha\rt/\rho_\alpha\rt$ and the adiabatic and
superadiabatic force fields acting on species $\alpha$ are given,
respectively, by
\begin{align}
  \fv_{\rm ad}^\alpha(\rv,t)&=
  -\nabla\frac{\delta F_{\rm exc}
    [\{\rho_{\alpha'}\}]}{\delta\rho_\alpha(\rv,t)},\\
  \fv_{\rm sup}^\alpha(\rv,t) &=
  -\frac{\delta P_t^{\rm exc}[\{\rho_{\alpha'},\Jv_{\alpha'}\}]}
  {\delta\Jv_\alpha(\rv,t)}.
\end{align}
We return to the physics of mixtures below in
Sec.~\ref{SECsuperdemixing} when we consider differential and total
motion.

Applying this general framework to the dynamics of the van Hove
function in its test-particle representation gives the two-body
equation of motion:
\begin{align}
  \frac{\partial}{\partial t} G_{\rm vH}^\alpha\rt
  &= D\nabla^2 G_{\rm vH}^\alpha(\rv,t)
  \\&\quad
  -\gamma^{-1}\nabla\cdot G_{\rm vH}^\alpha(\rv,t)
   (\fv_{\rm ad}^\alpha\rt+\fv_{\rm sup}^\alpha\rt),
  \notag
\end{align}
where the species index $\alpha= s,d$ labels the self and the distinct
part and $D=k_BT/\gamma$ is the diffusion constant, with identical
friction constant of self and distinct particles
($\gamma\equiv\gamma_s=\gamma_d$).

Results obtained from the adiabatic approximation, $\fv_{\rm
  sup}^\alpha\rt=0$, i.e., from DDFT, for the repulsive Gaussian core
model fluid \cite{archer2007dtpl} indicate that for the chosen
statepoint, the DDFT gives very good account of the simulation
data. This is not true in general, as was shown on the basis of BD
simulation results for the van Hove current for a dense Lennard-Jones
bulk liquid by \citet{schindler2016dynamicPairCorrelations}. They
split the total van Hove current, via explicitly constructing the
adiabatic state in simulations, into adiabatic and superadiabatic
contributions. The results indicate that both contributions are of
comparable magnitude and that they are distinctly different in form,
i.e., in the variation with distance. Recently, a comparison of DDFT
results with experimental data, obtained in a quasi two-dimensional
hard sphere dispersion using video microscopy was performed by
\citet{stopper2018dtpl}. The authors have modified the ``bare'' DDFT
and have obtained results in very good agreement with the experimental
data.

The test-particle concept was applied to investigate self-diffusion in
a model system of rod-like particles in the smectic (or lamellar)
phase by \citet{grelet2008}. A corresponding experimental system was a
colloidal suspension of filamentous {\it fd} virus particles, which
allowed the direct visualization at the scale of the single particle
of mass transport between the smectic layers. The authors found that
self-diffusion takes place preferentially in the direction normal to
the smectic layers and occurs in steps of one rod length, which is
reminiscent of a hopping-type of transport. The probability density
function was obtained experimentally at different times and found to
be in qualitative agreement with theoretical predictions based on a
dynamical density functional theory.  Closely related DDFT work was
carried out by \citet{bier2007pre,bier2008pre} and
\citet{bier2008prl}.

Most of the previously mentioned empirical corrections to the DDFT
dynamics in effect replace the bare diffusion constant $D$ (and
$\gamma$ accordingly) by a reduced value, which is an input to the
theory. In their recent investigation for the hard sphere van Hove
dynamics, \citet{treffenstaedt2021dtpl} proceeded differently. They
rather showed that the superadiabatic force contributions generate the
slow down of the dynamics. Their quantitative power functional
description of the superadiabatic force contributions yields results
that are in very good agreement with BD simulation data. Crucially, by
considering the total, but also the differential motion of the van
Hove function, i.e., the difference of self and distinct parts, they
were able to uniquely identify force contributions that also arise in
nonequilibrium.

\subsection{Custom flow algorithm}
\label{SECcustomFlow}

\begin{figure*}
  \includegraphics[width=1.67\columnwidth,angle=0]{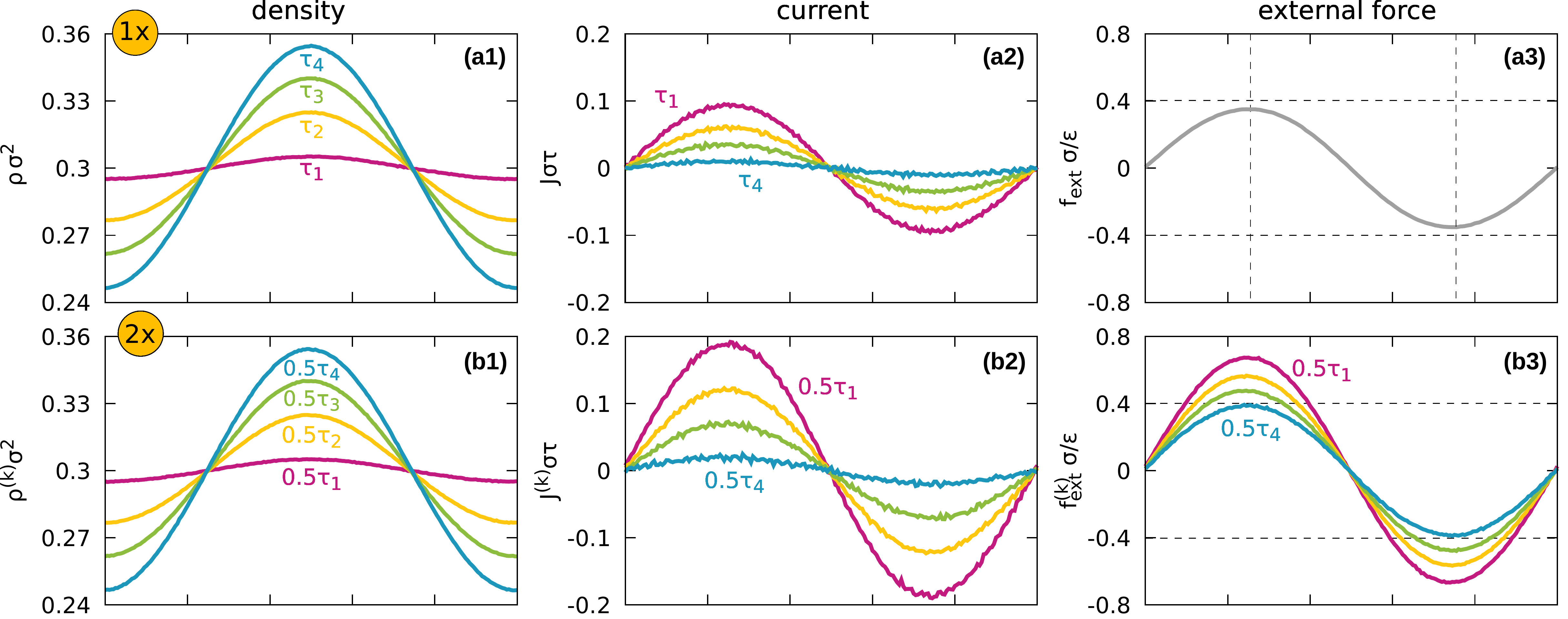}
  \caption{Custom flow method applied to a fluid of steeply repulsive
    particles of size $\sigma$ and strength of repulsion
    $\epsilon$. Shown are the density profile (left column), the local
    current (middle colunm), and the external force field (right
    colunm) at four consequtive times
    $0<\tau_1<\tau_2<\tau_3<\tau_4$. The system is initially in a
    homogeneous bulk fluid state (constant density and vanishing
    external force). In the original system, at times $t \geq 0$, a
    density inhomogeneity (a1) and corresponding flow (a2) develop,
    due to the action of a sinusoidal external force (a3) that is
    instantaneously switched on at $t=0$. Eventually the system
    reaches practically a new equilibrium at $t=\tau_4$ with almost
    zero current. In the fast forward system (second row) the density
    and the current are prescribed to evolve twice as fast as in the
    original system. The resulting sequence of density profiles (b1)
    is identical to that in the original system (a1), when rescaling
    the time label by $0.5$ (as indicated). Correspondingly, the
    amplitude of the current (b2) is twice as large as in the original
    system, but it is otherwise unchanged. The superscript $(k)$
    indicates the iteration step. The custom flow method finds the
    external force field (b3) that generates the prescribed (slow)
    time evolution. The resulting external force (b3) varies in the
    course of time, in contrast to the behaviour in the original
    system (a3). Adapted from \citet{delasheras2019customFlow}.}
  \label{FIGcustomFlow}
\end{figure*}

The custom flow algorithm constitutes a method to create a desired
spatio-temporal pattern of density and velocity, by constructing
(iteratively) the necessary external force field that creates this
prescribed motion.  As a causal relationship, one would view the
external forces as being at the origin of the motion, $\fv_{\rm
  ext}\rt\to\{\rho\rt,\Jv\rt\}$.  However, from power functional
theory \cite{schmidt2013pft}, the functional map is
\begin{align}
  \{\rho\rt,\Jv\rt\} \to \fv_{\rm int}\rt.
\end{align}
In the force balance relation the flow is given by
\begin{align}
  \gamma\vel\rt =
  -k_BT\nabla\ln\rho\rt + \fv_{\rm int}\rt + \fv_{\rm ext}\rt,
\end{align}
where the external force field in general consists of conservative and
nonconservative contributions [$\fv_{\rm ext}\rt=-\nabla V_{\rm
    ext}\rt+\fv_{\rm nc}\rt$]. As $\fv_{\rm int}\rt=\fv_{\rm
  int}(\rv,t,[\rho,\Jv])$, the force balance implies
\begin{align}
  \{\rho\rt,\Jv\rt\}\to\fv_{\rm ext}\rt,
\end{align}
which constitutes a reversal of the causal relationship. One might
wonder whether this has consequences and whether it can be exploited,
say, on the level of BD simulations.  To investigate this point, we
re-order the equation of motion trivially as
\begin{align}
  \fv_{\rm ext}\rt &= k_BT\nabla\ln\rho\rt
  -\fv_{\rm int}\rt + \gamma \vel\rt,
\end{align}
where the microscopic velocity field is $\vel\rt=\Jv\rt/\rho\rt$.

We first consider steady states, where there is no explicit time
dependence in the one-body fields. Hence,
\begin{align}
  \fv_{\rm ext}(\rv) &= k_BT\nabla\ln\rho(\rv)
  -\fv_{\rm int}(\rv) + \gamma \vel(\rv).
  \label{EQfextCustomFlow}
\end{align}
Here we know from power functional theory that the right-hand side
only depends on the density profile and current, and not directly on
the external force field. In a BD scheme however, one needs to
implement this relationship in a computational way. This can be
performed using the custom flow iteration scheme by
\citet{delasheras2019customFlow} to solve for $\fv_{\rm ext}(\rv)$.
Consider $\rho(\rv),\Jv(\rv)$ to be fixed target fields and search for
the form of $\fv_{\rm ext}(\rv)$ that generates, in BD simulations,
these targets in steady state. The targets need to be physical,
including the condition $\nabla\cdot\rho(\rv)\vel(\rv)=0$ for a steady
state. The iteration step is
\begin{align}
  \fv_{\rm ext}^{(k)}(\rv) &=
  k_BT\nabla\ln\rho(\rv)
  -\fv_{\rm int}^{(k-1)}(\rv)
  +\gamma\vel(\rv),
\end{align}
where $k$ labels the steps, $\rho(\rv)$ and $\vel(\rv)$ are the known
targets. The internal force field is sampled as
\begin{align}
  \fv_{\rm int}^{(k-1)}(\rv) &= -\frac{1}{\rho(\rv)}
  \Big\langle\sum_i\delta(\rv-\rv_i)\nabla_i u(\rv^N)\Big\rangle,
  \label{EQfintCustomFlow}
\end{align}
where the system is exposed to the action of the external force field
$\fv_{\rm ext}^{(k-1)}(\rv)$.  The iteration can be started with the
ideal gas ansatz as follows:
\begin{align}
  \fv_{\rm ext}^{(0)}(\rv) &=
  k_BT\nabla\ln\rho(\rv) + \gamma\vel(\rv).
\end{align}

\citet{delasheras2019customFlow} formulated convergence criteria and
showed that in practice convergence is fast and reliable and to a
unique solution. This demonstrates that for steady states
Eq.~\eqref{EQfextCustomFlow} performs the functional inversion from
$\fv_{\rm ext}(\rv)\to\{\rho(\rv),\vel(\rv)\}$ to
$\{\rho(\rv),\vel(\rv)\}\to\fv_{\rm ext}(\rv)$. Note that the
splitting of the internal force field into adiabatic and
superadiabatic contributions, $\fv_{\rm int}(\rv)=\fv_{\rm
  ad}(\rv)+\fv_{\rm sup}(\rv)$, has not been used here. [As a striking
  example of the method, one can keep $\rho(\rv)$ fixed and vary
  $\vel(\rv)$ only; see \citet{delasheras2019customFlow}.]

As a special case, for inverting in equilibrium, where $\vel(\rv)=0$,
the force balance \eqref{EQfextCustomFlow} simplifies to
\begin{align}
  \fv_{\rm ext}(\rv) = k_BT\nabla\ln\rho(\rv) - \fv_{\rm int}(\rv).
\end{align}
Here the iteration step is
\begin{align}
  \fv_{\rm ext}^{(k)}(\rv) &= k_BT\nabla\ln\rho(\rv)
  -\fv_{\rm int}^{(k-1)}(\rv),
\end{align}
with the iteration start at e.g.\ the ideal gas solution, $\fv_{\rm
  ext}^{(0)}(\rv) = k_BT\nabla\ln\rho(\rv)$. The sampling provides
$\fv_{\rm int}^{(k-1)}(\rv)$ from carrying out
Eq.~\eqref{EQfintCustomFlow} under the action of $\fv_{\rm
  ext}^{(k-1)}(\rv)$. This provides a powerful method to perform the
adiabatic constrution in practice (Sec.~\ref{SECadiabaticState}).  If
the sampling is performed in Monte Carlo, then one needs the external
potential which in effective one-dimensional situations of planar
geometry (space coordinate~$x$) can be obtained from integration
\begin{align}
  V_{\rm ext}^{(k)}(x) &=
  -k_BT\ln[\rho(x)\Lambda^d] + \int dx f_{{\rm int},x}^{(k-1)}(x).
\end{align}
In more general situations, one needs to use an inverse $\nabla^{-1}$
operator to perform the integration, see
e.g.~\citet{delasheras2018forceSampling,borgis2013}.

From the DFT context (Sec.~\ref{SECsketchDFT}), we know that
the internal force field can be expressed as
\begin{align}
  \fv_{\rm int}(\rv) &= -\nabla
  \frac{\delta F_{\rm exc}[\rho]}{\delta \rho(\rv)}.
\end{align}
Hence, the force balance is
\begin{align}
  \fv_{\rm ext}(\rv) &= k_BT\nabla\ln\rho(\rv)
  + \nabla \frac{\delta F_{\rm exc}[\rho]}{\delta \rho(\rv)},
\end{align}
where the right-hand side is a density functional. Hence, the
Mermin-Evans map $\rho(\rv)\to \fv_{\rm ext}(\rv)$, with $\fv_{\rm
  ext}(\rv)=-\nabla V_{\rm ext}(\rv)$.  The method of
\citet{fortini2014prl} can be derived from the present scheme, as
shown in \citet{delasheras2019customFlow}.  They also summarize three
different methods to sample the current distribution in BD
simulations. This includes (i) using the force density balance; (ii)
the method of centered finite time difference, where the velocity of
particle $i$ is given as $\vel_i(t) = (\rv_i(t+\Delta
t)+\rv_i(t-\Delta t))/(2\Delta t)$, with $\Delta t$ being the time
step in the BD algorithm; and (iii) using the continuity
equation. These methods can be implemented separately, and hence
provide valuable consistency checks.

Having restricted ourselves to steady states in the previous
description of custom flow, the method is amenable to time-dependent
problems, see \citet{delasheras2019customFlow}. Here a coarse-graining
time step is introduced and the steady state strategy described above
is performed in each step.  Figure \ref{FIGcustomFlow} shows
corresponding illustrative results of a density peak that grows in
time (top row), which is then sped up to proceed at twice its original
speed (bottom row).\footnote{The Supplemental Material for this work
  contains a video with speed up by a factor of 3, as also reflected
  in the accompanying shred guitar soundtrack.}  Very recently,
\citet{renner2021customFlowMD} have considered the problem of
prescribed flow in the context of Molecular Dynamics. They propose a
generic formulation of iterative custom flow methods, and demonstrate
that a particularly simple variant indeed allows to generate
tailor-made flow.

\subsection{Viscous and structural forces} 
\label{SECstructuralForces}
\begin{figure}
    \includegraphics[width=0.9\columnwidth,angle=0]{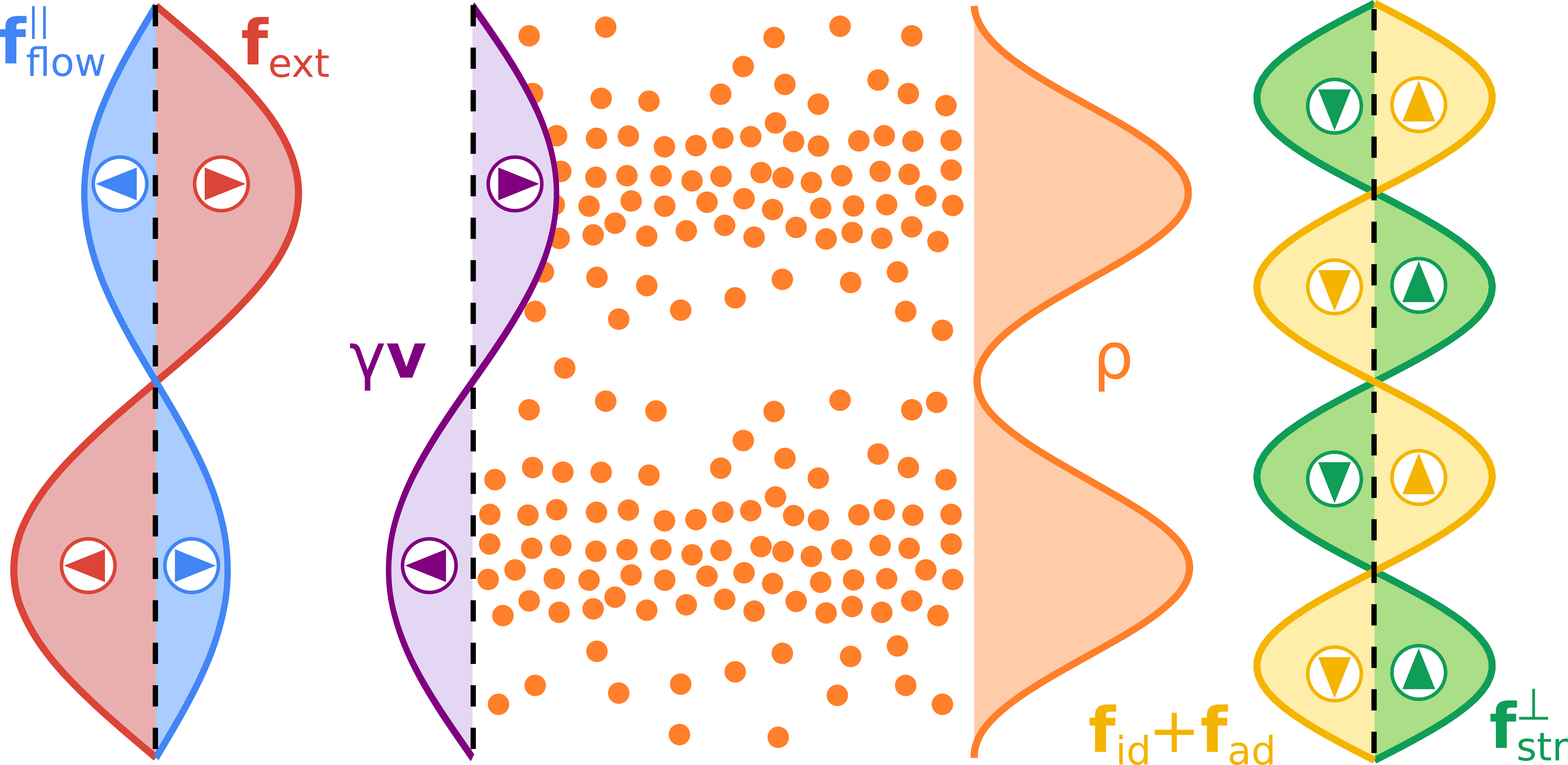}
  \caption{Prototypical inhomogeneous flow situation in which both
    viscous (blue) and structural nonequilibrium forces (green) occur
    in a fluid of repulsive particles (orange dots) that undergo
    Brownian dynamics. The system is in steady state that results from
    the action of an external force field $\fv_{\rm ext}$ (red) that
    is parallel to $\ev_x$ (horizonal) and varies its magnitude
    sinusoidally in $\ev_y$ (vertical). The induced flow (violet) is
    directed in the (horizontal) $\ev_x$-direction. Due to the
    structural forces, the density profile (orange) becomes
    inhomogeneous. The inhomogeneous density induces ideal and
    adiabatic forces (yellow) that tend to homogenize the system, but
    are counter acted by the structural nonequilibrium force $\fv_{\rm
      struc}$ (green). For each vector field its magnitude (curves)
    and direction (arrows) are indicated. Taken from
    \citet{delasheras2020fourForces}; see also
    \cite{stuhlmueller2018structural,jahreis2019shear}.  }
  \label{FIGfourForces}
\end{figure}

The rheology of colloidal systems is a rich and diverse subject, see
the excellent account given by \citet{brader2010}. In his seminal
treatment \citet{dhont1996} provided detailed background. Here we
consider inhomogeneous shear flow that is spatially oscillatory; see
Fig.~\ref{FIGfourForces} for an illustration. The external force field
that generates the flow is assumed to have the form
\begin{align}
  \fv_{\rm ext}(y) &= f_0 \sin(2\pi y/L) \ev_x,
\end{align}
where $f_0=\rm const$ controls the strength of the driving, $y$ is the
coordinate perpendicular to the driving, $L$ is the system size in the
$y$-direction, and $\ev_x$ is the unit vector in the $x$-direction;
note that the field acts along the $x$-direction, but varies its
strength in the orthogonal $y$-direction, which constitutes a generic
shear situation, where no unique flow potential exists.

The steady state force balance relation is
\begin{align}
  \gamma \vel(\rv) &= -k_BT\nabla\ln\rho(\rv) + \fv_{\rm ad}(\rv)
  +\fv_{\rm sup}(\rv) + \fv_{\rm ext}(\rv).
  \label{EQshearForceBalanceTotal}
\end{align}
The induced flow will in steady state also be along the $x$-direction
and the system will be homogeneous in $y$. We can hence split
Eq.~\eqref{EQshearForceBalanceTotal} into its vector components in the
flow ($\ev_x$) and gradient ($\ev_y$) directions, respectively given
by
\begin{align}
  \gamma \vel(\rv) &= \fv_{\rm visc}(\rv) + \fv_{\rm ext}(\rv),
  \label{EQshearForceBalanceFlow}\\
  0 &= -k_BT \nabla\ln\rho(\rv)
  + \fv_{\rm ad}(\rv) + \fv_{\rm struc}(\rv),
  \label{EQshearForceBalanceGradient} 
\end{align}
In Eqs.~\eqref{EQshearForceBalanceFlow} and
\eqref{EQshearForceBalanceGradient} we have split the superadiabatic
force field into two mutually orthogonal contributions according to
$\fv_{\rm sup}(\rv)=\fv_{\rm visc}(\rv)+\fv_{\rm struc}(\rv)$, where
the viscous contribution is parallel to the flow [$\fv_{\rm
    visc}(\rv)\parallel \ev_x$] and the ``structural'' force
contribution is perpendicular to the flow [$\fv_{\rm struc}(\rv)
  \parallel \ev_y$].

We assume the superadiabatic free power functional to consist of two
parts that correspond to viscous and structural effects,
\begin{align}
  P_t^{\rm exc}[\rho,\vel] &=
  P_t^{\rm visc}[\rho,\vel] + P_t^{\rm struc}[\rho,\vel].
\end{align}
The viscous contribution \eqref{EQPtexcViscous} is given by
\begin{align}
  P_t^{\rm visc}[\rho,\vel] &= \frac{1}{2}\int d\rv
  \big[
    \eta (\nabla\times\vel)^2 + \zeta (\nabla\cdot\vel)^2
    \big],
\end{align}
where $\eta$ and $\zeta$ are related to the shear viscosity and volume
(or ``bulk'') viscosity, respectively. The present geometry has no
compressional flow component, i.e., $\nabla\cdot\vel=0$, and hence
only the shear contribution contributes. Hence, the functional
derivative with respect to the velocity field $-\rho^{-1}\delta
P_t^{\rm visc}/\delta \vel$, gives a viscous superadiabatic force
field corresponding to that in Stokes flow,
\begin{align}
  \fv_{\rm visc}(y) &= \eta \nabla^2 \vel(y)
  \approx \frac{\partial^2}{\partial y^2}
  \frac{\eta f_0}{\gamma} \sin(ky)\ev_x
  \label{EQfviscApproximation}\\
  &= -\frac{\eta f_0 k^2}{\gamma} \sin(ky)\ev_x
  = -\frac{\eta k^2}{\gamma} \fv_{\rm ext}(y),
\end{align}
where we have made the approximation $\vel(\rv)=\fv_{\rm
  ext}(\rv)/\gamma$ in Eq.~\eqref{EQfviscApproximation} and have
introduced the wave number $k=2\pi/L$ that characterizes the
oscillatory shear field.

In the gradient direction, as described by
Eq.~\eqref{EQshearForceBalanceGradient}, we neglect the adiabatic
(interparticle interaction) contribution over the ideal diffusive
part, as is appropriate at low densities. Hence, $\fv_{\rm
  ad}(\rv)\approx 0$ and we obtain
\begin{align}
  k_BT\nabla\ln\rho(\rv) &= \fv_{\rm struc}(\rv),
  \label{EQshearStructuralForceBalance}
\end{align}
where the structural force field is necessarily a kinematic
functional, i.e., in general $\fv_{\rm struc}(\rv,t,[\rho,\vel])$.

We next assume the following form [Eq.~\eqref{EQPtexcStructural}] of
the structural contribution to the superadiabatic functional,
\begin{align}
  P_t^{\rm struc}[\rho,\vel] &= -\int d\rv \int_0^tdt' \int_0^tdt''
  \\&\qquad\qquad
  m_{tt't''} (\nabla\cdot\vel)
  (\nabla\times\vel')\cdot(\nabla\times\vel''), \notag
\end{align}
where the primed (double primed) velocity depends on $t'$ ($t''$) and
the position argument $\rv$ has been omitted for clarity. We obtain
the structural force density distribution $F_{\rm
  struc}\rt=\rho\rt\fv_{\rm struc}\rt$ via differentitation,
\begin{align}
  \Fv_{\rm struc}\rt &= -\frac{\delta P_t^{\rm struc}[\rho,\vel]}
     {\delta \vel\rt}\\
  &= -\int_0^t\!\!dt'\int_0^t\!\!dt'' \nabla m_{tt't''}
  (\nabla\times\vel')\cdot(\nabla\times\vel'')\\
  &= -\chi \nabla (\nabla\times\vel(\rv))^2,
  \label{EQshearStructuralForceField}
\end{align}
where in the last step we have assumed that $\rho\approx\rho_{\rm
  b}=\rm const$ and that the system is in steady state. The amplitude
of the structural force is given by the moment
\begin{align}
  \chi &= \lim_{t\to\infty} \int_0^tdt'\int_0^tdt''m_{tt't''}.
\end{align}
We next apply Eq.~\eqref{EQshearStructuralForceField} to the form of
the velocity field $\vel \approx \fv_{\rm ext}/\gamma=
f_0\sin(ky)\ev_x/\gamma$, which is straightforward to do as follows:
\begin{align}
  -\chi\nabla(\nabla\times\vel(\rv))^2 &=
  -\chi\frac{\partial}{\partial y}
  \Big( \frac{\partial}{\partial y}\frac{f_0}{\gamma} \sin(ky)  \Big)^2\ev_y\\
  &= -\frac{\chi f_0^2 k^2}{\gamma^2}
  \frac{\partial}{\partial y} \cos^2(ky)\ev_y\\
  &= \frac{\chi f_0^2 k^3}{\gamma^2} \sin(2ky)\ev_y.
\end{align}
The result for the structural (``migration'') force displays a
striking period doubling effect; the force tends to push particles
into the (two) regions of low shear rate. We can obtain the steady
state density profile from Eq.~\eqref{EQshearStructuralForceBalance},
again under the assumption of the density profile having only small
deviations from its bulk value, as follows:
\begin{align}
  \rho(y) &= \rho_{\rm b} -
  \frac{\chi f_0^2 k^2}{2\gamma^2 k_BT} \cos(2ky).
  \label{EQshearDensityProfileResult}
\end{align}
The shape of the density profile
[Eq.~\eqref{EQshearDensityProfileResult}] agrees very well with
results both from Brownian dynamics computer simulations and with
low-density results from exact numerical solution of the Smoluchowski
equation \cite{stuhlmueller2018structural}. From fitting the amplitude
to the simulation data, one can obtain the value of the migration
force amplitude $\chi$. Note that the structural force field causes no
dissipation, as $\vel\cdot\fv_{\rm struc}=0$, i.e., the force field is
orthogonal to the flow direction.

\citet{delasheras2020fourForces} have gone further in systematically
splitting the force balance relationship into flow and structural
parts, given respectively by
\begin{align}
  \gamma \vel\rt &= \fv_{\rm flow}\rt + \fv_{\rm ext,f}\rt,
  \\
  0 &= \fv_{\rm id}\rt + \fv_{\rm ad}\rt
  + \fv_{\rm struc}\rt + \fv_{\rm ext,s}\rt,
\end{align}
where $\fv_{\rm id}\rt=-k_BT \nabla\ln\rho\rt$. Here both the
superadiabatic and the external force field are split into flow and
structural contibutions: $\fv_{\rm sup}\rt = \fv_{\rm flow}\rt +
\fv_{\rm struc}\rt$ and $\fv_{\rm ext}\rt=\fv_{\rm ext,f}\rt+\fv_{\rm
  ext,s}\rt$. The different contributions are characterized by their
symmetry properties under motion reversal, which then also determines
the corresponding analytical form of the power functional
approximation. Furthermore, a vectorial decomposition yields force
components parallel and perpendicular to the flow field, which then
can be rationalized separately; see \citet{delasheras2020fourForces}
for details.

\subsection{Viscoelasticity and memory}
\label{SECviscoelasticity}
In light of significant recent interest in the study of memory kernels
as fundamental objects for collective dynamics
\cite{jung2016,jung2017,lesnicki2016}, \citet{treffenstaedt2019shear}
considered the hard sphere fluid exposed to transient switching
phenomena under shear. Both the spatial variation of the shear field
and its time dependence were highly idealized and chosen to trigger
strong response of the system. The results were obtained with
event-driven BD simulations \cite{scala2007} and the output from
simulation was rationalized on the basis of an approximative form of
the superadiabatic free power functional. As both the flow profile and
the force profile are available from the simulations, this strategy
allows for an unambiguous test of the theory. The shear protocol is
specified by an external force that varies as a step function in
space, i.e., with an infinite gradient at the shear plane(s)
perpendicular to the flow direction. This shear force field is first
instantaneously switched on, starting from a quiescent fluid, and
secondly instantaneously switched off after the system has reached a
steady shear state. Both transient processes, that after switching on
and that after switching off, were analyzed on the basis of the same
viscoelastic power functional approximation. (Recall that the sole
input to any superadiabatic force is the history of the kinematic
fields, which are known in the present case.)
Figures~\ref{FIGsedimentationAndShear}(b) and (c) show data from BD
compared to the results from the power functional approximation.

Different types of model forms for the memory kernel were considered.
The following first form constitutes a simple reference and it is
local in space and with a purely exponential temporal decay:
$K_L(\Delta \rv,\Delta t) = \delta(\Delta
\rv)\tau_M^{-1}\exp\left(-\Delta t/\tau_M\right)\Theta(\Delta t)$,
with $\tau_M$ indicating the memory time and $\Theta(\cdot)$ denoting
the Heaviside step function.  Here $\Delta \rv$ is the spatial
difference between the two coupled spacetime points, and $\Delta t$ is
their temporal difference.  The second version is spatially non-local
and hence can account for spatial correlation effects. This memory
kernel is assumed to have a diffusing form,
\begin{equation}
    K_D(\Delta\rv,\Delta t) = 
    \frac{e^{-\Delta \rv^2/(4 D_M \Delta t)-\Delta t/\tau_M}}
         {(4\pi D_M \Delta t)^{3/2} \tau_M} \Theta(\Delta t),
\end{equation}
with memory diffusion coefficient $D_M$.  The memory time
$\tau_\mathrm{M}$ again sets the time scale for the decay. In
principle, the parameters $\tau_\mathrm{M}$ and $D_\mathrm{M}$ are
determined by the underlying interparticle interactions. Adjusting
these parameters to match the simulation data results in very good
agreement with the BD results, see
Fig.~\ref{FIGsedimentationAndShear}(b, c). In particular, the global
motion reversal after switching off the shear force field is captured
correctly. The theory hence provides an explanation for the
effect. The superadiabatic forces that oppose the externally driven
current arise due to memory after switching off. The behaviour is of
genuinely viscoelastic nature: in the sheared steady state, viscous
forces oppose the current, but they elastically generate an opposing
current after switch-off. We refer the reader to
\citet{treffenstaedt2019shear} for the details of the theoretical
treatment and for further results and comparisons. The concepts were
also used by \citet{treffenstaedt2021dtpl} to investigate the dynamics
of the van Hove dynamical pair correlation function, where again
memory was found to play an important role. Furthermore, the self and
distinct splitting was complemented by splitting into total and
differential motion, as laid out in the following in a different
context.

\subsection{Superdemixing and laning}
\label{SECsuperdemixing}

\begin{figure*}
  \includegraphics[width=1.8\columnwidth,angle=0]{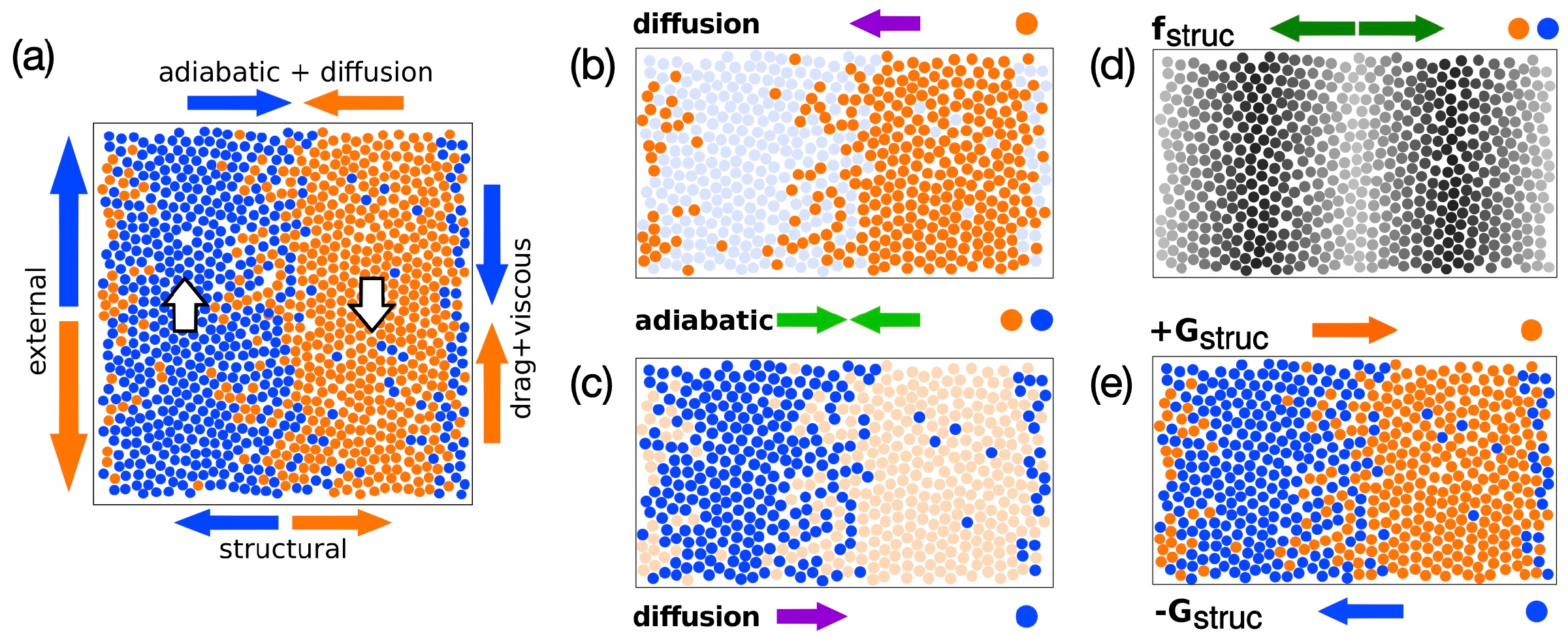}
  \caption{Lane formation in a counter driven binary mixture of purely
    repulsive particles.  The blue (red) particles are driven in the
    upward (downward) direction by a constant external force. (a) As a
    result two lanes that move against each other (white arrows) form
    spontaneously. The blue (red) arrows indicate forces that act on
    the blue (red) species; green arrows indicate forces that act
    irrespective of the colour.  (a) Each species experiences drag and
    viscous forces that tend to be directed against the species' local
    direction of motion. The adiabatic and diffusive forces tend to
    mix the system (top arrows). The nonequilibrium structural forces
    (bottom arrows) generate a superdemixing effect that stabilizes
    the laned state. (b) The diffusive force that acts on the red
    species tends to expand the red lane. (c) The corresponding effect
    for the blue particles acts in the opposite direction. Both
    species are pushed towards the interface by a (small) adiabatic
    force field (green arrows). (d) A structural nonequilibrium force,
    $\fv_{\rm struc}$, acts on both species and sustains a density
    depletion zone at the interface (light gray). (e) The differential
    superadiabatic force density $\Gv_{\rm struc}$ keeps each species
    inside of its lane and hence stabilizes the inhomogeneous steady
    state.  Adapted from \citet{geigenfeind2019laning}.  }
  \label{FIGlaning}
\end{figure*}

The formation of lanes is a prominent effect that occurs generically
once two (or more) different species are driven against each
other. Such situations arise e.g.\ in a binary colloidal mixture under
sedimentation, where light particles cream up and heavy particles
settle down, or when oppositely charged particles are exposed to a
uniform electric field. \citet{dzubiella2002} have presented an early
pioneering study of the effect. Experimental systems of magnetic
colloids that are placed above suitably patterned substrates offer
much sophisticated control of driving protocols
\cite{loehr2016,loehr2018}. \citet{geigenfeind2019laning} have
analyzed the forces that occur in the Brownian dynamics of a generic
two-dimensional binary repulsive sphere model. They recast the
internal force density for binary mixtures (as laid out in
Sec.~\ref{SECdtpl}) as
\begin{align}
  \Fv_{\rm int}^{(\alpha)}\rt &= 
  \rho_\alpha\rt \fv_{\rm int}\rt \pm \Gv_{\rm int}\rt,
\end{align}
where the two cases $\pm$ refer to species $\alpha=1,2$ and the force
field $\fv_{\rm int}$ acts on the total density $\rho_1+\rho_2$ and
the differential force density $\Gv_{\rm int}$ acts on the density
difference $\rho_2-\rho_1$. If the internal interactions are
independent of $\alpha$ (ideal mixture), then the species-resolved
superadiabatic force field was shown to have the structure
\begin{align}
  \fv_{\rm sup}^{(\alpha)}\rt&=
  \fv_{\rm visc}\rt \pm \frac{\Gv_{\rm drag}\rt}{\rho_\alpha\rt}
  \notag\\&\quad
  +\fv_{\rm struc}\rt \pm \frac{\Gv_{\rm struc}\rt}{\rho_\alpha\rt},
\end{align}
where the viscous force field $\fv_{\rm visc}\rt$ and the differential
drag force density $\Gv_{\rm drag}\rt$ both act parallel to the flow
direction; the total structural force field $\fv_{\rm struc}\rt$ and
the differential structural force density $\Gv_{\rm struc}\rt$ act
perpendicularly to the flow.  All four superadiabatic terms were
modelled by explicit kinematic functionals, which were shown to
reproduce the bare simulation data well. The force splitting concept
allows one to uniquely identify the physical mechanism that generates
the lane formation. It is the superadiabatic demixing force density
$\Gv_{\rm struc}\rt$ that drives the two species apart and stabilizes
the lanes. Figure \ref{FIGlaning} gives an illustration of this
effect, as well as of the action of all further forces in the driven
system. For the details of the analytical treatment of the problem,
together with the explicit power functional approximation, we refer
the reader to \citet{geigenfeind2019laning}.

\subsection{Active Brownian particles}
\label{SECactiveBrownianParticles}

\begin{figure*}
  \includegraphics[width=1.49\columnwidth,angle=0]{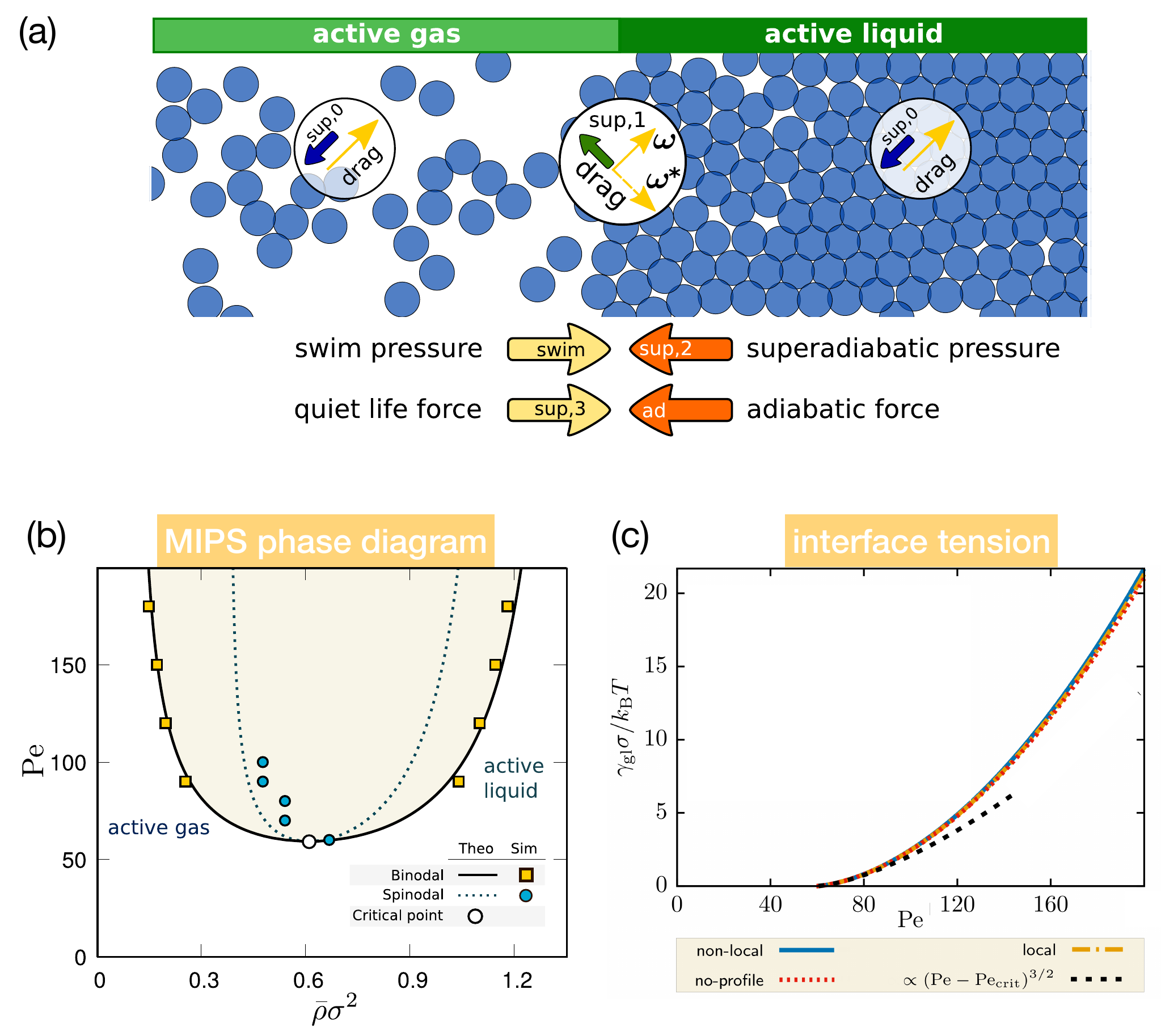}
\caption{Motility-induced phase separation of active Brownian
  particles. (a) Repulsive spheres that swim along an intrinsic
  orientation $\ov$ phase separate spontaneously into a dilute (active
  gas) and a dense (active liquid) phase. Drag forces hinder the
  swimming and slow down the motion (white circles). Thick arrows
  indicate different types of forces: The swim pressure arises from
  the interface polarization and it is balanced by the (intrinsic)
  superadiabatic pressure $\Pi_2$. The quiet life force compresses the
  liquid and balances the adiabatic force. The interparticle
  interactions are of is purely repulsive Lennard-Jones type, with
  length scale $\sigma$ and energy scale $\epsilon$.  (b) Phase
  diagram as a function of scaled density, $\bar\rho\sigma^2$, and
  Peclet number $\rm Pe$. Shown are the theoretical results for the
  binodal and the spinodal compared against simulation data (symbols)
  for the binodal \cite{paliwal2018chemicalPotential} and the spinodal
  \cite{stenhammar2013prl}.  (c) Interface tension $\gamma_{\rm gl}$
  as a function of Peclet number $\rm Pe$, as obtained from the
  present nonequilibrium square-gradient power functional treatment of
  the problem. Here $\gamma_{\rm gl}\geq 0$ in contrast to the
  findings by \citet{bialke2015}. Three methods (non-local,
  no-profile, local) give a unique result; the scaling with
  (mean-field) exponent $3/2$ is shown near the critical point.
  Adapted from \citet{hermann2019acif} (a,b) and
  \citet{hermann2019tension} (c).}
  \label{FIGabpForces}
\end{figure*}

Active Brownian particles have become a prototypical model for the
study of collective nonequilibrium phenomena. The power functional
framework is inherently set up to describe such driven systems, once
the generalization to the angular degrees of freedom (which describe
the direction of the active swimming) is performed. We detail the
theoretical layout in the following.

We first follow \citet{krinninger2016prl} and
\citet{krinninger2019jcp} who considered the drag effect in active
Brownian particles.  Here the $i$-th particle is described by position
$\rv_i$ and (unit vector) orientation $\ov_i$. We consider
two-dimensional systems in a volume $V$; the orientation $\ov$ is
parameterized by $\ov=(\cos\varphi,\sin\varphi)$, where $\varphi$ is
the angle of the particle orientation against the $x$-axis. The set of
orientations $\ov_1,\ldots,\ov_N\equiv \ov^N$ constitutes internal
degrees of freedom. The interparticle interaction potential $u(\rv^N)$
is that of spherical particles, independent of $\ov^N$. The particles
are self-propelled with a body force field,
\begin{align}
  \fv_{\rm swim}(\rv,\ov,t) &= \gamma s \ov,
  \label{EQswimForceField}
\end{align}
which is formally an addition to the external one-body force field.
The force \eqref{EQswimForceField} propels the particles into their
forward direction with strength $\gamma s$; the propulsion is
homogeneous in space and time. The parameter $s$ has the
interpretation of the speed of free swimming (i.e., without the
effects of collisions with other particles). The one-body fields
depend also on orientation $\ov$; hence, we have the density
distribution $\rho(\rv,\ov,t)$, the (translational) current
distribution $\Jv(\rv,\ov,t)$ and the orientational current
distribution $\Jv^\omega(\rv,\ov,t)$. As there are no explicit torques
acting in the system, the orientational motion is purely diffusive,
\begin{align}
  \Jv^\omega\one &=  
  -\frac{k_BT}{\gamma^\omega} \nabla^\omega \rho\one,
  \label{EQabpRotationalCurrent}
\end{align}
where $1\equiv\rv,\ov,t$ is a shorthand, $\gamma^\omega$ is the
friction constant for the (overdamped) rotational motion and
$\nabla^\omega$ is the orientational derivative, which in two spatial
dimensions is simply $\nabla^\omega \to \partial/\partial\varphi$. The
rotational current features in the continuity equation, which has the
form
\begin{align}
  \dot\rho\one &=
  -\nabla\cdot \Jv\one - \nabla^\omega\cdot \Jv^\omega\one.
  \label{EQabpContinuityEquation}
\end{align}
We consider steady states, and hence $\dot\rho\one=0$. Furthermore for
isotropic, homogeneous fluid steady states $\rho(\rv,\ov,t)=\rho_{\rm
  b}=\rm const$ and the current distribution is $\Jv(\rv,\ov,t)=J_{\rm
  b}\ov$, where $\rho_{\rm b}$ is the bulk fluid density, and $J_{\rm
  b}$ is the strength of the ``forward'' current (of the swimming
motion into the $\ov$ direction). For constant density, the rotational
current \eqref{EQabpRotationalCurrent} clearly vanishes
[$\Jv^\omega\one=0$], and hence the continuity equation
\eqref{EQabpContinuityEquation} is satisfied.

The task is to find a relationship $J_{\rm b}(\rho_{\rm b})$ that
would act like a dynamical equation of state and determine the average
current, given the fluid density of the system. We assume that the
superadiabatic free power contains a dissipative contribution,
\begin{align}
  P_t^{\rm exc}\rhoJ &= \frac{\gamma}{2}
  \int \!\!d1 \!\!\int \!\!d2 \rho(1) \rho(2) (\vel(1)-\vel(2))^2 M(1,2),
  \label{EQabpPtexcDrag}
\end{align}
where $1\equiv \rv,\ov$ and $2\equiv\rv',\ov'$ are again short-hand
notations and $M(1,2)$ is a (density-dependent) correlation kernel
that couples the two ``configurational'' points to each other and the
microscopic velocity field $\vel$ is defined as usual
[$\vel(1)=\Jv(1)/\rho(1)$]. Note that the squared velocity difference
in Eq.~\eqref{EQabpPtexcDrag} is a scalar measure of the crossflow
that occurs in the system. (Particles with different orientations tend
to collide, given a suitable spatial setup.) The squared velocity
difference is in Eq.~\eqref{EQabpPtexcDrag} multiplied by the density
distribution at both points, in order to give a statistical weight to
the actual occurrence of such collisions. Hence, besides the fact that
Eq.~\eqref{EQabpPtexcDrag} constitutes a simple low-order power series
term, we can find a quite clear physical interpretation of an
interflow dissipation measure.

\citet{hermann2019acif} have shown that motility-induced phase
separation into active gas and liquid phases is described when taking
into account further superadiabatic force contributions, besides drag,
as previously considered. Primarily, these consist of pressure and
``quiet life'' chemical potential terms. \citet{hermann2020longActive}
give much further background for the theory, and in particular of the
angular Fourier decomposion methods, as also used for the exact
solution of ideal active sedimentation in two dimensions
\cite{hermann2018activeSedimentation}. The theory yields the
interfacial tension in a natural way, as demonstrated by
\citet{hermann2019tension} in a nonequilibrium generalization of the
classical square-gradient interfacial theory
\cite{RowlinsonWidomBook}. Four different types of superadiabatic
force contributions are shown to be relevant [$\Fv_{\rm
    sup}(\rv,\ov)=\sum_{\alpha=0}^3 \Fv_{\rm sup,\alpha}(\rv,\ov)$] in
a full position- and orientation-resolved description.

Briefly, both $\Fv_{\rm sup,0}$ and $\Fv_{\rm sup,1}$ describe drag,
with the isotropic component leading to the reduction of the mean swim
speed in bulk.  The anisotropic drag force density $\Fv_{\rm sup,1}$
occurs in inhomogeneous situations and is smaller in magnitude than
$\Fv_{\rm sup,0}$. The force density $\Fv_{\rm sup,2}=-\nabla \Pi_2$
is naturally expressed as the gradient of a spherical superadiabatic
pressure $\Pi_2$.  This intrinsic term balances the swim pressure
$P_{\rm swim}$, with corresponding force density $-\nabla P_{\rm
  swim}$ that occurs due to the self propulsion force. The intrinsic
quiet life force density $\Fv_{\rm sup,3}=-\rho\nabla\nu_3$ originates
naturally from a nonequilibrium chemical potential $\nu_3$ and it is
this term that balances the strong adiabatic repulsion that
(primarily) occurs in the liquid phase. Ultimately $\nu_3$ drives the
motility-induced phase separation into dilute and dense steady states.

Figure \ref{FIGabpForces} summarizes these results, including an
illustration of the nonequilibrium phase coexistence and direction of
the relevant forces [Fig.~\ref{FIGabpForces}(a)], the theoretical bulk
phase diagram [Fig.~\ref{FIGabpForces}(b)], as compared to simulation
data, and the behaviour of the interfacial tension $\gamma_{\rm gl}$
[Fig.~\ref{FIGabpForces}(c)] of the free interface between the phase
separated bulk phases.  The theory yields $\gamma_{\rm gl}\geq 0$ in
contrast to the findings by \citet{bialke2015}; see also
\citet{speck2020} and
\citet{delasheras2021tension}. \citet{hermann2020polarization} have
derived an exact sum rule that links the total interface polarization
$M_{\rm tot}$ (i.e., the overall degree of orientational order that
particles near the interface exhibit) to the value of the swim current
in the adjacent bulk phases: $M_{\rm tot}/L=(J_g-J_l)/(2D_{\rm rot})$,
where $L$ is the length of interface, $J_g$ ($J_l$) is the current in
the active gas (liquid), and $D_{\rm rot}$ is the rotational diffusion
constant. Hence, \citet{hermann2020polarization} conclude that the
interface polarization is a state function; the power functional
approximation respects this exact property.  The sum rule itself was
verified with a light-controlled Janus-type swimmer in the vicinity of
an abrupt activity step, both in an experimental setup and using
numerical solution of the underlying Smoluchowski equation
\cite{auschra2020,soeker2020}.  On the basis of Noether's Theorem
\citet{hermann2021noether,hermann2021noetherPopular} have clarified
the role of interfacial forces in motility-induced phase separation.

\section{Conclusions and outlook}
\label{SECconclusions}

In conclusion, I have described approaching the dynamics of coupled
many-body systems in a functional setting. The functional point of
view allows for systematic coarse-graining, or synonymously
integrating out degrees of freedom, while retaining a microscopically
sharp description, both in space and in time. The fundamental
variational variables are one-body fields that depend on time and one
a single space coordinate. The existence of a generating functional
ensures that the description is complete, i.e., that two- and
higher-body correlation functions, again microscopically resolved in
space and in time, are contained in the treatment. The benefits of
this variational setup are the relative ease of carrying out practical
calculations as well as the direct access to physical effects, because
the one-body fields, corresponding to local density, velocity and
acceleration, are both manageable in terms of numerical
representability and they also admit direct physical intuition to be
exploited. The kinematic fields represent the dynamical behaviour of
the system directly. No coarse-graining in a hydrodynamical sense of
smoothening is implied. Rather correlations on the particle scale are
accessible. Concretely, for the case of two-body correlation
functions, both the dynamical test-particle limit as well as the
nonequilibrium OZ relations allow access.

While the existence of a generating functional guarantees and defines
the overarching theoretical structure, the approach is not a merely
formal one. This is, in particular, due to the description of the
dynamics as stemming both from adiabatic and superadiabatic
contributions, where the former are treated within the
well-established and powerful framework of density-functional
theory. The latter, superadiabatic effects are generated from a unique
and increasingly well-characterized object, the free power
functional. For the dynamical system the free power plays a role
similar to that of the free energy in equilibrium.

While the power functional variational principle may appear formal at
first glance, it has deep roots in the Gibbs-Appell-Gauss formulation
of classical mechanics. The Euler-Lagrange equation that results from
the free power minimization principle has the direct physical
interpretation of a force balance relationship (or Newton's second
law). This situation renders the task of constructing approximations
physically intuitive, as we have shown for a number of nonequilibrium
phenomena. Moreover, the concept allows straightforward access to the
physics via particle-based simulation work. Here we have covered
techniques such as custom flow that allow one to access relevant
information for the genuine structure and inner working of theory,
i.e., for the challenging task of findings approximations for the
superadiabatic free power functional.

I have described the essentials of density functional theory as the
appropriate technique to describe the adiabatic state, which is a
reference concept that allows to uniquely define those contributions
to the dynamics that can be understood on the basis of an equilibrium
free energy. This possibility might seem surprising to some readers,
given that the system is genuinely driven out of equilibrium and no
near-equilibrium, linear response or similar restriction applies. But
within the functional setting the adiabatic contribution is uniquely
identified as that part of the free power that is an instantaneous
density functional.  The superadiabatic contributions to the dynamics
are of genuine nonequilibrium character, as the corresponding
functional generator depends on the motion of the system, as
characterized by the flow, acceleration and time-dependent density
profile. The functional dependence is nonlocal in space and time,
where the latter dependence is causal and, as described in
e.g.\ dynamical test-particle and transient shear settings, the
dependence can be modelled by memory kernel techniques.

The custom flow concept and simulation algorithm, both for BD and MD,
allows one to implement directly the functional point of view of
many-body dynamics. Here the reversed map from the motion of the
system to the corresponding external driving force field is readily
explicitly constructed. The technique solves the inherent inverse
problem in an algorithmically straightforward and computationally
efficient way. It allows both for scrutiny of the functional concepts
as well as giving a powerful means for testing and developing concrete
power functional approximations.  I have described a range of such
concrete power functional approximations for dynamical phenomena,
ranging from the van Hove function to characterizes the equilibrium
dynamics of a quiescent bulk liquid, to viscoelastic, structural and
memory effects of sheared fluids, as well as to nonequilibrium
structure formation, such as laning in counterdriven mixtures and
motility-induced phase separation of active Brownian particles.

The underlying power functional approximations are based on the
unifying concept of kinematic dependence on the flow, and both local,
semi-local (i.e., via gradients), but also genuine spatio-temporal
nonlocal functional dependencies. As shown in the description of the
adiabatic state, these mathematical tools are formally akin to those
used in classical DFT, but with the import physical distinction of
representing kinematic functionals that operate on the motion of the
system, rather than mere functional dependence on the density profile.
The kinematic dependence is grounded in the many-body foundation of
the theory, where the central minimization principle is akin to the
Gibbs-Appell-Gaussian formulation of classical mechanics. Levy's
search method then facilitates the construction of the reduced
one-body description. The theory requires a modest amount of
functional calculus; Appendixes \ref{SECHamiltonsActionPrinciple} and
\ref{SECfunctionalDerivativeAppendix} give the necessary background
for readers who want to increase their knowledge on the topic.

Much of the concrete power functional work has been carried out to
date for the case of overdamped Brownian dynamics and tightly
interwoven with simulation methodology. These model dynamics are
simple and they represent, in a simple fashion, the motion of
colloids. The generality of the power functional approach is apparent
through the quantum and classical inertial dynamics, as represented by
the Schr\"odinger and Liouville equation, respectively. We have
presented significant background for readers who want to access the
original publications. The presented power functional approximations
both validate the variational concept and they also demonstrate that
the functional point of view allows to gain insights into the genuine
physics at play. The involved objects carry profound physical meaning
well beyond the functional nature. Examples thereof include the
flow-structure splitting of forces in Brownian dynamics
\cite{delasheras2020fourForces}, the total-differential decomposition
of forces in mixtures \cite{geigenfeind2019laning}. Furthermore, the
memory structure is such that very specific combinations of the
kinematic fields occur in the memory integral
\cite{treffenstaedt2021dtpl}.

The functional point of view both offers insights into the deep
structure of nonequilibrium dynamics, as well as being an excellent
candidate for providing a computational framework for the systematic
and comprehensive description and classification of many-body dynamics
out of equilibrium, similar to the role that DFT plays as the gold
standard for the behaviour in equilibrium. As no universal consensus
on the description of nonequilibrium dynamics has been reached, the
material covered in this review puts forward the power functional
theory as a competitive contender. Significant potential exists for
elucidating the mechanisms that govern the crossover regimes between
the considered types of dynamics, i.e., from quantum to classical, and
from inertial to overdamped dynamics. There is much potential for
cross fertilisation between these regimes.  Furthermore, it is highly
relevant to address from a general point of view open problems such as
nonequilibrium phase transitions [see \citet{lips2018basep} for a
  recent striking case] and the foundations of nonequilibrium
thermodynamics. On a practical level, it is easier to construct
approximations for scalars ($P_t^{\rm exc}$) than for force
fields. Having a reliable theoretical scheme also allows numerical
results to be obtained faster than what is possible in simulation
work.

Significant work lies ahead, both in conceptual terms, of a practical
nature, and for concrete systems and physical phenomena. We describe
several potentially very fruitful directions in the following.  It
would be very interesting to relate power functional theory to the
concept of quasi-universality of simple fluids, as pioneered by Dyre
and coworkers, see \citet{dyre2016} for a review, as well as to the
\citet{rosenfeld1977entropyScaling} entropy scaling; see
\citet{mittal2006prl} and \citet{mittal2008prl} for applications and
\citet{dyre2018} for a recent review. It would be worthwhile to
investigate superadiabatic effects in processes such as spinodal
decomposition \cite{evans1979,archer2004ddft} and in ``gravitational''
collapse of colloidal monolayers \cite{bleibel2014sm}.  Exploring
possible connections to generalized Langevin equations (see
\citet{amati2019} for recent work) and to the use of rate equations
(see \citet{dixit2018} for recent work) would be valuable.  It would
be interesting to see how power functional theory can be useful in
going beyond classical nucleation theory
\cite{lutsko2018extendingCNT}, investigating the ``dynamical barrier''
in many-body correlations in hard spheres \cite{robinson2019}, and
making connections to stochastic thermodynamics
\cite{seifert2010,seifert2012review,leonard2013}, and in particular to
the concept of the entropy production rate, see the insightful work by
\citet{speck2016pre}.

Furthermore, investigating nonisothermal conditions would be very
worthwhile, where recent work addressed the effects of fluctuating
hydrodynamics on Brownian motion \cite{falasco2016}.  Relating to the
behavior of memory kernels in Molecular Dynamics
\cite{lesnicki2016,jung2016,jung2017} is worthwhile.  Identifying the
superadiabatic force contributions in driven mixtures of spheres and
spherocylinders could be revealing for the observed ordering phenomena
in this system \cite{lueders2019}. Applying power functional theory to
the task of classifying new states of active matter
\cite{menzel2016perspective} lends an exciting perspective.  Even very
simple types of external fields, such as gravity, can induce complex
phenomena. Obtaining a predicitive quantitative framework for the
sedimentation dynamics of colloidal mixtures
\cite{delasheras2012sciRep,delasheras2013phaseStacking,
  delasheras2014sedimentationLiquid2014,
  hermann2018activeSedimentation,eckert2021} would be valuable.

It would be interesting to explore the consequences of the
nonequilibrium Ornstein-Zernike relation
\cite{brader2013nozOne,brader2014nozTwo} and in particular to apply it
to concrete problems. Given the central role of the equilibrium
Ornstein-Zernike equation for liquid state theory, one would expect
its nonequilibrium version to play a similar crucial future role in
the description of dynamical phenomena in complex liquids. When it is
flanked by the dynamical test-particle limit
\cite{archer2007dtpl,hopkins2010dtpl,brader2015dtpl}, one appears to
be well equipped for a fresh view on the dynamical two-body structure
of complex systems, and to explore fundamental links to, say,
mode-coupling theory [see \citet{janssen2018} for a recent review] in
order to investigate transient dynamics of colloidal liquids
\cite{zausch2008}. The relationship to recent progress beyond the
usual hydrodynamic description obtained in the Zwanzig-Mori
projection-operator formalism \cite{vogel2020} is worth
exploring. Furthermore the dynamical sum rules that follow from
Noether's theorem \cite{hermann2021noether} provide valuable
resources.

Lastly, and arguably most importantly and implicit in all of the
previously mentioned items, we point out the importance of future
developments of first-principles microscopically-based approximations
to the excess power functional. This task surely is highly
challenging, but with significant potential for ultimate high reward.

\vspace{5mm} {\bf Acknowledgments.} I thank Joseph Brader, Daniel de
las Heras and Sophie Hermann for their input and great scientific
contributions. Daniel and Sophie are acknowledged for careful
proof-reading and significant feedback on the article, for which I
also thank the editors and the referees. All remaining errors are
mine. I thank Elias Bernreuther, Moritz Br\"utting, Brams Dwandaru,
Tobias Eckert, Andrea Fortini, Thomas Geigenfeind, Paul Hopkins,
Nikolai Jahreis, Philip Krinninger, Jonas Landgraf, Johannes Renner,
Florian Samm\"uller, Thomas Schindler, Nico Stuhlm\"uller, Lucas
Treffenst\"adt, and Thomas Trepl, for their many and significant
scientific contributions.  I~am grateful for my stimulating and useful
exchange with Bob Evans, Thomas Fischer, and Roland Roth. This work is
supported by the German Research Foundation (DFG) via Project
No.~436306241.

\appendix
\section{Functional calculus}
\subsection{Variations and Hamilton's principle}
\label{SECHamiltonsActionPrinciple}
Consider a classical mechanical system with~$M$ degrees of freedom
represented by generalized coordinates $q_1,\ldots,q_M\equiv\qv$ and
by corresponding generalized velocities $\dot q_1,\ldots,\dot
q_M\equiv \dot\qv$. The Lagrangian $L(\qv,\dot\qv,t)$ specifies the
system via the difference of total kinetic and potential energy.

The action integral is then defined as
\begin{align}
  S &= \int_{t_1}^{t_2} dt L(\qv(t),\dot\qv(t),t),
  \label{EQactionIntegral}
\end{align}
for given start and end configurations, $\qv_1=\qv(t_1)$ and
$\qv_2=\qv(t_2)$, i.e., as specfied by all values of the generalized
coordinates, at an intial time $t_1$ and a final time~$t_2$.

There are several equivalent ways to carry out the variation of the
action integral. Often one introduces an auxiliary function
$\boldsymbol\epsilon(t)$ that ``perturbs'' the path, via
$\qv(t)\to\qv(t)+\boldsymbol\epsilon(t)$ and corresponding change in
velocity. Taylor expanding in the perturbation to first order then
gives the desired result.

The method via functional differentiation is more formal. In order to
calculate functional derivatives, $\delta/\delta f(x)$, with respect
to a function $f(x)$ one needs in practice often two rules: First, the
rules of differentiation in several variables apply (upon replacing
sums over indices of variables by integrals over argument
$x$). Secondly, $\delta f(x)/\delta f(x')=\delta(x-x')=\delta(x'-x$),
where $\delta(\cdot)$ is the Dirac distribution (which is even). The
method of functional differentiation might seem less intuitive at
first glance, but as it is entirely algebraic it is powerful in
practice. Our account is fully explicit in spelling out all function
arguments. In practice, this can be advantageous over commonly used,
more compact notation; see \citet{hansen2013}.

As a demonstration of functional differentiation, we apply the
functional derivative to the action integral as follows:
\begin{align}
  \frac{\delta S}{\delta \qv(t')} &=
  \frac{\delta}{\delta \qv(t')}
  \int_{t_1}^{t_2} dt L(\qv(t),\dot\qv(t),t)\\
  =&\int_{t_1}^{t_2}\! dt 
  \frac{\delta}{\delta \qv(t')}
  L(\qv(t),\dot\qv(t),t)\\
  =& \int_{t_1}^{t_2} \!dt
  \left(
  \frac{\partial L}{\partial \qv(t)}\cdot
  \frac{\delta \qv(t)}{\delta \qv(t')}
  +\frac{\partial L}{\partial \dot\qv(t)}\cdot
  \frac{\partial\dot\qv(t)}{\partial \qv(t')}
  \right)\\
  =&\int_{t_1}^{t_2} \!dt
  \left(
  \frac{\partial L}{\partial\qv(t)}\cdot{\bf 1}\delta(t-t')
  +\frac{\partial L}{\partial \dot\qv(t)}\cdot
  \frac{d}{dt}\frac{\delta \qv(t)}{\delta \qv(t')}
  \right)\\
  =&\int_{t_1}^{t_2} \!dt
  \left(
  \frac{\partial L}{\partial \qv(t)}\delta(t-t')
  -\Big(\frac{d}{dt}\frac{\partial L}{\partial \dot\qv(t)}
  \Big)\!\cdot\!{\bf 1} \delta(t-t')\!\!
  \right)\notag\\
  &\qquad + \frac{\partial L}{\partial \dot \qv}\delta(t-t')\Big|_{t_1}^{t_2}
  \label{EQactionWithBoundaryTerm}\\
  =& \frac{\partial L}{\partial \qv(t')}
    -\frac{d}{dt}\frac{\partial L}{\partial \dot\qv}\Big|_{t=t'},
\end{align}
where $\bf 1$ is the $M\times M$ unit matrix. The boundary term in
Eq.~\eqref{EQactionWithBoundaryTerm} vanishes for $t_1<t'<t_2$.
Multiplication with $\bf 1$ yields the vector to its left in
Eq.~\eqref{EQactionWithBoundaryTerm}.  Renaming the time variable $t'$
as $t$ and requesting stationarity, i.e., a vanishing derivative,
leads to the followsin Lagrange equations of motion:
\begin{align}
  \frac{d}{dt}\frac{\partial L}{\partial \dot\qv} 
  -\frac{\partial L}{\partial \qv} = 0.
\end{align}

To illustrate the method further, we also derive Hamilton's equations
of motion. Rewrite the action integral \eqref{EQactionIntegral} as
\begin{align}
  S &= \int_{t_1}^{t_2} dt
  \Big( \sum_i \dot q_ip_i - H(\qv,\pv,t) \Big),
\end{align}
where $H(\qv,\pv,t)$ is the Hamiltonian and $p_1,\ldots, p_M\equiv
\pv$ are the generalized momenta corresponding to $\qv$.  We vary the
action both in momentum and in coordinates, independently of each
other (using, as above and as is common, ``partial'' functional
derivatives). We start as follows with the variation in the $j$-th
generalized momentum:
\begin{align}
  \frac{\delta S}{\delta p_j(t')} &=
  \int_{t_1}^{t_2} dt
  \frac{\delta}{\delta p_j(t')}
  \Big(\sum_i\dot q_ip_i - H(\qv,\pv,t)
  \Big)\\
  &= \int_{t_1}^{t_2} dt \Big(
  \sum_i \dot q_i(t)\frac{\delta p_i(t)}{\delta p_j(t')}
  -\frac{\delta H(\qv,\pv,t)}{\delta p_j(t')}
  \Big)\\
  &=\int_{t_1}^{t_2}dt 
  \sum_i\Big(\dot q_i(t)\delta_{ij}\delta(t-t')
  -\frac{\partial H}{\partial p_i(t)}
  \frac{\delta p_i(t)}{\delta p_j(t')}
  \Big)\\
  &=\dot q_j(t') - \frac{\partial H}{\partial p_j} \Big|_{t'},
\end{align}
where we have used $\delta p_i(t)/\delta
p_j(t')=\delta(t-t')\delta_{ij}$, with $\delta_{ij}$ being the
Kronecker symbol.  We rename $t'\to t$ and from requesting $\delta
S/\delta \pv(t)=0$ we obtain
\begin{align}
  \dot \qv &= \frac{\partial H}{\partial \pv},
  \label{EQHamiltonsEquationsOfMotionsOne}
\end{align}
which is one part of Hamilton's equations of motion.

We next vary in coordinates:
\begin{align}
  \frac{\delta S}{\delta q_j(t')} &=
  \int_{t_1}^{t_2} dt
  \frac{\delta}{\delta q_j(t')}
  \Big(\sum_i \dot q_i p_i - H(\qv,\pv,t)
  \Big)\\
  &=\int_{t_1}^{t_2} dt
  \Big(
  -\frac{\delta}{\delta q_j(t')} \sum_i q_i \dot p_i
  -\frac{\partial H}{\partial q_j(t')}\delta(t-t')
  \Big)\notag\\&\quad
  +\frac{\delta}{\delta q_j(t')} \sum_i q_i p_i\Big|_{t_1}^{t_2}\\
  &= -\frac{\partial H}{\partial q_j}\Big|_{t'}
  -\dot p_j(t')
  +p_j\delta(t-t')\Big|_{t_1}^{t_2},
\end{align}
where the boundary term\footnote{For modern work on the status of the
  boundary values in Hamilton's principle, see, e.g.,
  \citet{galley2013prl}.} vanishes for $t_1<t'<t_2$.  From imposing
that $\delta S/\delta\qv(t)=0$ we conclude that
\begin{align}
  \dot\pv &= -\frac{\partial H}{\partial \qv},
  \label{EQHamiltonsEquationsOfMotionsTwo}
\end{align}
which together with Eq.~\eqref{EQHamiltonsEquationsOfMotionsOne} forms
the complete set of Hamilton's equations of motion.

\subsection{Spatiotemporal and time-slice functional derivatives} 
\label{SECfunctionalDerivativeAppendix}

Functional dependencies can be on functions of a single argument, such
as on time as described above in
Appendix~\ref{SECHamiltonsActionPrinciple} for the action
integral. The more general case involves functions of several
variables and we describe the generalization, which is
straightforward, in the following. We take position $\rv=(x,y,z)$ as
an example, where $x,y,z$ are the Cartesian components.  Consider a
generic functional $A[f]$, where $f(\rv)$ is its argument function.
The rules of functional differentiation laid out in
Sec.~\ref{SECHamiltonsActionPrinciple} then continue to hold,
including the functional chain rule, etc. When building the functional
derivative of $f(\rv)$ with respect to itself, $f(\rv')$, one has to
take account of the increased dimensionality of the arguments $\rv$
and $\rv'$.  The result is $\delta f(\rv)/\delta
f(\rv')=\delta(\rv-\rv')$, where $\delta(\cdot)$ indicates the Dirac
distribution in three dimensions. The dimensionality is implicit in
the notation, as the argument $\rv-\rv'$ is a three-dimensional
vector. [More explicit notation is $\delta^{(3)}(\rv) \equiv
  \delta^{(1)}(x)\delta^{(1)}(y)\delta^{(1)}(z)$ where the superscript
  indicates the dimension.]

Spatio-temporal functional derivatives with respect to a function
$f(\rv,t)$ follow the same scheme, with $\delta f(\rv,t)/\delta
f(\rv',t')=\delta(\rv-\rv')\delta(t-t')$, where the spatial part of
the result is again a three-dimensional Dirac distribution (multiplied
by a one-dimensional Dirac distribution in time). Note that two
distinct positions, $\rv$ and $\rv'$, as well as two distinct times,
$t$ and $t'$, are involved. Physically speaking, given some functional
$A[f]$ of $f(\rv,t)$, functionally differentiating, $\delta
A[f]/\delta f(\rv,t)$, monitors the response of $A[f]$ to a change in
argument function at spacetime point $\rv,t$.

An important special case involves only a single time
argument. Consider again the situation of a time-dependent function
$f(\rv,t)$, but we wish to disregard the temporal dependence and only
allow position changes. A simple example is an instantaneous position
integral, $A_t[f]=\int d\rv f(\rv,t)$, where $t$ is fixed and hence
treated as a constant. Then the ``time-slice'' functional derivative
yields $\delta A_t[f]/\delta f(\rv,t)=\int d\rv \delta f(\rv,t)/\delta
f(\rv',t)=\int d\rv \delta(\rv-\rv')=1$ for an appropriate integration
domain. Hence, the fundamental rule for the time slice derivative is
$\delta f(\rv,t)/\delta f(\rv',t)=\delta(\rv-\rv')$ with no time
dependence on the right-hand side. Here the time slice derivative is
notated by the same time argument $(t)$ appearing twice on the
left-hand side.

\subsection{Gibbs-Appell-Gaussian classical mechanics}
\label{SECgagOne}
We describe the Gibbs-Appell-Gaussian formulation of classical
mechanics following the presentation by \citet{EvansMorriss}; an
excellent pedagogical account is given by \citet{Desloge1988}.
Consider Newton's second law, using Cartesian coordinates in a system
with no constraints,
\begin{align}
  m_i\ddot \rv_i(t) = \fv_i(t)
\end{align}
where $\fv_i$ is the total force acting on particle $i$. Introduce the
acceleration
\begin{align}
  \av_i(t) = \ddot \rv_i(t),
 \end{align}
at time $t$. The task is to determine $\av_i(t)$.  In order to do so,
keep all positions $\rv_i(t)$ and all velocities
$\vel_i(t)=\dot\rv_i(t)$ fixed at time $t$. Then there are then two
alternatives.

\begin{itemize}
\item[(i)] Determine the acceleration from Newton's second law
  according to
\begin{align}
  \av_i(t) &= \frac{\fv_i(t)}{m_i},
  \label{EQaccerationsFixed}
\end{align}
where the right-hand side is (and must be) known. This fixes the
dynamics and constitutes Newton's version of classical mechanics.

\item[(ii)] The alternative is to construct a scalar (cost) function
  $G_t(\rv^N,\vel^N,\av^N,t)$, such that at the minimum with respect
  to all $\av_i(t)$ Newton's second law holds. Here the accelerations
  $\av_i(t)$ are considered to be trial functions. (Conceptually, this
  is a signifiant step in addition to the thinking behind Hamilton's
  principle.) Define the Gibbs-Appell-Gaussian as
\begin{align}
  G_t &= \sum_i \Big(
  \frac{m_i}{2}\av_i^2(t) - \fv_i(t)\cdot\av_i(t)
  \Big).
  \label{EQdefinitionGAGfunction}
\end{align}
At the minimum,
\begin{align}
  \frac{\partial G_t}{\partial \av_i(t)} &= 0 \quad \rm (min),
\end{align}
from which we conclude
\begin{align}
  m_i\av_i(t)-\fv_i(t) &= 0,
\end{align}
as desired, i.e., Eq.~\eqref{EQaccerationsFixed}. Hence, the result is
analogous to that of method i) above. In contrast to method i), here
the accelerations have the status of trial variables, i.e., they do
not have to possess the correct physical values at the stage of
Eq.~\eqref{EQdefinitionGAGfunction}.
\end{itemize}

When one puts things into context, classical mechanics features three
alternative variational principles, attributable to d'Alembert:
$\sum_i (m_i\ddot\rv_i-\fv_i)\cdot\delta \rv_i=0$, Jourdain: $\sum_i
(m_i\ddot\rv_i-\fv_i)\cdot\delta \dot \rv_i=0$, and Gibbs, Appell, and
Gauss: $\sum_i (m_i\ddot\rv_i-\fv_i)\cdot\delta \ddot \rv_i=0$. Here
the variations are performed respectively in position $\delta\rv_i$,
in velocity $\delta \dot\rv_i$, or in acceleration
$\delta\ddot\rv_i$. See \citet{EvansMorriss} for a thorough account,
including the treatment of contraints.


\begin{thebibliography}{31}

\bibitem[Amati \etal, 2019]{amati2019}
Amati, G.,  H. Meyer, and T. Schilling,
2019,
``Memory effects in the Fermi-Pasta-Ulam model,''
\href{https://doi.org/10.1007/s10955-018-2207-6}
{J. Stat. Phys. {\bf 174}, 219}.

\bibitem[Anero \etal, 2013]{anero2013}
Anero, J. G., P. Espa\~nol, and P. Tarazona,
2013,
``Functional thermo-dynamics: A generalization of dynamic
density functional theory to non-isothermal situations,''
\href{https://doi.org/10.1063/1.4811655}
{J. Chem. Phys. {\bf 139}, 034106}.

\bibitem[Angioletti-Uberti \etal, 2014]{angioletti2014}
Angioletti-Uberti, S., M. Ballauff, and J. Dzubiella,
2014,
``Dynamic density functional theory of protein adsorption
on polymer-coated nanoparticles,''
\href{https://doi.org/10.1039/c4sm01170h}
{Soft Matter {\bf 10}, 7932}.

\bibitem[Angioletti-Uberti \etal, 2018]{angioletti2018}
Angioletti-Uberti, S., M. Ballauff, and J. Dzubiella,
2018,
``Competitive adsorption of multiple proteins to nanoparticles:
the Vroman effect revisited,''
\href{https://doi.org/10.1080/00268976.2018.1467056}
{Mol. Phys. {\bf 116}, 3154.}

\bibitem[Archer and Evans, 2001]{archer2001}
  Archer, A. J., and R. Evans,
  2001,
  ``Binary Gaussian core model: Fluid-fluid phase separation,
  and interfacial properties,''
  \href{https://doi.org/10.1103/PhysRevE.64.041501}
  {Phys. Rev. E {\bf 64}, 041501.}

\bibitem[Archer \etal, 2002]{archer2002}  
  Archer, A. J., R. Evans, and R. Roth,
  2002,
  ``Microscopic theory of solvent-mediated long-range forces: Influence of wetting,''
  \href{https://doi.org/10.1209/epl/i2002-00137-2}
  {EPL (Europhys. Lett.) {\bf 59}, 526.}

\bibitemESSENTIAL{Archer and Evans, 2004}{archer2004ddft}
Archer, A. J., and R. Evans, 
2004,
``Dynamical density functional theory and its application to
spinodal decomposition,''
\href{https://doi.org/10.1063/1.1778374}
{J. Chem. Phys. {\bf 121}, 4246.}

\bibitem[Archer and Rauscher, 2004b]{archer2004rauscher}
Archer, A. J., and M. Rauscher,
2004,
``Dynamical density functional theory for interacting Brownian
particles: stochastic or deterministic?''
\href{https://doi.org/10.1088/0305-4470/37/40/001}
{J. Phys. A: Math. Gen. {\bf 37}, 9325}.

\bibitemORIGINAL{Archer \etal, 2007}{archer2007dtpl} 
Archer, A. J., P. Hopkins, and M.~Schmidt,
2007,
``Dynamics in inhomogeneous liquids and glasses via the test particle limit,''
\href{https://doi.org/10.1103/PhysRevE.75.040501}
{Phys. Rev. E {\bf 75}, 040501(R).}

\bibitem[Auschra \etal, 2020]{auschra2020}
  Auschra, S., V. Holubec,  N. A. S\"oker, F. Cichos, and K. Kroy,
  2020,
  ``Polarization-density patterns of active particles in motility gradients,''
  \href{https://arxiv.org/abs/2010.16234}
       {arXiv:2010.16234.}

\bibitem[Balucani and Zoppi, 1994]{balucani1994}
  Balucani, U., and M. Zoppi,
  1994,
  {\it Dynamics of the Liquid State} (Clarendon Press, Oxford).

\bibitem[Bechinger \etal, 2016]{loewen2016}
Bechinger, C., R. Di Leonardo, H. L\"owen,
C. Reichhardt, G. Volpe, and G. Volpe,
2016,
``Active particles in complex and crowded environments,''
\href{https://doi.org/10.1103/RevModPhys.88.045006}
  {Rev. Mod. Phys. {\bf 88}, 045006.}

\bibitem[Berner \etal, 2018]{berner2018}
  Berner, J., B. M\"uller, J. Ruben Gomez-Solano, M. Kr\"uger, and C. Bechinger,
  2018,
  ``Oscillating modes of driven colloids in overdamped systems,''
  \href{https://doi.org/10.1038/s41467-018-03345-2}
  {Nat. Comm. {\bf 9}, 999.}

\bibitemORIGINAL{Bernreuther and Schmidt, 2016}{bernreuther2016gcm} 
Bernreuther, E., and M. Schmidt, 
2016,
``Superadiabatic forces in the dynamics of the one-dimensional Gaussian core model,''
\href{https://doi.org/10.1103/PhysRevE.94.022105}
{Phys. Rev. E {\bf 94}, 022105.}

\bibitem[Bialk\'e \etal, 2015]{bialke2015}
  Bialk\'e, J., J. T. Siebert, H. L\"owen, and T. Speck, 
  2015,
``Negative interfacial tension in phase-separated active Brownian particles,''
  \href{https://doi.org/10.1103/PhysRevLett.115.098301}
  {Phys. Rev.  Lett. {\bf 115}, 098301.}

\bibitem[Bier and van Roij, 2007]{bier2007pre}
Bier, M., and R. van Roij,
2007,
``Relaxation dynamics in fluids of platelike colloidal particles,''
\href{https://doi.org/10.1103/PhysRevE.76.021405}
{Phys. Rev. E {\bf 76}, 021405}.

\bibitem[Bier \etal, 2008]{bier2008prl}
Bier, M., R. van Roij, M. Dijkstra, and P. van der Schoot,
2008,
``Self-diffusion of particles in complex fluids:
temporary cages and permanent barriers,''
\href{https://doi.org/10.1103/PhysRevLett.101.215901}
{Phys. Rev. Lett. {\bf 101}, 215901}.

\bibitem[Bier and van Roij, 2008]{bier2008pre}
Bier, M., and R. van Roij,
2008,
``Nonequilibrium steady states in fluids of platelike colloidal particles,''
\href{https://doi.org/10.1103/PhysRevE.77.021401}
{Phys. Rev. E {\bf 77}, 021401}.

\bibitem[Bird \etal, 1987]{bird1987}
  Bird, R.B., R.C.  Armstrong, and O. Hassager,
  1987
  {\it Dynamics of polymeric liquid / 1. Fluid mechanics}, 2nd ed.
  (John Wiley, New York).

\bibitem[Bleibel \etal, 2014]{bleibel2014sm}
Bleibel, J., A. Dom\'inguez, M. Oettel, and S. Dietrich,
2014,
``Capillary attraction induced collapse of colloidal monolayers at fluid interfaces,''
\href{https://doi.org/10.1039/c3sm53070a}
{Soft Matter {\bf 10}, 4091}.

\bibitem[Bleibel \etal, 2016]{bleibel2016}
Bleibel, J., A. Dom\'inguez, and M. Oettel,
2016,
``A dynamic DFT approach to generalized diffusion equations in a system
with long-ranged and hydrodynamic interactions,''
\href{https://doi.org/10.1088/0953-8984/28/24/244021}
{J. Phys.: Condens. Matter {\bf 28}, 244021}.

\bibitem[Borgis \etal, 2013]{borgis2013}
Borgis, D., R. Assaraf, B. Rotenberg, and R. Vuilleumier,
2013,
``Computation of pair distribution functions and three-dimensional
densities with a reduced variance principle,''
\href{https://doi.org/10.1080/00268976.2013.838316}
{Mol. Phys. {\bf 111}, 3486}.

\bibitem[Brader, 2010]{brader2010}
   Brader,  J. M.,
   2010,
   ``Nonlinear rheology of colloidal dispersions,''
   \href{https://doi.org/10.1088/0953-8984/22/36/363101}
        {J. Phys.: Condens. Matter {\bf 22}, 363101.}

\bibitemESSENTIAL{Brader \etal, 2012}{brader2012pre}
Brader, J.~M., M.~E. Cates, and M.~Fuchs,
2012,
``First-principles constitutive equation for suspension rheology,''
\href{https://doi.org/10.1103/PhysRevE.86.021403}
{Phys. Rev. E {\bf 86}, 021403.}
%
The appendix gives an excellent account of operator methods for
time-dependent correlators, and in particular describes essential
properties of time-ordered exponentials, as is relevant for
Sec.~\ref{SECnoz}.

\bibitem[Brader and Krueger, 2011]{brader2011shear}
  Brader, J. M., and M. Krueger,
  2011,
  ``Density profiles of a colloidal liquid at a wall under shear flow,''
  \href{https://doi.org/10.1080/00268976.2010.541889}
       {Mol. Phys. {\bf 109}, 1029.}

\bibitemORIGINAL{Brader and Schmidt, 2013}{brader2013nozOne} 
Brader, J.~M., and M. Schmidt, 
2013,
``Nonequilibrium Ornstein-Zernike relation for Brownian many-body dynamics,''
\href{https://doi.org/10.1063/1.4820399}
{J.~Chem. Phys. {\bf 139}, 104108.}

\bibitemORIGINAL{Brader and Schmidt, 2014}{brader2014nozTwo} 
Brader, J.~M., and M. Schmidt, 
2014,
``Dynamic correlations in Brownian many-body systems,''
  \href{https://doi.org/10.1063/1.4861041}
       {J. Chem. Phys. {\bf 140}, 034104.}

\bibitemORIGINAL{Brader and Schmidt, 2015}{brader2015dtpl} 
  Brader, J. M., and M. Schmidt,
  2015,
  ``Power functional theory for the dynamic test particle limit,''
  \href{https://doi.org/10.1088/0953-8984/27/19/194106}
       {J. Phys.: Condens. Matter {\bf 27}, 194106.}

\bibitemORIGINAL{Brader and Schmidt, 2015b}{brader2015functionalLineIntegration} 
  Brader, J. M., and M. Schmidt,
  2015,
  ``Free power dissipation from functional line integration,''
  \href{https://doi.org/10.1080/00268976.2015.1042086}
       {Mol. Phys. {\bf 113}, 2873.}
       (Special issue in honour of Jean-Pierre Hansen).

\bibitemORIGINAL{Br\"utting \etal, 2019}{bruetting2019viscosity} 
   Br\"utting, M., T. Trepl, D. de las Heras, and M. Schmidt,
   2019,
  ``Superadiabatic forces via the acceleration gradient in quantum many-body 
  dynamics,''
  \href{https://doi.org/10.3390/molecules24203660}
       {Molecules {\bf 24}, 3660.}

\bibitem[Cats \etal, 2021]{cats2021}
  Cats, P., S. Kuipers, S. de Wind, R. van Damme, G. M. Coli, 
  M. Dijkstra, and R. van Roij,
  2021,
  ``Machine-learning free-energy functionals using density
  profiles from simulations'',
  \href{https://arxiv.org/abs/2101.01942}
  {arXiv:2101.01942.}

\bibitem[CECAM, 2019]{CECAM2019}
  CECAM workshop, 2019
  ``Fundamentals of Density Functional Theory for $T>0$: Quantum meets Classical'',
  Lausanne, Switzerland, May 20, 2019 - May 23, 2019.
  \href{https://www.cecam.org/workshop-details/113}
       {https://www.cecam.org/workshop-details/113}

\bibitem[Chakrabarti \etal, 2003]{chakrabarti2003epl}
Chakrabarti, J., J. Dzubiella, and H. L\"owen,
2003,
``Dynamical instability in driven colloids,''
\href{https://doi.org/10.1209/epl/i2003-00193-6}
{EPL (Europhys. Lett.) {\bf 61}, 415}.

\bibitem[Chakrabarti \etal, 2004]{chakrabarti2004pre}
Chakrabarti, J., J. Dzubiella, and H. L\"owen,
2004,
``Reentrance effect in the lane formation of driven colloids,''
\href{https://doi.org/10.1103/PhysRevE.70.012401}
{Phys. Rev. E {\bf 70}, 012401}.

\bibitem[Chan and Finken, 2005]{chan2005}
Chan, G. K.-L., and R. Finken,
2005,
``Time-dependent density functional theory of classical fluids,''
\href{https://doi.org/10.1103/PhysRevLett.94.183001}
{Phys. Rev. Lett. {\bf 94}, 183001}.

\bibitem[Chacko \etal, 2017]{chacko2017}
  Chacko, B., R. Evans, and A. J. Archer,
  2017,
  ``Solvent fluctuations around solvophobic, solvophilic, and patchy
  nanostructures and the accompanying solvent mediated interactions,''
  \href{https://doi.org/10.1063/1.4978352}
  {J. Chem. Phys. {\bf 146}, 124703.}


\bibitem[Clerk Maxwell, 1874]{maxwell1874}
Clerk-Maxwell, J.,
1874,
``van der Waals on the continuity of the gaseous and liquid states,''
\href{https://doi.org/10.1038/010477a0}
{Nature {\bf 10}, 477}.

\bibitem[Coe \etal, 2020]{coe2022}
  Coe, M. K., R. Evans, and N. B. Wilding, 2022,
    ``Density depletion and enhanced fluctuations in water near hydrophobic solutes: identifying the underlying physics,''
  \href{https://doi.org/10.1103/PhysRevLett.128.045501}
  {Phys. Rev. Lett. {\bf 128}, 045501.}


\bibitem[Davidchack \etal, 2016]{davidchack2016}
  Davidchack, R. L., B. B. Laird, and R. Roth,
  2016,
  ``Hard spheres at a planar hard wall: Simulations and density functional theory,''
  \href{https://doi.org/10.5488/CMP.19.23001}
  {Condens. Matt. Phys. {\bf 19}, 23001.}

\bibitem[de las Heras \etal, 2011]{delasheras2011patchy}
de las Heras, D., J. M. Tavares, and M. M. Telo da Gama,
2011,
``Phase diagrams of binary mixtures of patchy colloids with distinct
numbers of patches: the network fluid regime,''
\href{https://doi.org/10.1039/C0SM01493A}
{Soft Matter {\bf 7}, 5615}.

\bibitem[de las Heras \etal, 2005]{delasheras2005capillary}
de las Heras, D., E. Velasco, and L. Mederos,
2005,
``Capillary smectization and layering in a confined liquid crystal,''
\href{https://doi.org/10.1103/PhysRevLett.94.017801}
{Phys. Rev. Lett. {\bf 94}, 017801}.

\bibitem[de las Heras \etal, 2012]{delasheras2012sciRep}
de las Heras, D., N. Doshi, T. Cosgrove, J. Phipps, 
D. I. Gittins, J. S. van Duijneveldt, and M. Schmidt, 
2012,
``Floating nematic phase in colloidal platelet-sphere mixtures,''
\href{https://doi.org/10.1038/srep00789}
{Sci. Rep. {\bf 2}, 789}.

\bibitemORIGINAL{de las Heras and Schmidt, 2013}{delasheras2013phaseStacking} 
de las Heras, D., and M. Schmidt, 
2013,
``Phase stacking diagram of colloidal mixtures under gravity,''
  \href{https://doi.org/10.1039/C3SM51491A}
       {Soft Matter {\bf 9}, 8636.}

\bibitemORIGINAL{de las Heras and Schmidt, 2014}{delasheras2014sedimentationLiquid2014} 
de las Heras,  D., and M. Schmidt,
2015,
  ``Sedimentation stacking diagram of binary colloidal mixtures and bulk phases
in the plane of chemical potentials,''
  \href{https://doi.org/10.1088/0953-8984/27/19/194115}
       {J. Phys.: Condens. Matter {\bf 27}, 194115.}
       (Special Issue for Liquids 2014).

\bibitemORIGINAL{de las Heras and Schmidt, 2014}{delasheras2014fullCanonical} 
de las Heras, D., and M. Schmidt, 
2014,
  ``Full canonical information from grand potential density functional theory,''
  \href{https://doi.org/10.1103/PhysRevLett.113.238304}
       {Phys. Rev. Lett. {\bf 113}, 238304.}

\bibitemORIGINAL{de las Heras \etal, 2016}{delasheras2016particleConservation} 
de las Heras,  D., J. M. Brader, A. Fortini, and M.~Schmidt,
2016,
  ``Particle conservation in dynamical density functional theory,''
  \href{https://doi.org/10.1088/0953-8984/28/24/244024}
       {J. Phys.: Condens. Matter {\bf 28}, 244024.}

\bibitemORIGINAL{de las Heras and Schmidt, 2018}{delasheras2018velocityGradient} 
 de las Heras,  D., and M. Schmidt,
2018,
  ``Velocity gradient power functional for Brownian dynamics,''
  \href{https://doi.org/10.1103/PhysRevLett.120.028001}
       {Phys. Rev. Lett. {\bf 120}, 028001.}

\bibitemORIGINAL{de las Heras and Schmidt, 2018b}{delasheras2018forceSampling} 
 de las Heras,  D., and M. Schmidt,
2018,
  ``Better than counting: Density profiles from force sampling,''
  \href{https://doi.org/10.1103/PhysRevLett.120.218001}
       {Phys. Rev. Lett. {\bf 120}, 218001}
       (selected as PRL Editors' Suggestion); see also \cite{borgis2013}.

\bibitemORIGINAL{de las Heras \etal, 2019}{delasheras2019customFlow} 
de las Heras,  D., J. Renner, and M. Schmidt,
2019,
  ``Custom flow in overdamped Brownian dynamics,''
  \href{https://doi.org/10.1103/PhysRevE.99.023306}
       {Phys. Rev. E {\bf 99}, 023306.}

\bibitemORIGINAL{de las Heras and Schmidt, 2020}{delasheras2020fourForces} 
 de las Heras, D., and M. Schmidt,
2020,
  ``Flow and structure in nonequilibrium Brownian many-body systems,''
  \href{https://doi.org/10.1103/PhysRevLett.125.018001}
       {Phys. Rev. Lett. {\bf 125}, 018001.}

\bibitem[de~las~Heras \etal, 2021]{delasheras2021tension} 
  de las Heras, D., S. Hermann, and M. Schmidt,
  2021
  (to be published).

\bibitemESSENTIAL{Desloge, 1998}{Desloge1988} 
Desloge, E.~A., 
1988,
``The Gibbs-Appell equation of motion,''
\href{https://doi.org/10.1119/1.15463}
{Am. J. Phys. {\bf 56}, 841.}

\bibitem[Dhont, 1996]{dhont1996}
Dhont, J.K.G.,
1996,
{\it An introduction to the dynamics of colloids}
(Elsevier, Amsterdam).

\bibitem[Dixit \etal, 2018]{dixit2018}
Dixit, M., T. Schilling, and M. Oettel,
2018,
``Growth of films with anisotropic particles: simulations and rate equations,''
\href{https://doi.org/10.1063/1.5031217}
{J. Chem. Phys. {\bf 149}, 064903}.

\bibitemORIGINAL{Dwandaru and Schmidt, 2011}{dwandaru2011levy} 
Dwandaru, W. S. B., and M. Schmidt, 
2011,
``Variational principle of classical density-functional theory via 
Levy's constrained search method,''
\href{https://doi.org/10.1103/PhysRevE.83.061133}
{Phys. Rev. E {\bf 83}, 061133.}

\bibitem[Dyre, 2006]{dyre2006}
Dyre, J. C.,
2006,
``Colloquium: The glass transition and elastic models of glass-forming liquids,''
\href{https://doi.org/10.1103/RevModPhys.78.953}
{Rev. Mod. Phys. {\bf 78}, 953.}

\bibitem[Dyre, 2016]{dyre2016}
Dyre, J. C.,
2016,
``Simple liquids' quasiuniversality and the hard-sphere paradigm,''
\href{https://doi.org/10.1088/0953-8984/28/32/323001}
{J. Phys.: Condens. Matter {\bf 28}, 323001.}

\bibitem[Dyre, 2018]{dyre2018}
Dyre, J. C.,
2018,
``Perspective: excess-entropy scaling,''
\href{https://doi.org/10.1063/1.5055064}
{J. Chem. Phys. {\bf 149}, 210901}.

\bibitem[Dzubiella \etal, 2002]{dzubiella2002}
Dzubiella, J.,  G. P. Hoffmann, and H.  L\"owen,
2002,
``Lane formation in colloidal mixtures driven by an external field,''
\href{https://doi.org/10.1103/PhysRevE.65.021402}
{Phys. Rev. E {\bf 65}, 021402.}

\bibitem[Dzubiella \etal, 2003]{dzubiella2003mfddft}
Dzubiella, J., and C. N. Likos,
2003,
``Mean-field dynamical density functional theory,''
\href{https://doi.org/10.1088/0953-8984/15/6/102}
{J. Phys.: Condens. Matter {\bf 15}, L147.}

\bibitemORIGINAL{Eckert \etal, 2020}{eckert2020auxiliaryFields} 
Eckert,  T., N.~C.~X. Stuhlm\"uller, F. Samm\"uller, and M. Schmidt,
2020,
  ``Fluctuation profiles in inhomogeneous fluids,''
\href{https://doi.org/10.1103/PhysRevLett.125.268004}
{Phys. Rev. Lett. {\bf 125}, 268004.}

\bibitem[Eckert \etal, 2021]{eckert2021}
  Eckert, T., M. Schmidt, and D. de las Heras, 
  2021,
  ``Gravity-induced phase phenomena in plate-rod colloidal mixtures,''
  \href{https://doi.org/10.1038/s42005-021-00706-0}
         {Commun. Phys. {\bf 4}, 202.}

\bibitemESSENTIAL{Evans, 1979}{evans1979}
Evans, R.,
1979,
``The nature of the liquid-vapour interface and other topics in the statistical
mechanics of non-uniform, classical fluids,''
\href{https://doi.org/10.1080/00018737900101365}
{Adv. Phys. {\bf 28},  143.} This is the classical
text for classical DFT, as relevant for Sec.~\ref{SECdft}; 
  for an overview of new developments see \cite{evans2016specialIssue}.

\bibitem[Evans, 1992]{evans1992}
  Evans, R.
  1992,
  ``Density functionals in the theory nonuniform fluids,''
  in {\it Fundamentals of Inhomogeneous Fluids},
  edited by D. Henderson (Dekker, New York, 1992).

\bibitemESSENTIAL{Evans \etal, 2016}{evans2016specialIssue}
Evans, R., M. Oettel, R. Roth, and G. Kahl,
2016,
``New developments in classical density functional theory,''
\href{https://doi.org/doi:10.1088/0953-8984/28/24/240401}
{J. Phys.: Condens. Matter {\bf 28}, 240401}.
%
This is the foreword of the {\it Special issue on new developments in 
classical density functional theory}
\href{https://iopscience.iop.org/issue/0953-8984/28/24}
{J. Phys.: Condens. Matter {\bf 28}(24) (2016)}.

\bibitem[Evans and Stewart, 2015]{evans2015jpcm}
  Evans, R., and M. C. Stewart,
  2015,
  ``The local compressibility of liquids near non-adsorbing substrates: a
  useful measure of solvophobicity and hydrophobicity?''
  \href{http://doi.org/10.1088/0953-8984/27/19/194111}
  {J. Phys.: Condens. Matter {\bf 27}, 194111.}

\bibitem[Evans and Wilding, 2015]{evans2015prl}
  Evans, R.,  and N. B. Wilding,
  2015,
  ``Quantifying density fluctuations in water at a hydrophobic surface:
  evidence for critical drying,''
  \href{http://doi.org/10.1103/PhysRevLett.115.016103}
  {Phys. Rev. Lett. {\bf 115}, 016103.}

\bibitem[Evans \etal, 2016]{evans2016prl}
  Evans, R., M. C. Stewart, and N. B. Wilding,
  2016,
  ``Critical Drying of Liquids,''
  \href{http://doi.org/10.1103/PhysRevLett.117.176102}
  {Phys. Rev. Lett. {\bf 117}, 176102.}

\bibitem[Evans \etal, 2017]{evans2017}
  Evans, R., M. C. Stewart, and N. B. Wilding,
  2017,
  ``Drying and wetting transitions of a Lennard-Jones fluid: Simulations
  and density functional theory,''
  \href{https://doi.org/10.1063/1.4993515}
  {J. Chem. Phys. {\bf 147}, 044701.}

\bibitem[Evans \etal, 2019]{evans2019pnas}
  Evans, R., M. C. Stewart, and N. B. Wilding,
  2019,
  ``A unified description of hydrophilic and superhydrophobic surfaces in terms of 
  the wetting and drying transitions of liquids,''
  \href{https://doi.org/10.1073/pnas.1913587116}
  {Proc. Nat. Acad. Sci. {\bf 116}, 23901.}

\bibitemESSENTIAL{Evans \etal, 2019}{evans2019physicsToday}
Evans, R., D. Frenkel, and M. Dijkstra, 
2019,
``From simple liquids to colloids and soft matter,''
\href{https://doi.org/10.1063/PT.3.4135}
{Physics Today {\bf 72}, 38.}

\bibitemESSENTIAL{Evans and Morriss, 2013}{EvansMorriss} 
Evans, D. J., and G. P. Morriss, 
2013,
{\it Statistical Mechanics of Nonequilibrium Liquids}, 2nd ed.\
(Cambridge University Press, Cambridge). 

\bibitem[Espa\~nol and L\"owen, 2009]{espanol2009}
Espa\~nol, P., and H. L\"owen,
2009,
``Derivation of dynamical density functional theory using
the projection operator technique,''
\href{https://doi.org/10.1063/1.3266943}
{J. Chem. Phys. {\bf 131}, 244101}.


\bibitem[Esztermann \etal, 2006]{esztermann2006}
Esztermann, A., H. Reich, and M. Schmidt, 
2006,
``Density functional theory for colloidal mixtures of hard platelets, rods, and spheres,''
\href{https://doi.org/10.1103/PhysRevE.73.011409}
{Phys. Rev. E. 73, 011409.}

\bibitem[Falasco and Kroy, 2016]{falasco2016}
Falasco, G., and K. Kroy,
2016,
``Nonisothermal fluctuating hydrodynamics and Brownian motion,''
\href{https://doi.org/10.1103/PhysRevE.93.032150}
{Phys. Rev. E {\bf 93}, 032150}.

\bibitem[Farage \etal, 2015]{farage2015}
Farage, T. F. F., P. Krinninger, and J. M. Brader,
2015,
``Effective interactions in active Brownian suspensions,''
\href{https://doi.org/10.1103/PhysRevE.91.042310}
{Phys. Rev. E {\bf 91}, 042310}.

\bibitemORIGINAL{Fortini \etal, 2014}{fortini2014prl} 
Fortini, A., D. de las Heras, J.~M. Brader, and M.~Schmidt, 
2014,
``Superadiabatic forces in Brownian many-body dynamics,''
\href{https://doi.org/10.1103/PhysRevLett.113.167801}
{Phys. Rev. Lett. {\bf 113}, 167801.}

\bibitemESSENTIAL{Galley, 2013}{galley2013prl}
Galley, C. R., 
2013,
``Classical mechanics of nonconservative systems,''
\href{https://doi.org/10.1103/PhysRevLett.110.174301}
{Phys. Rev. Lett. {\bf 110}, 174301.}

\bibitem[Geigenfeind and de las Heras, 2017]{geigenfeind2017sampleHeight}
  Geigenfeind, T. and D. de las Heras,
  2017,
  ``The role of sample height in the stacking diagram of
  colloidal mixtures under gravity''
  \href{https://doi.org/10.1088/1361-648X/aa4e04}
       {J. Phys: Condens. Matter {\bf 29}, 064006.}

\bibitemORIGINAL{Geigenfeind \etal, 2020}{geigenfeind2019laning} 
Geigenfeind, T., D. de las Heras, and M. Schmidt,
2020,
``Superadiabatic demixing in nonequilibrium colloids,''
\href{https://doi.org/10.1038/s42005-020-0287-5}
  {Commun. Phys. {\bf 3}, 23.}

\bibitem[Giacomello \etal, 2016]{giacomello2016}
  Giacomello, A., L. Schimmele, S. Dietrich, and M. Tasinkevych,
  2016,
  ``Perpetual superhydrophobicity,''
  \href{https://doi.org/10.1039/C6SM01727D}
  {Soft Matter {\bf 12}, 8927.}

\bibitem[Giacomello \etal, 2019]{giacomello2019}
  Giacomello, A., L. Schimmele, S. Dietrich, and M. Tasinkevych,
  2019,
  ``Recovering superhydrophobicity in nanoscale and macroscale surface textures,''
  \href{https://doi.org/10.1039/C9SM01049A}
  {Soft Matter {\bf 15}, 7462.}

\bibitem[Goddard \etal, 2012]{goddard2012prl}
  Goddard, B. D., A. Nold, N. Savva, G. A. Pavliotis, and S. Kalliadasis,
  2012,
  ``General dynamical density functional theory for classical fluids,''
  \href{https://doi.org/10.1103/PhysRevLett.109.120603}
       {Phys. Rev. Lett. {\bf 109}, 120603}.

\bibitem[Gonz\'alez \etal, 1997]{gonzalez1997}
   Gonz\'alez, A., J. A. White, F. L. Rom\'an, S. Velasco, and R. Evans,
   1997,
   ``Density functional theory for small systems: hard spheres in a
   closed spherical cavity,''
   \href{https://doi.org/10.1103/PhysRevLett.79.2466}
        {Phys. Rev. Lett. {\bf 79}, 2466.}

\bibitem[G\"otze, 2008]{goetze2008}
  G\"otze, W.,
  2008,
  {\it Complex dynamics of glass forming liquids}
  (Oxford University Press, Oxford).

\bibitem[Grelet \etal, 2008]{grelet2008}
Grelet, E., M. P. Lettinga, M. Bier, R. van Roij, and P. van der Schoot,
2008,
``Dynamical and structural insights into the smectic phase of rod-like particles,''
\href{https://doi.org/10.1088/0953-8984/20/49/494213}
{J. Phys.: Condens. Matter {\bf 20}, 494213}.

\bibitemESSENTIAL{Hansen and McDonald, 2013}{hansen2013}
Hansen, J.~P., and I.~R. McDonald,
2013,
{\it Theory of Simple Liquids}, 4th ed. (Academic Press, London).

\bibitem[Haertel \etal, 2012]{haertel2012}
  Haertel, A., M. Oettel, R. E. Rozas,  S. U. Egelhaaf, J. Horbach, 
  and H. L\"owen,
  2012,
  ``Tension and stiffness of the hard sphere crystal-fluid interface,''
  \href{http://doi.org/10.1103/PhysRevLett.108.226101}
  {Phys. Rev. Lett. {\bf 108}, 226101.}

\bibitem[Hansen-Goos and Roth, 2006]{hansengoos2006}
Hansen-Goos, H., and R. Roth,
2006,
``Density functional theory for hard-sphere mixtures: 
the White Bear version mark II,''
\href{https://doi.org/10.1088/0953-8984/18/37/002}
{J. Phys.: Condens. Matter {\bf 18}, 8413}.

\bibitem[Hansen-Goos and Mecke, 2009]{hansengoos2009prl}
Hansen-Goos, H., and K. Mecke,
2009,
``Fundamental measure theory for inhomogeneous fluids of non-spherical
hard particles,''
\href{https://doi.org/10.1103/PhysRevLett.102.018302}
{Phys. Rev. Lett. {\bf 102}, 018302}.

\bibitemORIGINAL{Hermann and Schmidt, 2018}{hermann2018activeSedimentation} 
Hermann, S., and M. Schmidt,
2018,
``Active ideal sedimentation: Exact two-dimensional steady states,''
\href{https://doi.org/10.1039/C7SM02515G}
  {Soft Matter {\bf 14}, 1614.}

\bibitemORIGINAL{Hermann \etal, 2019}{hermann2019acif} 
Hermann, S., P. Krinninger, D. de las Heras, and M. Schmidt,
2019,
``Phase coexistence of active Brownian particles,''
\href{https://link.aps.org/doi/10.1103/PhysRevE.100.052604}
{Phys. Rev. E {\bf 100}, 052604.}

\bibitemORIGINAL{Hermann \etal, 2019b}{hermann2019tension} 
Hermann, S., D. de las Heras, and M. Schmidt,
2019,
``Non-negative interfacial tension in phase-separated active Brownian particles,''
\href{https://doi.org/10.1103/PhysRevLett.123.268002}
    {Phys. Rev. Lett. {\bf 123}, 268002.}

\bibitemORIGINAL{Hermann and Schmidt, 2020}{hermann2020polarization} 
Hermann, S., and M. Schmidt, 
2020,
 ``Active interface polarization as a state function,''
 \href{https://doi.org/10.1103/PhysRevResearch.2.022003}
 {Phys. Rev. Research {\bf 2}, 022003(R).}

\bibitemORIGINAL{Hermann \etal, 2020b}{hermann2020longActive}
Hermann, S., D. de las Heras, and M. Schmidt, 
2021,
``Phase separation of active Brownian particles in two dimensions: 
 Anything for a quiet life,''
 Mol. Phys.
 \href{https://doi.org/10.1080/00268976.2021.1902585}{e1902585}; see also:
 \href{https://arxiv.org/abs/2103.03585}{arxiv:2103.03585}.

\bibitemORIGINAL{Hermann and Schmidt, 2021}{hermann2021noether}
Hermann, S., and M. Schmidt, 
2021,
``Noether's Theorem in Statistical Mechanics,''
\href{https://doi.org/10.1038/s42005-021-00669-2}
{Commun. Phys. {\bf 4}, 176.}

\bibitemORIGINAL{Hermann and Schmidt, 2021b}{hermann2021noetherPopular}
Hermann, S., and M. Schmidt, 
2021b,
``Why Noether's Theorem applies to Statistical Mechanics,''
\href{https://arxiv.org/abs/2109.10283}
     {arxiv:2109.10283}.

\bibitem[Hern\'andez-Mu\~noz \etal, 2019]{hernandez-munoz2019}
   Hern\'andez-Mu\~noz, J., E. Chac\'on, and P. Tarazona,
   2019,
  ``Density functional analysis of atomic force microscopy in a dense fluid,''
   \href{https://doi.org/10.1063/1.5110366}
   {J. Chem. Phys. {\bf 151}, 034701.}

\bibitemESSENTIAL{Hohenberg and Kohn, 1964}{HK1964}
Hohenberg, P., and W. Kohn,
1964,
``Inhomogeneous electron gas,''
\href{https://doi.org/10.1103/PhysRev.136.B864}
{Phys. Rev. {\bf 136}, B864.}

\bibitemORIGINAL{Hopkins \etal, 2010}{hopkins2010dtpl} 
Hopkins,  P., A. Fortini, A.~J. Archer, and M.~Schmidt, 
2010,
``The van Hove distribution function for Brownian hard spheres:
Dynamical test particle theory and computer simulations for bulk dynamics,''
\href{https://doi.org/10.1063/1.3511719}
{J. Chem. Phys. {\bf 133}, 224505.}

\bibitemESSENTIAL{Irving and Kirkwood, 1950}{IrvingKirkwood}
Irving, J.~H., and J.~G. Kirkwood,
1950,
``The statistical mechanical theory of transport processes. 
IV. The equations of hydrodynamics,''
\href{https://doi.org/10.1063/1.1747782}
{J. Chem. Phys. {\bf 18}, 817.}

\bibitemORIGINAL{Jahreis and Schmidt, 2020}{jahreis2019shear} 
Jahreis, N., and M. Schmidt,
2020,
  ``Shear-induced deconfinement of hard disks,''
  \href{https://doi.org/10.1007/s00396-020-04644-1}
       {Col. Pol. Sci. {\bf 298}, 895.}

\bibitem[Janssen, 2018]{janssen2018}
Janssen, L. M. C.,
2018,
``Mode-coupling theory of the glass transition: a primer,''
\href{https://doi.org/10.3389/fphy.2018.00097}
{Front. Phys. {\bf 6}, 97}.

\bibitem[Jeanmairet \etal, 2013]{jeanmairet2013jpcl}
Jeanmairet, G., M. Levesque, R. Vuilleumier, and D. Borgis,
2013,
``Molecular density functional theory of water,''
\href{https://doi.org/10.1021/jz301956b}
{J. Phys. Chem. Lett. {\bf 4}, 619}.

\bibitem[Jeanmairet \etal, 2013 b]{jeanmairet2013jcp}
  Jeanmairet, G., M. Levesque, and D. Borgis,
  2013,
  ``Molecular density functional theory of water describing 
  hydrophobicity at short and long length scales,''
  \href{https://doi.org/10.1063/1.4824737}
  {J. Chem. Phys. {\bf 139}, 154101.}

\bibitem[Jeanmairet \etal, 2019]{jeanmairet2019capacitor}
  Jeanmairet, G., B. Rotenberg, D. Borgis, and M. Salanne,
  2019,
  ``Study of a water-graphene capacitor with molecular density functional theory,''
  \href{https://doi.org/10.1063/1.5118301}
  {J. Chem. Phys. {\bf 151}, 124111.}


\bibitem[Jones and Gunnarsson, 2015]{jones1989}
  Jones, R.O., and O. Gunnarsson, 
  1989,
  ``The density functional formalism, its applications and prospects,''
  \href{https://doi.org/10.1103/RevModPhys.61.689}
       {Rev. Mod. Phys. {\bf 61}, 689.}

\bibitem[Jones, 2015]{jones2015}
  Jones, R.O.,
  2015,
  ``Density functional theory: Its origins, rise to prominence, and future,''
  \href{https://doi.org/10.1103/RevModPhys.87.897}
       {Rev. Mod. Phys. {\bf 87}, 897.}


\bibitem[Jung and Schmid, 2016]{jung2016}
Jung, G., and F. Schmid,
2016,
``Computing bulk and shear viscosities from simulations of fluids with
dissipative and stochastic interactions,''
\href{http://dx.doi.org/10.1063/1.4950760}
{J. Chem. Phys. {\bf 144}, 204104}.

\bibitem[Jung \etal, 2017]{jung2017}
Jung, G., M. Hanke, and F. Schmid,
2017,
``Iterative reconstruction of memory kernels,''
\href{http://dx.doi.org/10.1021/acs.jctc.7b00274}
{J. Chem. Theory Comput. {\bf 13}, 2481}.

\bibitem[Kierlik and Rosinberg, 1990]{kierlik1990}
Kierlik, E., and M. L. Rosinberg,
1990,
``Free-energy density functional for the inhomogeneous
hard-sphere fluid: Application to interfacial adsorption,''
\href{https://doi.org/10.1103/PhysRevA.42.3382}
{Phys. Rev. A {\bf 42}, 3382}. 

\bibitem[Klopotek \etal, 2017]{klopotek2017}
Klopotek, M., H. Hansen-Goos, M. Dixit, T. Schilling, 
F. Schreiber, and M. Oettel, 
2017,
``Monolayers of hard rods on planar substrates. II. Growth,''
\href{https://doi.org/10.1063/1.4976308}
{J. Chem. Phys. {\bf 146}, 084903}.

\bibitem[Kohn, 1999]{kohn1999nobel}
Kohn, W.,
1999,
``Nobel Lecture: Electronic structure of matter--wave functions 
and density functionals,''
\href{https://doi.org/10.1103/RevModPhys.71.1253}
{Rev. Mod. Phys. {\bf 71}, 1253}.

\bibitemORIGINAL{Krinninger \etal, 2016}{krinninger2016prl}
Krinninger, P., M. Schmidt, and J.~M. Brader, 
2017,
``Nonequilibrium phase behaviour from minimization of free power dissipation,''
\href{https://doi.org/10.1103/PhysRevLett.117.208003}
  {Phys. Rev. Lett. {\bf 117}, 208003 (2016);}
    \href{https://doi.org/10.1103/PhysRevLett.119.029902}
    {Erratum {\bf 119}, 029902 (2017).}

\bibitemORIGINAL{Krinninger and Schmidt, 2019}{krinninger2019jcp} 
Krinninger, P., and M. Schmidt, 
2019,
``Power functional theory for active Brownian particles:
general formulation and power sum rules,''
\href{https://doi.org/10.1063/1.5061764}
{J. Chem. Phys. {\bf 150}, 074112.}

\bibitem[Lafuente and Cuesta, 2004]{lafuente2004}
Lafuente, L., and J. A. Cuesta,
2004,
``Density functional theory for general hard-core lattice gases,''
\href{https://doi.org/10.1103/PhysRevLett.93.130603}
{Phys. Rev. Lett. {\bf 93}, 130603}.

\bibitemORIGINAL{Landgraf \etal, 2020}{landgraf2019torques} 
Landgraf, J.,  M. Schmidt, and D. de las Heras,
2021,
``Superadiabatic torques in the orientational dynamics of 
two-dimensional repulsive rods,''
(to be published).

\bibitem[Lesnicki \etal, 2016]{lesnicki2016}
Lesnicki, D., R. Vuilleumier, A. Carof, and B. Rotenberg,
2016,
``Molecular hydrodynamics from memory kernels,''
\href{https://doi.org/10.1103/PhysRevLett.116.147804}
{Phys. Rev. Lett. {\bf 116}, 147804}.

\bibitem[Levesque \etal, 2012]{levesque2012jcp}
  Levesque, M., R. Vuilleumier, and D. Borgis,
  2012,
  ``Scalar fundamental measure theory for hard spheres in three
  dimensions: Application to hydrophobic solvation,''
  \href{https://doi.org/10.1063/1.4734009}
  {J. Chem. Phys. {\bf 137}, 034115.}

\bibitemESSENTIAL{Levy, 1979}{levy1979}
Levy, M.,
1979,
``Universal variational functionals of electron densities, first-order
density matrices, and natural spin-orbitals and solution of the
$v$-representability problem,''
\href{https://doi.org/10.1073/pnas.76.12.6062}
{Proc. Nat. Acad. Sci. {\bf 76}, 6062.}

\bibitem[Leonard \etal, 2013]{leonard2013}
Leonard, T., B. Lander, U. Seifert, and T. Speck,
2013,
``Stochastic thermodynamics of fluctuating density fields:
Non-equilibrium free energy differences under coarse-graining,''
\href{https://doi.org/10.1063/1.4833136}
{J. Chem. Phys. {\bf 139}, 204109}.

\bibitem[Lin and Oettel, 2019]{lin2019ml}
Lin, S.-C., and M. Oettel,
2019,
``A classical density functional from machine learning and a
convolutional neural network,''
\href{https://doi.org/10.21468/SciPostPhys.6.2.025}
{SciPost Phys. {\bf 6}, 25}.

\bibitem[Lin \etal, 2020]{lin2020ml}
Lin,  S.-C., G. Martius and M. Oettel,
2020,
``Analytical classical density functionals from an equation learning network,''
\href{https://doi.org/10.1063/1.5135919}
{J. Chem. Phys. {\bf 152}, 021102.}

\bibitem[Lips \etal, 2018]{lips2018basep}
  Lips, D., A. Ryabov, and P. Maass,
  2018,
  ``Brownian asymmetric simple exclusion process,''
  \href{https://doi.org/10.1103/PhysRevLett.121.160601}
       {Phys. Rev. Lett. {\bf 121}, 160601}.

\bibitem[Loehr \etal, 2016]{loehr2016}
  Loehr, J. , M. Loenne, A. Ernst, D. de las Heras, and T. M. Fischer,
  2016,
  ``Topological protection of multiparticle dissipative transport,''
  \href{https://doi.org/10.1038/ncomms11745}
  {Nat. Commun. {\bf 7}, 11745.}

\bibitem[Loehr \etal, 2018]{loehr2018}
  Loehr, J., D. de las Heras, A. Jarosz, M. Urbaniak, F. Stobiecki, A. Tomita, R. Huhnstock, I. Koch, A. Ehresmann, D. Holzinger, and T. M. Fischer, 
  2018,
  ``Colloidal topological insulators,''
  \href{https://doi.org/10.1038/s42005-017-0004-1}
  {Commun. Phys. {\bf 1}, 4.}

\bibitem[L\"uders \etal, 2019]{lueders2019}
L\"uders, A., U. Siems, and P. Nielaba,
2019,
``Dynamic ordering of driven spherocylinders in a nonequilibrium
suspension of small colloidal spheres,''
\href{https://doi.org/10.1103/PhysRevE.99.022601}
{Phys. Rev. E {\bf 99}, 022601}.

\bibitem[Lutsko and Lam, 2018]{lutsko2018}
Lutsko, J. F., and J. Lam,
2018,
``Classical density functional theory, unconstrained crystallization,
and polymorphic behavior,''
\href{https://doi.org/10.1103/PhysRevE.98.012604}
{Phys. Rev. E {\bf 98}, 012604}.

\bibitem[Lutsko, 2018]{lutsko2018extendingCNT}
Lutsko, J. F.,
2018,
``Systematically extending classical nucleation theory,''
\href{https://doi.org/10.1088/1367-2630/aae174}
{New J. Phys. {\bf 20} 103015}.

\bibitemESSENTIAL{Lutsko, 2010}{lutsko2010review}
Lutsko, J.~F., 
2010,
``Recent developments in classical density functional theory,''
\href{https://doi.org/10.1002/9780470564318.ch1}
{Adv. Chem. Phys. {\bf 144}, 1}.

\bibitem[Lutsko, 2020]{lutsko2020stable}
  Lutsko, J. F.,
  2020,
  ``Explicitly stable fundamental-measure-theory models 
  for classical density functional theory,''
  \href{https://doi.org/10.1103/PhysRevE.102.062137}
  {Phys. Rev. E {\bf 102}, 062137.}


\bibitem[Lutsko and Oettel, 2021]{lutsko2021}
  Lutsko, J. F., and M. Oettel,
  2021,
  ``Reconsidering power functional theory,''
  \href{https://doi.org/10.1063/5.0055288}
       {J. Chem. Phys. {\bf 155}, 094901.}

\bibitem[Maes, 2020]{maes2020}
  Maes, C., 
  2020,
  ``Fluctuating motion in an active environment,''
  \href{https://doi.org/10.1103/PhysRevLett.125.208001}
  {Phys. Rev. Lett. {\bf 125}, 208001.}

\bibitem[Marchetti \etal, 2013]{liverpool2013}
Marchetti,  M. C., J. F. Joanny, S. Ramaswamy,
T. B. Liverpool, J. Prost, M. Rao, and R. A. Simha, 
2013,
``Hydrodynamics of soft active matter,''
\href{https://doi.org/10.1103/RevModPhys.85.1143}
{Rev. Mod. Phys. {\bf 85}, 1143.}

\bibitemESSENTIAL{Marconi and Tarazona, 1999}{marconi1999ddft}
Marconi, U. M. B., and P. Tarazona,
1999, 
``Dynamic density functional theory of fluids,''
\href{https://doi.org/10.1063/1.478705}
{J. Chem. Phys. {\bf 110}, 8032.}


\bibitem[Martin-Jimenez \etal, 2016]{martinjimenez2017natCom} 
   Martin-Jimenez, D., E. Chac\'on, P. Tarazona, and R. Garcia,
   2016,
  ``Atomically resolved three-dimensional structures of
  electrolyte aqueous solutions near a solid surface,''
  \href{https://doi.org/10.1038/ncomms12164}
  {Nat. Commun. {\bf 7}, 12164.}

\bibitem[Menzel \etal, 2016]{menzel2016}
Menzel, A. M., A. Saha, C. Hoell, and H. L\"owen,
2016,
``Dynamical density functional theory for microswimmers,''
\href{https://doi.org/10.1063/1.4939630}
{J. Chem. Phys. {\bf 144}, 024115}.

\bibitem[Menzel, 2016]{menzel2016perspective}
Menzel, A. M.,
2016,
``On the way of classifying new states of active matter,''
\href{https://doi.org/doi:10.1088/1367-2630/18/7/071001}
{New J. Phys. {\bf 18}, 071001}.

\bibitemESSENTIAL{Mermin, 1965}{mermin1965}
Mermin, N.~D., 
1965,
``Thermal properties of the inhomogeneous electron gas,''
\href{https://doi.org/10.1103/PhysRev.137.A1441}
{Phys. Rev. {\bf 137}, A1441.}

\bibitem[Mittal \etal, 2006]{mittal2006prl}
Mittal, J., J. R Errington, and T. M. Truskett,
2006,
``Thermodynamics predicts how confinement modifies the dynamics of the
equilibrium hard-sphere fluid,''
\href{https://doi.org/10.1103/PhysRevLett.96.177804}
{Phys. Rev. Lett. {\bf 96}, 177804}.

\bibitem[Mittal \etal, 2008]{mittal2008prl}
Mittal, J., T. M. Truskett, J. R. Errington, and G. Hummer,
2008,
``Layering and position-dependent diffusive dynamics of confined fluids,''
\href{https://doi.org/10.1103/PhysRevLett.100.145901}
{Phys. Rev. Lett. {\bf 100}, 145901}.

\bibitem[Moncho-Jorda \etal, 2019]{monchojorda2019acsnano}
Moncho-Jorda, A., A. Germ\'an-Bellod, S. Angioletti-Uberti, 
I. Adroher-Ben\'itez, and J. Dzubiella,
2019,
``Nonequilibrium uptake kinetics of molecular
cargo into hollow hydrogels tuned by electrosteric interactions,''
\href{https://doi.org/10.1021/acsnano.8b07609}
{ACS Nano {\bf 13}, 1603}.

\bibitem[Nagel, 2017]{nagel2017}
Nagel,  S. R., 
2017,
``Experimental soft-matter science,''
\href{https://doi.org/10.1103/RevModPhys.89.025002}
  {Rev. Mod. Phys. {\bf 89}, 025002.}

\bibitem[Nakatsukasa \etal, 2016]{nakatsukasa2016}
Nakatsukasa,  T., K. Matsuyanagi, M. Matsuo, and Y. Yabana,
2016,
``Time-dependent density-functional description of nuclear dynamics,''
\href{https://doi.org/10.1103/RevModPhys.88.045004}
  {Rev. Mod. Phys. {\bf 88}, 045004.}

\bibitem[Oettel \etal, 2016]{oettel2016}
  Oettel, M., M. Klopotek, M. Dixit,  E. Empting, T. Schilling, and H. Hansen–Goos,
  2016,
  ``Monolayers of hard rods on planar substrates. I. Equilibrium,''
  \href{https://doi.org/10.1063/1.4960618}
       {J. Chem. Phys. {\bf 145}, 074902 (2016).}


\bibitem[Onida \etal, 2002]{onida2002}
Onida, G., L. Reining, and A. Rubio, 
2002,
 ``Electronic excitations: density-functional versus many-body
 Green’s-function approaches,''
 \href{https://doi.org/10.1103/RevModPhys.74.601}
 {Rev. Mod. Phys. {\bf 74}, 601.}

\bibitem[Onsager, 1949]{onsager1949}
Onsager, L.,
1949,
``The effects of shape on the interaction of colloidal particles,''
\href{https://doi.org/10.1111/j.1749-6632.1949.tb27296.x}
{Ann. NY Acad. Sci. {\bf 51}, 627}.
See \cite{vanroij2005pedagogical} for a modern and pedagogical account.

\bibitem[Paliwal \etal, 2018]{paliwal2018chemicalPotential}
Paliwal, S., J. Rodenburg, R. van Roij, and M. Dijkstra,
2018,
``Chemical potential in active systems: predicting phase equilibrium 
from bulk equations of state?''
\href{https://doi.org/10.1088/1367-2630/aa9b4d}
{New J. Phys. {\bf 20}, 015003}.

\bibitem[Percus and Yevick, 1958]{percus1958}
Percus, J. K., and G. J. Yevick
1958,
``Analysis of Classical Statistical Mechanics by Means of Collective Coordinates''
\href{https://doi.org/10.1103/PhysRev.110.1}
{Phys. Rev. {\bf 110}, 1.}

\bibitem[Percus, 1962]{percus1962}
Percus, J. K.,
1962,
``Approximation methods in classical statistical mechanics,''
\href{https://dx.doi.org/10.1103/PhysRevLett.8.462}
{Phys. Rev. Lett. {\bf 8}, 462}.

\bibitem[Percus, 1976]{percus1976}
Percus, J. K.,
1976,
``Equilibrium state of a classical fluid of hard rods in an external-field,''
\href{https://doi.org/10.1007/BF01020803}
{J. Stat. Phys. {\bf 15}, 505.}

\bibitem[Qi and Schmid, 2017]{qi2017}
Qi, S., and F. Schmid,
2017,
``Hybrid particle-continuum simulations coupling
Brownian dynamics and local dynamic density functional theory,''
\href{https://doi.org/10.1039/c7sm01749a}
{Soft Matter {\bf 13}, 7938}.

\bibitem[Ramakrishnan and Yussouff, 1979]{ramakrishnan1979}
Ramakrishnan, T. V., and M. Yussouff,
1979,
``First-principles order-parameter theory of freezing,''
\href{https://doi.org/10.1103/PhysRevB.19.2775}
{Phys. Rev. B {\bf 19}, 2775}.

\bibitem[Reiss \etal, 1959]{reiss1959}
 Reiss, H., H.~L. Frisch, and J.~L. Lebowitz, 
 1959,
 ``Statistical Mechanics of Rigid Spheres,''
 \href{https://doi.org/10.1063/1.1730361}
 {J. Chem. Phys. {\bf 31}, 369.}

\bibitem[Remsing, 2019]{remsing2019pnas}
  Remsing, R. C.,
  2019,
  ``Commentary: Playing the long game wins the cohesion-adhesion rivalry,''
  \href{https://doi.org/10.1073/pnas.1916911116}
  {Proc. Nat. Acad. Sci. {\bf 116}, 23874.}

\bibitemORIGINAL{Renner \etal, 2021}{renner2021customFlowMD} 
Renner, J., M. Schmidt, and D. de las Heras,
2021,
``Custom flow in molecular dynamics,''
\href{https://doi.org/10.1103/PhysRevResearch.3.013281}
{Phys. Rev. Res. {\bf 3}, 013281.}


\bibitemORIGINAL{Renner \etal, 2022}{renner2022acceleration} 
Renner, J., M. Schmidt, and D. de las Heras,
2022,
``Shear and bulk acceleration viscosities in simple fluids,''
Phys. Rev. Lett. (to appear).

\bibitem[Rex and L\"owen, 2009]{rex2009epje}
Rex, M., and H. L\"owen,
2009,
``Dynamical density functional theory for colloidal
dispersions including hydrodynamic interactions,''
\href{https://doi.org/10.1140/epje/i2008-10363-x}
{Eur. Phys. J. E {\bf 28}, 139}.


\bibitem[Robinson \etal, 2019]{robinson2019}
Robinson, J. F., F. Turci, R. Roth, and C. P. Royall,
2019,
``Morphometric approach to many-body correlations in hard spheres,''
\href{https://doi.org/10.1103/PhysRevLett.122.068004}
{Phys. Rev. Lett. {\bf 122}, 068004}.

\bibitem[Robledo and Varea, 1981]{robledo1981}
  Robledo, A., C. and Varea,
  ``On the relationship between the density functional formalism and the 
  potential distribution theory for nonuniform fluids,''
  \href{https://doi.org/10.1007/BF01011432}
  {J. Stat. Phys. {\bf 26}, 513 (1981).}

\bibitem[Rosenfeld, 1977]{rosenfeld1977entropyScaling}
Rosenfeld, Y., 
1977,
``Relation between the transport coefficients and the
internal entropy of simple systems,''
\href{https://doi.org/10.1103/PhysRevA.15.2545}
{Phys. Rev. A {\bf 15}, 2545}.

\bibitem[Rosenfeld and Ashcroft, 1979]{rosenfeld1979mhnc}
Rosenfeld, Y., and N. W. Ashcroft,
1979,
``Theory of simple classical fluids: Universality in the short-range structure,''
\href{https://doi.org/10.1103/PhysRevA.20.1208}
{Phys. Rev. A {\bf 20}, 1208}.

\bibitem[Rosenfeld, 1988]{rosenfeld1988}
Rosenfeld, Y.,
1988,
``Scaled field particle theory of the structure and the
thermodynamics of isotropic hard particle fluids,''
\href{https://doi.org/10.1063/1.454810}
{J. Chem. Phys. {\bf 89}, 4272}.

\bibitem[Rosenfeld, 1989]{rosenfeld1989}
Rosenfeld, Y.,
1989,
``Free-energy model for the inhomogeneous hard-sphere fluid mixture and
density-functional theory of freezing,''
\href{https://doi.org/10.1103/PhysRevLett.63.980}
{Phys. Rev. Lett. {\bf 63}, 980.}

\bibitem[Rosenfeld, 1993]{rosenfeld1993testparticle}
Rosenfeld, Y., 
1993,
``Free-energy model for inhomogeneous fluid mixtures: Yukawa-charged
hard-spheres, general interactions, and plasmas,''
\href{https://doi.org/10.1063/1.464569}
{J. Chem. Phys. {\bf 98}, 8126}.

\bibitem[Rosenfeld, 1994]{rosenfeld1994nonspherical}
Rosenfeld, Y.,
1994,
``Density functional theory of molecular fluids: Free-energy model for
the inhomogeneous hard-body fluid,''
\href{https://doi.org/10.1103/PhysRevE.50.R3318}
{Phys. Rev. E {\bf 50}, R3318(R)}.

\bibitem[Rosenfeld \etal, 1997]{rosenfeld1997RSLT}
Rosenfeld, Y., M. Schmidt, H. L\"owen, and P. Tarazona, 
1997,
``Fundamental-measure free energy density functional for hard spheres:
Dimensional crossover and freezing,''
\href{https://doi.org/10.1103/PhysRevE.55.4245}
{Phys. Rev. E {\bf 55}, 4245}.

\bibitem[Rosenfeld \etal, 2000]{rosenfeld2000ps}
Rosenfeld, Y., M. Schmidt, M. Watzlawek, and H. L\"owen,
2000,
``Fluid of penetrable spheres: Testing the universality of the bridge
functional,''
\href{https://doi.org/10.1103/PhysRevE.62.5006}
{Phys. Rev. E {\bf 62}, 5006}.

\bibitem[Rotenberg, 2020]{rotenberg2020}
Rotenberg, B.,
2020,
``Use the force! Reduced variance estimators for densities,
radial distribution functions, and local mobilities in molecular simulations,''
\href{https://doi.org/10.1063/5.0029113}
{J. Chem. Phys. {\bf 153}, 150902.}

\bibitem[Roth \etal, 2002]{roth2002}
Roth, R., R. Evans, A. Lang, and G. Kahl,
2002,
``Fundamental measure theory for hard-sphere mixtures revisited: the
White Bear version,''
\href{https://doi.org/10.1088/0953-8984/14/46/313}
{J. Phys.: Condens. Matter {\bf 14}, 12063}.

\bibitemESSENTIAL{Roth, 2010}{roth2010review}
Roth, R.,
2010,
``Fundamental measure theory for hard-sphere mixtures: A review,''
\href{https://doi.org/10.1088/0953-8984/22/6/063102}
{J. Phys.: Condens. Matter {\bf 22}, 063102.}

\bibitemESSENTIAL{Rowlinson and Widom, 2002}{RowlinsonWidomBook}
Rowlinson, J.~S., and B. Widom, 
2002,
{\it  Molecular Theory of Capillarity}
  (Dover, New York).

\bibitemORIGINAL{Royall \etal, 2007}{royall2007dynamicSedimentation} 
Royall, C.~P., J. Dzubiella, M. Schmidt, and A.~van Blaaderen, 
2007,
  ``Non-equilibrium sedimentation of colloids on the particle scale,''
  \href{https://doi.org/10.1103/PhysRevLett.98.188304}
       {Phys. Rev. Lett. {\bf 98}, 188304.}

\bibitem[Runge and Gross, 1984]{runge1984}
Runge, E., and E. K. U. Gross,
1984,
``Density-functional theory for time-dependent systems,''
\href{https://doi.org/10.1103/PhysRevLett.52.997}
{Phys. Rev. Lett. {\bf 52}, 997}.

\bibitem[Scacchi \etal, 2017]{scacchi2017pre} 
Scacchi, A., A. J. Archer, and J. M. Brader,
2017,
``Dynamical density functional theory analysis of
the laning instability in sheared soft matter,''
\href{https://doi.org/10.1103/PhysRevE.96.062616}
{Phys. Rev. E {\bf 96}, 062616}.

\bibitem[Samm\"uller and Schmidt, 2021]{sammueller2021} 
  Samm\"uller F., and Schmidt, M.
  2021,
  ``Adaptive Brownian dynamics,''
  \href{https://doi.org/10.1063/5.0062396}
       {J. Chem. Phys. {\bf 155}, 134107 (2021);} (featured on the cover).

\bibitem[Scacchi and Brader, 2018]{scacchi2018}
 Scacchi, A., and J. M. Brader,
 2018,
 ``Local phase transitions in driven colloidal suspensions,''
 \href{https://doi.org/10.1080/00268976.2017.1393117}
 {Mol. Phys. {\bf 116}, 378.}

\bibitem[Scala \etal, 2007]{scala2007}
  Scala, A., T. Voigtmann, and C. De Michele,
  2007,
  ``Event-driven Brownian dynamics for hard spheres,''
  \href{https://doi.org/10.1063/1.2719190}
  {J. Chem. Phys. {\bf 126}, 134109.}

\bibitem[Schilling, 2021]{schilling2021}
  Schilling, T.,
  2021
  ``Coarse-Grained Modelling Out of Equilibrium,''
  \href{https://arxiv.org/abs/2107.09972}
  {arxiv:2107.09972}

\bibitem[Sergiievskyi \etal, 2014]{sergiievskyi2014}
Sergiievskyi, V. P., G. Jeanmairet, M. Levesque, and D. Borgis,
2014,
``Fast Computation of Solvation Free Energies with Molecular 
Density Functional Theory: Thermodynamic-Ensemble Partial
Molar Volume Corrections,''
\href{https://doi.org/10.1021/jz500428s}
{J. Phys. Chem. Lett. {\bf 5}, 1935.}


\bibitemORIGINAL{Schindler and Schmidt, 2016}{schindler2016dynamicPairCorrelations} 
Schindler,  T., and M. Schmidt, 
2016,
``Dynamic pair correlations and superadiabatic forces in a dense Brownian liquid,''
\href{https://doi.org/10.1063/1.4960031}
{J. Chem. Phys. {\bf 145}, 064506.}

\bibitem[Schindler \etal, 2019]{schindler2019}
  Schindler, T., R. Wittmann, and J. M. Brader,
  2019,
  ``Particle-conserving dynamics on the single-particle level,''
  \href{https://doi.org/10.1103/PhysRevE.99.012605}
  {Phys. Rev. E {\bf 99}, 012605.}


\bibitem[Schmidt, 1999]{schmidt1999ps}
Schmidt, M., 
1999,
``Ab-initio density-functional theory for penetrable spheres,''
\href{https://doi.org/10.1088/0953-8984/11/50/309}
{J. Phys.: Condens. Matter {\bf 11}, 10163}.

\bibitem[Schmidt \etal, 2000]{schmidt2000ao}
Schmidt, M., H. L\"owen, J. M. Brader, and R. Evans,
2000,
``Density functional for a model colloid-polymer mixture,''
\href{https://doi.org/10.1103/PhysRevLett.85.1934}
{Phys. Rev. Lett. {\bf 85}, 1934}.

\bibitem[Schmidt, 2001]{schmidt2001rsf}
Schmidt, M., 
2001,
``Density functional theory for colloidal rod-sphere mixtures,''
\href{https://doi.org/10.1103/PhysRevE.63.050201}
{Phys. Rev. E {\bf 63}, 050201(R)}. 

\bibitem[Schmidt, 2001b]{schmidt2001wr}
Schmidt, M.,
2001,
``Density functional for the Widom-Rowlinson model,''
\href{https://doi.org/10.1103/PhysRevE.63.010101}
{Phys. Rev. E {\bf 63}, 010101(R)}.

\bibitem[Schmidt, 2002]{schmidt2002qa}
Schmidt, M., 
2002,
``Density functional theory for fluids in porous media,''
\href{https://doi.org/10.1103/PhysRevE.66.041108}
{Phys. Rev. E {\bf 66}, 041108.}

\bibitem[Schmidt \etal, 2002]{schmidt2002versus}
Schmidt, M., E. Sch\"oll-Paschinger, J. K\"ofinger, and G. Kahl,
2002,
``Model colloid-polymer mixtures in porous matrices:
density functional versus integral equations,''
\href{https://doi.org/10.1088/0953-8984/14/46/315}
{J. Phys.: Condens. Matter {\bf 14}, 12099}.

\bibitem[Schmidt, 2004]{schmidt2004nahs}
Schmidt, M.,
2004,
``Rosenfeld functional for non-additive hard spheres,''
\href{https://doi.org/10.1088/0953-8984/16/30/L01}
{J. Phys.: Condens. Matter {\bf 16}, L351}.

\bibitemORIGINAL{Schmidt \etal, 2008}{schmidt2008dynamicSedimentation} 
Schmidt, M., C. P. Royall, A. van Blaaderen, and J. Dzubiella, 
2008,
``Non-equilibrium sedimentation of colloids: Confocal microscopy
and Brownian dynamics simulations,''
\href{https://doi.org/10.1088/0953-8984/20/49/494222}
{J. Phys.: Condens. Matter {\bf 20}, 494222.}

\bibitem[Schmidt, 2011]{schmidt2011tnas}
Schmidt, M., 
2011,
``Density functional for ternary non-additive hard sphere mixtures,''
\href{https://doi.org/10.1088/0953-8984/23/41/415101}
{J. Phys.: Condens. Matter {\bf 23}, 415101}.

\bibitemORIGINAL{Schmidt, 2011b}{schmidt2011internalEnergyFunctional} 
Schmidt, M.,
2011,
``Statics and dynamics of inhomogeneous liquids via the internal-energy functional,''
\href{https://doi.org/10.1103/PhysRevE.84.051203}
{Phys. Rev. E {\bf 84}, 051203.}

\bibitemORIGINAL{Schmidt and Brader, 2013}{schmidt2013pft} 
Schmidt, M., and J.~M. Brader,
2013,
``Power functional theory for Brownian dynamics,''
\href{https://doi.org/10.1063/1.4807586}
{J. Chem. Phys. {\bf 138}, 214101.}

\bibitemORIGINAL{Schmidt, 2015}{schmidt2015qpft} 
Schmidt, M., 
2015,
``Quantum power functional theory for many-body dynamics,''
\href{https://doi.org/10.1063/1.4934881}
{J. Chem. Phys. {\bf 143}, 174108.}

\bibitemORIGINAL{Schmidt, 2018}{schmidt2018md} 
Schmidt, M.,
2018,
``Power functional theory for Newtonian many-body dynamics,''
\href{https://doi.org/10.1063/1.5008608}
{J. Chem. Phys. {\bf 148}, 044502.}

\bibitemESSENTIAL{Schofield and Henderson, 1982}{SchofieldHenderson}
Schofield, P., and J.~R. Henderson, 
1982,
``Statistical mechanics of inhomogeneous fluids,''
\href{https://doi.org/10.1098/rspa.1982.0015}
{Proc. R. Soc. A {\bf 379}, 231.}
Important demonstration of the nonuniqueness of the stress tensor
distribution.

\bibitem[Seifert and Speck, 2010]{seifert2010}
Seifert, U., and T. Speck,
2010,
``Fluctuation-dissipation theorem in nonequilibrium steady states,''
\href{https://doi.org/10.1209/0295-5075/89/10007}
{EPL  (Europhys. Lett.) {\bf 89}, 10007}.

\bibitem[Seifert, 2012]{seifert2012review}
Seifert, U.,
2012,
``Stochastic thermodynamics, fluctuation theorems and molecular machines,''
\href{https://doi.org/doi:10.1088/0034-4885/75/12/126001}
{Rep. Prog. Phys. {\bf 75}, 126001}.

\bibitem[S\"oker \etal, 2020]{soeker2020}
  S\"oker, N. A., S. Auschra, V. Holubec, K. Kroy, and F. Cichos,
  2020,
  ``Active-particle polarization without alignment forces,''
  \href{https://arxiv.org/abs/2010.15106}
       {arXiv:2010.15106.}

\bibitem[Speck, 2016]{speck2016pre}
Speck, T.,
2016,
``Thermodynamic formalism and linear response theory for nonequilibrium
steady states,''
\href{http://dx.doi.org/10.1103/PhysRevE.94.022131}
{Phys. Rev. E {\bf 94}, 022131}.

\bibitem[Speck, 2020]{speck2020}
  Speck, T.,
  2020,
  ``Collective forces in scalar active matter,''
  \href{https://doi.org/10.1039/D0SM00176G}
  {Soft Matter {\bf 16}, 2652.}

\bibitem[Stewart and Evans, 2012]{stewart2012pre}
  Stewart, M. C., and R. Evans,
  2012,
  ``Phase behavior and structure of a fluid confined between competing
  (solvophobic and solvophilic) walls,''
  \href{http://dx.doi.org/10.1103/PhysRevE.86.031601}
  {Phys. Rev. E {\bf 86}, 031601.}

\bibitem[Stewart and Evans, 2014]{stewart2014jcp}
  Stewart, M. C., and R. Evans,
  2014,
  ``Layering transitions and solvation forces in an asymmetrically confined fluid,''
  \href{https://doi.org/10.1063/1.4869868}
  {J. Chem. Phys. {\bf 140}, 134704.}

\bibitem[Stopper \etal, 2018]{stopper2018dtpl}
Stopper, D., A. L. Thorneywork, R. P. A. Dullens, and R. Roth,
2018,
``Bulk dynamics of Brownian hard disks: Dynamical density functional
theory versus experiments on two-dimensional colloidal hard spheres,''
\href{https://doi.org/10.1063/1.5019447}
{J. Chem. Phys. {\bf 148}, 104501.}


\bibitem[Squires and Quake, 2005]{squires2005}
Squires, T. M., and S. R. Quake, 
2005,
``Microfluidics: Fluid physics at the nanoliter scale,''
\href{https://doi.org/10.1103/RevModPhys.77.977}
{Rev. Mod. Phys. {\bf 77}, 977.}

\bibitem[Stenhammar \etal, 2013]{stenhammar2013prl}
Stenhammar, J., A. Tiribocchi, R. J. Allen, D. Marenduzzo, and M. E. Cates, 
2013,
``Continuum theory of phase separation kinetics for active Brownian particles,''
\href{https://doi.org/10.1103/PhysRevLett.111.145702}
{Phys. Rev. Lett. {\bf 111}, 145702.}


\bibitemORIGINAL{Stuhlm\"uller \etal, 2018}{stuhlmueller2018structural} 
Stuhlm\"uller, N.~C.~X., T. Eckert, D. de las Heras, and M. Schmidt,
2018,
``Structural nonequilibrium forces in driven colloidal systems,''
\href{https://doi.org/10.1103/PhysRevLett.121.098002}
{Phys. Rev. Lett. {\bf 121}, 098002.}

\bibitem[Tarantino and Ullrich, 2021]{tarantino2021}
  Tarantino, W., and C. A. Ullrich,
  2021,
  ``A reformulation of time-dependent Kohn– Sham theory in terms of the 
  second time derivative of the density,''
  \href{https://doi.org/10.1063/5.0039962}
       {J. Chem. Phys. {\bf 154}, 204112.}

\bibitem[Tarazona and Evans, 1984]{tarazona1984}
Tarazona, P., and R. Evans,
1984,
``A simple density functional theory for inhomogeneous liquids:
Wetting by gas at a solid-liquid interface,''
\href{https://doi.org/10.1080/00268978400101601}
{Mol. Phys. {\bf 52}, 847}.

\bibitem[Tarazona, 2000]{tarazona2000}
Tarazona, P.,
2000,
``Density functional for hard sphere crystals: a fundamental
measure approach,''
\href{https://doi.org/10.1103/PhysRevLett.84.694}
{Phys. Rev. Lett. {\bf 84}, 694}.

\bibitemESSENTIAL{Tarazona \etal, 2008}{tarazona2008review}
Tarazona, P., J. A. Cuesta, and Y. Mart\'inez-Rat\'on,
2008,
``Density functional theories of hard particle systems,''
\href{https://doi.org/10.1007/978-3-540-78767-9_7}
{Lect. Notes Phys. {\bf 753}, 247.}

\bibitem[Tchenkoue \etal, 2019]{tchenkoue2019}
  Tchenkoue, M.~M., M. Penz, I. Theophilou, M. Ruggenthaler, and A. Rubio,
  2019,
  ``Force balance approach for advanced approximations in 
  density functional theories,''
  \href{https://doi.org/10.1063/1.5123608}
       {J. Chem. Phys. {\bf 151}, 154107.}

\bibitem[te~Vrugt \etal, 2020]{tevrugt2020}
te~Vrugt, M., H. L\"owen, and R. Wittkowski,
2020,
``Classical dynamical density functional theory: from fundamentals to applications,''
\href{https://doi.org/10.1080/00018732.2020.1854965}
{Adv. Phys. {\bf 69}, 121.}

\bibitem[Thorneywork \etal, 2014]{thorneywork2014}
Thorneywork, A. L., R. Roth, D. G. A. L. Aarts, and R. P. A. Dullens,
2014,
``Communication: Radial distribution functions in a two-dimensional
binary colloidal hard sphere system.''
\href{https://doi.org/10.1063/1.4872365}
{J. Chem. Phys. {\bf 140}, 161106.}

\bibitem[Tonks, 1936]{tonks1936}
Tonks, L.,
1936,
``The complete equation of state of one, two and 
three-dimensional gases of hard elastic spheres,''
\href{https://doi.org/10.1103/PhysRev.50.955}
{Phys. Rev. {\bf 50}, 955}.

\bibitemORIGINAL{Treffenst\"adt and Schmidt, 2020}{treffenstaedt2019shear} 
Treffenst\"adt, L. L., and M. Schmidt,
2020,
``Memory-induced motion reversal in Brownian liquids,''
\href{https://doi.org/10.1039/C9SM02005E}
{Soft Matter {\bf 16}, 1518.}

\bibitemORIGINAL{Treffenst\"adt and Schmidt, 2021}{treffenstaedt2021dtpl} 
Treffenst\"adt, L. L., and M. Schmidt,
2021,
``Universality in driven and equilibrium hard sphere liquid dynamics,''
\href{https://doi.org/10.1103/PhysRevLett.126.058002}
{Phys. Rev. Lett. {\bf 126}, 058002.}

\bibitemORIGINAL{Treffenst\"adt \etal, 2021}{treffenstaedt2021asymptotic} 
Treffenst\"adt, L. L., T. Schindler, and M. Schmidt,
2021,
``Dynamic decay and superadiabatic forces in the van Hove dynamics of bulk hard sphere fluids'' (to be published).

\bibitem[Tschopp \etal, 2020]{tschopp2020}
  Tschopp, S.~M., H.~D. Vuijk, A. Sharma, and J.~M. Brader,
  2020,
  ``Mean-field theory of inhomogeneous fluids,''
  \href{https://doi.org/10.1103/PhysRevE.102.042140}
  {Phys. Rev. E {\bf 102}, 042140.}

\bibitem[Tschopp and Brader, 2021]{tschopp2021}
  Tschopp, S.~M., and J.~M. Brader,
  2021,
  ``Fundamental measure theory of inhomogeneous two-body correlation functions,''
  \href{https://doi.org/10.1103/PhysRevE.103.042103}
  {Phys. Rev. E {\bf 103}, 042103}.

\bibitem[Turci and Wilding, 2021]{turci2021}
  Turci, F., and N. B. Wilding,
  2021,
  ``Phase separation and multibody effects in three-dimensional active 
  Brownian particles,''
  \href{https://doi.org/10.1103/PhysRevLett.126.038002}
  {Phys. Rev. Lett. {\bf  126}, 038002}.

\bibitemESSENTIAL{van der Waals, 2004}{vanderWaals1873} 
van der Waals, J.~D., 
2004,
``On the continuity of the gaseous and liquid states,''
edited by J.~S. Rowlinson, (Dover, New York).

\bibitemESSENTIAL{van der Waals, 1893}{vanderWaals1893}
van der Waals, J.~D.,
1894,
``The thermodynamik theory of capillarity under the hypothesis of a 
continuous variation of density (translation),''
\href{https://doi.org/10.1515/zpch-1894-1338}
  {Z. Phys. Chem. {\bf   13}, 657} 
  (German translation by J. J. van Laar);
  see also: J.~S. Rowlinson, 1979,
  \href{https://doi.org/10.1007/BF01011514}
       {J. Stat. Phys. {\bf 20}, 197}
       (English translation by J.~S. Rowlinson).

\bibitemESSENTIAL{van Roij, 2005}{vanroij2005pedagogical}
van Roij, R., 
2005,
``The isotropic and nematic liquid crystal phase of colloidal rods,''
\href{https://doi.org/10.1088/0143-0807/26/5/S07}
{Eur. J. Phys. {\bf 26}, S57.}

\bibitem[Vanderlick \etal, 1989]{vanderlick1989}
Vanderlick, T. K., H. T. Davis, and J. K. Percus,
1989,
``The statistical mechanics of inhomogeneous hard rod mixtures,''
\href{https://doi.org/10.1063/1.457329}
{J. Chem. Phys. {\bf 91}, 7136}.

\bibitem[Vogel and Fuchs, 2020]{vogel2020}
  Vogel, F., and M. Fuchs,
  2020,
  ``Stress correlation function and linear response of Brownian particles,''
  \href{https://doi.org/10.1140/epje/i2020-11993-4}
  {Eur. Phys. J. E {\bf 43}, 70.}

\bibitem[W\"achtler \etal, 2016]{waechtler2016}
W\"achtler, C. W., F. Kogler, and S. H. L. Klapp,
2016,
``Lane formation in a driven attractive fluid,''
\href{https://doi.org/10.1103/PhysRevE.94.052603}
{Phys. Rev. E {\bf 94}, 052603.}

\bibitem[White \etal, 2000]{white2000prl}
  White, J.A., and A. Gonz\'alez, F. L. Rom\'an, and S. Velsasco,
  ``Density-functional theory of inhomogeneous fluids in the canonical ensemble,''
  2000,
  \href{https://doi.org/10.1103/PhysRevLett.84.1220}
       {Phys. Rev. Lett. {\bf 84}, 1220.}

\bibitem[White and Gonz\'alez, 2002]{white2002jpcm}
  White, J.A., and A. Gonz\'alez,
  ``The extended variable space approach to density functional theory in the 
  canonical ensemble,''
  2002,
  \href{https://doi.org/10.1088/0953-8984/14/46/302}
  {J. Phys.: Condens. Matter {\bf 14}, 11907.}


\bibitem[Wittkowski \etal, 2012]{wittkowski2012}
Wittkowski, R., H. L\"owen, and H. R. Brand,
2012,
``Extended dynamical density functional theory for colloidal mixtures
with temperature gradients,''
\href{https://doi.org/10.1063/1.4769101}
{J. Chem. Phys. {\bf 137}, 224904}.

\bibitem[Wittmann \etal, 2015]{wittmann2015}
Wittmann, R., M. Marechal, and K. Mecke,
2015,
``Fundamental mixed measure theory for non-spherical colloids,''
\href{https://doi.org/10.1209/0295-5075/109/26003}
{EPL (Europhys. Lett.) {\bf 109}, 26003}.

\bibitem[Wittmann \etal, 2021]{wittmann2021}
  Wittmann, R., H. L\"owen, and J. M. Brader,
  2021,
  ``Order-preserving dynamics in one dimension -- single-file
  diffusion and caging from the perspective of dynamical density
  functional theory,''
  \href{https://doi.org/10.1080/00268976.2020.1867250}
  {Mol. Phys. e1867250.}

\bibitem[Zausch \etal, 2008]{zausch2008}
Zausch, J., J. Horbach, M. Laurati, S. U. Egelhaaf, J. M. Brader,
T. Voigtmann, and M. Fuchs,
2008,
``From equilibrium to steady state: the transient dynamics of colloidal
liquids under shear,''
\href{http://doi.org/10.1088/0953-8984/20/40/404210}
{J. Phys.: Condens. Matter {\bf 20}, 404210.}

\bibitem[Zwanzig, 2001]{zwanzig2001}
  Zwanzig, R.,
  2001,
 {\it Nonequilibrium Statistical Mechanics}
  (Oxford University Press, Oxford).

\end{thebibliography}
\end{document}